\documentclass[a4paper,12pt,twoside]{report}

\usepackage[utf8]{inputenc}
\usepackage[T1]{fontenc}
\usepackage[]{graphicx}
\usepackage[]{float}
\usepackage[font=footnotesize]{caption} 
\usepackage[]{lmodern}
\usepackage[]{amsmath}
\usepackage[]{amssymb}
\usepackage[]{amsthm}
\usepackage[]{mathrsfs}
\usepackage[]{fancyhdr}
\usepackage[]{tabularx}
\usepackage[usenames,dvipsnames]{xcolor}
\usepackage[]{tikz}
\usepackage[]{subfig}
\usepackage[]{url}
\usepackage[]{braket}
\usepackage[]{psfrag}
\usepackage[]{textcomp}
\usepackage[top=2.8cm, bottom=2.8cm, left=2.6cm, right=2.6cm]{geometry}
\usepackage[]{cite}
\usepackage[]{ifthen}
\usepackage[clearempty]{titlesec}
\usepackage[]{bm}
\usepackage[]{xspace}
\usepackage{color}
\definecolor{linkcol}{rgb}{0,0,0.4} 
\definecolor{citecol}{rgb}{0.5,0,0} 
\usepackage[dvips,ps2pdf,colorlinks=true,citecolor=linkcol,linkcolor=linkcol,linktocpage,pdfhighlight=/P]{hyperref}

\usepackage[font=small,labelfont={bf,sf}]{caption} 
\usetikzlibrary{calc,arrows,decorations.pathmorphing,decorations.markings,backgrounds,positioning,fit,petri,patterns,matrix}
\usepackage{minitoc}
\usepackage{titletoc}
\usepackage{parskip}
\usepackage[vcentermath]{youngtab}

\usepackage[export]{adjustbox}
\usepackage{subfig}

\usepackage{bm}
\usepackage{rotating}
\usepackage{chngcntr} 

\numberwithin{equation}{section}
\numberwithin{figure}{section}

\renewcommand{\thechapter}{\Roman{chapter}}	 

\newcommand{\dd}{\ensuremath{\mathrm{d}}} 
\newcommand{\dr}{\ensuremath{\partial}} 
\newcommand{\Hrule}{\rule{\linewidth}{1.5pt}}

\newcommand{\Tr}{\ensuremath{\mathrm{Tr}}} 
\newcommand{\Id}{{\mbox{l\hspace{-0.52em}1}}} 
\newcommand{\Yb}{\ensuremath{{}^{173}\mathrm{Yb}}} 
\newcommand{\clearemptydoublepage}{\newpage{\pagestyle{empty}\cleardoublepage}}
\newcommand{\derivee}[3][1]{
      \ifthenelse{\equal{#1}{1}}{%
	    \ensuremath{\frac{\dd#2}{\dd#3}}
	  }{%
	    \ensuremath{\frac{\dd^{#1}#2}{\dd#3^{#1}}}
	   }
}
\newcommand{\deriveepart}[3][1]{
      \ifthenelse{\equal{#1}{1}}{%
	    \ensuremath{\frac{\dr#2}{\dr#3}}
	  }{%
	    \ensuremath{\frac{\dr^{#1}#2}{\dr#3^{#1}}}
	   }
}
\newcommand{\integrale}[4][1]{
	  \ifthenelse{\equal{#1}{1}}{
	      \int_{#2}^{#3} \dd#4 \, %
	  }{%
	      \int_{#2}^{#3} \dd^{#1}#4 \, %
	   }
}

\titleformat{\chapter}[display]	
{\Huge}%
{\vspace{-5ex}\filcenter\sc\huge%
\chaptertitlename\ \thechapter\vspace{-0.3cm}}%
{0.1pt}
{\Hrule\\\sffamily\Huge\bfseries\filcenter}

\titleformat{name=\chapter,numberless}[display] 
{\huge}%
{\vspace{-10ex}}%
{0pt}
{\Hrule\\[1ex]\sffamily\Huge\bfseries\filcenter}[\Hrule]

\titleformat{\section}[block]
    {\Large\bf\sffamily}%
    {\thesection.}%
    {5pt}{}
\titleformat{\subsection}[block]{\large\bf\sffamily}{\thesubsection.}{0.7em}{} 
\titleformat{\subsubsection}[block]{\normalsize\bf\sffamily}{\thesubsubsection.}{0.7em}{} 
\titleformat{\paragraph}[hang]{\normalsize\bf\sffamily}{}{0.7em}{}

\renewcommand{\chaptermark}[1]{\markboth{Chap \thechapter \ -\ #1}{#1}} 

\titlecontents{chapter}
[1pt] 
{\addvspace{1pc}}
{\large\textsc{Chapter \thecontentslabel} --- \bf\sffamily}
{\large \bf\sffamily} 
{\dotfill \contentspage}
[\addvspace{0.5pc}]

\titlecontents{section}
[3.8em] 
{\addvspace{0.5pc}}
{\contentslabel{2.3em} \bf\sffamily}
{\large \bf\sffamily} 
{\dotfill \contentspage}
[\addvspace{0.5pc}]

\titleformat{\part}[display]
{\Large\bfseries\sffamily\center}
{\partname\nobreakspace\thepart}
{0mm}
{\vspace{3ex}\huge\bfseries\sffamily}

\newcounter{myc}
\newcounter{mye}
\newenvironment{partintro}{%
\markboth{Introduction}{Introduction}

\renewenvironment{figure}{\addtocounter{myc}{1} \begin{ofigure}}{\end{ofigure}}

\renewenvironment{equation}{\addtocounter{mye}{1}\begin{oequation}}{\end{oequation}}
}{%

\counterwithin{figure}{chapter}

\counterwithin{equation}{chapter}
}

\hypersetup{
    pdfauthor={Jean Decamp},
    pdfsubject={Thesis Manuscript},
    pdftitle={Symmetries and Correlations in Strongly Interacting One-dimensional Quantum Gases}
} 

\begin{document}

\thispagestyle{empty}


~\vspace{4cm} 

\begin{center}
{\sffamily 

{\bfseries{\Huge Symmetries and Correlations in Strongly Interacting One-dimensional Quantum Gases}} 
\vspace{2.5cm}

{\bfseries{\LARGE
Jean DECAMP
}
} \\

\vspace{0.5cm}
{\large Institut de Physique de Nice } \\

\vfill

\begin{minipage}{4.5cm}
 {\bfseries Submitted for the title of Doctor of} Physics 
 {\bfseries of} Université Côte d'Azur \\[0.4cm]
 {\bfseries Supervised by} Patrizia Vignolo and Mathias Albert \\[0.4cm]
 {\bfseries Defended on:} September 25, 2018 
\end{minipage}
\hfill
\begin{minipage}{10.5cm}
 {\bfseries Before the jury composed of: } \\[0.2cm]
    \textbf{Mathias Albert}, {\small{Associate professor, Institut de Physique de Nice (Université Côte d'Azur)}} \\
     \textbf{George Batrouni}, {\small{Professor, Institut de Physique de Nice (Université Côte d'Azur)}} \\
      \textbf{Leonardo Fallani}, {\small{Professor, University of Florence}} \\
  \textbf{Maxim Olchanyi}, {\small{Professor, University of Massachusetts at Boston}} \\
 \textbf{Ludovic Pricoupenko}, {\small{Professor, Laboratoire de Physique Th\'{e}orique de la Matière Condens\'{e}e (Université Pierre et Marie Curie)}}\\
\textbf{Patrizia Vignolo}, {\small{Professor, Institut de Physique de Nice (Université Côte d'Azur)}}\\
  \textbf{Nikolaj T. Zinner}, {\small{Associate professor, Aarhus Institute of Advanced Studies (Aarhus University)}}

\end{minipage}

} 

\end{center}

\clearemptydoublepage
\pagestyle{empty}

~
\vfill
\begin{minipage}{\textwidth}

{\sffamily

\begin{center}
 {\bfseries \huge Symmetries and Correlations in Strongly Interacting One-dimensional
Quantum Gases}\\[1cm]
 {\textit{ \Large Sym\'{e}tries et corr\'{e}lations dans les gaz quantiques fortement interagissants \`{a} une dimension}}\\[4cm]
\end{center}

Jury:\\[0.2cm]

Reviewers

\textbf{Maxim Olchanyi}, Professor, University of Massachusetts at Boston\\
\textbf{Ludovic Pricoupenko}, Professor, Laboratoire de Physique Th\'{e}orique de la Matière Condens\'{e}e (Université Pierre et Marie Curie) \\

Examiners

\textbf{George Batrouni}, Professor, Institut de Physique de Nice (Université Côte d'Azur) \\
\textbf{Leonardo Fallani}, Professor, University of Florence \\
\textbf{Nikolaj T. Zinner}, Associate professor, Aarhus Institute of Advanced Studies (Aarhus University) \\

Thesis supervisors

\textbf{Mathias Albert}, Associate professor, Institut de Physique de Nice (Université Côte d'Azur)\\
\textbf{Patrizia Vignolo}, Professor, Institut de Physique de Nice (Université Côte d'Azur)

}

\end{minipage}
\vfill
~
\clearemptydoublepage
\pagestyle{empty}

\newgeometry{top=2cm, bottom=2cm, left=2cm, right=2cm}

{\sffamily

~
\vfill


\small

{\bfseries\large Sym\'{e}tries et corr\'{e}lations dans les gaz quantiques fortement interagissants \`{a} une dimension} \\[0.1cm]
{\bfseries R\'{e}sum\'{e}}

L'objectif principal de cette thèse est l'\'{e}tude th\'{e}orique de m\'{e}langes quantiques fortement interagissants \`{a} une dimension et soumis \`{a} un potentiel externe harmonique.
De tels systèmes fortement corr\'{e}l\'{e}s peuvent être r\'{e}alis\'{e}s et test\'{e}s dans des exp\'{e}riences d'atomes ultrafroids.
Leurs propri\'{e}t\'{e}s de sym\'{e}trie par permutation non triviales sont \'{e}tudi\'{e}es, ainsi que leurs effets sur les corr\'{e}lations.
\\
Exploitant une solution exacte pour des interactions fortes, nous extrayons des propri\'{e}t\'{e}s g\'{e}n\'{e}rales des corr\'{e}lations  encod\'{e}es dans la matrice densit\'{e} \`{a} un corps et dans les distributions des impulsions associ\'{e}es, dans les m\'{e}langes fermioniques et de Bose-Fermi. En particulier, nous obtenons des r\'{e}sultats substantiels sur
le comportement \`{a} courtes distances, et donc les
queues \`{a} haute impulsions, qui suivent des lois en $k^{-4}$ typiques. Les poids de ces queues, d\'{e}not\'{e}s contacts de Tan, sont li\'{e}s \`{a} de nombreuses propri\'{e}t\'{e}s thermodynamiques
des systèmes telles que les corr\'{e}lations \`{a} deux corps, la d\'{e}riv\'{e}e de l'\'{e}nergie par rapport \`{a} la longueur de diffusion unidimensionnelle, ou le facteur de structure statique. Nous montrons que ces contacts universels de Tan permettent \'{e}galement de caract\'{e}riser la sym\'{e}trie spatiale
des systèmes, et constitue donc une connexion profonde entre les corr\'{e}lations et les sym\'{e}tries. En outre, la sym\'{e}trie d'\'{e}change est extraite en utilisant
une m\'{e}thode de th\'{e}orie des groupes, \`{a} savoir la m\'{e}thode de la somme des classes (\textit{class-sum method} en anglais), qui provient \`{a} l'origine de la physique nucl\'{e}aire. De plus, nous montrons que ces systèmes suivent une
version g\'{e}n\'{e}ralis\'{e}e du fameux th\'{e}orème de Lieb-Mattis.
Souhaitant rendre nos r\'{e}sultats aussi pertinents exp\'{e}rimentalement que possible, nous d\'{e}rivons des lois d'\'{e}chelle pour le contact de Tan en fonction de l'interaction, de la
temp\'{e}rature et du confinement transverse.
Ces lois pr\'{e}sentent des effets int\'{e}ressants li\'{e}s aux fortes corr\'{e}lations et \`{a} la dimensionnalit\'{e}.

{\bfseries Mots cl\'{e}s : } Gaz quantiques, atomes ultrafroids, dimension un, mixtures quantiques, sym\'{e}trie d'\'{e}change, th\'{e}orie des groupes, m\'{e}thode de la somme des classes, fermionisation, corr\'{e}lations \`{a} un corps, contact de Tan, lois d'\'{e}chelle

\vspace{0.4cm}

{\bfseries\large Symmetries and Correlations in Strongly Interacting One-dimensional Quantum Gases} \\[0.1cm]
{\bfseries Abstract}

The main focus of this thesis is the theoretical study of strongly interacting quantum mixtures confined in one dimension and subjected to a harmonic external potential. 
Such strongly correlated systems can be realized and tested in ultracold atoms experiments.
Their non-trivial permutational symmetry properties are investigated, as well as their interplay with correlations.
\\
Exploiting an exact solution at strong interactions, we  extract general correlation properties encoded in the one-body density 
matrix and in the associated momentum distributions, in fermionic and Bose-Fermi mixtures. In particular, we obtain substantial results about the 
short-range behavior, and therefore the 
high-momentum tails, which display typical $k^{-4}$ laws. The weights of these tails, denoted as Tan's contacts, are related to numerous thermodynamic 
properties
of the systems such as the two-body correlations, the derivative of the energy with respect to the one-dimensional scattering length, or the static structure factor. We show that these universal Tan's contacts also allow to characterize the spatial symmetry
of the systems, and therefore is a deep connection between correlations and symmetries. Besides, the exchange symmetry is extracted using
a group theory method, namely the class-sum method, which comes originally from nuclear physics. Moreover, we show that these systems follow a 
generalized
version of the famous Lieb-Mattis theorem.
Wishing to make our results as experimentally relevant as possible, we derive scaling laws for Tan's contact as a function of the interaction, 
temperature and transverse confinement.
These laws display interesting effects related to strong correlations and dimensionality.


\vfill

}

\restoregeometry
\clearemptydoublepage
\pagestyle{empty}

\chapter*{Acknowledgements}

There are many people I should thank in these acknowledgments, and for many different reasons. I will certainly not be exhaustive in what follows, but I hope that those who made this adventure possible realize how grateful I am for this. 

First of all, I would like to thank George Batrouni, Leonardo Fallani, Maxim Olshanii, Ludovic Pricoupenko and Nikolaj T. Zinner for agreeing to be part of my jury and for their enthusiasm for this work. In particular, I thank the two people who guided me during these three years of PhD, my supervisors Patrizia and Mathias. Both of you have always managed to be available when I needed it, and to share enthusiastic discussions about Physics, but not only. I feel lucky that I had you as my supervisors, and I thank you for that.

I should thank the other permanent members of the \textit{Institut de Physique de Nice} (INPHYNI) for bringing such a good atmosphere to the lab. Thank you to the researchers with whom I had the chance to have interesting discussions during a lunch, a coffee break, or in the corridors: Mario, Fr\'{e}d\'{e}ric, Thierry, Robin, William, among others. A special thanks to Guillaume and Xavier, with whom I have learned so much about the Magnus effect (let's put it this way). I do not forget the members of the administrative and computer staffs: thank you Nathalie, Isabelle, François-R\'{e}gis and Christian for your kindness and patience, you needed it with me! 

I also thank all the PhD students, postdocs and interns of the INPHYNI for being such nice traveling companions: Abdoulaye, Bruno, François, Thibaut, Ali, Simona, Antoine, Samir, Tao, Cristina, Michelle, Patrice, Aurélien, Guido, Axel, Guillaume \textit{1145B}, Pierre, Romain, Patrizia, Marius, Anna, Julian, Vittorio, Florent, Alexis, and of course my compatriot Antonin. While writing your names I am thinking about the good moments I have shared with all of you, thank you for this! In particular, I thank Thibaut for the nice discussions and memorable moments we have shared together. Simona, for all the laughs, surrealistic debates about dragons, and for having enough imagination for both of us. Guillaume 1145B, for the passionate discussions about Science and for the \textit{petites sœurs}. My buddies of the "G. crew", Patrice and Aurélien, for all the games we have invented during breaks and for never leaving room for boredom in our office. I especially thank our "daddy" Patrice for taking care of Aurélien and I, who really needed it --- I honestly wonder how we could have survived without you. And Aurélien, if Patrice is our daddy, you are with no doubt my brother: I feel like we have experienced the same things, good or bad, at the same moments during these three years. And every time, you were there when I needed to. More than thank you, I want to say: \textit{bien joué Aurélien !}

There are also people who were there before my doctoral studies, who are still here now, and whom I would like to thank for their support and just for being a part of my life. First, the very good friends I have met during my studies, starting with my \textit{vieux frère} Manu and the \textit{trois frères} Hugo, Mickael and Romain during my unforgettable years of \textit{classe prépa}, followed by my \textit{grand Ami} Etienne (who is not only a great friend but also was kind enough to read and make spelling corrections to part of this manuscript) and my \textit{copain} Will at the \textit{magistère}, without forgetting my friends Audrey, Nathaniel, Jérémy and Anne-Charlotte (among others) of the \textit{prépa agreg}. Besides, I thank Roch, Philippe, and all the great teachers I was fortunate to have in \textit{classe prépa} and after, to be largely responsible for the choices that led me to do this thesis today.

I also thank my weird family\footnote{admit it, you guys are a bit weird!}, for teaching me to always be curious about the world. In particular, I thank my brother François, whom I nicknamed Science-Boy at the time, for having forced me to sit in front of our white board and to learn too-advanced-for-my-age science when I was young. I also want to thank my other brother Manu, who taught me the importance of having a balanced life. Moreover, I am grateful to my partner's extravagant family to have welcomed me with such kindness among theirs.

Last but not least, I thank the two loves of my life, my partner Lorène and my daughter Ambre, for bringing me joy every day. Lorène, needless to say that a "thank you" is far from being enough, seeing all that you have done for me, especially during this last year. And Ambre, starting each day by seeing your smile is the best gift I could have dreamed of.
\clearemptydoublepage
\pagestyle{empty}

%
%
~
\vspace{4cm}
\begin{flushright} 
\textit{We were on a walk and somehow began to talk about space. I had just read Weyl's book \textit{Space, Time and Matter}, and under its influence was proud to declare that space was simply the field of linear operations.\\
"Nonsense," said Heisenberg, "space is blue and birds fly through it."}

\vspace{0.3cm}
\rule{2cm}{1.5pt}
\vspace{0.3cm}

Felix Bloch

\textit{Heisenberg and the early days of quantum mechanics}
\end{flushright}

%
%
%
\vfill
\clearemptydoublepage
\pagestyle{empty}
%

\dominitoc
\tableofcontents

\setcounter{mtc}{1} 

\clearemptydoublepage
\pagestyle{fancy}
\thispagestyle{empty}
\markboth{Introduction}{Introduction}
\begin{partintro}
\chapter*{Introduction}
\addcontentsline{toc}{chapter}{Introduction}

One of the greatest challenges of modern physics is to understand the so-called \textit{strongly correlated systems}, where particles have so much influence over each other that completely new paradigms have to be involved in order to describe them. Indeed, free models are pretty well understood, and in many systems the correlations can be treated as a \textit{perturbation} of the non-interacting model. However, in other ones, interactions are so strong that they cannot be treated perturbatively and lead to totally different properties than their weak-coupling counterparts. In solid state physics for instance, many phenomena are associated or believed to be associated with electron-electron interactions, including the very debated \textit{high-temperature superconductivity} \cite{Auerbach1994}. The realization of a common framework in order to treat strong correlations appears to be such a complicate but groundbreaking task that it is sometimes referred as a \textit{third quantum revolution}.

Among the possible classes of strongly correlated systems that one could think of, there is one whose strongly correlated nature is present even for very weak interactions, which is the class of one-dimensional systems \cite{Giamarchi_book}. This can be understood pretty easily by the fact that, because of the dimensional constraint, particles cannot avoid each other. Thus, even the slightest particle excitation will directly turn into a collective one. Therefore, physics in one dimension must be addressed in a whole different way than in higher dimensions, and is rich of counter-intuitive and interesting phenomena.

Inseparable from the notion of quantum correlations is the notion of \textit{indistinguishability}, a counter-intuitive aspect of quantum physics which makes the quantum many-body problem even harder to treat, at least conceptually, than the classical $n$-body problem. In quantum physics, particles with the same intrinsic properties (mass, charge, spin...) are said identical and cannot be distinguished with each other. Therefore, permuting two identical particles should not change the physical properties of the system, and can thus only change the many-body wave function describing it by a phase factor. Depending on the spin of these identical particles, one can show that this phase factor is either $+1$ or $-1$, corresponding respectively to a \textit{symmetric} and an \textit{anti-symmetric} exchange \cite{Schwinger1951}. This discriminates identical particles into two classes, the so-called \textit{bosons} and \textit{fermions}, depending on their symmetrical or anti-symmetrical nature.

The consequences of the symmetrization postulate of identical particles are striking. On a purely conceptual point of view, it implies that all identical particles are, in a certain way, correlated, even if they are not interacting. For fermions, it implies that two identical particles cannot be in the same quantum state --- for instance, all the electrons of the Universe are in different states! This fact, known as the \textit{Pauli exclusion principle}, explains a huge variety of properties of everyday life, such as for instance the fact that ordinary matter does not collapse \cite{Dyson1967}. The implications are even more spectacular for bosons. They have indeed the tendency to accumulate in their lowest energy state, at the origin of the celebrated \textit{Bose-Einstein condensation} phenomenon at ultracold temperatures, where millions of particles behave as a single macroscopic wave. The bosonic symmetry is also related to other surprising quantum phases of matter, such as \textit{superfluids}, a type of fluids which has  zero viscosity, or \textit{superconductors}, a type of materials with zero resistance where electrons form bosonic \textit{Cooper pairs}.

Thus, the study of strongly correlated systems is extremely intricate. As we have previously stated, they cannot be treated perturbatively, and the exponential growth of their complexity with increasing number of particles makes it extremely difficult to access by exact analytical or numerical calculations. In order to face this problem, Feynman suggested in the eighties to make use of \textit{quantum simulators} \cite{Feynman1982,Feynman1986}. The idea is to create a clean and controllable experimental system with a given  Hamiltonian coming from other branches of physics such as condensed matter, quantum chemistry or high-energy physics.

Following Feynman's direction, huge progresses have been done in the field of ultracold atomic physics after the realization of the first atomic Bose-Einstein condensate in the nineties \cite{Tollett1995,Petrich1995,Davis1995,Anderson1995,Bradley1995}. Using only lasers and magnetic fields, experimentalists are now able to prepare their systems in various external potentials, to tune the interactions between the strongly attractive to the strongly repulsive regimes, with almost no coupling to the environment \cite{Bloch2008,Bloch2012}. They have then been able to simulate a large number of strongly correlated systems, such as the \textit{Bose-Hubbard model} \cite{Greiner2002}, the quantum Hall effect \cite{Lin2011} (making use of \textit{artificial gauge fields} in order to circumvent the fact that atoms are neutral particles), and even cosmological models such as a black hole-like system in a Bose-Einstein condensate \cite{Lahav2010} or Universe's expansion \cite{Eckel2018}, among many others.

Moreover, by making the external trapping potential very anisotropic, experimentalists have been able to access the one-dimensional regime, allowing to test some of surprising predictions of low-dimensional quantum physics \cite{Cazalilla2011}. Many experiments have been performed on spinless bosons, and the exceptional control over interactions has allowed to observe, for instance, the \textit{fermionization} of bosons at very large repulsions \cite{Paredes04,Kinoshita2004}. However, a lot of typically one-dimensional phenomena, such as the \textit{spin-charge separation} between the spin and density excitations, are expect to happen in one-dimensional fermionic spin mixtures \cite{Voit1995}. 

Recently, a one-dimensional fermionic mixture with up to six spin-components was realized in the experimental group of Leonardo Fallani in the LENS \cite{Pagano2014}, paving the way for the verification of many hitherto untested theoretical predictions. Their experiment was performed using fermionic Ytterbium atoms, whose ground-state has a purely nuclear spin. This implies that particles are subjected to the same external and interaction potentials regardless of their spin-orientation, and that there are no spin-flipping collisions. This confers the so-called $SU(\kappa)$ symmetry to the system, where $\kappa$ is the number of spin-components \cite{Gorshkov2010}, making it an ideal quantum simulator for the Yang-Mills gauge theories involved in the standard model of elementary particles \cite{Banerjee2013,Zohar2013,Tagliacozzo2013}.

This thesis is devoted to the theoretical study of one-dimensional quantum mixtures (fermionic, bosonic, or mixed), in the fermionized regime of very strong repulsions. As in the LENS experiment, the particles have the same mass, are subjected to the same (harmonic) external potential and ($\delta-$type) interaction potential whatever their species, and the number of particle per species is fixed. As one can see, many aspects of strongly correlated systems are tackled, namely the strong interactions, low-dimensionality, and the question of exchange symmetry and quantum statistics. During this work, we have tried to link these concepts together and analyze their effects, always keeping in mind the experimental aspects. The central question that we  address is the following: How to characterize, both theoretically and experimentally, the exchange symmetry in our system? Indeed, although the \textit{total} symmetry of the particles is fixed by their fermionic or bosonic nature, the fact that they have different spin orientations allows to obtain other kinds of \textit{spatial} exchange symmetries.

This manuscript is organized as follows.

Chapter \ref{Exactsol} is an introduction to the concepts and techniques related to strongly interacting one-dimensional quantum mixtures. After describing some of the theoretical peculiarities and experimental aspects of one-dimension, we  explain the effects of strong interactions and the so-called fermionization. In particular, we  describe the method we implemented in order to obtain exact analytical results for few-body systems, which is based on a mapping to a non-interacting fermionic problem combined with a perturbative expansion performed over the inverse of the coupling strength. Besides, we give an interpretation of this method in terms of graph theory. 

In chapter \ref{chap:sym}, we explain how, given an exact solution obtained by the aforementioned perturbative method, we are able to characterize its exchange symmetry. To do so, we adapt the so-called \textit{class-sum method}, which is originally due to Dirac. We try to present it in a pedagogical but mathematically rigorous way, with the hope that this manuscript can serve as a good introduction to this method. The exchange symmetry of various few-body mixtures are then analyzed. We show in particular that the class-sum method can serve to generalize the so-called \textit{Lieb-Mattis theorem}, which allows to compare the ordering of the energy levels associated to certain symmetry classes.

In chapter \ref{chap:1bcor}, we study the correlations in our system, and more precisely the \textit{one-body correlations}, which embed the density distributions of the particles in real and momentum space and are easily accessible in a cold atom experiment.  First, we analyze the effects of interactions and symmetries on few-body systems, and show that the density and momentum profiles can be qualitatively deduced from symmetry arguments. Second, we focus on the so-called \textit{Tan's contact}, an observable that governs the high-momentum behavior of shortly-interacting quantum gases. We show in particular that a measurement of Tan's contact allows to deduce uniquely the exchange symmetry of the system. Then, in order to be as experimentally relevant as possible, we derive scaling laws for Tan's contact, as a function of the interaction strength, number of particles and components, temperature, and transverse confinement.

The so-called \textit{coordinate Bethe ansatz}, which allows to obtain exact results in the absence of an external potential but for any value of the interaction strength, is explained in details in appendix \ref{secbethe}. Appendix \ref{listpub} contains the list of publications of the author of this thesis.

\end{partintro}
\clearemptydoublepage

\clearemptydoublepage
\pagestyle{fancy}
\thispagestyle{empty}
\chapter{Strongly interacting one-dimensional quantum gases}
\label{Exactsol}

\minitoc
\newpage


What makes one-dimensional systems extremely interesting for theoretical physicists is divided in two main reasons. First, due to the extreme complexity of many-body physics, 
these systems are usually easier to solve analytically than many-body systems in higher dimensions. Second, and perhaps more interestingly, phenomena in one-dimension 
are often very different from what our three-dimensional intuition may suggests us. Indeed, many of the theories that work extremely well in higher dimensions 
have to be completely modified in order to tackle their one-dimensional counterparts. Finally, besides these two historical reasons, what makes this  research area even 
more exciting is the huge progresses realized in ultracold atom physics, that allow to engineer and probe various one-dimensional systems with an incredible precision.

This chapter is devoted to the study of the models describing strongly interacting one-dimensional ultracold atomic gases. First, in section \ref{secgen1D}, we provide a 
very general description of one-dimensional systems, from their theoretical peculiarities and modelization to their main experimental realization techniques. 
Then, in section \ref{secstrong}, we focus on the case of strongly interacting one-dimensional quantum gases in the experimentally relevant case of a harmonic external
potential, and describe an exact analytic solution that we will extensively use throughout this thesis.

\section{Generalities on one-dimensional quantum gases}
\label{secgen1D}
In this introductory section, we discuss some general aspects of ultracold atomic gases in one dimension. First, we are going to review some of the very peculiar 
theoretical features of one dimension in \ref{pec1d}. Then, we briefly present how experimental physicists are able to create highly clean and tunable ultracold atomic systems 
that can be modeled by one-dimensional theories in \ref{expasp}. Finally, in \ref{2body}, we focus on the theoretical description of interactions in one-dimensional ultracold atomic gases, whose diluted nature 
reduce the problem to two-body interactions. This last part will serve as a preliminary for the rest of the chapter, by justifying precisely the form of the interaction 
potential and its consequence on the many-body wave function.

\subsection{Some theoretical peculiarities of one dimension}
\label{pec1d}

\begin{figure}\centering
\begin{tikzpicture}
\draw (0,4) rectangle (4,0);
\draw (6,2) -- (12,2);
\draw[fill] (7,2) circle(0.1);
\draw[fill] (7.5,2) circle(0.1);
\draw[fill] (8.5,2) circle(0.1);
\draw[fill] (10,2) circle(0.1);
\draw[fill] (10.3,2) circle(0.1);
\draw[fill] (10.8,2) circle(0.1);
\draw[fill] (11.5,2) circle(0.1);
\draw[fill] (1.2,2) circle(0.1);
\draw[fill] (0.5,1.2) circle(0.1);
\draw[fill] (2,3) circle(0.1);
\draw[fill] (2.1,1.5) circle(0.1);
\draw[fill] (3,0.5) circle(0.1);
\draw[fill] (0.6,3.5) circle(0.1);
\draw[fill] (3.1,3.1) circle(0.1);
\draw[fill] (3.6,2) circle(0.1);
\draw[>=latex,->] (2.3,1.7) -- (2.7,2.1) ;
\draw[>=latex,->] (0.6,3.2) -- (0.6,2.7) ;
\draw[>=latex,->] (8.5,2.3) -- (9,2.3) ;
\draw[>=latex,<-] (9.5,2.3) -- (10,2.3) ;
\end{tikzpicture}
\caption{\label{1dtop}Classical interpretation of the dimensional constraint. In the 2D gas (left), particles are free to avoid each other, which is not the case in 1D (right).}
\end{figure}
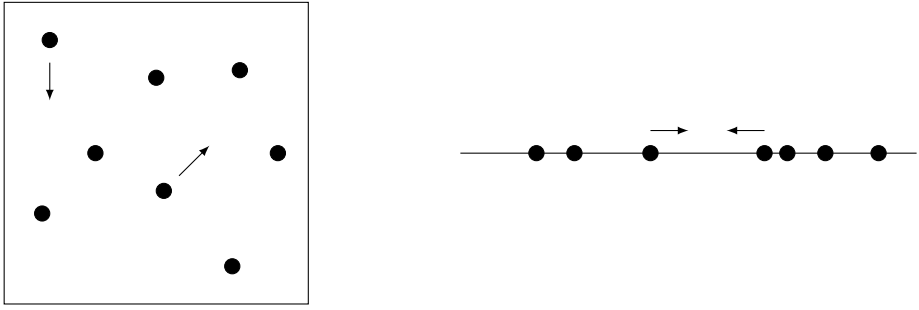

The one-dimensional (1D) world is, on many aspects, profoundly distinct from higher dimensions \cite{Giamarchi_book}. Many quantum theories that have proven to be very efficient 
in higher dimensions, such as the Landau-Fermi liquid theory describing interacting electrons \cite{Landau1957} simply breakdown in 1D. It could naively seem strange that models that work perfectly well in
two (2D), three (3D) and, formally, any dimension $d\ge2$, fail so dramatically in 1D. On a mathematical point of view, this is a reflect of the very peculiar topological 
properties of low-dimensional spaces. Physically, for a many-body one-dimensional system, the dimensional constraint is actually very simple to understand: contrary 
to higher dimensions, particles do not have the possibility to avoid each other (see Fig. \ref{1dtop}).

This very simple observation has dramatic impacts. First, 1D systems are strongly correlated, even if the interactions between particles are weak. It seems clear then 
that naive mean-field approaches will not succeed to describe it. Second, a single-particle excitation will automatically lead to a collective excitation. 
This explains why the Landau-Fermi liquid paradigm fails to describe 1D systems. Indeed, Landau's description is based on the idea 
that a Fermi liquid behaves essentially as a free Fermi gas, but with "dressed" particles, that is whose dynamical properties such as the mass or the magnetic
moment are renormalized due to interactions. This so-called \textit{adiabatic connection} implies that the elementary excitations of a Fermi liquid can be treated very similarly to 
the individual excitations of a free Fermi gas, as quasi-particles with a very long life time $\tau$. Due to the collective nature of excitations in 1D, such a description by essentially 
free quasi-particles clearly becomes irrelevant. 

The 1D counterpart of the Landau-Fermi liquid universality class is the so-called Tomonaga-Luttinger liquid universality class \cite{Haldane1981}. The Tomonaga-Luttinger liquid theory describes the low-energy excitations of a large 
number of 1D models, by mean of the so-called \textit{bosonization} method. It consists in linearizing the energy spectrum $\epsilon_k$ around the two Fermi 
points $+k_F$ and $-k_F$ writing $\epsilon_k\simeq\pm\hbar v_F(k-k_F)$. This implies the following low-energy effective 
description for the Hamiltonian $\hat{H}_0$ of the free Fermi gas:
\begin{equation}
\label{h0lutt}
\hat{H}_0=\sum_k\hbar v_Fk\left(\hat{c}_k^{R\dagger}\hat{c}_k^{R}-\hat{c}_k^{L\dagger}\hat{c}_k^L\right),
\end{equation}
where $\hat{c}_k^{R/L\dagger}$, $\hat{c}_k^{R/L}$ are the usual fermionic creation/annihilation operator, the $R/L$ exponents standing for right/left moving particles. 
Motivated by the long wavelength collective behavior of low-energy excitations, we define the density fluctuation operators as
\begin{equation}
\hat{\rho}_q^{R/L\dagger}=\sum_k c_{k+q}^{R/L\dagger}c_k^{R/L}
\end{equation}
and
\begin{equation}
\hat{\rho}_q^{R/L}=\sum_k c_{k-q}^{R/L\dagger}c_k^{R/L}.
\end{equation}
We then use the associated field operators $\hat{\rho}^{R/L}(x)$ to define the current and density fields as the following 
bosonic fields\footnote{The following expressions are only true in the thermodynamic limit $L\to\infty$. Finite-size definitions are more involved and make use of a cut-off parameter.}:
\begin{equation}
\begin{split}
&\phi(x)=-\pi\left(\hat{\rho}^{R}(x)+\hat{\rho}^{L}(x)\right),\\
&\Pi(x)=\hat{\rho}^{R}(x)-\hat{\rho}^{L}(x).
\end{split}
\end{equation}
Then, Eq.~\eqref{h0lutt} can be re-written
\begin{equation}
\label{h0dc}
\hat{H}_0=\frac{\hbar}{2}\int dx ~v_F\left[\pi\Pi(x)^2+\frac{1}{\pi}\left(\partial_x\phi(x)\right)^2\right].
\end{equation}
When interactions are taken into account, Haldane has shown that the total Hamiltonian $\hat{H}$ is very similar to Eq.~\eqref{h0dc}:
\begin{equation}
\label{lutham}
\hat{H}=\frac{\hbar}{2}\int dx ~u\left[\pi K\Pi(x)^2+\frac{1}{\pi K}\left(\partial_x\phi(x)\right)^2\right],
\end{equation}
where $u$ is the sound velocity and the so-called \textit{Luttinger parameter} $K$ controls the long-distance behavior of the correlation functions. 
More generally, we can define the Tomonaga-Luttinger liquids as the class of 1D liquids described by an Hamiltonian
of the form of Eq.~\eqref{lutham}. It is then sufficient to determine the model-dependent parameters $K$ and $u$  to obtain precious information about 
the low-energy and low-momentum behavior of the system, hence the power of this method. Moreover, although derived at first for spinless fermions, it also 
allows to describe spinless bosonic systems as well as spin mixtures.

If we consider a mixture of different spin species, another striking effect of the dimensionality occurs. In the case of a spin-$\frac{1}{2}$ mixture, we can
define respectively the charge and spin fields as the following spin-symmetric and spin-anti-symmetric operators:
\begin{equation}
\begin{split}
&\phi_{c/s}(x)=\frac{1}{\sqrt{2}}\left[\phi_{\uparrow}(x)\pm\phi_{\downarrow}(x)\right],\\
&\Pi_{c/s}(x)=\frac{1}{\sqrt{2}}\left[\Pi_{\uparrow}(x)\pm\Pi_{\downarrow}(x)\right],
\end{split}
\end{equation}
where $c/s$ are respectively associated with $+/-$. It can the been shown that the Hamiltonian can be written:
\begin{equation}
\hat{H}=\hat{H}_c+\hat{H}_s=\sum_{\nu=c,s}\frac{\hbar}{2}\int dx ~u_{\nu}\left[\pi K_{\nu}\Pi_{\nu}(x)^2+\frac{1}{\pi K_{\nu}}\left(\partial_x\phi_{\nu}(x)\right)^2\right].
\end{equation}
To put it in words, the excitations in 1D are completely decoupled between charge and spin excitations! This counter-intuitive effect is known as the \textit{spin-charge separation}. It is intimately connected with 
the collective nature of excitations.

Another very surprising property of 1D systems is that they are more interacting at low densities. This can be understood with a simple dimensional argument: 
if we denote the atomic lineic density by $n$ and the interaction strength by $g$, the typical interaction and kinetic energies per particle are respectively 
given by $E_{int}\sim ng$ and $E_{K}\sim \hbar^2n^2/2m$. Thus, the ratio of these energies is $\gamma\sim E_{int}/E_K\sim g/n$. It is then clear that the interactions 
will have a more important influence on the system for low $n$.

Let us finish this introduction to the beauty of one-dimensional physics by a very profound and important theorem known as the Mermin-Wagner-Hohenberg theorem \cite{Mermin1966, Hohenberg1967}. It states 
that, because of the too important long-range fluctuations induced by the dimensional constraint, there is no phase transition with spontaneous breaking of a continuous symmetry at non-zero temperatures (see section \ref{importancesym}). This 
implies, in particular, the absence of the celebrated Bose-Einstein condensation in 1D, which is associated with the breakdown of the phase symmetry $U(1)$ \cite{Pitaevskii_book}.

\subsection{Experimental aspects}
\label{expasp}

%
%

Although very interesting on a purely theoretical point of view, it would be a little bit unsatisfying if one-dimensional physics was only a set of 
mathematical \textit{toy models} without any experimental significance. Fortunately, wonderful progress has been achieved in the domain of ultracold gases 
(\cite{Bloch2008} and references therein), allowing to access the 1D world with an incredible degree of experimental control \cite{Moritz2003,Stoferle2004}. This motivates a very stimulating interplay between theoretical 
and experimental physicists. We thus briefly take the time to present some of the  
experimental techniques used to address 1D physics with ultracold gases. The probing techniques will be exposed in the chapter devoted to one-body correlations 
(chapter \ref{chap:1bcor}). Remark that other physical experimental systems can be modelized by 1D theories, ranging from edge states in the quantum hall effect \cite{Milliken1996}, 
to carbon nanotubes \cite{Bockrath1999} or quantum wires \cite{Auslaender2005} and spin chains \cite{Lake2005}. However these systems don't offer the experimental 
control of ultracold atomic physics. We focus on the latter in the following.

\subsubsection{Cooling atoms}

The main tool in order to cool atoms is the so-called \textit{magneto-optical trap} (MOT), whose principle was suggested by Jean Dalibard 
in the eighties, and whose first experimental realization was reported in \cite{Raab1987}. It allows to cool atoms at the order of a few $\mu$K. 

\begin{figure}\centering
\begin{tikzpicture}
	\draw[very thick] (-4,0) -- (3,0);
	\draw (-7,0) node{\footnotesize $|\mathrm{F} = 0\rangle$};
	\draw(-5.5,0) node{\footnotesize $m = 0$};
	
	\draw[very thick] (-4,4) -- (3,4);
	\draw[very thick] (-4,3) -- (3,5);
	\draw[very thick] (-4,5) -- (3,3);
	\draw (-7,4) node{\footnotesize $|\mathrm{F}^{\prime} =1\rangle$};
	\draw (-5.5,4) node{\footnotesize $m = 0$};
	\draw (-5.5,5.2) node{\footnotesize $m = +1$};
	\draw (-5.5,2.8) node{\footnotesize $m = -1$};

	\draw[<-,latex-,very thick] (0,5) -- (1,5);
	\draw (0.5,5) node[above]{\footnotesize $\vec{\nabla}\cdot\vec{B}$};

	\draw[red,very  thick, ->,-latex] (-2,0) -- (-2,3.565); \draw[red,very  thick, ->,-latex] (1.,0) -- (1.,3.565);
	\draw[fill=black] (-2,0) circle(0.15); \draw[fill=black] (1.,0) circle(0.15);

	\draw[red,fill = red] (-4.2,1.5) rectangle (-3.5,1.75);
	\draw[red,fill=red] (-3.5,2) -- (-3,1.625) -- (-3.5,1.25);
	\draw (-3.8,1) node{\footnotesize $\sigma^{-}$};

	\draw[red,fill = red] (3.2,1.5) rectangle (2.5,1.75);
	\draw[red,fill=red] (2.5,2) -- (2,1.625) -- (2.5,1.25);
	\draw (2.9,1) node{\footnotesize $\sigma^{+}$};
\end{tikzpicture}
\caption{\label{fig:MOT}Schematic diagram of a MOT. The magnetic gradient generates a Zeeman splitting of the $|\mathrm{F}^{\prime} =1\rangle$ state. The red-detuned counter-propagating lasers 
have opposite circular polarizations. Because of the atomic selection rules, each laser can only induce a transition to a single Zeeman level. Thus, the atoms are slowed down 
and pushed toward the center of the MOT.}
\end{figure}
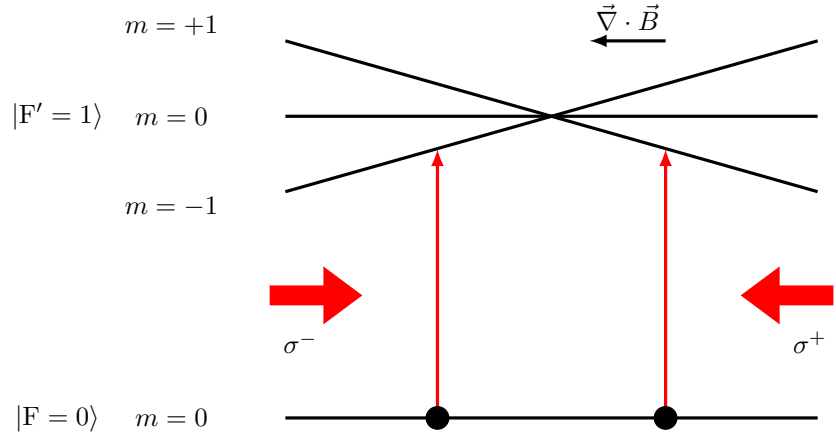

There are two main ingredients in a MOT: 
first, the atoms are illuminated by counter-propagating \textit{red-detuned} laser beams, i.e. with a lower frequency than the resonant frequency of the atoms. Because of the Doppler effect,
when an atom moves, it will absorb a photon coming from the opposite direction and carrying a momentum $\vec{p}$, thus reducing the momentum of the atom. The second ingredient of the MOT is a spatially varying magnetic field generated by magnetic coils in anti-Helmholtz configuration. This causes a Zeeman splitting of the 
energy levels of the atoms, which increases with the distance from the center of the trap, and therefore shifts the atomic resonance closer to the 
frequency of the lasers. Thus, atoms are more likely to absorb a "photon kick" when far from the trap center. Moreover, by choosing the laser polarizations so that photons interact with the correct energy levels, 
these photon kicks are always pushing the atoms toward the center. A schematic diagram of a MOT is given in Fig.~\ref{fig:MOT}.

However, the cooling induced by a MOT is limited by the fact that atoms spontaneously emit the photons previously absorbed during the cooling process, which in result will heat the atoms. In order to 
reach the ultracold limit (around 100 $n$K), the next step is the \textit{evaporative cooling}. It consists in trapping the atoms in a magnetic trap of finite depth $\epsilon$. The "hot" atoms with energies 
higher than $\epsilon$ will then escape from the trap, resulting in a reduced average temperature of the atoms remaining in the trap. It is the exact analog of cooling a hot drink by blowing on it! This technique was employed 
in order to obtain the first Bose-Einstein condensate with ultracold atoms \cite{Tollett1995,Petrich1995,Davis1995,Anderson1995,Bradley1995}.

\subsubsection{Trapping ultracold atoms}

\label{trapuco}

The basic way of generating 1D atomic traps is to manipulate the atomic potential in order to make it very anisotropic. In a typical harmonic oscillator potential 
characterized by frequencies $\omega_x=\omega_y=\omega_{\perp}$ and $\omega_z$, we want the \textit{aspect ratio} $\lambda=\omega_z/\omega_{\perp}$ to be sufficiently small, so
that the typical energy of a particle is smaller than the energy of the first transverse excited state. This way, the so-called \textit{quasi-1D} regime is achieved, with no 
transverse excited modes and all the dynamics occurring in the $z$ direction \cite{Olshanii1998}. The main ways to generate atomic potentials are discussed below.

\begin{figure}\centering
\begin{tikzpicture}
\begin{scope}[scale=0.8]
\foreach \x in {0,1.2,...,7}
    \foreach \y in {0,0.9,...,4} \draw[fill=gray!50,rotate around={45:(\x,\y)}] (\x,\y) ellipse (2cm and 0.1cm); 
\draw[very thick,->] (-3,4) -- (-2,4);\draw[very thick,->] (-3,4) -- (-3,5);\draw[very thick,->] (-3,4) -- (-2.3,4.7);
\draw (-2,4) node[below]{$y$};\draw (-3,5) node[left]{$x$};\draw (-2.3,4.7) node[right]{$z$};
\tikzstyle{fleche}=[->,>=latex,line width=3mm,red];\draw[fleche] (-4,2)--(-2,2);\draw[fleche] (10,2)--(8,2);
\draw[fleche] (3,7.5)--(3,5.5);\draw[fleche] (3,-4)--(3,-2);
\end{scope}
\end{tikzpicture}
\caption{\label{fig:optlat}A 2D optical lattice: the superposition of two standing waves generates 1D traps. The hopping time between these tubes has to be significantly greater than the characteristic time of the experiment.}
\end{figure}
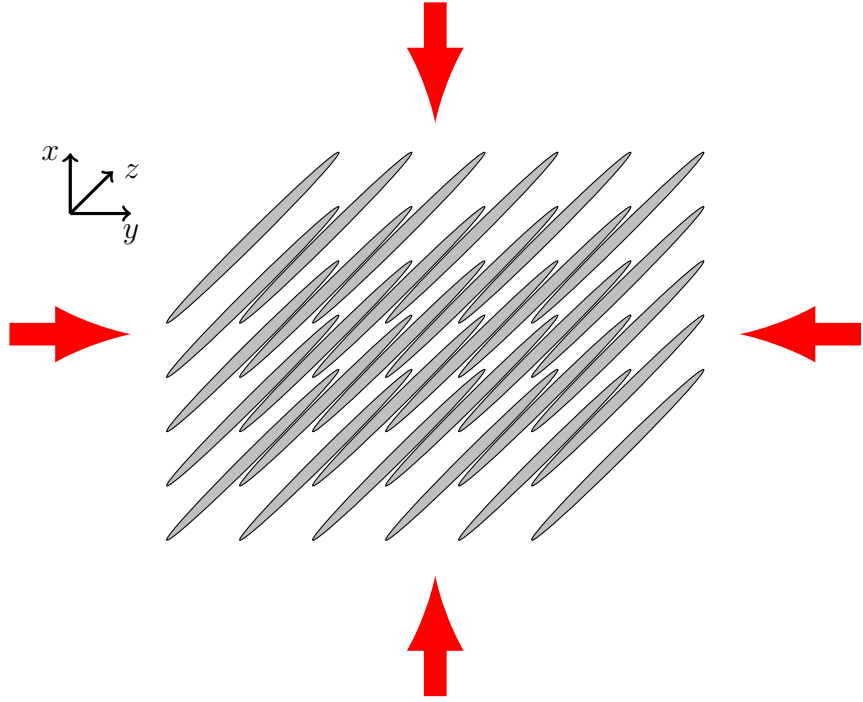

There are two main methods in order to trap ultracold atoms: \textit{optical} and \textit{magnetic} trapping. The optical trapping is based on 
interference patterns created by a superposition of laser beams. More precisely, the electric fields generated by the light
will interact with the atom and generate a small dipole moment. The resulting dipolar force  $\vec{F}$ between the atom and the laser of frequency 
$\omega_L$ has the form \cite{Grimm2000}:
\begin{equation}
\label{opttrap}
\vec{F}(\vec{r})=\frac{1}{2}\alpha(\omega_L)\nabla\left[\left|\vec{E}(\vec{r})\right|^2\right],
\end{equation}
where $\alpha(\omega_L)$ is the polarizability. If $\omega_L$ is close to the resonance frequency $\omega_0$ between the ground $\left|g\right\rangle$ and excited state $\left|e\right\rangle$ 
of the atom, it can be shown that the polarizability has the form:
\begin{equation}
\alpha(\omega_L)\simeq\frac{\left|\left\langle e\left|\hat{d}_{\vec{E}}\right|g\right\rangle\right|^2}{\hbar(\omega_0-\omega_L)},
\end{equation}
where $\hat{d}_{\vec{E}}$ is the dipole operator in the direction of the field. Therefore, we see that the atoms will be attracted to regions of high light intensity 
when the laser is \textit{red detuned} ($\omega_L<\omega_0$) and to regions of weak light intensity when the laser is \textit{blue detuned} ($\omega_L>\omega_0$).
An example of 2D optical lattice allowing to generate 1D traps is given in figure \ref{fig:optlat}.

The magnetic trapping techniques are based on the Zeeman coupling between an external magnetic field $\vec{B}(\vec{r})$ and the total spin $\vec{S}$ of the atom.
In 1D, the magnetic field is generally generated by a so-called \textit{atom chip}, which consists in a surface where wires were deposited with high-precision micro-fabrication techniques \cite{Reichel1999,Jacqmin2011}.
If we denote the magnetic moment of the atom by $\vec{\mu}$, the atomic potential is given by \cite{Folman2002}:
\begin{equation}
V_{mag}(\vec{r})=-\vec{\mu}\cdot\vec{B}(\vec{r}),
\end{equation}
which is simply proportional to $||\vec{B}(\vec{r})||$ when $\vec{B}(\vec{r})$ is sufficiently slowly varying. Similarly to the case 
of the optical trapping, there are two cases, depending if $\vec{\mu}$ is in the same 
direction as $\vec{B}$ ($V_{mag}<0$) or in the opposite direction ($V_{mag}>0$). In these two so-called \textit{strong field seeking} (resp. \textit{weak field seeking}) states,
the atoms are attracted by maxima (resp. minima) of $||\vec{B}({r})||$. Due to a theorem by Earnshaw which states that there cannot be maxima of the magnetic field in free space, the 
source of the field must be inside the trapping region in the strong field seeking case \cite{Ketterle1992}. This is the reason  why the weak field seeking state is more used. However, it has  the inconvenient 
of preventing the use of strong magnetic fields, which can be useful in order to tune the interactions (see section \ref{expint}).

\subsubsection{Tuning the interactions}
\label{expint}

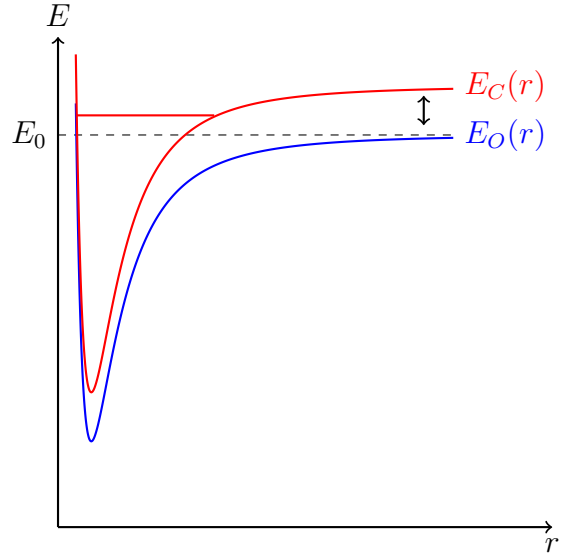
\begin{figure}\centering
\begin{tikzpicture}
\begin{scope}[scale=1.3]
\draw[thick,->] (0,0) -- (5,0);
\draw[dashed] (0,4) -- (4,4);
\draw[thick,->] (0,0) -- (0,5);
\draw (5,0) node[below]{$r$};
\draw (0,4) node[left]{$E_0$};
\draw (4,4) node[right,blue]{$E_O(r)$};
\draw (4,4.5) node[right,red]{$E_C(r)$};
\draw (0,5) node[above]{$E$};
\draw[thick,red] (0.18,4.2) -- (1.58,4.2);
\draw[thick,<->] (3.7,4.1) -- (3.7,4.4);
\draw[blue,samples=1000,thick,domain=0.18:4] plot (\x,{4-2.5/(\x+0.4)^3+0.5/(\x+0.4)^6});
\draw[red,samples=1000,thick,domain=0.18:4] plot (\x,{4.5-2.5/(\x+0.4)^3+0.5/(\x+0.4)^6});

\end{scope}
\end{tikzpicture}

\caption{\label{fig:feshb}Two-channel model for a Feshbach resonance. When varying an external magnetic field $\vec{B}_{ext}$, the incident energy $E_0$ of the particle in the open channel $E_O$
  can be in resonance at short inter-atomic distance $r$ with the energy of a bound state of the closed channel $E_C$.}
\end{figure}

Ultracold gases offer a very powerful tool that allows to tune the interactions very precisely, and even to change the sign of the interactions, namely 
the Feshbach resonances \cite{Feshbach1958,Fano1961}. The basic idea is to generate, via an external magnetic field\footnote{Alternatively, one could also use 
an external optical field \cite{Fedichev1996}.}, a resonance between a bound-state in a close channel and the scattering continuum of an open channel. Typically,
given two atoms, these two states correspond to different two-spin configurations. When the two atoms are scattering together, this induced resonance will result 
in a quasi-bound state which will considerably modify the (3D) scattering length $a_{3D}$ (a precise definition of the scattering length will be given in section \ref{2body}). 
Phenomenologically, this resonance can be described for the scattering length by \cite{Bloch2008}:
\begin{equation}
\label{feshphen}
a_{3D}(B)=a_{bg}\left[1-\frac{\Delta B}{B-B_0} \right],
\end{equation}
where $a_{bg}$ is the background scattering length in the absence of the external magnetic field $B$, and $\Delta B$ and $B_0$ are respectively the width and position of the magnetic resonance.
We can see in Eq.~\eqref{feshphen} that a Feshbach resonance even allows to change the sign of $a_{3D}$, and to tune it from $-\infty$ to $+\infty$ just by changing 
the external magnetic field \cite{Tiesinga1993}.
For a graphical interpretation of a two-channel Feshbach resonance, see Fig.~\ref{fig:feshb}.

There are alternative methods in order to tune the interactions that are specific to 1D. First, as seen in section \ref{pec1d}, one way to change 
the interaction regime in 1D is given by the atomic density $n$: the lower the density, the stronger the interactions. Moreover, it can been shown \cite{Olshanii1998} that when 
the system is in the quasi-1D regime (see section \ref{trapuco}), the scattering processes can be described by an effective interaction potential  $U(x)=g_{1D}\delta(x)$ 
where the effective coupling parameter is
\begin{equation}
\label{g1d}
g_{1D}=\frac{-2\hbar^2}{ma_{1D}}=\frac{2\hbar^2a_{3D}}{ma_{\perp}^2}\left(1-C\frac{a_{3D}}{a_{\perp}}\right)^{-1},
\end{equation}
where $a_{1D}$ is the effective 1D scattering length, $a_{\perp}=\sqrt{\hbar/m\omega_{\perp}}$ is the transverse harmonic oscillator length and $C=|\zeta(1/2)|/\sqrt{2}\simeq1.0326$, 
$\zeta$ being the Riemann zeta function. Then, by varying $\omega_{\perp}$, we see that we can approach a \textit{confined induced resonance} when $a_{\perp}\simeq Ca_{3D}$. 
This method was achieved experimentally, see e.g. \cite{Peano2005}.

The theoretical aspects of two-body interactions in 1D quantum gases will be discussed in more detail in section \ref{2body}.

\subsubsection{Spin mixtures}
\label{florexp}

In this thesis, we will mainly focus on quantum mixtures, i.e. ultracold atomic gases with different spin components. On the experimental side, a huge breakthrough 
was made in an experiment made in the LENS (European Laboratory for Nonlinear Spectroscopy) in Florence by the team of L. Fallani \cite{Pagano2014}. We briefly describe it here, as a paradigmatic realization of the system we studied in this thesis.   

L. Fallani and coworkers realized a 1D fermionic 
mixture of $\Yb$ atoms with a tunable number $\kappa\in\{1,\cdots,6\}$ of components. The ground state of $\Yb$ atoms has a nuclear spin $I=5/2$ and a zero electronic angular momentum $J=0$. This last property, which is also common to all alkaline-earth atoms \cite{Gorshkov2010}, implies that the electronic wave function 
is independent of the nuclear spin state --- there is no hyperfine structure. Therefore, the interactions between atoms of different nuclear spin states $\sigma\neq\sigma'$, which depends only on their electronic 
wave functions, are the same regardless of the choice of $\sigma\neq\sigma'$. Moreover, it implies the absence of spin-flipping collisions. 
Thus, their system is invariant by any permutation of the $\kappa\in\{1,\cdots,6\}$ spin populations, which is referred as the $SU(\kappa)$ symmetry (see chapter \ref{chap:sym}).

Moreover, the system can be initialized with a given number $\kappa<6$ of spin-components with an equal number of atoms per component, starting from 
an equally populated six-component mixture. The idea is the following: the first excited state $(I=5/2,J=1)$ is Zeeman-splitted by mean of a magnetic field.
Then, they can optically pump out the population of a given spin state $\sigma$ to another spin state $\sigma\pm1$ through polarized beams resonant on a specific 
Zeeman component of the excited state. The unwanted populations are then put in an optically closed transition and removed by evaporation.

Because the spin of $\Yb$ is purely nuclear, spin-resolved detection cannot be done by mean of a standard magnetic Stern-Gerlach procedure \cite{Gerlach1922}. In order to 
circumvent this problem, an \textit{optical} Stern-Gerlach experiment is performed \cite{Taie2010}, by misaligning a laser beam on the atomic cloud. The Gaussian profile 
of the laser generates an electric gradient. The atoms will then interact with the laser in a similar way as described in expression \ref{opttrap}, but with a polarizability depending 
on the spin-state, thus splitting the populations.

It is important to note that when $\kappa>2$, interactions cannot be tuned using Feshbach resonances if one wants to preserve the $SU(\kappa)$ symmetry. Experimentalists 
have to rely on the other methods (decreasing the atomic density or generating confined induced resonances) in order to reach the strongly interacting regime. 
In the LENS experiment, interactions are typically in the intermediate regime.

\subsection{The two-body problem}
\label{2body}
In this section we discuss the properties of interactions in ultracold quantum gases. We will justify the form of the interaction Hamiltonian on the one hand,
and on the other hand we will discuss its consequence on the many-body wave function, known as the \textit{cusp condition}.

\subsubsection{Scattering length and pseudo-potential}

Since atoms are neutral particles, they interact via van der Waals forces \cite{Dzyaloshinskii1961}. Typically, at large distance $r$ the interaction potential $U(r)$
is of the type $\propto -1/r^6$ and at distances lower than the so-called \textit{van der Waals contact distance} $r_0$, the electronic clouds will 
lead to a strong contact repulsion. The systems we consider are ultracold and diluted: therefore, $r_0$ can be considered negligible as compared to the de 
Broglie wavelength $\lambda_{\mathrm{dB}}$ and the inter-atomic separation $n^{-1/3}$ (where $n$ is the atomic density). This property, together with the fact that the temperatures are  very low, explains why the system 
can be effectively described by low energy two-body collisions, and by a single parameter, namely the $s$-wave scattering length $a_{3D}$ \cite{Pitaevskii_book}. 
Considering spinless particles, the two-body problem can be written:
\begin{equation}
\label{2bschro}
\left(-\frac{\hbar^2}{2\mu}\Delta+U(r)-E\right)\psi(\vec{r})=0,
\end{equation}
where $\vec{r}=\vec{r_1}-\vec{r_2}$ and $\mu=m/2$ is the reduced mass. In the asymptotic region $r\gg r_0$, the solution can be written as the superposition 
of an incident plane wave in the $x$ direction and a spherical scattered wave:
\begin{equation}
\label{psiscat}
\psi(\vec{r})\simeq e^{ikx}+f(\theta,k)\frac{e^{ikr}}{r},
\end{equation}
where 
\begin{equation}
k=\sqrt{\frac{2\mu E}{\hbar^2}},
\end{equation}
$\theta$ is the angle between $\vec{r}$ and the $x$-axis, and $f$ is defined as the \textit{scattering amplitude}. Expanding $\psi$ in the natural basis of 
Legendre polynomials $P_l$ yields:
\begin{equation}
\psi(\vec{r})=\sum_{l=0}^{\infty}P_l(\cos \theta)\frac{\chi_{kl}(r)}{kr},
\end{equation}
where $\chi_{kl}(r)$ satisfies
\begin{equation}
\frac{d^2\chi_{kl}(r)}{dr^2}-\frac{l(l+1)}{r^2}\chi_{kl}(r)+\frac{2\mu}{\hbar^2}\left(E-U(r)\right)\chi_{kl}(r)=0.
\end{equation}
In the asymptotic region $r\gg r_0$, one can neglect the $\propto 1/r^2$ term, which yields the simple expression:
\begin{equation}
\chi_{kl}(r)=A_l\sin \left(kr-\frac{\pi l}{2}+\delta_l(k)\right),
\end{equation}
where $\delta_l(k)$ are defined as \textit{phase shifts}. Then, an appropriate choice for $A_l$ gives after some simple algebra:
\begin{equation}
\label{scatamp}
f(\theta,k)=\frac{1}{2ik}\sum_{l=0}^{\infty}(2l+1)P_l(\cos \theta)(e^{2i\delta_l}-1).
\end{equation}
Thus, the scattering cross-section $\sigma$, related to $f(\theta,k)$ through $\sigma=\frac{4\pi}{k}\mathrm{Im}\left[ f(0,k)\right]$, is given by:
\begin{equation}
\sigma(k)=\frac{4\pi}{k^2}\sum_{l=0}^{\infty}(2l+1)\sin^2\delta_l(k).
\end{equation}
For low energy collisions and momenta $k\ll 1/r_0$, one can prove that only the $s$-wave $l=0$ term will be relevant. Therefore, Eq.~\eqref{scatamp} simply becomes:
\begin{equation}
f(\theta,k)\simeq\frac{e^{2i\delta_0}-1}{2ik}=\frac{1}{k\cot \delta_0(k)-ik}.
\end{equation}
We define the $s$-wave scattering length $a_{3D}$ as
\begin{equation}
a_{3D}=-\lim_{k\to 0}\frac{\tan \delta_0(k)}{k},
\end{equation}
which implies that $\sigma\simeq 4\pi a_{3D}^2$. Intuitively, this can be interpreted as if the system had the same low energy scattering properties as a hard sphere of radius
$a_{3D}$.

In the case where we consider the spin-statistics, we can observe an important effect. Indeed, because of the (anti-)symmetrization of the two-body wave function 
for identical (fermions) bosons, it implies that only the (odd) even $l$ terms in Eq.~\eqref{scatamp} will contribute to the cross-section $\sigma$. Thus, 
in the $s$-wave scattering limit of low energy and for distances $r\ll r_0$, identical fermions are not interacting.

In the previous discussion, we have never used a precise expression for the interaction potential $U(r)$ in Eq.~\eqref{2bschro}, but instead showed that, in 
our regime, all the physics is captured by the $s$-wave scattering asymptotic behavior and thus $a_{3D}$. It is then sufficient to replace $U(r)$ by a simpler
\textit{pseudopotential} $U_{\mathrm{pseudo}}(r)$ which has the same long distance $s$-wave scattering properties than $U(r)$. It was shown in \cite{Huang1957}
that the following pseudo-potential has the right properties:
\begin{equation}
U_{\mathrm{Huang}}(r)=g_{3D}\delta(\vec{r})\frac{\partial}{\partial r}(r\cdot),
\end{equation}
where $g_{3D}=4\pi\hbar^2a_{3D}/m$ is the coupling constant. The $\frac{\partial}{\partial r}(r\cdot)$ is useful in order to remove the $1/r$ divergence in 
Eq.~\eqref{psiscat}.

\subsubsection{Interactions in one dimension}

In the quasi-1D case where the atoms are in a highly elongated trap in the $x$ direction, Eq.~\eqref{2bschro} becomes, in the pseudo-potential approximation:
\begin{equation}
\left(-\frac{\hbar^2}{2\mu}\frac{\partial^2}{\partial x^2}+U_{\mathrm{Huang}}(r)+\hat{H}_{\perp}-E\right)\psi(\vec{r})=0,
\end{equation}
where 
\begin{equation}
\hat{H}_{\perp}=-\frac{\hbar^2}{2\mu}\left(\frac{\partial^2}{\partial y^2}+\frac{\partial^2}{\partial z^2}\right)+\frac{1}{2}\mu\omega_{\perp}^2(y^2+z^2)
\end{equation}
and $\omega_{\perp}$ is the radial confinement frequency, which is such that $\hbar\omega_{\perp}$ is 
sufficiently higher than the other typical energies so that the quasi-1D regime can be achieved. This case was studied 
in \cite{Olshanii1998}. Then, considering that the incident wave is in the ground state $\phi_0(y,z)$ of $\hat{H}_{\perp}$,
the analogue of Eq.~\eqref{psiscat} for the asymptotic form of the scattered wave function is:
\begin{equation}
\Psi(\vec{r})\simeq \left(e^{ik_x x}+f_{even}(k)e^{ik_x |x|}+f_{odd}(k) ~\mathrm{sign} (z)e^{ik_x |x|}\right)\phi_0(y,z).
\end{equation}
The scattering amplitudes can then be calculated analytically, by correctly taking into account 
the \textit{virtual} transverse excited states during the collision process. This gives $f_{odd}=0$ and, in the $k_xa_{1D}\ll 1$ limit:
\begin{equation}
f_{even}(k)=-\frac{1}{1+ik_xa_{1D}+\mathcal{O}((k_xa_{1D})^3)},
\end{equation}
where the 1D effective scattering length is defined as in Eq.~\eqref{g1d} by
\begin{equation}
\label{a1dexpress}
a_{1D}=-\frac{a_{\perp}^2}{a_{3D}}\left(1-C\frac{a_{3D}}{a_{\perp}}\right).
\end{equation}
As stated in section \ref{expint}, the analogue of Huang's pseudo-potential that captures the correct even wave scattering behavior in 1D is 
\begin{equation}
\label{upseudo}
U_{pseudo}(x)=g_{1D}\delta(x),
\end{equation}
where $g_{1D}=-\frac{2\hbar^2}{ma_{1D}}$.

\subsubsection{The cusp condition}
\label{secusp}

Here we prove a very important corollary to the fact that the 1D two-body interactions can be described by the potential \eqref{upseudo}.
Let us consider a strictly 1D Schr\"{o}dinger equation for two atoms of equal masses $m$ and coordinates $x_1,x_2$:
\begin{equation}
\label{sch1D}
\left(-\frac{\hbar^2}{2m}\left(\frac{\partial^2}{\partial x_1^2}+\frac{\partial^2}{\partial x_2^2}\right)+g_{1D}\delta(x_1-x_2)+V_{ext}(x_1,x_2)-E\right)\psi(x_1,x_2)=0,
\end{equation}
where $V_{ext}$ is an arbitrary (continuous) external potential. We then define the 
relative coordinate by $x_{12}=x_1-x_2$ and the center-of-mass coordinate by $X_{12}=\frac{x_1+x_2}{2}$. Eq.~\eqref{sch1D} then becomes:
\begin{equation}
\left(-\frac{\hbar^2}{2M}\frac{\partial^2}{\partial X_{12}^2}-\frac{\hbar^2}{2\mu}\frac{\partial^2}{\partial x_{12}^2}+g_{1D}\delta(x_{12})+V_{ext}(x_{12},X_{12})-E\right)\psi(X_{12},x_{12})=0,
\end{equation}
with $M=2m$ and $\mu=m/2$. Let us now integrate this equation in the vicinity $[-\epsilon,+\epsilon]$ of $x_{12}=0$. If we suppose that $\psi$ is continuous, the only relevant terms
in the $\epsilon\to 0$ limit are:
\begin{equation}
-\frac{\hbar^2}{2\mu}\int_{-\epsilon}^{\epsilon}\frac{\partial^2\psi(X_{12},x_{12})}{\partial x_{12}^2}\,dx_{12}+g_{1D}\int \delta(x_{12})\psi(X_{12},x_{12})\,dx_{12}.
\end{equation}
Thus, we obtain the \textit{cusp condition} for the two-body wave function in $x_1=x_2$:
\begin{equation}
\label{cusp2b}
\frac{\partial\psi}{\partial x_{12}}(X_{12},0^{+})-\frac{\partial\psi}{\partial x_{12}}(X_{12},0^{-})=\frac{2\mu g_{1D}}{\hbar^2}\psi(X_{12},0).
\end{equation}

This simple discontinuity condition on the wave function has extremely important consequences, and will be used crucially throughout this thesis.

On a side note, it is interesting to remark that if the two atoms are identical fermions, the anti-symmetrization requirement on Eq.~\eqref{cusp2b} will imply that $\psi(X_{12},0)=0$\footnote{Note that in this case, there is no cusp on the many-body wave-function.}.
This means that, even if identical fermions do not interact in experiments (see section \ref{2body}), it will not change the physics to consider a $\delta$-type 
interaction between them \textit{a priori}, as the symmetry properties will naturally cancel its effect.


\section{Strongly-interacting systems}
\label{secstrong}
When we consider the more realist case where the system is confined by an external harmonic potential $V_{ext}(x)=\frac{1}{2}m\omega^2x^2$ along the longitudinal direction, quantum integrability 
breaks down and the system can no longer be directly solved by Bethe ansatz. Indeed, since the system is not translation invariant anymore, the scattering 
events will also depend on where they took place in the trap, so that the system is no longer \textit{integrable} (c.f. appendix \ref{secbethe}). In the limit of very strong repulsion however, one can obtain 
exact analytical solutions using the so-called \textit{fermionization} property, as first pointed out in \cite{Girardeau1960} for a gas of impenetrable bosons, or Tonks-Girardeau 
gas\footnote{Tonks studied the classical gas of hard spheres in \cite{Tonks1936}.}. 
In this section, we will first explain the notion of fermionization for this simple model in \ref{tggas}. Then, we will turn to a more general method developed in \cite{Volosniev2014} that is valid 
for any choice of the mixture in \ref{secvolosniev}. The last one has a central role in this thesis, as it is the one we used in order to obtain exact analytic expressions 
for the many-body wave functions.

\subsection{The Tonks-Girardeau gas}
\label{tggas}

\subsubsection{Model}

As a pedagogical example of a simple strongly interacting system, we briefly present here the solution of the so-called Tonks-Girardeau gas. It consists in a 
gas of identical impenetrable bosons. It can be described, for an arbitrary external potential $V_{ext}$, by the following equation:
\begin{equation}
\label{schrodtg}
\left[\sum_{j=1}^{N}\left(-\frac{\hbar^2}{2m}\frac{\partial^2}{\partial x_j^2}+V_{ext}(x_j)\right)-E\right]\psi_B=0,
\end{equation}
together with the boundary conditions:
\begin{equation}
\label{bouncondtg}
\psi_B(x_1,\ldots,x_N)=0\quad \text{if}\quad x_i=x_j,\quad 1\le i<j\le N.
\end{equation}
Alternatively, one can say that the Tonks-Girardeau gas is the hardcore limit $g_{1D}\to\infty$ of the (trapped) Lieb-Liniger gas:
\begin{equation}
\left[\sum_{j=1}^{N}\left(-\frac{\hbar^2}{2m}\frac{\partial^2}{\partial x_j^2}+V_{ext}(x_j)\right)+g_{1D}\sum_{i<j}\delta(x_i-x_j)-E\right]\psi_B=0.
\end{equation}
Notice that the resulting cusp condition (see e.g. Eq.~\eqref{cusp2b}) implies Eq.~\eqref{bouncondtg} in the $g_{1D}\to\infty$ limit. Experimentally, observation of a 1D Tonks-Girardeau gas was first reported in \cite{Paredes04,Kinoshita2004}.

\subsubsection{Bose-Fermi mapping}

Girardeau's idea 
in order to solve this problem was to remark that the many-body wave function of a spinless gas of fermions $\psi_F$ that satisfies Eq.~\eqref{schrodtg} also 
satisfies Eq.~\eqref{bouncondtg}. Because of the 
Bose statistics, the solution for the Tonks-Girardeau gas can then be written \cite{Girardeau1960}:
\begin{equation}
\label{bfmapping}
\psi_B(x_1,\ldots,x_N)=A(x_1,\ldots,x_N)\psi_F(x_1,\ldots,x_N),
\end{equation}
where $A(x_1,\ldots,x_N)=\prod_{i>j}\mathrm{sign}(x_i-x_j)=\pm 1$ compensates the anti-symmetrization of $\psi_F$. This observation, known as the \textit{Bose-Fermi mapping} or 
more generally as the \textit{fermionization} of bosons, maps a strongly interacting problem onto a (simpler!) non-interacting one.

Since $\psi_B$ and $\psi_F$ both satisfy Eq.~\eqref{schrodtg}, their energy spectrum are identical. In the homogeneous case $V_{ext}=0$ on a ring of size $L$ with periodic boundary conditions, one finds 
for the ground-state many-body wave function:
\begin{equation}
\psi_B^0(x_1,\ldots,x_N)=\sqrt{\frac{2^{N(N-1)}}{N!L^N}}\prod_{i<j}\sin\left(\frac{\pi}{L}|x_i-x_j|\right),
\end{equation}
with the associated ground-state energy:
\begin{equation}
E_0=\left(N-\frac{1}{N}\right)\frac{\hbar^2\pi^2n^2}{6m},
\end{equation}
with $n=N/L$ the particle density. Note that in the thermodynamic limit this relation corresponds to the $\gamma\to\infty$ limit in Eq.~\eqref{gsell}.
In the case where the particles are trapped in a harmonic potential, Eq~\eqref{schrodtg} becomes:
\begin{equation}
\label{tgharm}
\left[\sum_{j=1}^N\left(-\frac{\hbar^2}{2m} \frac{\partial^2}{\partial x_j^2}+\frac{1}{2}m\omega^2x_j^2\right)-E\right]\psi_B=0
\end{equation}
This case is solved in \cite{Girardeau2001} and \cite{Forrester2003}. The fermionic solution $\psi_F$ of Eq.~\eqref{tgharm} is given by the well-known Slater determinant:
\begin{equation}
\label{slaterdet}
\psi_F(x_1,\ldots,x_N)=\frac{1}{\sqrt{N!}}\det\left[\phi_j(x_i/a_0)\right]_{i\in\{1,\dots,N\},~j\in\{0,\ldots,N-1\}},
\end{equation}
where the eigenstates $\phi_j$ of the single-particle Hamiltonian $\hat{H}_1=-\frac{\hbar^2}{2m} \frac{\partial^2}{\partial x^2}+\frac{1}{2}m\omega^2x^2$ are defined by:
\begin{equation}
\phi_j(x/a_0)=\frac{e^{-(x/a_0)^2/2}H_j(x/a_0)}{\sqrt{\sqrt{\pi}2^jj!}},
\end{equation}
where $a_0=\sqrt{\hbar/m\omega}$ is the harmonic oscillator length and $H_j$ is the $j$th Hermite polynomial. Using a Vandermonde determinant formula and Eq.~\eqref{bfmapping}, Forrester \textit{et al.} deduced the expression for the ground-state 
of the harmonically trapped Tonks-Girardeau gas:
\begin{equation}
\psi_B^0(x_1,\ldots,x_N)=\frac{1}{a_0^{N/2}\sqrt{N!\prod_{m=0}^{N-1}2^{-m}\sqrt{\pi}m!}}\prod_{k=1}^Ne^{-(x_k/a_0)^2/2}\prod_{1\le j<k\le N}|x_j-x_k|.
\end{equation}

The Bose-Fermi mapping can be further exploited  by noticing than $|\psi_B|=|\psi_F|$. Thus, the density profiles $n(x)$ defined by
\begin{equation}
n(x)=N\int dx_2\cdots dx_N |\psi(x,x_2,\ldots,x_N)|^2
\end{equation}
and measuring the probability (normalized to $N$) of finding a particle at a point $x$ are the same for the Tonks-Girardeau gas and spinless fermions. More explicitely, the density profile in the harmonic trap is given by \cite{Vignolo2000,Kolomeisky2000}:
\begin{equation}
\label{spinlessdensprof}
n_F(x)=\frac{1}{\sqrt{\pi a_0}}\sum_{k=0}^{N-1}\frac{1}{2^kk!}H_k^2(x/a_0)e^{-(x/a_0)^2}. 
\end{equation}
This analogy between hardcore bosons and non-interacting fermions is also true, for example, for the pair distribution functions defined by
\begin{equation}
D(x,y)=N(N-1)\int dx_3\cdots dx_N |\psi(x,y,x_3,\ldots,x_N)|^2,
\end{equation}
which measures the joint probability (normalized to $N(N-1)$) of finding one atom in $x$ and another in $y$ \cite{Vignolo2001}.

However, if we consider the off-diagonal correlations or the momentum distributions, the symmetry plays an important role, and the analogy is no longer true. 
The one-body correlations will be discussed in details in chapter \ref{chap:1bcor}.

\subsection{Strongly interacting multi-component systems in atomic traps}
\label{secvolosniev}

Here we generalize the previous considerations and turn to the main Hamiltonian that we studied in this thesis.

\subsubsection{Model}

We consider a strongly interacting system of $N$ particles divided in $\kappa$ different spin species 
with populations $N_1,\ldots,N_{\kappa}$. The 
species can be either fermions or bosons. We impose that all the particles:
\begin{enumerate}
\item Have the same mass $m$;
\item Interact via a $\delta$-type potential of same strength $g_{1D}$;
\item Are submitted to the same external potential $V_{ext}(x)=\frac{1}{2}m\omega^2x^2$.
\end{enumerate}

In the case of a fermionic mixture, note that Florence's experiment with $\Yb$ atoms \cite{Pagano2014} described in section \ref{florexp} fulfills these 
conditions, with $\kappa\in\{2,\ldots,6\}$ and $N_1=\cdots=N_{\kappa}\simeq10^4$. In the case of Bose-Fermi mixtures, these assumptions are demanding, but can 
however be considered as good approximations in the case of isotopes, as realized in \cite{Fukuhura2009} with an $\Yb-{}^{174}\mathrm{Yb}$ mixture.

The stationary Schr\"{o}dinger equation for this system is given by:
\begin{equation}
\label{maineq}
\left[\sum_{j=1}^N\left(-\frac{\hbar^2}{2m} \frac{\partial^2}{\partial x_j^2}+\frac{1}{2}m\omega^2x_j^2\right)+g_{1D}\sum_{i<j}\delta(x_i-x_j)-E\right]\psi=0.
\end{equation}
As stated in previous sections (see e.g. section \ref{secusp}), it is not necessary to specify the mixture in this equation, since the Pauli principle for identical 
fermions is naturally obtained by Eq.~\eqref{cusp2b}, that we recall here for any couple of particles $i,j$:
\begin{equation}
\label{maincusp}
\left.\left(\frac{\partial\psi}{\partial x_i}-\frac{\partial\psi}{\partial x_j}\right)\right|_{x_{ij}=0^+}-
\left.\left(\frac{\partial\psi}{\partial x_{i}}-\frac{\partial\psi}{\partial x_{j}}\right)\right|_{x_{ij}=0^-}=\frac{mg_{1D}}{\hbar^2}\left.\psi\right|_{x_{ij}=0},
\end{equation}
with $x_{ij}=x_{i}-x_{j}$. Eq.~\eqref{maineq} is the main equation studied in this thesis.

As in the Tonks-Girardeau gas of section \ref{tggas}, in the impenetrable limit the system will fermionize, since the $g_{1D}\to\infty$ limit in the cusp 
condition \eqref{maincusp} implies that $\psi=0$ whenever two particles are at the same point. The system can thus be again mapped onto a spinless fermionic gas. 
In particular, the Slater determinant $\psi_F$ of Eq.~\eqref{slaterdet} has the right nodes and energy $E_0$ as the ground state. Since all the systems have the same 
eigenspectrum as the free Fermi gas regardless of their composition and symmetry when $g_{1D}=\infty$, these points of the energy spectrum are often called 
\textit{degenerate manifolds} \cite{Harshman2014}. The question is, if we consider that $g_{1D}$ is in the vicinity of the $g_{1D}=\infty$ point, how will this degeneracy  be removed as a function 
of the mixture's symmetry?

\subsubsection{A perturbative ansatz}
\label{volansatz}

\paragraph{Method}

The method developped in \cite{Volosniev2013,Volosniev2014} allows to answer last paragraph's question. Although it works for any confining potential $V_{ext}$, 
I focus here on the harmonic case. We switch to natural units of $a_0=\sqrt{\hbar/m\omega}$ for length and $\hbar\omega$ for energy.

The first idea is to use the aforementioned fact that the Slater determinant $\psi_F$ has the right nodes in the $g_{1D}\to\infty$ limit. Then, in order to 
have a solution $\psi$ that respects the symmetry of the considered mixture, Volosniev \textit{et al.} proposed the following ansatz:
\begin{equation}
\label{Psivolo}
\psi(x_1,\ldots,x_N)=\sum_{P\in \mathfrak S_N}a_P\theta(x_{P1}<\cdots<  x_{PN})\psi_F(x_1,\ldots,x_N),
\end{equation}
where $\mathfrak S_N$ is the permutation group of $\{1,\dots,N\}$ (see section \ref{subsecpermut}) and $\theta(x_1<\cdots<  x_N)$, the indicator function of the sector $\{x_1<\cdots<  x_N\}\subset\mathbb{R}^N$, is equal to 1 is $x_1<\cdots<  x_N$ and 0 otherwise.
This ansatz is very much in the spirit of the Bethe ansatz (cf Eq.~\eqref{bethe}), except that the asymptotic plane wave basis in the last one is replaced with 
the fermionic Slater determinant here. 

Because of the (anti-)symmetrization constraint imposed for identical (fermions) bosons, the choice of a given mixture 
$N_1,\dots,N_{\kappa}$ will reduce the number of independent $a_P$ coefficient in Eq~\eqref{Psivolo} to the multinomial coefficient:
\begin{equation}
\label{degeneracy}
D_{N_1,\dots,N_{\kappa}}={N \choose N_1, N_2, \ldots, N_{\kappa}}=\frac{N!}{N_1!N_2!\cdots N_{\kappa}!}.
\end{equation}
The number $D_{N_1,\dots,N_{\kappa}}$ corresponds to the number of linearly independent states that can be written in terms of $\psi_F$. Since $\psi_F$ is 
associated with the ground state energy $E_0$, $D_{N_1,\dots,N_{\kappa}}$ is in fact the dimension of the degenerate manifold at $g_{1D}=\infty$. Instead 
of writing the solution in one of the $N!$ sectors of $\mathbb{R}^N$, this observation encourages to write the solution in one of the $D_{N_1,\dots,N_{\kappa}}$ 
so-called \textit{snippets} \cite{Deuretzbacher,Fang2011} of $\text{Sect}(\mathbb{R}^N)/\mathcal{R}$, where $\text{Sect}(\mathbb{R}^N)/\mathcal{R}$ is the quotient set of all sectors of $\mathbb{R}^N$ by the equivalence relation 
$\mathcal{R}$: 'two sectors are equivalent if they are equal up to permutations of \textit{identical} particles'. This operation considerably reduces 
the dimension of the problem.

In order to determine these $a_P$ coefficients, the idea is to remove the degeneracy by analyzing the system in the vicinity of the degenerate point $g_{1D}=\infty$. 
In other words, one has to consider a linear perturbation in the energy around $1/g_{1D}=0$:
\begin{equation}
\label{perten}
E(1/g_{1D})=_{+\infty}E_0+\frac{1}{g_{1D}}(-K)+o(1/g_{1D}).
\end{equation}
The \textit{energy slope} $K=-\lim_{g_{1D}\to\infty}\frac{\partial E}{\partial g_{1D}^{-1}}=g_{1D}^2\frac{\partial E}{\partial g_{1D}}$ will be different 
for different states of the degenerate manifold, and  is then a functionnal of the $a_P$ coefficients. Using the Hellmann-Feynman theorem \cite{Feynman1939} on the Hamiltonian 
of Eq.~\eqref{maineq}, one gets:
\begin{equation}
K=\lim_{g_{1D}\to\infty}g_{1D}^2\frac{\displaystyle{\sum}_{i<j}\displaystyle{\int} dx_1\ldots dx_N~\delta(x_i-x_j)|\psi|^2}{\displaystyle{\int} dx_1\ldots dx_N~|\psi|^2},
\end{equation}
where the $g_{1D}$-dependence is removed by the cusp condition \eqref{maincusp}, yielding
\begin{equation}
K=\frac{1}{4}\frac{\displaystyle{\sum_{i<j}\int} dx_1\ldots dx_N~\delta(x_i-x_j)\left|\left(\left.\left(\frac{\partial}{\partial x_i}-\frac{\partial}{\partial x_j}\right)\right|_{x_{ij}=0^+}-
\left.\left(\frac{\partial}{\partial x_{i}}-\frac{\partial}{\partial x_{j}}\right)\right|_{x_{ij}=0^-}\right)\psi\right|^2}{\displaystyle{\int} dx_1\ldots dx_N~|\psi|^2},
\end{equation}
with $x_{ij}=x_i-x_j$. Then, using Eq.~\eqref{Psivolo} and normalizing $\psi$ to unity, one gets:
\begin{equation}
\label{kvolo}
K=\sum_{P,Q\in\mathfrak S_N}(a_P-a_Q)^2\alpha_{P,Q},
\end{equation}
where $\alpha_{P,Q}$ is defined by
\begin{equation}
\label{alphavolo}
\alpha_{P,Q}\equiv\alpha_k=\int dx_1\ldots dx_N~\theta(x_1<\cdots<x_N)\delta(x_k-x_{k+1})\left|\frac{\partial\psi_F}{\partial x_k}\right|^2
\end{equation}
if $P=Q(k,k+1)$ (that is to say $P$ and $Q$ are equal up to a transposition in $k$ and $k+1$), where the particles in $k$ and $k+1$ are not identical fermions, and $\alpha_{P,Q}=0$ otherwise. These $\alpha_k$ coefficients can be seen as the energy 
cost of an exchange of particles at positions $k$ and $k+1$ in the trap, and are sometimes called \textit{nearest-neighbor exchange constants} \cite{Deuretzbacher2014,Laird2017}. Note that the parity invariance of the harmonic potential implies that 
$\alpha_k=\alpha_{N-k}$, reducing the number of different $\alpha_k$ coefficients to $\lfloor N/2\rfloor$.
As seen in section \ref{tggas}, expression for $\psi_F$ is given by \cite{Girardeau2001,Forrester2003}:
\begin{equation}
\label{psivander}
\psi_F(x_1,\ldots,x_N)=\frac{1}{\sqrt{N!\prod_{m=0}^{N-1}2^{-m}\sqrt{\pi}m!}}\prod_{k=1}^Ne^{-x_k^2/2}\prod_{1\le j<k\le N}(x_j-x_k).
\end{equation}
We then have after some algebra
\begin{equation}
\left[\frac{\partial\psi_F}{\partial x_k}\right]_{x_k=x_{k+1}}=\frac{1}{\sqrt{N!\prod_{m=0}^{N-1}2^{-m}\sqrt{\pi}m!}}e^{-x_k^2}\prod_{\substack{i=1 \\ i\neq k,k+1 }}^Ne^{-x_i^2/2}(x_i-x_k)^2\prod_{\substack{1\le j<\ell \le N \\j,l\neq k,k+1 }}(x_j-x_{\ell}),
\end{equation}
and thus, using a Vandermonde formula:
\begin{equation}
\left[\frac{\partial\psi_F}{\partial x_k}\right]_{x_k=x_{k+1}}^2=\frac{2^{2N-3}e^{-2x_k^2}}{\pi N!(N-1)!(N-2)!}\left[\det    \left[ \phi_{i-1}(x_j)\right]\right]^2_{N-2\times N-2}
\prod_{\substack{i=1 \\ i\neq k,k+1 }}^N(x_i-x_k)^4,
\end{equation}
We then make use of permutational and parity invariances, and obtain:
\begin{equation}
\label{alphak}
\begin{split}
\alpha_k=&\frac{2^{2N-3}}{\pi N!(N-1)!(N-2)!}\frac{1}{(k-1)!(N-k-1)!}\int_{-\infty}^{+\infty}dx_ke^{-2x_k^2}\sum_{P,Q\in \mathfrak S_{N-2}}\epsilon (P) \epsilon (Q)\\
&\times\prod_{\substack{i=1 \\ i\neq k,k+1}}^N\int_{L_k(i)}^{U_k(i)}dx_i(x_i-x_k)^4\phi_{Pi-1}(x_i)\phi_{Qi-1}(x_i),\\
\end{split}
\end{equation}
where 
\begin{equation}
(L_k(i),U_k(i))=\left\{\begin{array}{ll}
 (-\infty,x_k) & \mbox{if } i<k \\ (x_k,+\infty) & \mbox{if } i\ge k
\end{array}
\right. .
\end{equation}

Eq.~\eqref{alphak} was obtained in \cite{Decamp2016}. 
Alternatively, it is possible to obtain an approximate value for $\alpha_k$ by performing a local density approximation on the homogeneous results with 
periodic boundary counditions \cite{Matveev2008,Deuretzbacher2014}.

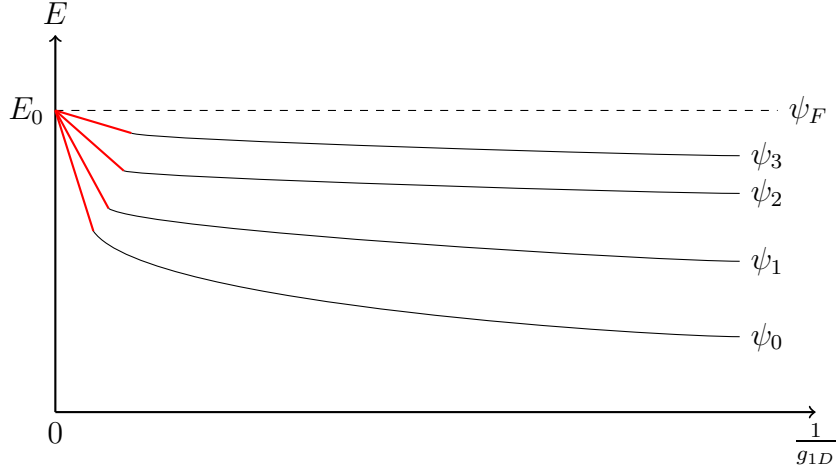
\begin{figure}\centering
\begin{tikzpicture}
\draw[thick,->] (0,0) -- (10,0);
\draw[dashed] (0,4) -- (9.5,4);
\draw[thick,->] (0,0) -- (0,5);
\draw (0,0) node[below]{$0$};
\draw (10,0) node[below]{$\frac{1}{g_{1D}}$};
\draw (0,4) node[left]{$E_0$};
\draw (9.5,4) node[right]{$\psi_F$};
\draw (0,5) node[above]{$E$};
\draw[thick,red] (0,4) -- (1,3.7);
\draw[thick,red] (0,4) -- (0.9,3.2);
\draw[thick,red] (0,4) -- (0.7,2.7);
\draw[thick,red] (0,4) -- (0.5,2.4);
\draw (1,3.7) .. controls (1.2,3.6) and (8,3.4) .. (9,3.4);
\draw (9,3.4) node[right]{$\psi_3$};
\draw (0.9,3.2) .. controls (1.2,3.1) and (8,2.9) .. (9,2.9);
\draw (9,2.9) node[right]{$\psi_2$};
\draw (0.7,2.7) .. controls (1.2,2.4) and (8,2) .. (9,2);
\draw (9,2) node[right]{$\psi_1$};
\draw (0.5,2.4) .. controls (1.2,1.4) and (8,1) .. (9,1);
\draw (9,1) node[right]{$\psi_0$};

\end{tikzpicture}

\caption{\label{figvolo}Graphic interpretation of the perturbative ansatz. The red lines correspond to the linear approximations of 
Eq.~\eqref{perten}, whose slopes are given by $-K$.}
\end{figure}

Once the $\alpha$ exchange constants computed in Eq.~\eqref{kvolo}, the next step is to determine the $a_P$ coefficients corresponding to each one of the states 
in the degenerated manifold. To do so, the idea is to notice that the lowest energy will be obtained by maximizing the energy slope functional $K\left[\{a_P\}\right]$ (because of the minus sign 
in Eq.~\eqref{perten}). Subsequently, solving $\partial K/\partial a_P=0$ for all $P\in\mathfrak S_N$ yields the following diagonalization problem:
\begin{equation}
\label{diagkeq}
V\vec{a}=K\vec{a},
\end{equation}
where $\vec{a}$ is the vector of the $D_{N_1,\dots,N_{\kappa}}$ independent $a_P$ coefficients and $V$ is a $D_{N_1,\dots,N_{\kappa}}\times D_{N_1,\dots,N_{\kappa}}$ matrix 
defined in the snippet basis by
\begin{equation}
\label{matrixvolo}
V_{ij}=\left\{
\begin{array}{ll}
 -\alpha_{i,j} & \mbox{if } i\ne j \\ \sum_{\mathrm{d},k\ne i}\alpha_{i,k}+2\sum_{\mathrm{b},k\ne i}\alpha_{i,k}& \mbox{if  } i=j
\end{array}
\right.,
\end{equation}
where the index $\mathrm{d}$ means that the sum has to be taken over snippets $k$ that transpose distinguishable particles as compared to snippet $i$, 
while $\mathrm{b}$ means that the sum is taken over sectors that transpose identical bosons. If the system contains only fermions, it reduces to the first sum. The parity invariance of the harmonic trap implies that $V$ is symmetric, and thus that it can be diagonalized in an orthogonal basis. 
Then, the highest eigenvalue $K_0$ and the corresponding eigenvector $\vec{a}_0$ determines the ground-state wave function $\psi_0$ through Eq.~\eqref{Psivolo}. Incidentally, all the eigenstates corresponding to the same 
degenerate manifold at $g_{1D}=\infty$ are given by the orthogonal eigenvectors of $V$.

I provide an illustration of the method in the next paragraph. It will be further implemented in the next chapters. A graphical illustration of the method is given in Fig. \ref{figvolo}.

\paragraph{A first example}

As a first pedagogical example, let us consider the case of a four-particle fermionic mixtures (two spin up, two spin down), as it is done in \cite{Volosniev2014}.
We will use this example in the next paragraph, where we give an interpretation of the method in terms of graph theory.
The number of snippets, i.e. the dimension of the degenerate manifold, is given by $D_{2,2}=4!/(2!2!)=6$. Explicitly, they are given by:
\begin{equation}
\begin{split}
&a_1\text{ : }\uparrow\uparrow\downarrow\downarrow\\
&a_2\text{ : }\uparrow\downarrow\uparrow\downarrow\\
&a_3\text{ : }\uparrow\downarrow\downarrow\uparrow\\
&a_4\text{ : }\downarrow\uparrow\uparrow\downarrow\\
&a_5\text{ : }\downarrow\uparrow\downarrow\uparrow\\
&a_6\text{ : }\downarrow\downarrow\uparrow\uparrow
\end{split}
\end{equation}
where we have written the six corresponding independent $a_P$ coefficients. For each configuration $i\in\{1,\ldots,6\}$ of the particles, the wave function 
will be then given by $\psi|_{\{i\}}=a_i\psi_F$ according to Eq.~\eqref{Psivolo}. The energy slope $K$ defined in Eq.~\eqref{kvolo} is then:
\begin{equation}
K=\alpha_2(a_1-a_2)^2+\alpha_3(a_2-a_3)^2+\alpha_1(a_2-a_4)^2+\alpha_1(a_3-a_5)^2+\alpha_3(a_4-a_5)^2+\alpha_2(a_5-a_6)^2,
\end{equation}
where the exchange coefficients defined in Eq.~\eqref{alphavolo} are:
\begin{equation}
\alpha_1=\alpha_3=\int_{x_1<x_2<x_3<x_4}dx_1dx_2dx_3dx_4\delta(x_1-x_2)\left|\frac{\partial\psi_F}{\partial x_1}\right|^2
\end{equation}
and
\begin{equation}
\alpha_2=\int_{x_1<x_2<x_3<x_4}dx_1dx_2dx_3dx_4\delta(x_2-x_3)\left|\frac{\partial\psi_F}{\partial x_2}\right|^2.
\end{equation}
They can be computed exactly using Eq.~\eqref{alphak}. The $V$ matrix that has to be diagonalized in order to obtain the ground and excited states (Eq.~\eqref{matrixvolo}) is:
\begin{equation}
\label{voloexa}
V=\begin{pmatrix}
\alpha_2 & -\alpha_2 & 0 & 0 & 0 & 0\\
\alpha_2 & 2\alpha_1+\alpha_2 & -\alpha_1 & -\alpha_1 & 0 & 0\\
0 & -\alpha_1 & 2\alpha_1 & 0 & -\alpha_1 & 0\\
0 & -\alpha_1 & 0 & 2\alpha_1 & -\alpha_1 & 0\\
0 & 0 &  -\alpha_1 & -\alpha_1 & 2\alpha_1+\alpha_2 & \alpha_2\\
0 & 0 & 0 & 0 & -\alpha_2 & \alpha_2 \\
\end{pmatrix}.
\end{equation}
The ground state energy slope $K_0$, given by the highest eigenvalue of $V$, 
is given by:
\begin{equation}
K_0=2\alpha_1+\alpha_2+\sqrt{4\alpha_1^2-2\alpha_1\alpha_2+\alpha_2^2}.
\end{equation}
The lower eigenspectrum of $V$ determines the energy slopes of the excited states.

\begin{figure}\centering
\begin{tikzpicture}
\draw (2,1) circle(1);
\draw (2,1) node{$\downarrow\downarrow\uparrow\uparrow$};
\draw (-1,3) circle(1);
\draw (-1,3) node{$\downarrow\uparrow\uparrow\downarrow$};
\draw (-1,6) circle(1);
\draw (-1,6) node{$\uparrow\downarrow\uparrow\downarrow$};
\draw (2,8) circle(1);
\draw (2,8) node{$\uparrow\uparrow\downarrow\downarrow$};
\draw (5,3) circle(1);
\draw (5,3) node{$\downarrow\uparrow\downarrow\uparrow$};
\draw (5,6) circle(1);
\draw (5,6) node{$\uparrow\downarrow\downarrow\uparrow$};
\draw[thick] (0,3) -- (4,3); 
\draw[thick] (-1,4) -- (-1,5); 
\draw[thick] (0,6) -- (4,6); 
\draw[thick] (5,4) -- (5,5); 
\draw[thick] (-0.16795,6.5547) -- (1.16795,7.4453); 
\draw[thick] (2.83205,1.5547) -- (4.16795,2.4453); 
\draw (2,3) node[above]{$\alpha_1$};
\draw (2,6) node[below]{$\alpha_1$};
\draw (-1,4.5) node[right]{$\alpha_1$};
\draw (5,4.5) node[left]{$\alpha_1$};
\draw (3,2) node[above]{$\alpha_2$};
\draw (1,7) node[below]{$\alpha_2$};
\end{tikzpicture}

\caption{\label{figgraphvolo}Weighted graph $\mathcal{G}_V$ associated with the $(2+2)$ fermionic example of the text. The laplacian matrix associated to this graph 
is exactly equal to $V$ of Eq.~\eqref{voloexa}.}
\end{figure}
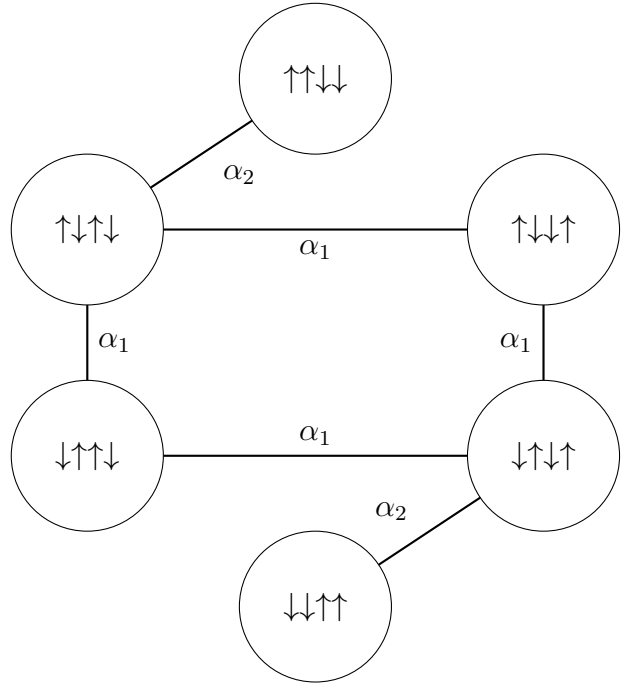

\paragraph{Graph theory interpretation}

In this paragraph, we give an interpretation of the perturbative ansatz in terms of graph theory. Although this observation has not been published yet,
I strongly believe this analogy can provide useful results about $V$'s eigenspectrum.

Let us first recall some basic definitions of graph theory \cite{Bondy2008}. A \textit{graph} 
is a pair $\mathcal{G}=(\mathcal{V},\mathcal{E})$ where $\mathcal{V}$ is a set of \textit{vertices} (or \textit{points}) and $\mathcal{E}$ is a set of \textit{edges} (or \textit{lines}) which are unordered pairs of elements of $\mathcal{V}$.
The \textit{degree} $\mathrm{deg}(v)$ of a vertex $v$ is the number of vertices that are connected by an edge to $v$.
If each element of  $\mathcal{E}$ is associated with a number, we say that $\mathcal{G}$ is a \textit{weighted graph}. 

Given a graph $\mathcal{G}$, it is possible to associate matrices to $\mathcal{G}$, allowing to use linear algebra tools in order to analyze $\mathcal{G}$. The simplest matrix that we can define 
is the \textit{adjacency matrix} $A$, whose elements $A_{ij}$ are equal to $1$ if vertex "$i$" is connected by an edge (or \textit{adjacent}) to vertex "$j$"  and $0$ otherwise. Another interesting matrix is the \textit{Laplacian matrix}
$L$ whose elements are given by:
\begin{equation}
\label{lapmat}
L_{ij}=\left\{
\begin{array}{lll}
 \mathrm{deg}(v_i) & \mbox{if } i=j \\ -1 & \mbox{if  } i\ne j \mbox{ and $v_i$ is adjacent to $v_j$}\\ 0 & \mbox{otherwise}
\end{array}
\right..
\end{equation}
In the case of a weighted graph, $L$ is defined similarly, where $\mathrm{deg}(v_i)$ is the sum of the weights of the edges connected to $v_i$, and the off-diagonal terms in -1 are replaced by minus the weight of the edge between $i$ and $j$. 
This matrix is extremely important in graph theory, and can be seen as a discrete version of the laplacian operator $\Delta$. In particular, the spectral properties of the laplacian matrices are well studied \cite{Brouwer2012}.

Let us now go back to our physical system and apply the aforementioned definitions. Given a mixture $N_1,\dots,N_{\kappa}$, we define a weighted graph $\mathcal{G}_V$ 
where each vertex is associated with a snippet, and where two snippets are adjacent with weight $\alpha_k$ if they are 
equal up to a transposition of particles in positions $k$ and $k+1$. Then, the laplacian matrix $L$ of $\mathcal{G}_V$ is equal to the matrix $V$ defined in Eq.~\eqref{matrixvolo}. 
An illustration based on the last paragraph's example is given in Fig. \ref{figgraphvolo}.

\paragraph{Implementation}

I have developed a \texttt{Mathematica} program that allows to obtain the $V$ matrix for any kind of mixture (fermionic, bosonic or mixed), 
and for any number of particles $N$. In practice, we have chosen to study in detail the case $N=6$, since the program's complexity increases extremely rapidly 
with increasing $N$ (as $\mathcal{O}(N!^2)$), and, as we will see, the case $N=6$ already allows to observe effects that are not present when $N\le 5$.

\begin{figure}
\begin{center}
  \includegraphics[width=0.6\linewidth]{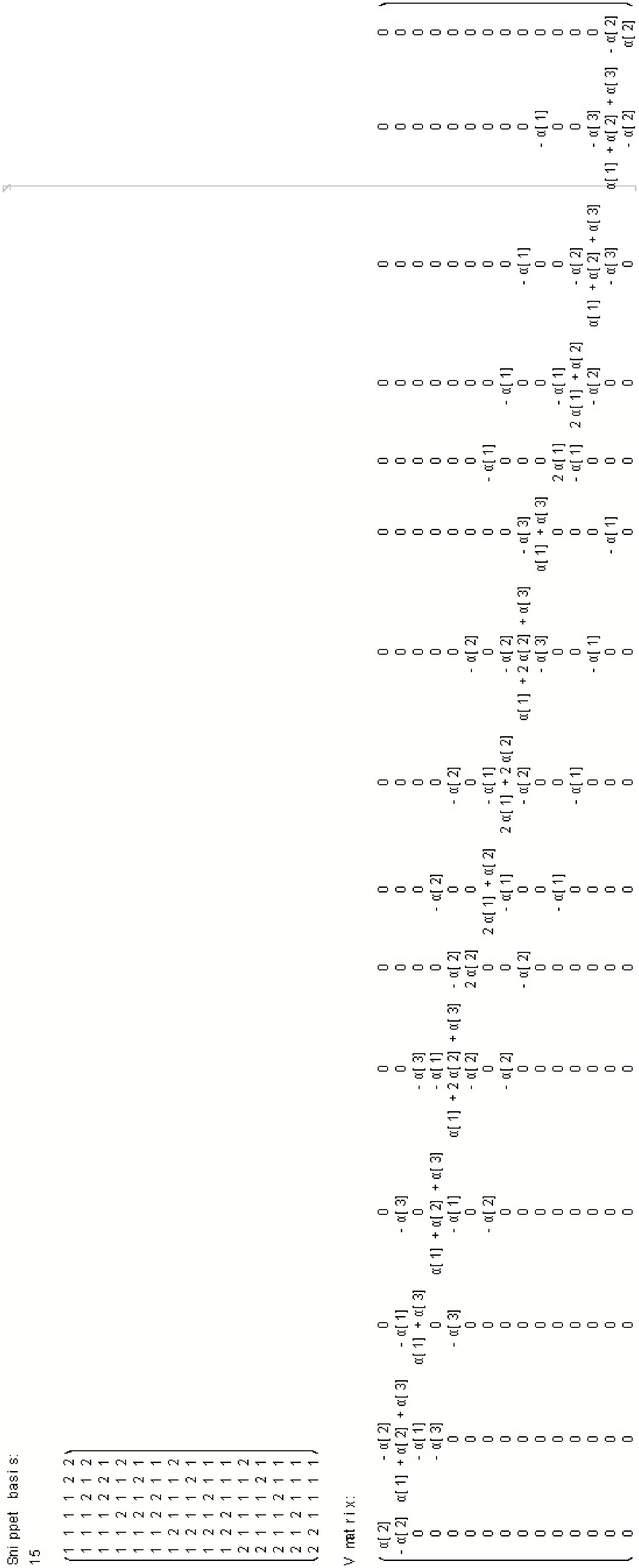}
  \caption{Output obtained from our \texttt{Mathematica} program in order to compute the $V$ matrix (Eq.~\eqref{voloexa}) associated with the 
  fermionic mixture $\uparrow\uparrow\uparrow\uparrow\downarrow\downarrow$: it returns the snippet basis and its length, and then the $V$ matrix. For $N=6$ 
  mixtures, the program typically runs for a few seconds, but it grows very rapidly for larger $N$'s. 
  \label{fig:exvol}}
  \end{center}
\end{figure}

For example, let us add two spin-up to our last example, so that we have a two-component fermionic mixture of the type 
$\uparrow\uparrow\uparrow\uparrow\downarrow\downarrow$. In order to compute the $V$ matrix associated with this quantum mixture (Eq.~\eqref{voloexa}), one just need to assign a number 
to each component, \textit{e.g.} $\uparrow=1$ and $\downarrow=2$, and to enter the following vector in the program: $\{1,1,1,1,2,2\}$. The program then directly 
computes the snippet basis and the $V$ matrix as a function of the exchange constants $\alpha_k$. The output given by the program for this example is shown 
in Fig.~\ref{fig:exvol}.

\clearemptydoublepage

\clearemptydoublepage
\pagestyle{fancy}
\thispagestyle{empty}
\chapter{Symmetry analysis}
\label{chap:sym}

\minitoc
\newpage

In chapter \ref{Exactsol}, we have seen that we can express exact solutions for strongly repulsive one-dimensional trapped quantum mixtures by mean of a summation over the permutations of all particles. It appears clearly that the permutation symmetry is of fundamental importance in our system. This chapter is  devoted to the precise symmetry analysis of the exact solutions obtained in the last chapter.

Section \ref{sec:gensym} is a very general introduction to the concept of symmetry, with a peculiar focus on the permutational symmetry we are interested in and the mathematical properties of the associated group, namely the \textit{symmetric group}. In section \ref{secsyman}, we will explain the method we used in order to characterize the symmetry of our solutions, with the will to justify it as precisely as possible in terms of group theory. We will then implement it for mixtures of $N=6$ particles, and analyze the results we can extract from it.

\section{Generalities}
\label{sec:gensym}

The concept of symmetry is one of the cornerstones of modern physics \cite{Weyl1952,Gross1996,Gieres1997}. It is first an extremely valuable tool in order to solve a physical problem: indeed, the fact 
that a classical or quantum Hamiltonian has a certain symmetry gives considerable information about the structure of the solution itself. Furthermore, the concept of 
symmetry appears to be so powerful and fundamental in Nature that many modern successful theories have been constructed almost only from symmetry considerations.
In this section, we first discuss some very general aspects of symmetries in \ref{syminp}. Then, we focus on the peculiar fundamental symmetry we are interested in, namely the
exchange symmetry, in \ref{secexcsym}.

\subsection{Symmetries in physics}
\label{syminp}

\subsubsection{What is symmetry?}

\begin{figure}
\begin{center}
  \includegraphics[width=0.3\linewidth]{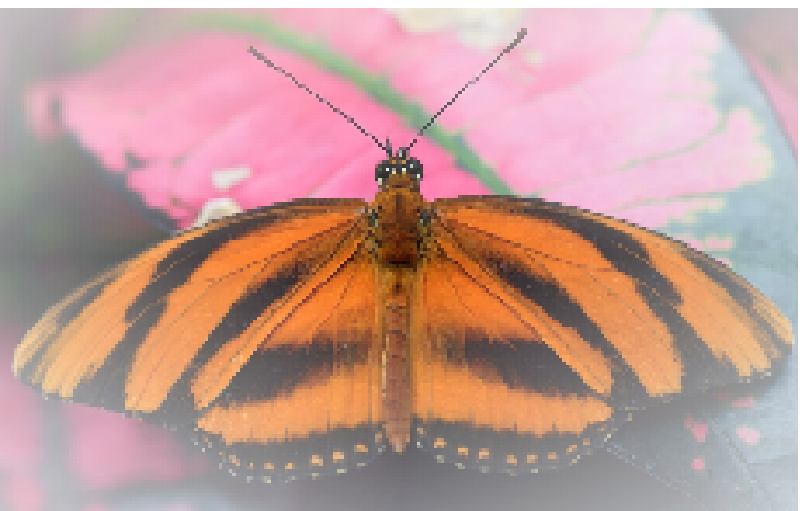}
  \includegraphics[width=0.317\linewidth]{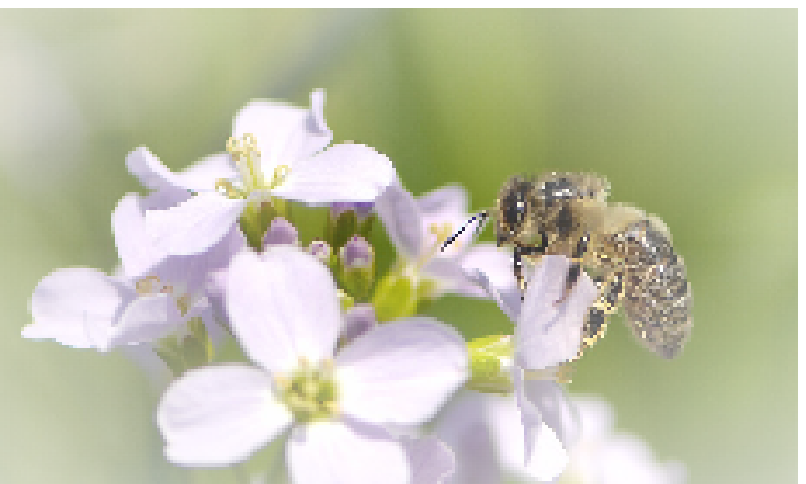}
  \includegraphics[width=0.283\linewidth]{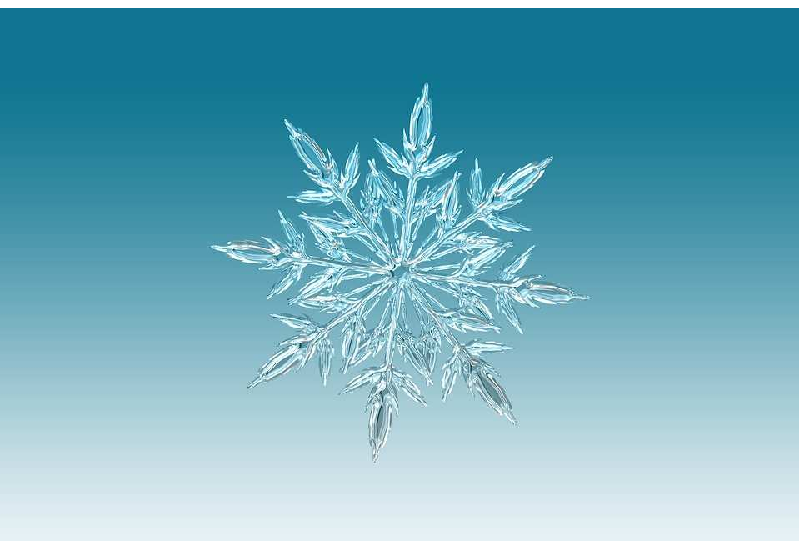} 
  \caption{Some examples of the symmetries encountered in Nature.
  \label{fig:exsym}}
  \end{center}
\end{figure}

The notion of symmetry is very intuitive, as it is omnipresent in everyday life: human beings and animals are left-right symmetric, flowers have a \textit{discrete} rotational 
symmetry, the sun has a \textit{continuous} rotational symmetry, snowflakes have very complex symmetry structure, and so on (see Fig.~\ref{fig:exsym}). Interestingly, on a more metaphysical 
point of view, it seems that  "\textit{Beauty} is bound up with symmetry" \cite{Weyl1952}.

\begin{figure}\centering
\begin{tikzpicture}
\draw[thick] (0,0) -- (2,0) -- (2,2) -- (1,2.5) -- (0,2) -- (0,0);
\draw[thick] (4,0) -- (7,0) -- (5.5,2.598) -- (4,0);
\draw[thick] (10,1.3) circle(1.3);

\end{tikzpicture}
\caption{Some examples of geometrical symmetries. From left to right, their symmetry group are respectively the \textit{discrete} diedral groups $D_2\simeq\mathbb{Z}_2$ and 
$D_3\simeq\mathbb{Z}_3\rtimes\mathbb{Z}_2$ of reflections and discrete rotations and the \textit{continuous} orthogonal group $O(2)$. The groups are "bigger and bigger" 
from left to right, which is equivalent to the fact that the objects are more and more symmetric.
  \label{fig:exsymgeo}}
\end{figure}
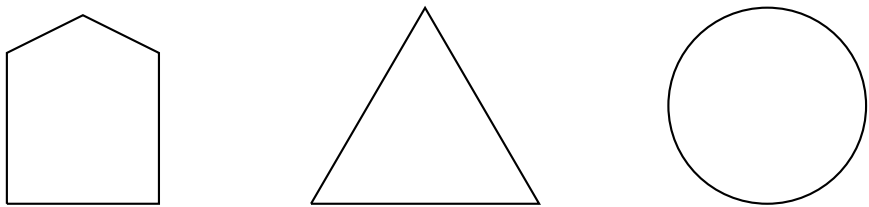

But how to precisely define the concept of symmetry? Roughly speaking, we can say that an object has a symmetry when you can apply some 
transformations to it which leaves the system invariant. More formally, this set of transformations is a group $G$, the \textit{symmetry group} of the system, and 
this group acts on the system through the mathematical notion of linear representation $\hat{D}_G$ \cite{Hamermesh_book}. The "bigger" is the group, the more symmetric is the object. Some examples of symmetric geometrical 
objects and their associated symmetry groups are given in Fig~\ref{fig:exsymgeo}. These geometrical symmetries are found in physics, for example in crystallography, where the crystals are defined 
by their discrete spatial periodicity and classified by their so-called \textit{space group symmetry} \cite{cryst}. 

\begin{figure}
\begin{center}
  \includegraphics[width=0.4\linewidth]{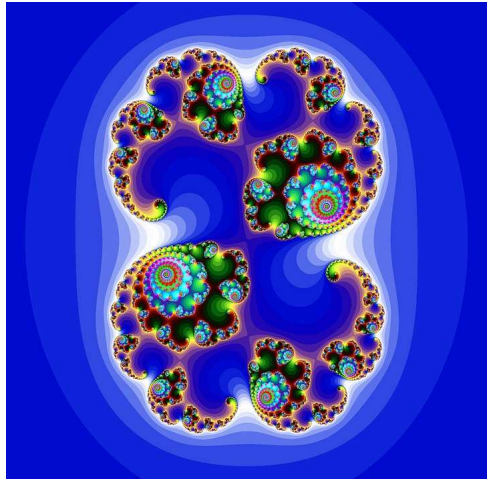}
  \caption{A fractal, characterized by its auto-similarity, or discrete scale invariance. Fractal theory is intimately linked to 
  the theory of critical phenomena and quantum field theories through the famous \textit{renormalization (semi-)group}.
  \label{fig:fractal}}
  \end{center}
\end{figure}

In a more fundamental way, a physical law has a symmetry when the equation describing the law, hence 
the Hamiltonian or equivalently the Lagrangian or the action, is invariant by a change of variables. In this case, the word \textit{covariant} is often used. These variables can be space-time coordinates: \textit{e.g}., if we require that a physical law is subjected to the 
relativity principle \cite{Einstein1905}, it implies that the Lagrangian must be invariant by the Lorentz group, that is the group of Lorentz transformations or equivalently the group $O(1,3)$ of isometries of Minkowski 
space. Another symmetry involving a geometrical transformation is given by the \textit{conformal group}, the group of transformations that locally conserve the angles. It involves, in particular, dilatations. Scale invariance, which 
is related to the mathematical notion of \textit{fractals}, 
is found in many theories such as critical phenomena and quantum field theories through the notion of \textit{renormalization} \cite{Wilson1975} (see Fig~\ref{fig:fractal}).

These symmetries can be more abstract in the sense that they do not affect the space-time coordinates, in the case of the so-called \textit{internal} symmetries. The most simple example of internal symmetry 
is the invariance of the electrodynamic model by a global phase-change $U(1)$. Another symmetry of peculiar importance in this thesis, that we will discuss more below, is 
the \textit{permutational} symmetry, namely the symmetry that exchanges the states of the particles in a many-body problem.

\subsubsection{Importance in physics}

\label{importancesym}

In classical physics, already, the symmetry properties of a Hamiltonian have extremely strong implications. For example, if we know that a classical Hamiltonian $H$
is invariant under rotation transformations and we have obtained a certain solution $s(t)$, we can deduce other solutions noticing that the spatially rotated $s(t)$ is also a solution. More profoundly, the notion of  
symmetry is intimately linked with the notion of \textit{conservation laws} through the celebrated Noether's theorem \cite{Noether1918}, which states that any continuous symmetry 
is associated with a conserved quantity. Some examples of this powerful correspondence are given in table \ref{tab:noether}.

In quantum physics, the importance of symmetries is even higher \cite{Landau1981}. The intrinsic linearity of quantum physics related to the superimposition 
principle implies that a lot of information can be obtained from representation theory. More precisely, suppose that a quantum Hamiltonian $\hat{H}$ has a certain symmetry group $G$,
\textit{i.e.} that, for every $g\in G$, 
\begin{equation}
\label{hamcomut}
\left[\hat{H},\hat{D}(g)\right]=\hat{H}\hat{D}(g)-\hat{D}(g)\hat{H}=0,
\end{equation}
where the representation $\hat{D}$ acts on the Hilbert space $V$ of the systems. Then, it follows from Maschke's theorem that the representation can be split\footnote{In generic cases.} as a direct sum 
of \textit{irreducible representations}, or \textit{irreps}, \textit{i.e.} non-zero representations that have no proper sub-representations:
\begin{equation}
\label{splitrep}
V=\bigoplus_k V^{(k)},\quad \hat{D}(g)=\bigoplus_k \hat{D}^{(k)}(g),\quad \hat{D}(g)V^{(k)}\subset V^{(k)}.
\end{equation}
Given Eqs.~\eqref{hamcomut} and \eqref{splitrep}, it follows that $\hat{H}$ can also be split according to the irreps of $G$:
\begin{equation}
\label{splith}
\hat{H}=\bigoplus_k \hat{H}^{(k)},~~~\hat{H}V^{(k)}\subset V^{(k)},
\end{equation}
and that the spectrum of $\hat{H}$ can be split into subspectra with degeneracies equal to the dimensions of the irreps. Thus, group theory provides \textit{good quantum numbers} 
and allows to classify the spectral properties of a system. Moreover, it can provides information about the form of the corresponding eigenstates. One of the most 
famous examples is Bloch's theorem \cite{Bloch1929}, which uses the translation invariance of crystals to determine the form of the wave functions.
Another important application of symmetry and representations in quantum mechanics is given by selection rules. Indeed, the probability of a transition between 
two states associated with different irreps can be obtained by the direct product of these irreps \cite{Hamermesh_book}. Finally, another interesting application in 
terms of spectral properties is given by perturbation theory: suppose our initial Hamiltonian $\hat{H}$ with a symmetry group $G$ is perturbed by a Hamiltonian 
$\epsilon\hat{H}_1$ with a symmetry group $G_1\subset G$, \textit{i.e.} with a lower symmetry. This case is referred as an \textit{explicit symmetry breaking}. Then, how the spectrum of $\hat{H}$, whose degeneracies are given 
by the irreps of $G$, is affected? If we consider an eigenvalue $E$ of $\hat{H}$ with a degeneracy corresponding to the dimension $m$ of the associated irrep $D$ 
of $G$, this irrep $\hat{D}$ is in general a reducible representation of $G'$, the symmetry group of the perturbed Hamiltonian $\hat{H}'=\hat{H}+\epsilon\hat{H}_1$. 
Then, by decomposing $\hat{D}$ into irreps of $G'$, we obtain how the degeneracy of $E$ is lifted by the perturbation.

\begin{table}
\centering
\begin{tabular}{l|r}
\textbf{Symmetry} & \textbf{Conserved quantity} \\\hline
Translation in time & Energy \\
Translation in space & Momentum \\
Rotation in space & Angular momentum \\
Global phase invariance $U(1)$ & Electric charge
\end{tabular}
\caption{\label{tab:noether}Examples of continuous symmetries and their corresponding conservation laws.}
\end{table}

The notion of symmetry breaking is also very useful in the case of a \textit{spontaneous} symmetry breaking. Here, the ground-state breaks the 
symmetry of the Hamiltonian. This is at the origin of numerous physical phenomena such as crystals (which break translation invariance), magnets (which 
breaks rotational invariance) or Bose-Einstein condensates (which breaks the global phase invariance). Each global spontaneous symmetry breaking  
is associated with low energy fluctuations, or \textit{Goldstone bosons} \cite{Goldstone1962}: sound waves or phonons for crystals, spin waves or magnons in magnets,
Bogoliubov quasi-particles in Bose-Einstein condensates. Spontaneous symmetry breaking is intimately linked with the notion of phase transition: when the temperature becomes 
larger than a certain temperature $T_c$, the symmetry is restored. The Mermin–Wagner–Hohenberg theorem that we briefly discussed in section \ref{pec1d} and that explains in 
particular why there is no Bose-Einstein condensation in 1D is based on the fact that the Goldstone bosons associated with a spontaneous symmetry breaking 
would have infrared diverging correlation functions for dimensions lower than 2 \cite{Mermin1966, Hohenberg1967}.

Finally, a lot of modern successful theories are \textit{constructed from symmetries}. This is the case of special 
and general relativity, which are respectively based on the global and the local Lorentz covariances \cite{Einstein1905,Einstein1917}. In quantum physics, and more precisely in quantum 
field theories, the fundamental interactions are dictated by local internal symmetries of the fields, or \textit{gauge} symmetries: $U(1)$ for quantum electrodynamics \cite{Tomonaga1946,Schwinger1948,Feynman1950},
$U(1)\times SU(2)$ for the electroweak theory \cite{Glashow1959,Salam1959,Weinberg1967}, $SU(3)$ for quantum chromodynamics (which describes strong interaction) \cite{Gross1973,Politzer1973}. The product of these gauge symmetries 
$U(1)\times SU(2) \times SU(3)$, together with the \textit{Higgs mechanism} which is nothing more than a spontaneous symmetry breaking explaining why the interaction 
bosons of the weak interaction are massive \cite{Englert1964,Higgs1964}, form the \textit{standard model}, which describes all fundamental interactions except gravity. 
The incredible success of these theories suggests that symmetry, 
more than being an extremely valuable tool for physicists, has in fact a more fundamental meaning and importance in Nature.

\subsection{Exchange symmetry}
\label{secexcsym}

\subsubsection{Identical particles}

\textit{Identical} particles are particles with the same intrinsic properties (mass, charge, spin...), and that are therefore impossible to differentiate. For example, all 
the electrons of the universe are identical, and so are all the hydrogen atoms. Then, if a system is composed of identical particles, its properties must be 
invariant when exchanging them. This \textit{exchange symmetry} is very important in quantum physics because it causes problems regarding its fundamental postulates \cite{Cohena,Cohenb}. Indeed,
let us consider a system of $N$ identical particles. If $\mathcal{E}$ is the Hilbert space associated with one particle, the total Hilbert space of this system 
can be obtained by taking the tensorial product
\begin{equation}
\mathcal{E}_{tot}\equiv\mathcal{E}(1) \otimes \mathcal{E}(2)\otimes\cdots\otimes\mathcal{E}(N),
\end{equation}
 where numbers were assigned 
to the particles in an arbitrary way. Let us now consider a set of $N$ commuting observables $(\mathcal{O}(i))_{i\in\{1,\ldots,N\}}$ associated with the $N$ 
particles, and with the same spectrum $\{\sigma_n;n=1,2,\ldots\}$. Suppose that in an experiment we have 
measured simultaneously $\mathcal{O}$ for the $N$ particles, and that we have obtained $\{\sigma_1,\sigma_2,\ldots,\sigma_N\}$. Then, because of the indiscernibility of the 
particles, it is \textit{a priori} impossible to know to which state of $\mathcal{E}_{tot}$ it corresponds: it can be either 
\begin{equation}
\ket{1:\sigma_1}\otimes\ket{2:\sigma_2}\otimes\cdots\otimes\ket{N:\sigma_N}\in\mathcal{E}_{tot}
\end{equation}
or any of its $N!$ permutations, \textit{e.g.} 
\begin{equation}
\ket{1:\sigma_2}\otimes\ket{2:\sigma_1}\otimes\ket{3:\sigma_3}\otimes\cdots\otimes\ket{N:\sigma_N}\in\mathcal{E}_{tot}.
\end{equation}
This is known as the \textit{exchange degeneracy}.

Let us consider the case of $N=2$ in the previous example, and let us define the permutation operator $\hat{P}_{12}$ such that 
\begin{equation}
\hat{P}_{12}\ket{1:\sigma_1}\otimes\ket{2:\sigma_2}=\ket{1:\sigma_2}\otimes\ket{2:\sigma_1}.
\end{equation}
It is clear then that $\hat{P}_{12}$ is an involution ($(\hat{P}_{12})^2=1$) and that it is self-adjoint ($\hat{P}_{12}^{\dagger}=\hat{P}_{12}$). Therefore, 
its only eigenvalues are $+1$ and $-1$. The corresponding orthogonal eigenstates are respectively \textit{symmetric} and \textit{anti-symmetric}.

In the general case, one can also define a permutation operator $\hat{P}$ associated with every $P\in\mathfrak S_N$, where $\mathfrak S_N$ is the permutation group 
of $N$ objects (see section \ref{subsecpermut}). The situation is however more complicate in this case, and one can not write $\mathcal{E}_{tot}$ as a direct 
sum of a completely symmetric and a completely anti-symmetric subspace. However, we can define the following projectors:
\begin{equation}
\begin{split}
&\hat{S}=\frac{1}{N!}\sum_{P\in\mathfrak S_N}\hat{P}\\
&\hat{A}=\frac{1}{N!}\sum_{P\in\mathfrak S_N}\epsilon(P)\hat{P},
\end{split}
\end{equation}
where $\epsilon(P)$ is the signature of permutation $P$, that project respectively onto a completely symmetric subspace $\mathcal{E}_S$ and a completely anti-symmetric 
subspace $\mathcal{E}_A$ of $\mathcal{E}_{tot}$. The symmetrization postulate states that, depending on the nature of the $N$ identical particles, the physical 
$N$-body states belong either to $\mathcal{E}_S$, and in this case they are called \textit{bosons}, or to $\mathcal{E}_A$, and then they are called \textit{fermions}\footnote{In principle, 
one can consider identical particles which are neither bosons nor fermions. It only happens in very rare occasion, a notable exception being given by the so-called
\textit{anyons} that one encounters in 2D materials and in particular in the fractionnal quantum Hall effect \cite{Wilczek1982}.}. This postulate completely removes the exchange degeneracy: indeed, 
depending on the nature of the identical particles in our example, a measure $\{\sigma_1,\sigma_2,\ldots,\sigma_N\}$ is associated with a single possible state,
which is $\hat{S}\ket{1:\sigma_1}\otimes\cdots\otimes\ket{N:\sigma_N}\in\mathcal{E}_S$ for bosons and $\hat{A}\ket{1:\sigma_1}\otimes\cdots\otimes\ket{N:\sigma_N}\in\mathcal{E}_A$ for fermions.

The fact that a particle is a boson or a fermion has extremely important physical consequences. To see this, let us go back to the case of two particles. In 
this case, $\hat{S}=\frac{1}{2}\left(1+\hat{P}_{12}\right)$ and $\hat{A}=\frac{1}{2}\left(1-\hat{P}_{12}\right)$. Let us consider the state $\ket{u}=\ket{1:\phi}\otimes\ket{2:\phi}$ 
where the two particles are in the same one-body state. Then, it is clear that $\hat{S}\ket{u}=\ket{u}$ and $\hat{A}\ket{u}=0$. In other words, it is impossible 
to put two identical fermions in the same quantum state. This well-known fact is known as the \textit{Pauli exclusion principle}. It implies in particular that electrons,
which are fermions, cannot be in the same state (and in particular at the same place), and therefore explains a lot of properties of matter. In contrast, there are 
no reason for bosons not to occupy the same quantum state. Particles will therefore have the tendency to accumulate in their individual states of lowest energy at very low 
temperatures, which is at the origin of spectacular quantum effects such as Bose-Einstein condensation, superfluidity and superconductivity (where electrons form bosonic 
\textit{Cooper pairs}). Even at non-zero temperatures, the statistical behaviors of fermions and bosons are completely different, where the first follows 
the \textit{Fermi-Dirac statistics} and the second the \textit{Bose-Einstein statistics}.

Finally, the statistics of identical particles and their spin are related through the so-called \textit{spin-statistics theorem} claims that particles with 
integer spins are bosons and particles with half-integer spins are fermions \cite{Schwinger1951}. This theorem can be intuitively understood as making the link between exchanging particles and rotating them --- since the spin labels the irreducible representations of $SU(2)$ (the projective group of $SO(3)$), it is then intimately related to the exchange symmetry.

\subsubsection{The symmetric group and its representations}
\label{subsecpermut}

We have already encountered several times the \textit{symmetric group}\footnote{A somehow confusing name!}, when dealing with strongly interacting mixtures (section \ref{secvolosniev}), or in appendix \ref{secbethe} with the Bethe ansatz. Here we recall some of the basic mathematical properties of this 
group and its irreps, without entering the details. For further readings, the interested reader can turn to one of the many excellent references on the subject 
\cite{Fulton2004,Liebeck,Hamermesh_book}.

\paragraph{Basic properties of $\mathfrak S_N$}

The symmetric group $\mathfrak S_N$ of index $N$ is the group of all bijections 
from $\{1,2,\ldots,N\}$ to itself, or \textit{permutations}, with the composition 
as an internal law. In physics, as we just saw, it has an extreme importance 
when dealing with identical particles. It is also very useful when studying 
\textit{unitary groups} \cite{Itzykson1966}, and therefore in gauge theories. Moreover, on a purely 
mathematical point of view, Cayley's theorem states that every finite group is isomorphic 
to a subgroup of $\mathfrak{S}_N$ \cite{Cayley1854}, which gives to the latter a central importance in group theory.

The group $\mathfrak S_N$ has order $N!$. It is generated by the 2-cycles, or \textit{transpositions}, of the form $\tau_i=(i,i+1)$ (that is 
which exchanges elements $i$ and $i+1$). The number $t$ of transpositions by which 
a permutation $P\in\mathfrak S_N$ can be decomposed defines the \textit{signature}  through $\epsilon(P)=(-1)^{t}$. The kernel of this morphism is denoted $\mathfrak A_N$, 
the \textit{alternating group}.

Any permutation $P\in\mathfrak S_N$ can be decomposed in disjoint cycles. Moreover, the \textit{conjugacy class} of $P$, that is the set
\begin{equation}
cc(P)=\{Q PQ^{-1}~|~Q\in\mathfrak S_N\},
\end{equation}
is given by the set of all permutations whose decomposition in disjoint cycles has the same structure as $P$ (same number of cycles of every length). If this structure consists in $k_1$ 1-cycles, $k_2$ 2-cycles, ... , $k_m$ $m$-cycles, the number $|cc(P)|$ of elements in $cc(P)$ is given by
\begin{equation}
\label{nbelcc}
|cc(P)|=\frac{N!}{1^{k_1}k_1!\ldots m^{k_m}k_m!}.
\end{equation}
The number of conjugacy classes of $\mathfrak S_N$ is then the number $p(N)$ of \textit{partitions} of the integer $N$. It can be obtained by expanding its \textit{generating function} as a geometric series \cite{0486612724}:
\begin{equation}
\label{partfunc}
\sum_{N=0}^{\infty}p(N)X^N=\prod_{k=1}^{\infty}\frac{1}{1-X^k}.
\end{equation}
An asymptotic expansion for $p(N)$ is given by \cite{doi:10.1112/plms/s2-17.1.75}:
\begin{equation}
p(n)\underset{\infty}{\sim}\frac{1}{4N\sqrt{3}}\exp\left(\pi\sqrt{\frac{2N}{3}}\right).
\end{equation}
As one can see, the number of conjugacy classes of $\mathfrak S_N$ grows rapidly with $N$, which makes its extensive study more adapted to low $N$.

\paragraph{Irreducible representations of $\mathfrak S_N$. Young formalism.}

The number of non-equivalent irreps of a group is given by the number of its conjugacy classes. Therefore, 
the number of irreps of $\mathfrak S_N$ is given by the number $p(N)$ of partitions of $N$. 

A convenient way of representing a partition of an integer 
is  by a so-called \textit{Young diagram}, which is defined as the following. Let us consider a partition of the form 
$\Lambda\equiv\left[\lambda_1,\lambda_2,\ldots,\lambda_n\right]$ with $\lambda_1\ge\lambda_2\ge\cdots\ge\lambda_n$ and $\lambda_1+\lambda_2+\cdots+\lambda_n=N$. Then, we represent this partition 
by a left-justified set of boxes with $n$ rows, where each row $i\in\{1,\ldots,n\}$ contains $\lambda_i$ boxes. For example, the Young diagram associated with the partition $\Lambda_0=\left[3,2,2,1\right]$ of 8 is given by
\begin{equation}
\label{exyoungd}
Y_{\Lambda_0}=\yng(3,2,2,1).
\end{equation}

Associated with Young diagrams are the \textit{Young tableaux}, where the boxes are filled 
with symbols taken from some totally ordered set (\textit{e.g.} integers, Latin alphabet...). If the entries are always increasing from left to right in every row and top to bottom in every column, the tableau is said to be \textit{standard}. In the case where several symbols appear more than once, 
the tableau is said \textit{semistandard} when the entries increase in the weak sense. Here are two examples of Young tableaux, one standard and one semistandard:
\begin{equation}
\young(135,26,47,8)\quad \text{ and }\quad \young(aab,bb,bc,c).
\end{equation}

Thus, there is a one to one correspondence between the Young diagrams with $N$ boxes and the irreps of $\mathfrak S_N$. This correspondence is done using the so-called 
\textit{Young symmetrizers} associated with the Young diagrams \cite{Fulton2004}. Intuitively, it indicates that the rows are symmetrized, while the column are anti-symmetrized. In the case of $N=3$ for example, 
there are three possible Young diagrams, which are given by
\begin{equation}
Y_{[3]}=\yng(3)~,\quad Y_{[1,1,1]}=\yng(1,1,1)~,\quad\text{ and }\quad Y_{[2,1]}=\yng(2,1)~.
\end{equation}
They correspond respectively to the \textit{trivial representation} (where every permutation is represented by the identity matrix $~\Id$), 
to the \textit{sign representation} (where every permutation is represented by its signature) and to the \textit{standard representation}. In terms of identical particles, $Y_{[3]}$ is the exchange symmetry of three identical bosons,
and $Y_{[1,1,1]}$ is the exchange symmetry of three identical fermions. $Y_{[2,1]}$ corresponds necessarily to a mixture of non-identical particles.

The dimension of a given irrep $D_{\Lambda}$ is given by the number of standard Young tableaux that one can construct from the corresponding Young diagram $Y_{\Lambda}$. This number can be obtained from the \textit{hook length formula}.
A \textit{hook length} $h_{\Lambda}(i,j)$ of a box of coordinates $(i,j)$ of $Y_{\Lambda}$ is given by the numbers of boxes below $(i,j)$ + the number of boxes at the right of the box $(i,j)$ + 1. For example, in $Y_{\Lambda_0}$ of Eq.~\eqref{exyoungd}, the hook lengths are given by
\begin{equation}
\young(641,42,31,1).
\end{equation}
Then, the hook length formula states that
\begin{equation}
\label{dimrep}
\dim D_{\Lambda}=\frac{N!}{\prod_{(i,j)\in\Lambda}h_{\Lambda}(i,j)}.
\end{equation}
In particular, in our example we have $\dim D_{\Lambda_0}=8!/(6*4^2*3*2)=70$. Moreover, more generally, the trivial and sign representations, which corresponds respectively to the horizontal and vertical Young diagrams, have always a dimension 1, whereas other irreps are multidimensional. 
This explains why the symmetrization postulate, which states that identical particles are either bosons or fermions, removes the exchange degeneracy (see section \ref{secexcsym}). Besides, two \textit{conjugate} irreps, that is irreps whose Young diagrams are transposed, have the same dimension.

Finally, for a given irrep $D_{\Lambda}$, each element of $\mathfrak S_N$ is represented by a $(\dim D_{\Lambda})$-dimensional matrix. Then, the function 
\begin{equation}
\chi_{\Lambda}~:\quad g\in\mathfrak S_N\quad\mapsto\quad \mathrm{Tr}\left(D_{\Lambda}(g)\right)  \in\mathbb{C}
\end{equation}
is called the \textit{character} of $D_{\Lambda}$. These functions have many important properties that greatly help the study of irreps. In particular, there are constant on the conjugacy classes of $\mathfrak S_N$, and 
isomorphic representations have the same characters.

\section{Symmetry analysis of strongly interacting quantum mixtures}
\label{secsyman}

In this section we turn to the symmetry analysis of the system that we mainly studied in this thesis, that is strongly interacting quantum mixtures in one-dimensional 
harmonic traps (c.f. section \ref{secvolosniev}). We will describe the problem and its relation with quantum magnetism in \ref{introsym}. Then, we will present the method that 
we used in order to solve it, namely the \textit{class-sum method}, in \ref{classsummethod}. Finally, we will analyze the ordering of energies as a function of the 
symmetries in \ref{lmandb}. The methods and results described in this section where published in \cite{Decamp2016,Decamp2016-2} for fermionic mixtures and \cite{Decamp2017} for Bose-Fermi mixtures.

\subsection{One-dimensional $SU(\kappa)$ quantum gases and quantum magnetism}
\label{introsym}

The systems we are interested in are mixtures of $N$ atoms divided in $\kappa$ different fermionic and/or bosonic spin species 
with populations $N_1,\ldots,N_{\kappa}$, and submitted to the Hamiltonian of Eq.~\eqref{maineq} that we recall here:
\begin{equation}
\label{mainham}
\left[\sum_{j=1}^N\left(-\frac{\hbar^2}{2m} \frac{\partial^2}{\partial x_j^2}+\frac{1}{2}m\omega^2x_j^2\right)+g_{1D}\sum_{i<j}\delta(x_i-x_j)-E\right]\psi=0.
\end{equation}
The key ingredient that allows to describe such mixtures with this Hamiltonian is that we are considering atoms whose ground-state has a purely nuclear spin. 
This ensures that the interactions between particles with different spins are independent of the relative spin orientations, that collisions are not spin-flipping, and that 
all particles are subjected to the same external potential. This property is common to Ytterbium atoms and to fermionic alkaline-earth atoms \cite{Gorshkov2010}.
Therefore, when exchanging fermionic (resp. bosonic) spin components, it does not affect the properties of the system. In the case of a purely fermionic or 
purely bosonic mixture, this is refered as the $SU(\kappa)$ symmetry, as it constitutes a generalization of the $SU(2)$ symmetry of spin-$\frac{1}{2}$ electronic mixtures \cite{Laird2017}. If the 
system is composed of $\mu$ bosonic  components and $\nu$ fermionic  component, the system is said to be $SU(\mu,\nu)$-symmetric \cite{Kato1995}.
$SU(\kappa)$ systems with $\kappa>2$ display exotic features and phase diagrams that has attracted the attention of theoretical physics \cite{Affleck1988,Marston1989,Read1989}, 
recently renewed by the experimental realization of a one-dimensional $SU(\kappa)$ fermionic mixture described in section \ref{florexp}.

The $SU(\kappa)$-symmetry is not transparent when just looking at Eq.~\eqref{mainham}. However, in the limit  of strongly repulsive 
interactions $g_{1D}\to\infty$ that we are considering, one can map this Hamiltonian onto a spin Hamiltonian \cite{Deuretzbacher2014,Massignan2015,MassignanPaarish}. Indeed, 
if we use the same notations as in section \ref{secvolosniev}, we can write in the vicinity of $1/g_{1D}=0$ an effective Hamiltonian of the form:
\begin{equation}
\label{spinmodel}
H_S=\left(E_0-\sum_{k=1}^{N-1}J_k\right)\Id\pm\sum_{k=1}^{N-1}J_k\hat{P}_{k,k+1}~,
\end{equation}
where the $+$ ($-$) sign is for fermions (resp. bosons),  the nearest-neighbor exchange constants $J_k=\alpha_k/g_{1D}$, and the permutation matrices $\hat{P}_{k,k+1}$ 
are defined in the snippet basis by:
\begin{equation}
\left(\hat{P}_{k,k+1}\right)_{ij}=\left\{ \begin{array}{ll}
 1 & \mbox{ if snippets } i \mbox{ and }j \mbox{ are equal up to a transposition } (k,k+1) \\ 0 & \mbox{ otherwise }
\end{array}\right..
\end{equation}
Then, the operator $\hat{P}_{k,k+1}$ can be re-written in terms of spin operators \cite{Deuretzbacher2014}: e.g. when $\kappa=2$, it can be expressed in terms of 
the Pauli vector $\vec{\sigma}$ by 
\begin{equation}
\hat{P}_{k,k+1}=\frac{1}{2}\left(\vec{\sigma}^{(k)}\cdot\vec{\sigma}^{(k+1)}+\Id\right),
\end{equation}
and one recovers a Heisenberg Hamiltonian. When $\kappa>2$, $\vec{\sigma}$ has to be replaced by a spin operator $\vec{S}$ associated with the generalized generators 
of $SU(\kappa)$ Lie algebra \cite{Bourbaki:2008:LGL:1502204,Gorshkov2010}. In any case, Eq.~\eqref{spinmodel} shows that our problem \eqref{mainham} is equivalent to an effective $SU(\kappa)$ spin model. 

Note that in the fermionized limit $g_{1D}\to\infty$, Eq.~\eqref{spinmodel} reduces to a trivial Hamiltonian
\begin{equation}
H_S^{\infty}=E_0~\Id.
\end{equation}
This Hamiltonian has then a bigger symmetry, namely a $SU(N)$ symmetry.  This emergent symmetry explains the existence of degenerate manifolds at $g_{1D}=\infty$, and why this degeneracy is lifted whenever $g_{1D}$ becomes finite \cite{Harshman2014,Harshman2015,Harshman2016}. Our system can then be seen as a perturbed Hamiltonian with an explicit symmetry breaking from $SU(N)$ to $SU(\kappa)$.

Thus, our system is a perfect model to study \textit{itinerant magnetism}, that is magnetism without a lattice. In the following, we will focus on the symmetry 
of the \textit{spatial} wave function. This symmetry is dual to its \textit{spin} symmetry (and thus magnetism). Indeed, the symmetry of the total wave function of identical particles 
is fixed by their bosonic or fermionic nature. However, if these identical particles have different spin orientations as it is the case in our model, although their total wave function 
still has to be totally symmetric or totally anti-symmetric, their spatial and spin wave function can be other representations of the symmetric group, as long as 
they are conjugate when identical fermions and equal when identical bosons. Thus, the problem that 
we are going to address in this section is the following: given a solution $\psi$ of Eq.~\eqref{mainham} in the $g_{1D}\to\infty$ limit, given by the perturbative method 
described in section \ref{secvolosniev} and hence by a vector of real coefficients $\vec{a}$, how to characterize its symmetry, or equivalently, to which  
representation of $\mathfrak S_N$ does it belong?

\subsection{The class-sum method}
\label{classsummethod}

Historically, the first description of the so-called \textit{class-sum method} is due Dirac, who used it to study the eigenstates of a many-electron 
system \cite{Dirac1929}. It was then developped in the context of nuclear physics \cite{Talmi,Novolesky1995}, and adapted to the study of one-dimensional 
quantum gases in \cite{Fang2011,Decamp2016,Decamp2016-2,Decamp2017}. After a description of the mathematical objects used in this method \ref{gendescls}, we will 
describe how this method can be adapted 
to our system 
of interest in order to answer the last paragraph's problem in \ref{csmeth}, and implement it in a $N=6$ system \ref{impcls}.

\subsubsection{Class-sums and central characters}
\label{gendescls}

Let us consider a Hamiltonian $\hat{H}$ which is invariant by permutation, as it is the case in our Hamiltonian of interest Eq.~\eqref{mainham}. More precisely, for every $P\in\mathfrak{S}_N$, if we define the linear operator $\hat{P}$ by
\begin{equation}
\hat{P}\psi(x_1,x_2,\ldots,x_N)=\psi(x_{P1},x_{P2},\ldots,x_{PN}),
\end{equation}
we then have, as in Eq.~\eqref{hamcomut},
\begin{equation}
\left[\hat{H},\hat{P}\right]=0.
\end{equation}

However, since the $N!$ permutation operators $\hat{P}$ do not commute with each other, we cannot diagonalize $\hat{H}$ and all the $\hat{P}$'s simultaneously 
in a common eigenbasis and use them directly in order to classify the states. We would like to define a set of operators $\hat{C}$ as a function of the $\hat{P}$'s that all commute with each other and also with $\hat{H}$, and that completely characterize the permutational symmetry of the system: i.e. a \textit{complete set of commuting observables} (CSCO) \cite{Cohena}. This is the same idea which leads to  the definition of the \textit{Casimir elements} $L^2=L_x^2+L_y^2+L_z^2$, the square angular momentum, when studying the states of the Hydrogen atom which are labeled by the eigenvalues $\{n,l,m\}$ of the CSCO $\{\hat{H}_{Hyd},L^2,L_z\}$.

For every conjugacy class $cc_{\Lambda}$ of $\mathfrak{S}_N$, where $\Lambda\equiv\left[\lambda_1,\lambda_2,\ldots,\lambda_n\right]$ is a partition of $N$, we now define the \textit{conjugacy class-sum} as:
\begin{equation}
\label{clasdef}
C_{\Lambda}=\sum_{P\in cc_{\Lambda}}P.
\end{equation}
This sum has to be understood in the context of the \textit{group algebra}  of $\mathfrak{S}_N$, that is the vector space whose basis is indexed by the elements of the group and that linearly extends the laws of the group \cite{Liebeck}.
It is clear then that every $Q\in\mathfrak{S}_N$ commutes with $C_{\Lambda}$ and thus with each other: by definition of a conjugacy class $QC_{\Lambda}Q^{-1}=C_{\Lambda}$ since it is the same sum in a different order. Moreover, if another linear combination of the $P$'s commutes with all the $P$'s, it is necessarily a linear combination of the $C_{\Lambda}$'s. Indeed, the condition
\begin{equation}
Q\left(\sum_{P\in\mathfrak{S}_N}b_PP\right)Q^{-1}=\sum_{P\in\mathfrak{S}_N}b_P'P
\end{equation}
for every $Q\in\mathfrak{S}_N$ implies that $b_P=b_P'$ whenever $P$ and $P'$ belong to the same conjugacy class. Thus we have shown that the $C_{\Lambda}$'s form a basis of the \textit{center} of the group algebra of $\mathfrak{S}_N$ \cite{Liebeck}, which is the that the associated permutation operators $\hat{C}_{\Lambda}$ form, together with $\hat{H}$, a CSCO.

The natural question that follows is then: what values are taken by the observables $\hat{C}_{\Lambda}$? One can prove that it takes one constant specific value for each irreductible representation $D_{\Gamma}$ of $\mathfrak{S}_N$, and that this value is related, up to a proportionality factor, to the irreductible character $\chi_\Gamma$ of $\mathfrak{S}_N$ \cite{Novolesky1995,book:491124}. More precisely, $D_{\Gamma}(C_\Lambda)$ commutes with $D_{\Gamma}(P)$ for every $P\in\mathfrak{S}_N$. Then, by Schur's lemma \cite{Fulton2004}, $D_{\Gamma}(C_\Lambda)$ is a \textit{homothety}, i.e. there is 
a real value $\omega_\Gamma^\Lambda$ such that:
\begin{equation}
\label{classumrep}
D_{\Gamma}(C_\Lambda)=\omega_\Gamma^\Lambda~\Id_{|\Gamma|},
\end{equation}
where $|\Gamma|\equiv\dim D_{\Gamma}$ is obtained with Eq.~\eqref{dimrep}. By taking traces in Eq.~\eqref{classumrep} we get
\begin{equation}
\label{principchar}
|\Lambda|\chi_\Gamma^\Lambda=|\Gamma|\omega_\Gamma^\Lambda,
\end{equation}
with $\chi_\Gamma^\Lambda$ the value taken by the irreducible character $\chi_\Gamma$ over the conjugacy class $cc_\Lambda$, and $|\Lambda|$ is the number of elements in $cc_\Lambda$, which is given by Eq.~\eqref{nbelcc}. Since, from Maschke's theorem, every representation $D$ can be written as a direct sum of irreps, the $\omega_\Gamma^\Lambda$ values of Eq.~\eqref{principchar} are the only eigenvalues of $D(C_{\Lambda})$, hence the only values that are taken by the observables $\hat{C}_{\Lambda}$. Because of their link with irreducible characters, they are called the \textit{central characters}. 

In the case where $\Lambda=\left[r,1,\ldots,1\right]\equiv\left[(r)\right]$, i.e. when the conjugacy class $cc_\Lambda$ consists in a single $r$-cycle and $(N-r)$ 1-cycles, there are explicit expressions for the central characters $\omega_\Gamma^{\left[(r)\right]}$. More precisely, if $\Gamma=[\gamma_1,\ldots,\gamma_m]$ (with $\gamma_1\ge\gamma_2\ge\cdots\ge\gamma_m$ and $\gamma_1+\gamma_2+\cdots+\gamma_m=N$), we can define \cite{Decamp2017}\footnote{The expression in \cite{Katriel1993} contains a small notation misprint.}
\begin{equation}
\mu_i=\gamma_i-i+m,\quad i=1,\ldots,m.
\end{equation}
and then obtain, using $\chi_{\left[(r)\right]}^\Lambda$'s expression \cite{book:491124,Katriel1993}:
\begin{equation}
\label{cychar}
\omega_\Gamma^{\left[(r)\right]}=\frac{1}{r}\sum_{i=1}^m\frac{\mu_i!}{(\mu_i-r)!}\prod_{j\neq i}\frac{\mu_i-\mu_j-r}{\mu_i-\mu_j},
\end{equation}
where to product is taken to be equal to 1 when $\Gamma$ contains only one line ($m=1$). When $r=2$, one can use one of these  more user-friendly formulas \cite{Novolesky1994}:
\begin{equation}
\label{userfriend}
\omega_\Gamma^{\left[(2)\right]}=
\left\{ \begin{array}{lll}
 \frac{1}{2}\sum_{i=1}^m \gamma_i\left(\gamma_i -2i+1\right)\\  \mbox{ or } \\ \sum_{(i,j)\in\Gamma}\left(j-i\right)  
\end{array}\right..
\end{equation}

Alternative ways of computing the central characters can be found in \cite{Katriel1993,Katriel1993b,KATRIEL1996149,Goupil2000}.

\subsubsection{Description of the method}
\label{csmeth}

Let us first precise our notations. The system $\left(N_1^B,\ldots,N_\mu^B,N_1^F,\ldots,N_\nu^F\right)$ we are considering is a mixture  of 
$N_B=N_1^B+\cdots+N_\mu^B$ bosons and $N_F=N_1^F+\cdots+N_\nu^F$  fermions divided respectively in $\mu$ and $\nu$ spin 
components, with $N=N_B+N_F$. Let us suppose that we have solved Eq.~\eqref{mainham} in the $g_{1D}\to\infty$ fermionized limit with the method of section \ref{secvolosniev}. Thus, we have obtained
a set of $D_{N_1^B,\ldots,N_\mu^B,N_1^F,\ldots,N_\nu^F}$ (c.f. Eq.~\eqref{degeneracy}) vectors  $\vec{a}$ of coefficients corresponding to solutions of Eq.~\eqref{mainham}
via Eq.~\eqref{Psivolo}. We want to determine to which representations of $\mathfrak S_N$ these vectors belong using the class-sums and central characters.

The method we have used in \cite{Decamp2016,Decamp2016-2,Decamp2017} can be decomposed into the following steps:

\paragraph{Step 1: Determine the irreps that are compatible with the mixture}

When identical bosons (resp. fermions) belong to the same spin component, their spatial wave function must be symmetric (resp. anti-symmetric) when exchanging 
their coordinates. This reduces the number of possible irreps that are compatible with a given mixture. In practice, one has to order the different species by 
decreasing order of population and to assign a letter in the alphabetic order to each one of the species. Then, one has to construct all the semistandard 
Young tableaux with $N$ boxes labeled by the species, imposing that no bosons (resp. fermions) belonging to the same spin component can be in the same column (resp. row).

For example, in a $N=4$ mixtures, there are \textit{a priori} $p(4)=5$ possible irreps (c.f. Eq.~\eqref{partfunc}), given by the following Young diagrams:
\begin{equation}
\yng(4)~,\quad\yng(3,1)~,\quad\yng(2,2)~,\quad\yng(2,1,1)~,\quad\yng(1,1,1,1)~.
\end{equation}
Suppose now that this mixture is given by $(2^F,2^F)$. Then, the only possible semistandard Young tableaux are:
\begin{equation}
\label{y2f2f}
\young(a,a,b,b)~,\quad\young(ab,a,b)~,\quad\young(ab,ab)~.
\end{equation}
If this mixture is instead $(2^B,2^F)$, there are only two possible symmetries:
\begin{equation}
\young(aab,b)~,\quad\young(aa,b,b)~.
\end{equation}

We now dispose of a set $\{Y_{\Gamma_1},\dots,Y_{\Gamma_k}\}$ of Young diagrams (with $k\le p(N)$). This initial step substantially reduces the complexity of the method.

\paragraph{Step 2: Compute the central characters}

We have seen in \ref{gendescls} that, for a given partition $\Lambda$ of $N$, the central characters $\omega_\Gamma^\Lambda$ have a constant value on each of 
the irreps $D_\Gamma$ of $\mathfrak S_N$ because of their relation with irreducible characters (Eq.~\eqref{principchar}). Analogously to a \textit{character table} 
$(\chi_\Gamma^\Lambda)_{\Lambda,\Gamma}$ \cite{Fulton2004}, we could construct a table $(\omega_\Gamma^\Lambda)_{\Lambda,\Gamma}$ for central characters and 
compute $\omega_\Gamma^\Lambda$ for all irrep $D_\Gamma$ that is compatible with the mixture and all partition $\Lambda$. Thus, we would be able to completely 
characterize the symmetry of a solution $\vec{a}$. This is, however, not necessary: it is sufficient to compute the central characters for $\Lambda$'s so that 
all the central characters belonging to different irreps have different values for at least one $\Lambda$.

More precisely, given our set $\{Y_{\Gamma_1},\dots,Y_{\Gamma_k}\}$ of Young diagrams, we start by computing the central characters 
$\omega_{\Gamma_1}^\Lambda,\ldots,\omega_{\Gamma_k}^\Lambda$ of the transposition class-sums, i.e. when  $\Lambda=\left[(2)\right]$, using 
Eq.~\eqref{cychar} or Eq.~\eqref{userfriend}. If the $\omega_{\Gamma_i}^{\left[(2)\right]}$'s are all different, it is sufficient in order to characterize the symmetry. If not, we compute 
the $\omega_{\Gamma_i}^{\left[(3)\right]}$'s using again Eq.~\eqref{cychar}, and so on, until a rank $r_{\max}$ where all the irreps $D_{\Gamma_i}$ have a 
different set of central characters $\omega_{\Gamma_i}^{\left[(2)\right]},\omega_{\Gamma_i}^{\left[(3)\right]},\ldots,\omega_{\Gamma_i}^{\left[(r_{\max})\right]}$.

In practice, when $N\le 5$, it is sufficient to compute the central characters of the transposition class-sums (see table \ref{tabcc}). For $N=6$, $r_{\max}=3$ (c.f. section \ref{impcls}).

\begin{table}
\renewcommand{\arraystretch}{3}
\centering
\subfloat[$N=2$]{
\begin{tabular}{|c|c|}\hline
$Y_\Gamma$ & $\omega^{\left[(2)\right]}_\Gamma$ \\\hline\hline
{\tiny\yng(2)} & 1 \\\hline
{\tiny\yng(1,1)} & -1 \\\hline
\end{tabular}
}\hfill
\subfloat[$N=3$]{
\begin{tabular}{|c|c|}\hline
$Y_\Gamma$ & $\omega^{\left[(2)\right]}_\Gamma$ \\\hline\hline
{\tiny\yng(3)} & 3 \\\hline
{\tiny\yng(2,1)} & 0 \\\hline
{\tiny\yng(1,1,1)} & -3 \\\hline
\end{tabular}
}\hfill
\subfloat[$N=4$]{
\begin{tabular}{|c|c|}\hline
$Y_\Gamma$ & $\omega^{\left[(2)\right]}_\Gamma$ \\\hline\hline
{\tiny\yng(4)} & 6 \\\hline
{\tiny\yng(3,1)} & 2 \\\hline
{\tiny\yng(2,2)} & 0 \\\hline
{\tiny\yng(2,1,1)} & -2 \\\hline
{\tiny\yng(1,1,1,1)} & -6 \\\hline
\end{tabular}
}\hfill
\subfloat[$N=5$]{
\begin{tabular}{|c|c|}\hline
$Y_\Gamma$ & $\omega^{\left[(2)\right]}_\Gamma$ \\\hline\hline
{\tiny\yng(5)} & 10 \\\hline
{\tiny\yng(4,1)} & 5 \\\hline
{\tiny\yng(3,2)} & 2 \\\hline
{\tiny\yng(3,1,1)} & 0 \\\hline
{\tiny\yng(2,2,1)} & -2 \\\hline
{\tiny\yng(2,1,1,1)} & -5 \\\hline
{\tiny\yng(1,1,1,1,1)} & -10 \\\hline
\end{tabular}
}
\caption{\label{tabcc}Central characters of the transposition class-sums $\omega_\Gamma^{\left[(2)\right]}$ for $N\le 5$. 
Since they are all different (at fixed $N$), there is no need to compute  $\omega_\Gamma^\Lambda$ for other $\Lambda$'s.
Note that conjugate diagrams have an opposite $\omega_\Gamma^{\left[(2)\right]}$.}
\end{table}

\paragraph{Step 3: Compute the class-sums in the sector basis representation}

The next thing to do is to construct the class-sums $C_{[(2)]},\ldots,C_{[(r_{\max})]}$ defined in Eq.~\eqref{clasdef} in the same basis as the $\vec{a}$ vectors, that is in the sector basis. We have to be careful to the fact that, in the definition of $\vec{a}$ (Eq.~\eqref{Psivolo}), the $a_P$ coefficients are the projection of ansatz $\psi$ over the totally anti-symmetric wave function $\psi_A$: therefore, in order to characterize the symmetry of $\psi$, we have to compensate this anti-symmetry when constructing the class-sums. Accordingly, let us define the matrix representation $D_{\mathrm{sec}}$ by:
\begin{equation}
\begin{array}{ccccc}
D_{\mathrm{sec}} &: & \mathfrak{S}_N & \to & GL_{N!}(\mathbb{R})\\
 & & g & \mapsto & (\epsilon(g)\delta_{P,gQ})_{P,Q}
\end{array},
\end{equation}
with $\epsilon$ the signature, which can be extended linearly to a morphism defined on the group algebra, where the class-sums are defined.
With this definition, the cycle class-sums are represented by:
\begin{equation}
D_{\mathrm{sec}}\left(C_{[(r)]}\right)=(-1)^{r-1}\sum_{\sigma\in cc_{[(r)]}}M_\sigma ,
\end{equation}
where $M_\sigma^r$ is a $N!\times N!$ matrix whose coefficients are $(M_\sigma)_{PQ}=\delta_{P,\sigma Q}$. 

For the sake of clarity, let us consider the example of $N=3$. There are $3!=6$ elements in the sector basis, that we order in the following way:
\begin{equation}
(e_{123},e_{132},e_{213},e_{312},e_{321}).
\end{equation} 
In this basis, 
\begin{equation}
D_{\mathrm{sec}}\left(C_{[(2)]}\right)=-\begin{pmatrix} 
0 & 1 & 1 & 0 & 0 & 1 \\
1 & 0 & 0 & 1 & 1 & 0 \\ 
1 & 0 & 0 & 1 & 1 & 0 \\
 0 & 1 & 1 & 0 & 0 & 1 \\
  0 & 1 & 1 & 0 & 0 & 1 \\
1 & 0 & 0 & 1 & 1 & 0 \\
\end{pmatrix},
\end{equation}
and
\begin{equation}
D_{\mathrm{sec}}\left(C_{[(3)]}\right)=\begin{pmatrix} 
0 & 0 & 0 & 1 & 1 & 0 \\
0 & 0 & 1 & 0 & 0 & 1 \\ 
0 & 1 & 0 & 0 & 0 & 1 \\
1 & 0 & 0 & 0 & 1 & 0 \\
1 & 0 & 0 & 1 & 0 & 0 \\
0 & 1 & 1 & 0 & 0 & 0 \\
\end{pmatrix}.
\end{equation}

\paragraph{Step 4: Project $\vec{a}$ on the eigenbasis of the class-sum representations}

As we have seen in section \ref{gendescls}, in each irrep $D_\Gamma$, the class-sums are homotheties whose ratios are given by the central characters (Eq.~\eqref{classumrep}). Although $D_{\mathrm{sec}}$ is \textit{a priori} not an irrep, it is (up to a sign) the \textit{regular representation} of $\mathfrak{S}_N$ \cite{Liebeck}. Then, $D_{\mathrm{sec}}$ is a direct sum of all the irreps of  $\mathfrak{S}_N$ (with multiplicities equal to their degrees). In other words, the spectrum of the cycle class-sum representations is equal to the set of all the associated central characters. Therefore, by diagonalizing the $D_{\mathrm{sec}}\left(C_{[(r)]}\right)$ matrices, we will get a set of eigenspaces corresponding to the irreps $D_\Gamma$ and with eigenvalues $\omega_{[(r)]}^\Lambda$.

Thus, the final step of this method is to diagonalize our set $D_{\mathrm{sec}}\left(C_{[(2)]}\right),\ldots,D_{\mathrm{sec}}\left(C_{[(r_{\max})]}\right)$ of matrices and project $\vec{a}$ in their eigenbasis. The symmetry of the associated wave function will then be completely characterized.

\paragraph{Alternative: the snippet representation}

Note that for a given mixture it is also possible to write the class-sums in the lower dimensional snippet representation $D_{\mathrm{snip}}$  (see section \ref{volansatz}) by 
summing (or subtracting when permuting same-component bosons) over the elements of the class-sums in the sector representation that belong to the same snippet. The only central characters that will appear when 
diagonalizing $D_{\mathrm{snip}}\left(C_{[(2)]}\right),\ldots,D_{\mathrm{snip}}\left(C_{[(r_{\max})]}\right)$ will then be the ones corresponding to the irreps 
allowed by the mixture (step 1). If we consider the fermionic mixture $(2^F,2^F)$ that we used as a first example in section \ref{volansatz}, we obtain in the same basis:
\begin{equation}
D_{\mathrm{snip}}^{(2^F,2^F)}\left(C_{[(2)]}\right)=-\begin{pmatrix} 
2 & 1 & 1 & 1 & 1 & 0 \\
2 & 2 & 1 & 1 & 0 & 1 \\
2 & 1 & 2 & 0 & 1 & 1 \\
2 & 1 & 0 & 2 & 1 & 1 \\
2 & 0 & 1 & 1 & 2 & 1 \\
0 & 1 & 1 & 1 & 1 & 2 \\
\end{pmatrix},
\end{equation}
which is similar to the diagonal matrix $\mathrm{diag}(-6,-2,-2,-2,0,0)$. The corresponding central characters are, as expected, associated with the three diagrams of Eq.~\eqref{y2f2f} allowed 
by the mixture (c.f. table \ref{tabcc}).

Although doing this adds a step between Step 3 and Step 4, it has the advantage of facilitating the diagonalization of Step 4 by reducing the size of the matrix 
representations. However, it has the inconvenient to be usable only for a given mixture, contrary to the sector representation which is usable for any mixture at 
given $N$.

\begin{table}
\renewcommand{\arraystretch}{3}
\centering
\begin{tabular}{|c|c|c|c|c|}\hline
$\Gamma$ & $Y_\Gamma$ & $\dim D_{\Gamma}$ & $\omega^{\left[(2)\right]}_\Gamma$ & $\omega^{\left[(3)\right]}_\Gamma$\\\hline\hline
$[6]$ & {\tiny\yng(6)} & 1 & 15 & 40  \\\hline
$[5,1]$  & {\tiny\yng(5,1)} & 5 & 9 & 16 \\\hline
 $[4,2]$  & {\tiny\yng(4,2)} & 9 & 5 & 0 \\\hline
  $[3,3]$  & {\tiny\yng(3,3)}  & 5 & \textbf{3} & -8  \\\hline
   $[4,1,1]$  & {\tiny\yng(4,1,1)} & 10 & \textbf{3} & 4 \\\hline
    $[3,2,1]$  & {\tiny\yng(3,2,1)} & 16 & 0 & -5 \\\hline
     $[3,1,1,1]$  & {\tiny\yng(3,1,1,1)} & 10 & \textbf{-3}  & 4  \\\hline
     $[2,2,2]$   & {\tiny\yng(2,2,2)} & 5 & \textbf{-3} & -8  \\\hline
     $[2,2,1,1]$    & {\tiny\yng(2,2,1,1)} & 9 & -5 & 0  \\\hline
       $[2,1,1,1,1]$   & {\tiny\yng(2,1,1,1,1)}  & 5 & -9 & 16  \\\hline
       $[1,1,1,1,1,1]$    & {\tiny\yng(1,1,1,1,1,1)} & 1 & -15 & 40  \\\hline
\end{tabular}
\caption{\label{tabcentch} Central characters $\omega^{\left[(2)\right]}_\Gamma$ and $\omega^{\left[(3)\right]}_\Gamma$ in the $N=6$ case, obtained from Eqs.~\eqref{cychar} and \eqref{userfriend}. For each partition $\Gamma$, the dimension of the associated irrep $D_{\Gamma}$, which corresponds to the multiplicity of the corresponding central characters as eigenvalues of the class-sums in the sector representation, is computed with the hook length formula (Eq.~\eqref{dimrep}).}
\end{table}

\subsubsection{Implementation for $N=6$ mixtures and first observations}
\label{impcls}

The $N=6$ case has interesting new features as compared to the $N\le 5$ ones. First, we will see very soon that here the central characters of the transposition class-sums $\omega_{[(2)]}^\Lambda$ will be degenerate for two irreps, hence the need to compute the three-cycle class sum. Moreover, it allows to study various kind of mixtures, \textit{e.g.} a completely imbalanced three-component mixtures of the type $(3^F,2^F,1^F)$. Thus, we can observe new features in these few-body systems that allow to better understand the large $N$ behaviors of the solutions of Eq.~\eqref{mainham}. We studied these $N=6$ systems in detail, in \cite{Decamp2016,Decamp2016-2} for fermionic mixtures and \cite{Decamp2017} for Bose-Fermi mixtures.

There are $p(6)=11$ non-isomorphic irreps of $\mathfrak{S}_N$. Since we will study various quantum mixtures, we will enumerate all the associated Young diagrams and compute all the corresponding central characters $\omega_{[(2)]}^\Lambda$ and $\omega_{[(3)]}^\Lambda$. The results are summarized in table \ref{tabcentch}.

\begin{figure}
\begin{center}
  \includegraphics[width=0.9\linewidth]{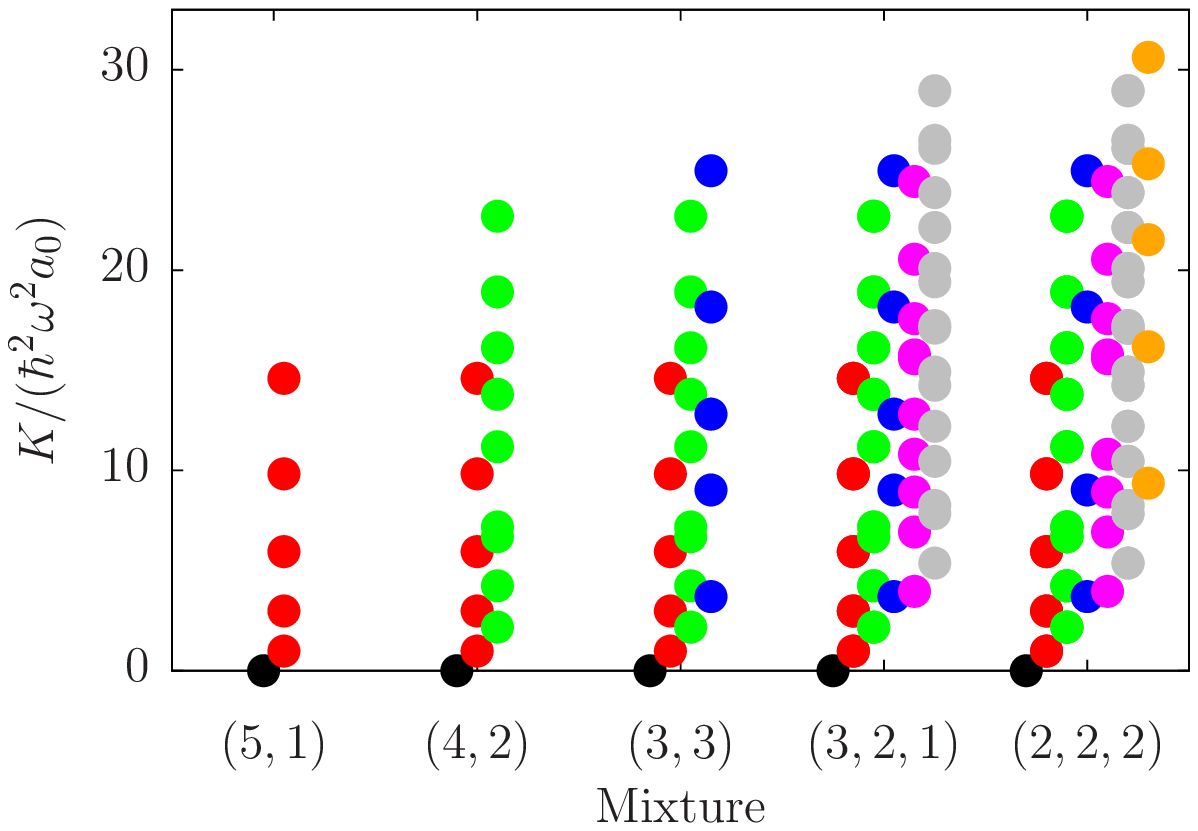}
      \begin{tabular}{p{0.2\linewidth}p{0.2\linewidth}p{0.2\linewidth}p{0.2\linewidth}}
      \textcolor{black}{$\bm{\scriptstyle{\yng(1,1,1,1,1,1)}}$} & \textcolor{red}{$\bm{\scriptstyle{\yng(2,1,1,1,1)}}$} & \textcolor{green}{$\bm{\scriptstyle{\yng(2,2,1,1)}}$}&\textcolor{blue}{$\bm{\scriptstyle{\yng(2,2,2)}}$}\\[0.3\linewidth]
      \\
      \textcolor{magenta}{$\bm{\scriptstyle{\yng(3,1,1,1)}}$} & \textcolor{gray}{$\bm{\scriptstyle{\yng(3,2,1)}}$} & \textcolor{orange}{$\bm{\scriptstyle{\yng(3,3)}}$}\\[0.2\linewidth]
          \end{tabular}
  \caption{Energy slopes $K$ for different kind of fermionic mixtures, with the associated symmetries given by their colors. The dots are shifted from left 
  to right with the same order of appearance than the symmetries.
  These results where published in \cite{Decamp2016-2}. 
  \label{fig:symferm}}
  \end{center}
\end{figure}

We observe that conjugate irreps have equal dimensions and three-cycle central characters, and opposite transposition central characters. 
Moreover, as we previously mentioned, we see that there are two degeneracies in the transposition central characters, namely 
$\omega^{\left[(2)\right]}_{[3,3]}=\omega^{\left[(2)\right]}_{[4,1,1]}=3$ and 
$\omega^{\left[(2)\right]}_{[3,1,1,1]}=\omega^{\left[(2)\right]}_{[2,2,2]}=-3$. This degeneracy is no longer present for the three-cycle central 
characters, and therefore we do not have to compute the four-cycle central characters.

The next step is then to compute $D_{\mathrm{sec}}\left(C_{[(2)]}\right)$ and $D_{\mathrm{sec}}\left(C_{[(3)]}\right)$. We have developed a \texttt{Mathematica} 
program similar to the one we created for the computing the $V$ matrix (see section \ref{volansatz}) that computes $D_{\mathrm{sec}}\left(C_{[(2)]}\right)$ and $D_{\mathrm{sec}}\left(C_{[(3)]}\right)$ 
$N!\times N!$ matrices for any $N$ (with a complexity of $\mathcal{O}(N!^2)$), and diagonalize them. In the $N=6$ case it consists in a $720\times 720$ matrix that we will of 
course not present here.  However, if we consider the example $(4^F,2^F)$
of section \ref{volansatz}, we can write in the snippet basis of Fig. \ref{fig:exvol}:
\begin{equation}
\setcounter{MaxMatrixCols}{15}
D_{\mathrm{snip}}^{(4^F,2^F)}\left(C_{[(2)]}\right)=-\begin{pmatrix} 
7 & 1 & 1 & 1 & 1 & 0 & 1 & 1 & 0 & 0 & 1 & 1 & 0 & 0 & 0 \\
1 & 7 & 1 & 1 & 0 & 1 & 1 & 0 & 1 & 0 & 1 & 0 & 1 & 0 & 0 \\
1 & 1 & 7 & 0 & 1 & 1 & 0 & 1 & 1 & 0 & 0 & 1 & 1 & 0 & 0 \\
1 & 1 & 0 & 7 & 1 & 1 & 1 & 0 & 0 & 1 & 1 & 0 & 0 & 1 & 0 \\
1 & 0 & 1 & 1 & 7 & 1 & 0 & 1 & 0 & 1 & 0 & 1 & 0 & 1 & 0 \\
0 & 1 & 1 & 1 & 1 & 7 & 0 & 0 & 1 & 1 & 0 & 0 & 1 & 1 & 0 \\
1 & 1 & 0 & 1 & 0 & 0 & 7 & 1 & 1 & 1 & 1 & 0 & 0 & 0 & 1 \\
1 & 0 & 1 & 0 & 1 & 0 & 1 & 7 & 1 & 1 & 0 & 1 & 0 & 0 & 1 \\
0 & 1 & 1 & 0 & 0 & 1 & 1 & 1 & 7 & 1 & 0 & 0 & 1 & 0 & 1 \\
0 & 0 & 0 & 1 & 1 & 1 & 1 & 1 & 1 & 7 & 0 & 0 & 0 & 1 & 1 \\
1 & 1 & 0 & 1 & 0 & 0 & 1 & 0 & 0 & 0 & 7 & 1 & 1 & 1 & 1 \\
1 & 0 & 1 & 0 & 1 & 0 & 0 & 1 & 0 & 0 & 1 & 7 & 1 & 1 & 1 \\
0 & 1 & 1 & 0 & 0 & 1 & 0 & 0 & 1 & 0 & 1 & 1 & 7 & 1 & 1 \\
0 & 0 & 0 & 1 & 1 & 1 & 0 & 0 & 0 & 1 & 1 & 1 & 1 & 7 & 1 \\
0 & 0 & 0 & 0 & 0 & 0 & 1 & 1 & 1 & 1 & 1 & 1 & 1 & 1 & 7 \\
\end{pmatrix},
\end{equation}
which is similar to 
\begin{equation}
\mathrm{diag}(-15, -9, -9, -9, -9, -9, -5, -5, -5, -5, -5, -5, -5, -5, -5). 
\end{equation}
As expected, it corresponds to the $\omega^{\left[(2)\right]}_\Gamma$'s 
with $\Gamma\in\{[2,2,1,1],[2,1,1,1,1],[1,1,1,1,1,1]\}$ and with multiplicities equal to the dimensions of the associated irreps (c.f. table \ref{tabcentch}). Note that 
for this mixture it is not necessary to compute $D_{\mathrm{snip}}^{(4^F,2^F)}\left(C_{[(3)]}\right)$, but it is \textit{e.g.} for $(2^F,2^F,2^F)$. If we diagonalize 
the $V$-matrix associated with this mixture (c.f. Fig. \ref{fig:exvol}), we obtain a set of energy slopes $K$ (Eqs.~\eqref{kvolo} and \eqref{diagkeq}) and $\vec{a}$ vectors whose 
symmetries are obtained by projecting the $\vec{a}$ vectors on the eigenbasis of $D_{\mathrm{snip}}^{(4^F,2^F)}\left(C_{[(2)]}\right)$.

Results for various fermionic mixtures are given in Fig.~\ref{fig:symferm}. By comparing it to table \ref{tabcentch}, we observe that the number of states 
that correspond to a given symmetry $\Gamma$ is equal to $\dim D_{\Gamma}$, and that, for a given mixture $(N_1,\ldots,N_{\kappa})$ with symmetries $\Gamma_1,\ldots,\Gamma_k$, we have
\begin{equation}
\sum_{i=1}^k\dim D_{\Gamma_i}=D_{N_1,\ldots,N_{\kappa}},
\end{equation}
where $D_{N_1,\ldots,N_{\kappa}}$ is the dimension of the degenerate manifold defined in Eq.~\eqref{degeneracy}. This is indeed expected from the discussion 
in section \ref{importancesym} (see Eqs.~\eqref{hamcomut}, \eqref{splitrep} and \eqref{splith} with $G=\mathfrak{S}_N$). Moreover, note that the set of energy slopes for a given irrep is independent of the choice of the mixture.

\subsection{Ordering of energy levels}
\label{lmandb}

\subsubsection{The Lieb-Mattis theorem}

The Lieb-Mattis theorem \cite{LiebMattisPR} is a fundamental theorem of many-body quantum physics that links the ordering of energy levels with their 
relative symmetries, and that has important consequences in condensed matter and the theory of magnetism. In this section we will enunciate and prove both 
versions of the theorem, following the original work of Lieb and Mattis, and then discuss some of their consequences.

\paragraph{The first Lieb-Mattis theorem (LMT I)}

The first and most well-known version of the Lieb-Mattis theorem, that we will refer as LMT I in the following, concerns spin-$1/2$ particles, or $SU(2)$ 
gases. Before enunciating it, let us first define our notations. Let us consider a system of $N$ particles divided into two fermionic components and 
subjected to a general Hamiltonian of the form (in units of $\hbar=2m=1$):
\begin{equation}
\label{hamlm}
\hat{H}=-\sum_{i=1}^N \frac{\partial^2}{\partial x_i^2}+U(x_1,\ldots,x_N),
\end{equation}
where $U$ is a real, permutational invariant potential that contains both the external and interaction potentials. The many-body wave function $\Psi$ can 
be subjected to various boundary conditions: $\Psi=0$ or $\partial \Psi/\partial x_i=0$ whenever $x_i=0$ or $x_i=L$ if particles are on a box of size $L$, 
or $\Psi\in L^2(\mathbb{R}^N)$ if particles are, for example, in a harmonic trap.

We will be interested in states ${}_M^S\Psi$ with a definite total spin $S$ and a definite total azimuthal quantum number $M$. These numbers are 
classically defined as
\begin{equation}
\vec{S}^2~{}_M^S\Psi=S(S+1)~{}_M^S\Psi,
\end{equation}   
and
\begin{equation}
\vec{S}_z~{}_M^S\Psi=M~{}_M^S\Psi,
\end{equation}
where $\vec{S}^2=\sum_{i=1}^N(\vec{S}^2)^i$ and $\vec{S}_z=\sum_{i=1}^N\vec{S}_z^i$ are the usual spin operators. For given $S$ and $M$, let us denote by 
$E(S)$ and $E(M)$ (resp.) the \textit{ground-state energies} of $S$ and $M$ (resp.), that is the minimum eigenvalues of $\hat{H}$ in Eq.~\eqref{hamlm} whose 
associated eigenstates have spin and azimuthal numbers $S$ and $M$ (resp.).

The first Lieb-Mattis theorem can be enunciated as follows:
\vspace{0.5cm}

\begin{center}
\fbox{\begin{minipage}{0.9\textwidth}
\textbf{Theorem (LMT I):} \textit{If $S>S'$, then $E(S)\ge E(S')$. Moreover, $E(S)=E(S')$  only if $U$ is pathological (in a sense that will be 
specified in the proof).}
\end{minipage}}
\end{center}
\vspace{0.5cm}

In order to prove this theorem, remark that it is sufficient to prove that, at fixed $M\ge 0$ (the opposite case is of course similar), 
the associate ground-state wave function ${}_M\Psi$ should have $S=M$, and hence $E(M)=E(S)$. Indeed, if we consider $S>S'$, the ground-state 
${}^S\Psi$ corresponding to $E(S)$ is degenerate and could have any azimuthal number $M\in\{-S,-S+1,\ldots,S-1,S\}$, and in particular ${}^S_{M=S'}\Psi$ has 
an energy $E(S)$. But since the ground-state energy of states with an azimuthal number equal to $S'$ should have a spin also equal to $S'$, we have  $E(S')\le E(S)$.

Let us consider $M\ge 0$. A typical total wave function ${}_M\Psi$ with an azimuthal number $M$ is given by:
\begin{equation}
{}_M\Psi=\sum_{P\in\mathfrak{S}_N}\epsilon(P)(\hat{P}{}_M\phi)(\hat{P}{}_MG),
\end{equation}
where the spin wave function ${}_MG$ has the form
\begin{equation}
{}_MG=(--\cdots--++\cdots++),
\end{equation}
with $p\equiv N-M$ spin down $(-)$ and $N-p$ spin up $(+)$, and the spatial wave function ${}_M\phi$ has the form
\begin{equation}
\label{phimform}
{}_M\phi=\phi(x_1,\ldots,x_p|x_{p+1},\ldots,x_N),
\end{equation}
where $\phi$ is anti-symmetric by permutation of $x_1,\ldots,x_p$ and $x_{p+1},\ldots,x_N$.

Then, we can affirm that a necessary and sufficient condition on $\phi$ so that $S=M$ is that $\phi$ cannot be anti-symmetrized with respect to 
$x_p,x_{p+1},\ldots,x_N$, \textit{i.e.} that the bar "$|$" in $\phi$ cannot be moved to the left. Indeed, $S$ can, \textit{a priori}, take all values 
between $M,M+1,\ldots,N/2$. Then, $S=M$ is equivalent to $S_+{}_M^M\Psi=0$, which means for ${}_MG$ that one cannot transform one spin down in one spin 
up, and for $\phi$ that the bar cannot be moved to the left.

Now we are going to study the properties of the ground-state wave function with azimuthal number $M$. We define the fundamental domain $R_M$ of $\mathbb{R}^N$ by
\begin{equation}
x_1\le\cdots\le x_p
\end{equation}
and
\begin{equation}
x_{p+1}\le\cdots\le x_N.
\end{equation}
Then, one can prove that the Schr\"{o}dinger equation $\hat{H}\varphi=E\varphi$ in $R_M$ with the boundary conditions $\varphi=0$ on the boundary of $R_M$ has 
a positive ground-state solution $\varphi_0$. In fact, one can show that $\varphi_0$ is strictly positive in the interior of $R_M$ unless $U$ is \textit{pathological} 
in the sense that it contains "sufficiently strong infinities". An example of pathological potential is given by the fermionized limit $g_{1D}\to\infty$ in our system Eq.~\eqref{mainham}, 
which explains why states with different symmetries have the same energy in this limit.

Given $\varphi_0$, we define the function $\Phi_0$ on $\mathbb{R}^N$ by
\begin{equation}
\Phi_0=\epsilon(P)\epsilon(Q)\hat{P}\hat{Q}~\varphi_0\quad\text{in}\quad PQ(R_M),
\end{equation}
where $P\in\mathfrak{S}_p$ and $Q\in\mathfrak{S}_{N-p},$ $\hat{P}$ and $\hat{Q}$ permute the variables $x_1,\ldots,x_p$ and $x_{p+1},\ldots,x_N$ (resp.) and the domain $PQ(R_M)$ is defined by
\begin{equation}
x_{P1}\le\cdots\le x_{Pp}
\end{equation}
and
\begin{equation}
x_{Q(p+1)}\le\cdots\le x_{QN}.
\end{equation}
Then, $\Phi_0$ is a solution of the Schr\"{o}dinger equation $\hat{H}\phi=E\phi$ in $\mathbb{R}^N$ (thanks to the boundary conditions of $\varphi_0$)
with an azimuthal number of $M$ (because it is of the form of Eq.~\eqref{phimform}). In particular, since $\varphi_0$ is the ground state in $R_M$, $\Phi_0$ 
is the ground state wave function with an azimuthal number of $M$.

Let us show that $\Phi_0$ verifies $S=M$. To do so, we define the following function on $\mathbb{R}^N$:
\begin{equation}
\mathcal{V}(x_1,\ldots,x_N)=
\det\begin{pmatrix}1 & x_1  & \dots & x_1^{p-1}\\
1 & x_2  & \dots & x_2^{p-1}\\
\vdots & \vdots  & \ddots &\vdots \\
1 & x_p &  \dots & x_p^{p-1}\end{pmatrix}
\det\begin{pmatrix}1 & x_{p+1} & \dots & x_{p+1}^{N-p-1}\\
1 & x_{p+2} & \dots & x_{p+2}^{N-p-1}\\
\vdots & \vdots & \ddots &\vdots \\
1 & x_N & \dots & x_N^{N-p-1}\end{pmatrix},
\end{equation}
a product of two Vandermonde determinant, which is then also equal to
\begin{equation}
\mathcal{V}(x_1,\ldots,x_N)=\prod_{1\le i<j\le p}(x_j-x_i)\prod_{p+1\le k<l\le N}(x_l-x_k).
\end{equation}
It is clear that $\mathcal{V}$ is totally anti-symmetric in the variables $x_1,\ldots,x_p$ and $x_{p+1},\ldots,x_N$ and then is a function of the 
form of Eq.~\eqref{phimform}, and moreover that "the bar cannot be moved to the left": hence $\mathcal{V}$ is characterized by an azimuthal number $M$ and by 
a spin $S=M$. Besides $\mathcal{V}$ is strictly positive in the fundamental domain $R_M$. Finally, we can define the following scalar product for functions 
of the type of Eq.~\eqref{phimform}:
\begin{equation}
\braket{{}_Mf,{}_Mg}_M=\int_{\mathbb{R}^N}{}_Mf~{}_Mg=p!(N-p)!\int_{R_M}{}_Mf~{}_Mg.
\end{equation}
If two functions ${}^S_Mf,{}^{S'}_Mg$ have definite spin values $S,S'$, one can easily see that $\braket{{}^S_Mf,{}^{S'}_Mg}_M\ne 0$ only if $S=S'$. We can 
say that the spin labels the irreps of $SU(2)$ and that non-equivalent irreps are orthogonal. Therefore, since $\varphi_0\ge0$ and 
$\mathcal{V}>0$ on $R_M$, we can conclude that $\Phi_0$ verifies $S=M$. Thus, we have completed the proof of LMT I.

\paragraph{Second Lieb-Mattis theorem (LMT II) and pouring principle}

The second Lieb-Mattis theorem is a direct generalization of LMT I to $SU(\kappa)$ systems. Although very powerful, we will see that it does not allow to 
compare all the ground states energies of the irreps of $\mathfrak{S}_N$. Here, more than giving a detailed proof, we will develop the analogy with the last paragraph. 

Instead of considering two-component fermionic particles subjected to the Hamiltonian defined in Eq.~\eqref{hamlm}, we do not specify the mixture and allow it to obey any statistics. In other words, it can belong to any irrep 
of $\mathfrak{S}_N$. 

For LMT I, particles could belong to any irrep of the type ${}^t\left[N/2+S,N/2-S\right]$, where the notation ${}^t\left[~\right]$ means that we specify 
the number of boxes in the columns of the corresponding Young diagram\footnote{Notice the difference of convention with section \ref{subsecpermut}, which is 
appropriate here since we considered fermionic species in LMT I.}. Then the generalization of $S$ is to specify the irrep $D_{{}^t\Gamma}$ associated 
with a partition ${}^t\Gamma$.

Moreover, functions with an azimuthal number $M$ where characterized by being of the type of Eq.~\eqref{phimform}. Therefore, instead of considering $M$, we 
directly consider functions of the form:
\begin{equation}
\label{phimfgen}
\phi(x_1,\ldots|\ldots,x_{N-N_1-N_2}|x_{N-N_1-N_2+1},\ldots,x_{N-N_1}|x_{N-N_1+1},\dots,x_N),
\end{equation}
where variables between two bars are anti-symmetrized. Instead of choosing $M\ge 0$, here we set $N_1\ge N_2\ge\cdots$. Besides, bars can always be moved to 
the right without changing of irrep (we can always lower $M$), and a necessary and sufficient condition for a function of the type of Eq.~\eqref{phimfgen} to belong to the irrep 
${}^t\left[N_1,N_2,\ldots\right]$ ($S=M$ in LMT I) is that the bars cannot be moved to the left.

The central idea of the proof of LMT I was that $E(S)$ was degenerate for $M$ and that therefore it was sufficient to prove that  the ground state wave function 
of azimuthal number $M$ should have $S=M$. Here this degeneracy is more subtle: for a partition ${}^t\Gamma$, the only functions of the type of Eq.~\eqref{phimfgen} 
that can belong to $D_{{}^t\Gamma}$ are the ones when the bars are not moved to the left. Thus, we cannot compare all the irreps.

Accordingly, we define a partial order $\succ$ on the irreps: considering $D_{{}^t\Gamma},D_{{}^t\Gamma'}$ with ${}^t\Gamma={}^t\left[N_1,N_2,N_3,\ldots\right]$ 
and ${}^t\Gamma'={}^t\left[N_1',N_2',N_3',\ldots\right]$, we say that $D_{{}^t\Gamma}$ can be \textit{poured} into $D_{{}^t\Gamma'}$ and write 
${}^t\Gamma\succ{}^t\Gamma'$ if $N_1\ge N_1'$, $(N_1-N_1')+N_2\ge N_2'$, $(N_1-N_1'+N_2-N_2')+N_3\ge N_3'$, etc. In other words, $Y_{{}^t\Gamma}$ can 
be transformed into $Y_{{}^t\Gamma'}$ by moving the boxes of the diagrams to the right. In the context of algebric combinatorics and representation theory, $\succ$ is known as the \textit{dominance order} \cite{book:491124}. For example, we have
\begin{equation}
\yng(3,3,2)\quad\succ\quad\yng(4,4)~,
\end{equation}
or
\begin{equation}
\yng(2,2,2)\quad\succ\quad\yng(5,1)~.
\end{equation}
However, it is impossible to compare these two diagrams:
\begin{equation}
\yng(3,2,2,1,1,1)\quad\text{and}\quad\yng(2,2,2,2,2)~.
\end{equation}

Writing $E({}^t\Gamma)$ the ground state energy associated with the symmetry class ${}^t\Gamma$, we can now enunciate the second Lieb-Mattis theorem:
\vspace{0.5cm}

\begin{center}
\fbox{\begin{minipage}{0.9\textwidth}
\textbf{Theorem (LMT II):} \textit{If ${}^t\Gamma\succ{}^t\Gamma'$, then $E({}^t\Gamma)\ge E({}^t\Gamma')$. Moreover, $E({}^t\Gamma)=E({}^t\Gamma')$ only if $U$ is pathological.}
\end{minipage}}
\end{center}
\vspace{0.5cm}

The proof of this theorem is now very similar to the proof of LMT I: considering the ground state wave function of the form of Eq.~\eqref{phimfgen}, we show 
that it is positive on the fundamental domain defined analogously to $R_M$. Then we consider an appropriate product of Vandermonde determinant similar to $\mathcal{V}$, 
which belongs to the irrep $D_{{}^t\Gamma}$ with ${}^t\Gamma={}^t\left[N_1,N_2,N_3,\dots\right]$ and is also positive on the aforementioned fundamental domain. 
Thus it is not orthogonal to the ground state wave function, which therefore belongs to $D_{{}^t\Gamma}$.

\begin{figure}\centering
\begin{tikzpicture}[scale=1.1]
\draw (2,0) node{\footnotesize{\yng(1,1,1,1,1,1)}};
\draw (2,2.9) node{\footnotesize{\yng(2,1,1,1,1)}};
\draw (2,5.5) node{\footnotesize{\yng(2,2,1,1)}};
\draw (-1,7) node{\footnotesize{\yng(3,1,1,1)}};
\draw (5,7) node{\footnotesize{\yng(2,2,2)}};
\draw (2,8.5) node{\footnotesize{\yng(3,2,1)}};
\draw (-1,10) node{\footnotesize{\yng(4,1,1)}};
\draw (5,10) node{\footnotesize{\yng(3,3)}};
\draw (2,11.8) node{\footnotesize{\yng(4,2)}};
\draw (2,13.6) node{\footnotesize{\yng(5,1)}};
\draw (2,15.4) node{\footnotesize{\yng(6)}};
\draw[line width = 1mm,gray] (2,1.2) -- (2,1.85);
\draw[line width = 1mm,gray] (2,3.9) -- (2,4.65);
\draw[line width = 1mm,gray] (1,6) -- (-1,6.5);
\draw[line width = 1mm,gray] (3,6) -- (4.8,6.3);
\draw[line width = 1mm,gray] (1,9) -- (-1,9.5);
\draw[line width = 1mm,gray] (3,9) -- (5,9.5);
\draw[line width = 1mm,gray] (1,8.3) -- (-1,7.9);
\draw[line width = 1mm,gray] (3,8.3) -- (5,7.7);
\draw[line width = 1mm,gray] (1,11.5) -- (-1,10.7);
\draw[line width = 1mm,gray] (3,11.5) -- (5,10.5);
\draw[line width = 1mm,gray] (2,12.3) -- (2,13.5);
\draw[line width = 1mm,gray] (2,14.1) -- (2,15.1);
\end{tikzpicture}
\caption{\label{domord}Hasse diagram of the partially ordered set $(I_6,\succ)$.}
\end{figure}
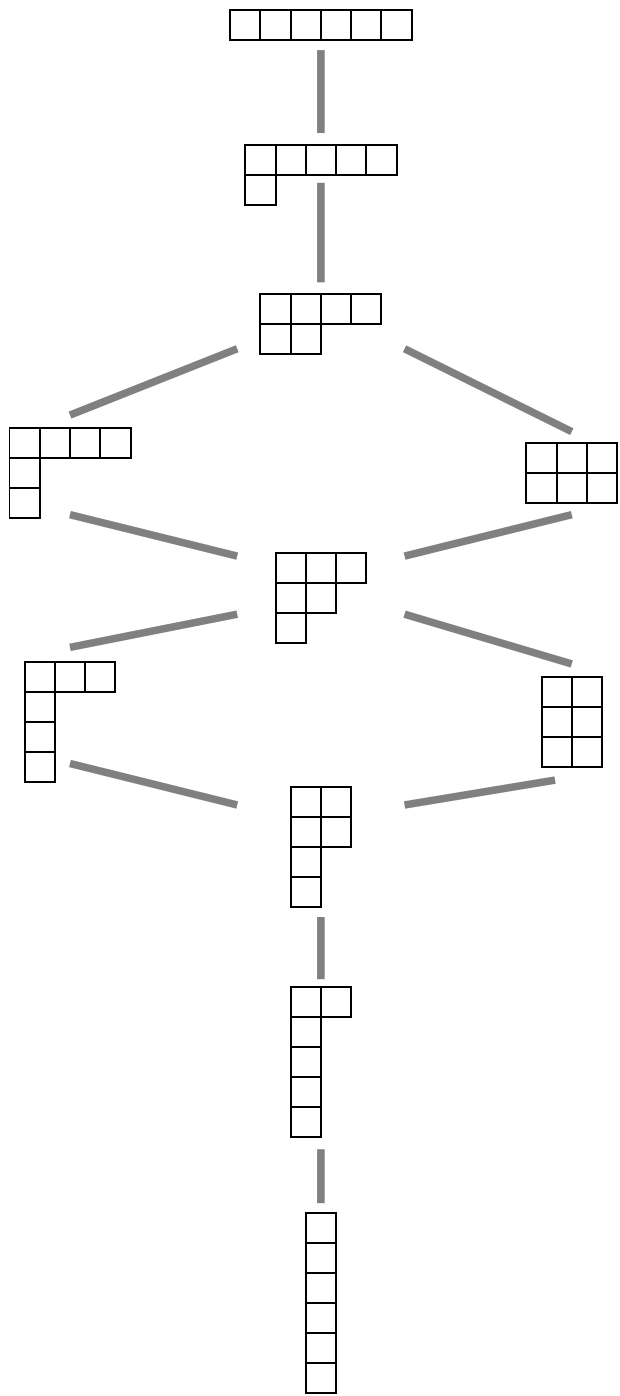

\paragraph{Consequences for the theory of magnetism}

Lieb and Mattis have discussed the consequences of LMT I for the theory of magnetism, by saying that a two-component fermionic system that obeys to the Hamiltonian in Eq.~\eqref{hamlm} has an non-ferromagnetic ground state. It is clear indeed that such a system will not have a non-zero $S$ value that grows proportionally with its size, since the ground state has a minimal spin.

In the case of a one-component Bose gas, on the contrary, since LMT II implies that the \textit{spatial wavefunction} is as symmetric as possible and since the total wavefunction of a Bose gas is totally symmetric, it implies that the spin wavefunction is also totally symmetric. Thus, a consequence of LMT II is that the ground state of a one-dimensional Bose gas is fully polarized. It was shown in \cite{Lieb2002} that even in three-dimensional systems at finite temperature, it is  also the case as long as there are no spin-dependent forces.

Moreover, Lieb and Mattis have shown, with the help of LMT II, that it is possible to extend LMT I to higher dimensions if the potential $U$ is \textit{separably symmetric} (\textit{e.g.} for a two-dimensional potential $U(x_1,\ldots,x_N;y_1,\ldots,y_N)$, $U$ is symmetric in the $x_i$ coordinates of the $N$ particles, and also in the $y_i$). Therefore, it does not apply, for example, to three-dimensional systems with central forces among the particles. 

More generally, the Lieb-Mattis theorems do not apply to systems with momentum-dependent or spin-dependent forces. It does apply, however, to one-dimensional quantum gases that can be simulated with cold-atom experiments such as the ones discussed in this thesis.

\subsubsection{Analysis in our system}

First of all, let us note that, although the $1/g_{1D}=0$ point corresponds to a pathological case of Lieb-Mattis theorem and is degenerate for the irreps, 
the set of energy slopes $K$ allows to deduce the ordering of energy levels in the vicinity of that point (see \textit{e.g.} Fig. \ref{figvolo}). Notice 
also that, because of the minus sign in $K$'s definition (Eq.~\eqref{perten}), the ground state energies will correspond to the maximum $K$ values. 
Accordingly, given a symmetry class $\Gamma$, we define $K(\Gamma)$ as the maximum $K$ whose corresponding state belongs to $D_{\Gamma}$. With this notation, 
LMT II becomes, in our context:
\begin{equation}
\Gamma\succ\Gamma'\quad\Rightarrow\quad K(\Gamma)< K(\Gamma'),
\end{equation}
where we have written "$<$" instead of "$\le$" since the interaction potential in Eq.~\eqref{mainham} is non-pathological in the vicinity of the degenerate 
manifold $1/g_{1D}=0$\footnote{Since the interaction potential is pathological in the degenerate manifold, the energetic counterpart should still be written 
$\Gamma\succ\Gamma'~\Rightarrow~E(\Gamma)\ge E(\Gamma')$.}.

Before analyzing our results, let us first see how the set $I_6$ of irreps of $\mathfrak{S}_6$ is partially ordered by the dominance order $\succ$. 
The conventional way of representing a partially ordered $(E,\le)$ set is through a \textit{Hasse diagram} \cite{Simovici2014}, \textit{i.e.} a graph 
(see section \ref{volansatz}) where each vertex is an element of $E$, and if an edge relies two vertices $A$ and $B$ where if $A$ is above $B$, then 
$A\ge B$. The Hasse diagram of $(I_6,\succ)$ is given in Fig. \ref{domord}.

\begin{table}\centering
\renewcommand{\arraystretch}{2.5}
\subfloat[Two-component Bose-Fermi mixtures.]{
\begin{tabular}{|c|c|c|}\hline
Mixture & $Y_{\Gamma}$ & $K(\Gamma)/(\hbar^2\omega^2a_0)$ \\\hline\hline
$(2^B,2^F)$ & {\tiny\yng(3,1)} &  10.66 \\
          & {\tiny\yng(2,1,1)} &   7.08 \\\hline
$(3^B,3^F)$ & {\tiny\yng(4,1,1)} &  30.37 \\
          & {\tiny\yng(3,1,1,1)} &  24.43 \\\hline
\end{tabular}
\label{tab:sym1}
}
\subfloat[Three-component Bose-Fermi mixtures.]{
\begin{tabular}{|c|c|c|}\hline
Mixtures & $Y_{\Gamma}$ &  $K(\Gamma)/(\hbar^2\omega^2a_0)$ \\\hline\hline
 & \textcolor{blue}{{\tiny\yng(5,1)}} & 33.35 \\
& {\tiny\yng(4,2)} &  32.16 \\
     \textcolor{blue}{$(2^B,2^B,2^F)$}     & \textcolor{blue}{{\tiny\yng(3,3)}} & 30.63 \\
          & {\tiny\yng(4,1,1)} & 30.37 \\
          & {\tiny\yng(3,2,1)} &  28.96 \\
          & \textcolor{red}{{\tiny\yng(2,2,2)}} & 24.97 \\
   \textcolor{red}{$(2^B,2^F,2^F)$}        & {\tiny\yng(3,1,1,1)} & 24.43 \\
          & {\tiny\yng(2,2,1,1)} & 22.69 \\
          & \textcolor{red}{{\tiny\yng(2,1,1,1,1)}} & 14.60 \\\hline
\end{tabular}
\label{tab:sym2}
}
\caption{\label{tab:symbf}Symmetry classes $\Gamma$ and corresponding ground-state energy slopes $K(\Gamma)$ for two- (\ref{tab:sym1}) and three- (\ref{tab:sym2}) 
component Bose-Fermi mixtures. In table \ref{tab:sym2}, the black diagrams are common to both mixtures, whereas the blue (resp. red) diagrams 
are specifically associated to $(2^B,2^B,2^F)$ (resp. $(2^B,2^F,2^F)$). These results where published in \cite{Decamp2017}.}
\end{table}

Now, a direct comparison of Fig. \ref{domord} and Fig. \ref{fig:symferm} shows that our system, in the fermionic case, verifies LMT II. In \cite{Decamp2017}, 
we extended this result to Bose-Fermi mixtures, as one can see in table \ref{tab:symbf}. On a side note, remark that three-component Bose-Fermi mixtures display 
much more complex symmetry structures than their two-component counterparts. Indeed, mixtures of the type $(N_1^B,N_1^F)$ have only two possible conjugate irreps 
associated with $\Gamma=[N_1^B,1,\ldots,1]$ and $\Gamma=[N_1^B+1,1,\ldots,1]$ \cite{Fang2011}. Besides, the $K(\Gamma)$ depend only on $\Gamma$ and are independent 
of the mixture, which shows the fundamental importance of symmetries in our system.

Moreover, we are able to compare energy levels that goes beyond the scope of LMT II and that are not comparable by the dominance order $\succ$ and the pouring 
principle. Indeed, we observe in Fig. \ref{fig:symferm} and table \ref{tab:symbf} that, although that the symmetry classes $[4,1,1]$ (resp. $[3,1,1,1]$) are not 
comparable with $[3,3]$ (resp. $[2,2,2]$) by the dominance order $\succ$ (c.f. Fig \ref{domord}), we have proved:
\begin{equation}
K\left([3,3]\right)>K\left([4,1,1]\right)\quad\text{and}\quad K\left([2,2,2]\right)>K\left([3,1,1,1]\right).
\end{equation}
This ordering was also obtained for fermionic mixtures using Bethe ansatz equations in the homogeneous case and local density approximation in the harmonic 
trap \cite{Lei2017}.

Interestingly, the two pairs of irreps that are not comparable by the dominance order are also the ones who have identical transposition central characters 
$\omega^{\left[(2)\right]}_\Gamma$ (c.f. table \ref{tabcentch}). Moreover, we observe that:
\begin{equation}
\label{conjcc}
\Gamma\succ\Gamma' \quad\Leftrightarrow\quad \omega^{\left[(2)\right]}_\Gamma < \omega^{\left[(2)\right]}_{\Gamma'}
\end{equation}
This suggests that a profound analysis of the central characters would allow to predict the energy ordering beyond LMT II. The idea would be to define an order "$\triangleright$" on the set of $\omega^{\left[\Lambda\right]}_\Gamma$'s that would, unlike $\succ$, be a \textit{total order}. For the moment, we only have conjectures of the type of Eq.~\eqref{conjcc}. A deeper analysis, using the full power of representation theory, is required and in progress.

Let us conclude this section by two remarks: first, the \textit{a priori} knowledge of the irrep $D_{\Gamma}$ of the ground-state of our system is extremely valuable information in terms of practical computing. Indeed, the dimension $\dim D_{\Gamma}$ of this irrep is much lesser than the dimension of the total space, which implies faster and less memory-expensive programs \cite{PhysRevLett.113.127204,PhysRevB.93.155134,PhysRevB.96.115159}. Second, although our analysis is done in the fermionized limit $g_{1D}\to\infty$, there are good reasons to think that the energy ordering is the same for all $0<g_{1D}<\infty$. Indeed, since the degeneracies of $\hat{H}$ are equal to the dimensions of the irreps of its symmetry group (c.f. section \ref{importancesym}), there are no energy level crossings unless there is an additional symmetry for a certain value of $g_{1D}$ \cite{Harshman2014}, which is unlikely (but would require however a rigorous proof). 


\clearemptydoublepage

\clearemptydoublepage
\pagestyle{fancy}
\thispagestyle{empty}
\chapter{One-body correlations. Tan's contact}
\label{chap:1bcor}

\minitoc
\newpage

%
%
%

Correlation functions are extremely important quantities in many-body quantum physics: Usually easier to compute than the total wave functions, they are involved in the characterization of many exotic properties of quantum matter, ranging for the celebrated Bose-Einstein condensation to the quasi-long range order in one dimension. Moreover, they can be extracted in a typical ultracold atom experiment. 

In this chapter, we focus on the properties of the first order correlation functions, from which a large number of properties of the system can be deduced. In particular, we analyze what are the effects of the strong interactions and of the permutational symmetry that we have characterized in chapter \ref{chap:sym}. The leitmotiv of this study is the following: How to extract uniquely the symmetry of the system from a measure of the one-body correlation? In section \ref{sec:1bdc}, we analyze in detail the exact density and momentum profiles of few body strongly interacting quantum mixtures from the exact solution computed in chapter \ref{Exactsol}, trying to extract some general features. Then, in section \ref{sectancont}, we focus on the so-called \textit{Tan's contact}, a quantity that governs the high-momentum behavior and has become a pivot in the study of short-range interacting quantum gases, showing especially that it allows to answer this chapter's main question. Furthermore, in order to be the more experimentally relevant as possible, we derive Tan's contact dependences on the interaction strength, temperature, and transverse confinement.

\section{One-body correlations}
\label{sec:1bdc}

In this section, we discuss the properties of the one-body correlations in strongly repulsive 
one-dimensional quantum mixtures. After theoretically defining this quantity and recalling 
the experimental ways that are used in order to measure it in cold atom set ups in \ref{gen1bdc}, we 
will then explain in \ref{sec:ex1bdcqualitative} how we obtained it from the exact solutions computed in section \ref{secstrong}. In particular, 
we compute and analyze the  density profiles and momentum distributions of $N=6$ mixtures, focusing on the effects 
of strong interactions and symmetries.

\subsection{Generalities}
\label{gen1bdc}

\subsubsection{Definitions}

The notion of correlation was first introduced in the mathematical theory of statistics \cite{Feller1947}. Intuitively, given several random/statistical 
variables, it allows to measure how these variables are linked together on average. This notion is extensively used in many areas of science, ranging 
from financial analysis and  sociology to optics and statistical physics. In the latter, the universal behaviors of correlation functions in the vicinity 
of critical points are at the heart of the theory of phase transitions. 

In quantum physics, which is an intrinsically probabilistic theory, quantum correlation functions have proven to be a very efficient theoretical tool. 
They are for example extensively used in quantum field theories\footnote{In this context, correlation functions are sometimes called "Green functions", 
which must not be confused with the Green functions defined in functional analysis as the inverts of differential operators!}, where the Feynman paradigm 
of path integrals and Feynman diagrams have proven to be a very efficient computational tool.

Now, let us give some general definitions. Suppose that a one-dimensional quantum system is described, in the second quantization formalism, by quantum 
field creation and annihilation operators $\hat{\Psi}^{\dagger}(x)$ and $\hat{\Psi}(x)$ (respectively), where $x$ is the spatial coordinate. Then, the 
\textit{first order} correlation function $G^{(1)}$, also called \textit{one-body density matrix} or \textit{one-point correlators}, is defined as 
\cite{Pitaevskii_book}:
\begin{equation}
\label{g1secondqdef}
G^{(1)}(x,x')=\braket{\hat{\Psi}^{\dagger}(x)\hat{\Psi}(x')}.
\end{equation}
If the system can be described by a normalized $N$-particle many-body wave function $\Psi(x_1,\ldots,x_N)$, the first quantization analogue of Eq.~\eqref{g1secondqdef} is
\begin{equation}
\label{defg1}
G^{(1)}(x,x')=N\int dx_2\ldots x_N~\Psi^*(x,x_2,\ldots,x_N)\Psi(x',x_2,\ldots,x_N).
\end{equation}
In a similar way we can define the second order correlation function as
\begin{equation}
\label{g2def2nd}
G^{(2)}(x,x')=\braket{\hat{\Psi}^{\dagger}(x)\hat{\Psi}^{\dagger}(x')\hat{\Psi}(x)\hat{\Psi}(x')},
\end{equation}
and so on.

A lot of information can be obtained just from $G^{(1)}$. The diagonal part $n(x)$ of $G^{(1)}$ is the \textit{density profile} and is  associated with the probability (normalized to $N$) of finding a particle at point $x$:
\begin{equation}
n(x)=G^{(1)}(x,x)=N\int dx_2\ldots x_N~\left|\Psi(x,x_2,\ldots,x_N)\right|^2.
\end{equation}
The off-diagonal part of $G^{(1)}$ can be related to the \textit{momentum distribution} $n(k)$, associated with the probability  of having a particle with a momentum $k$, by the following formula:
\begin{equation}
\label{mddef}
\begin{split}
n(k)&=\braket{\hat{\Psi}^{\dagger}(k)\hat{\Psi}(k)}\\
&=\frac{1}{2\pi}\iint dx dx'~ G^{(1)}(x,x')e^{-\frac{i}{\hbar}k(x-x')},
\end{split}
\end{equation} 
which is simply due to $\hat{\Psi}(k)=\frac{1}{2\pi}\int dx~ \hat{\Psi}(x)e^{-\frac{i}{\hbar}kx}$. Note that we have
\begin{equation}
\int dx~ n(x) = \int dk ~n(k) = N.
\end{equation}
Notice also that we can extract the total kinetic energy $E_K$ from the momentum distribution by
\begin{equation}
E_K=\int dk~ \frac{k^2}{2}n(k),
\end{equation}
which is convergent only if we have
\begin{equation}
n(k)=\underset{k\to\infty}{o}\left(\frac{1}{k^3}\right).
\end{equation}
We will see in section \ref{sectancont} than in fact $n(k)\underset{\infty}{\sim}\mathcal{C}/k^4$. This asymptotic behavior for $n(k)$ is associated with the \textit{short-range} correlations of the system. Conversely, the asymptotic behavior $|x-y|\to\infty$ of $G^{(1)}(x,y)$ is associated with the \textit{long-range order}. It can serve as a characterization of the coherence properties of the system, and e.g. intervenes in the definition of a Bose-Einstein condensate \cite{Penrose1956,Yang1962}. Since it is equivalent to the low-momentum properties, it is well described by the Luttinger liquid theory and the bosonization method  that we briefly mentioned is section \ref{pec1d}. In this thesis, we focus more on the short-range properties, which go beyond the scope of Luttinger paradigm.

\subsubsection{Experimental probes}
\label{expcorr}

Not only correlation functions have a great theoretical importance and are often easier to calculate than the total wave functions, they also happen to be very efficiently measurable in cold atoms experiments \cite{Bloch2008,Cazalilla2011}.

\begin{figure}\centering
\subfloat[Image obtained by absorption imaging of a Bose-Einstein condensate of Rubidium atoms (courtesy of Guillaume Labeyrie).
  \label{fig:absim}]{
  \includegraphics[width=0.35\linewidth]{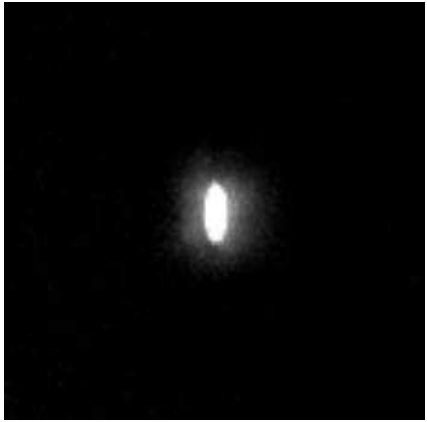}}\quad\quad
  \subfloat[\label{fig:bectof}Momentum distributions obtained by TOF and absorption imaging on a Rubidium gas. The two images on the right correspond to a Bose-Einstein condensate. From \cite{Anderson1995}.]{
  \includegraphics[width=0.53\linewidth]{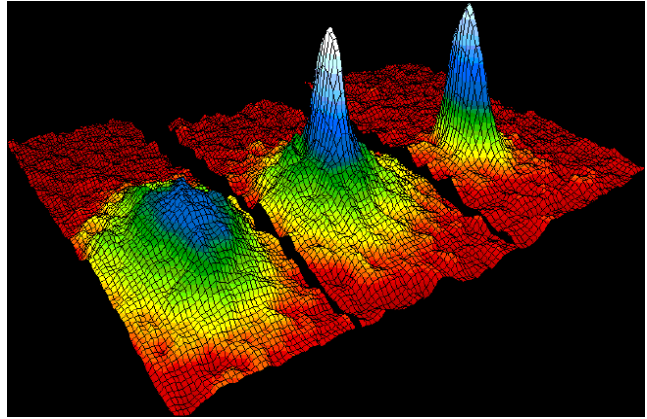}}
  \caption{\label{figabsimtof}Absorption imaging of Bose-Einstein condensates of Rubidium atoms.}
\end{figure}

The main method to measure the density profile of an atomic cloud is the \textit{in-situ} absorption imaging.  It basically consists in measuring the absorbed light with a CCD camera when the cloud is submitted to a resonant laser beam. Examples of images obtained with this method are given in Fig.~\ref{figabsimtof}.  

The usual way of measuring the momentum distribution $n(\vec{p})$ is called the \textit{time-of-flight} (TOF) technique. The idea is to suddenly release the trapping potential, and to measure with absorption imaging the density profile $n_{TOF}(\vec{r},\tau)$ after a time $\tau\gg mL^2/\hbar$, where $L$ is the linear size of the cloud before switching off the trap. If interactions can be safely neglected during the expansion so that it can be considered as ballistic, the momentum distribution in the trap $n(\vec{p})$ can be related to $n_{TOF}(\vec{r},\tau)$ by the simple formula:
\begin{equation}
n_{TOF}(\vec{r},\tau)=\left(\frac{m}{\hbar\tau}\right)^3n(\vec{p}(\vec{r})),
\end{equation}
where $\vec{p}(\vec{r})=m\vec{r}/\hbar\tau$. For example, in the case of a Bose-Einstein condensate, the absorption image after a TOF will display a typical central peak (see Fig~\ref{fig:bectof}).


\subsection{Exact one-body correlations for strongly repulsive systems}
\label{sec:ex1bdcqualitative}

\subsubsection{A formula for the one-body density matrix}

In this section we provide an expression we have obtained for the one-body density matrix $G^{(1)}(x,x')$ for the strongly interacting limit $g_{1D}\to +\infty$ of our model, using 
the notations of section \ref{volansatz}. This is the formula we used in order to plot the exact density profiles and exact momentum distributions in 
\cite{Decamp2016,Decamp2016-2,Decamp2017}. Here we will use units of $a_0=\sqrt{\hbar/m\omega}$ for length and $\hbar\omega$ for energy.

Suppose that our system is of the form $(N_1,\ldots,N_{\kappa})$ and that we want, for a given solution and thus a given vector $\vec{a}$ of size $N!$, and for a 
given spin-component $\sigma\in\{1\ldots,\kappa\}$, to compute the associated one-body density matrix:
\begin{equation}
\label{defg1sigma}
G^{(1)}_{\sigma}(x,x')=N_{\sigma}\int dx_2\ldots x_N~\Psi^*(x,x_2,\ldots,x_N)\Psi(x',x_2,\ldots,x_N),
\end{equation}
where "particle 1" belongs to component $\sigma$, $\Psi$ is given by Eq.~\eqref{Psivolo} and we have normalized $G^{(1)}_{\sigma}(x,x')$ to $N_{\sigma}$ so that the total one-body density verifies
\begin{equation}
G^{(1)}(x,x')=\sum_{\sigma=1}^{\kappa}G^{(1)}_{\sigma}(x,x').
\end{equation}
Let us organize the set of permutations $\mathfrak{S}_N$ in a convenient way: each permutation will be written
\begin{equation}
P_{\{i,k\}},\quad i\in\{1,\ldots,N\},~k\in\{1,\ldots,(N-1)!\},
\end{equation}
where $i$ denotes the position of particle 1 after permutation and $k$ labels the permutation of the $N-1$ other particles. With this notation, Eq.~\eqref{Psivolo} becomes
\begin{equation}
\label{Psivolg1}
\Psi(x_1,\ldots,x_N)=\sum_{i=1}^N\sum_{k=1}^{(N-1)!}a_{\{i,k\}}\theta_{\{i,k\}}(x_1,\ldots,  x_N)\psi_F(x_1,\ldots,x_N),
\end{equation}
where $a_{P_{\{i,k\}}}\equiv a_{\{i,k\}}$ and $\theta(x_{P_{\{i,k\}}1}<\cdots<  x_{P_{\{i,k\}}N})\equiv \theta_{\{i,k\}}(x_1,\ldots,x_N)$.

We  can now write $G^{(1)}_{\sigma}(x,x')$ by supposing $x\le x'$ (which is possible since $G^{(1)}_{\sigma}(x,x')=G^{(1)}_{\sigma}(x',x)$) and  using 
Eq.~\eqref{Psivolg1}:
\begin{equation}
\label{g1sig1}
\begin{split}
G^{(1)}_{\sigma}(x,x')=&N_{\sigma}\sum_{1\le i\le j \le N}\int dx_2\ldots dx_N\left[\sum_{k=1}^{(N-1)!}a_{\{i,k\}}\theta_{\{i,k\}}(x,x_2,\ldots,  x_N)\right.\\
&\times\Biggl.\psi_F(x,x_2,\ldots,x_N)\Biggr]\left[\sum_{l=1}^{(N-1)!}a_{\{j,l\}}\theta_{\{j,l\}}(x',x_2,\ldots,  x_N)\psi_F(x',x_2,\ldots,x_N)\right].
\end{split}
\end{equation}
Moreover, remark that 
\begin{equation}
\int dx_2\ldots dx_N \theta_{\{i,k\}}(x,x_2,\ldots,  x_N) \theta_{\{j,l\}}(x',x_2,\ldots,  x_N) (\cdots)\propto \delta_{kl},
\end{equation}
where $(\cdots)$ is an arbitrary function of $x,x',x_2,\ldots,x_N$. Notice also that the following product
\begin{equation}
\psi_F(x,x_2,\ldots,x_N)\psi_F(x',x_2,\ldots,x_N),
\end{equation}
is symmetric by any permutation of the variables $x_2,\ldots,x_N$. Therefore, Eq.~\eqref{g1sig1} becomes:
\begin{equation}
\label{g1sig2}
\begin{split}
G^{(1)}_{\sigma}(x,x')=&N_{\sigma}\sum_{1\le i\le j \le N}C_{ij}\int_{-\infty}^x dx_2\ldots dx_i \int_{x}^{x'}dx_{i+1}\ldots dx_j\\
&\times \int_{x'}^{+\infty}dx_{j+1}\dots dx_N \psi_F(x,x_2,\ldots,x_N)\psi_F(x',x_2,\ldots,x_N),
\end{split}
\end{equation}
where
\begin{equation}
C_{ij}=\frac{1}{(i-1)!(j-i)!(N-j)!}\sum_{k=1}^{(N-1)!}a_{\{i,k\}}a_{\{j,k\}}.
\end{equation}

The last step is to use again the Vandermonde trick that we used in order to obtain the exchange coefficient $\alpha_k$ (Eq.~\eqref{alphak})\footnote{This idea was suggested by Matteo Rizzi, and allows to consider sums over $\mathfrak{S}_{N-1}$ instead of $\mathfrak{S}_N$.}. More precisely, 
by noticing the Vandermonde determinant expression in Eq.~\eqref{psivander}, we get:
\begin{equation}
\label{g1sig3}
\begin{split}
\psi_F(x,x_2,\ldots,x_N)&=\frac{1}{\sqrt{N!\prod_{m=0}^{N-1}2^{-m}\sqrt{\pi}m!}}\prod_{k=2}^Ne^{-x_k^2/2}\prod_{2\le j<k\le N}(x_j-x_k)\prod_{l=2}^N(x_l-x)e^{-x^2/2}\\
&=\frac{2^{(N-1)/2}}{\sqrt{\pi}N!(N-1)!}\begin{vmatrix} \phi_0(x_2) & \cdots & \phi_{N-2}(x_2) \\ \vdots & \ddots & \vdots \\ \phi_0(x_N) & \cdots & \phi_{N-2}(x_N) \end{vmatrix}\prod_{l=2}^N(x_l-x)e^{-x^2/2}\\
&=\frac{2^{(N-1)/2}}{\sqrt{\pi}N!(N-1)!}\sum_{P\in\mathfrak{S}_{N-1}}\epsilon(P)\prod_{l=2}^N \phi_{P(l)-2}(x_l)~(x_l-x)e^{-x^2/2}.
\end{split}
\end{equation}
Finally, by putting this expression for $\psi_F$ into Eq.~\eqref{g1sig2}, we get
\begin{equation}
\label{g1sigf}
\begin{split}
G^{(1)}_{\sigma}(x,x')=&N_{\sigma}\frac{2^{(N-1)/2}}{\sqrt{\pi}N!(N-1)!}\sum_{1\le i\le j \le N}C_{ij}\sum_{P,Q\in\mathfrak{S}_{N-1}}\epsilon(P)\epsilon(Q)\\
&\times\prod_{l=2}^N\int_{L_{ij}(l)}^{U_{ij}(l)}dz(z-x)(z-x')\phi_{P(l)-2}(z)\phi_{Q(l)-2}(z),
\end{split}
\end{equation}
with the integration limits given by
\begin{equation}
(L_{ij}(l),U_{ij}(l))=\left\{\begin{array}{lll}
 (-\infty,x) & \mbox{if } l\le i \\ (x,x') & \mbox{if } i<l\le j  \\ (x',+\infty) & \mbox{if } l>j 
\end{array}
\right. .
\end{equation}

I implemented Eq.~\eqref{g1sigf} in a \texttt{Mathematica} program, that runs typically 20 minutes for $N=6$ systems. Notice that it has a complexity of 
$O(N^2((N-1)!)^2)$, so that $N>6$ systems are very time-consuming.

\subsubsection{Density profile analysis}
\label{dpanalysis}

\begin{figure}\centering
\includegraphics[width=1\linewidth]{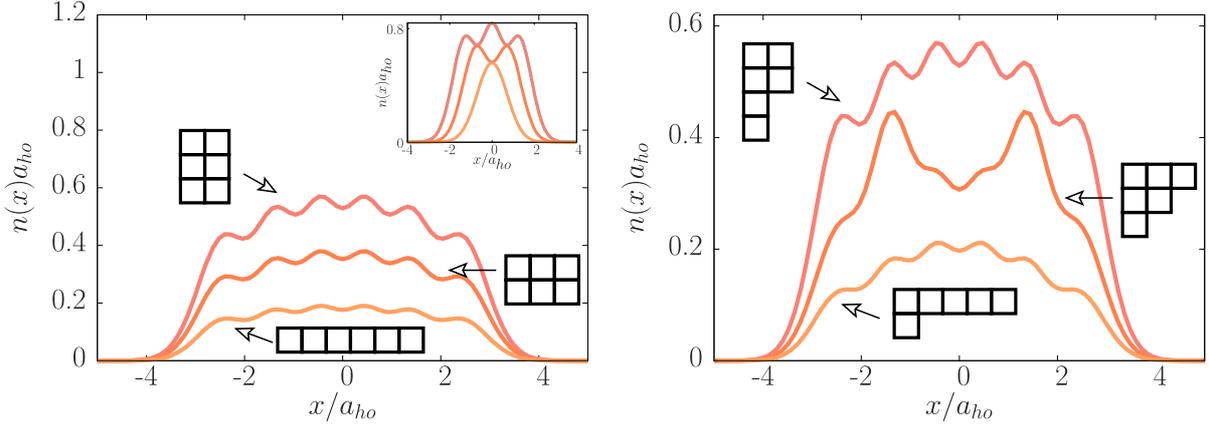}
\caption{\label{density_ferm_balanced}Ground state (left panel) and first excited state with a different symmetry (right panel) density profiles in units of $a_{ho}^{-1}=\sqrt{m\omega/\hbar}$ for three strongly interacting balanced mixtures of $N=6$ fermions, normalized 
to the number $N/\kappa$ of particles in each spin-component (from top to bottom: $\kappa={\color{Salmon}2},{\color{Peach}3},{\color{Apricot}6}$.). For a given balanced mixture, the $\kappa$ spin-components have the same density profile. The corresponding symmetries $Y_{\Gamma}$ that were determined in chapter \ref{chap:sym} are associated with each profile. In order to observe the effects of fermionization, the  
density profiles of the same mixtures in the non-interacting cases are shown in the insets. From \cite{Decamp2016}.}
\end{figure}
\begin{figure}\centering
\includegraphics[width=1\linewidth]{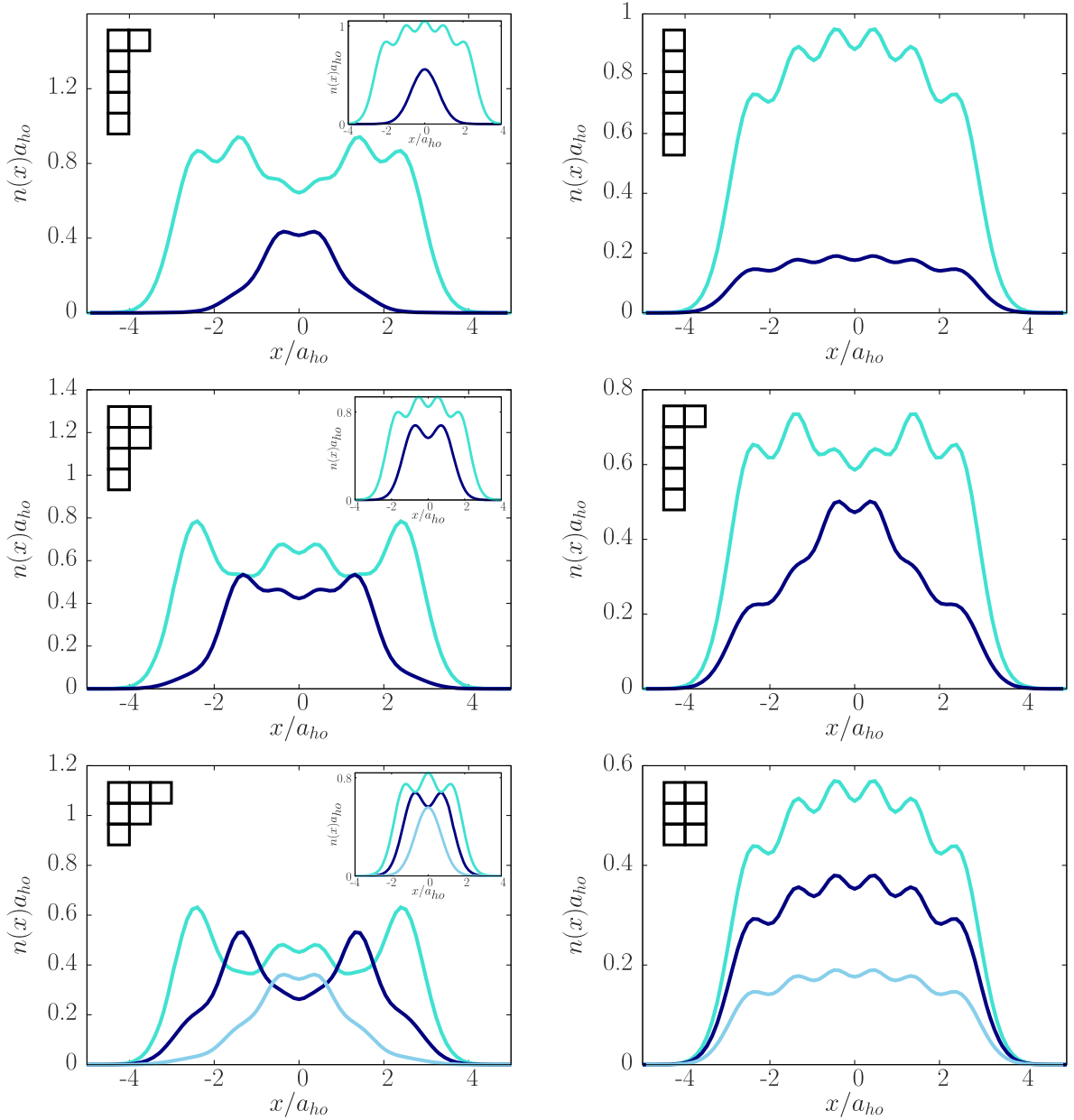}
\caption{\label{density_ferm_unbalanced}Ground state (left panel) and first excited state with a different symmetry (right panel) density profiles in units of $a_{ho}^{-1}=\sqrt{m\omega/\hbar}$ for three strongly interacting imbalanced mixtures of $N=6$ fermions, normalized 
to the number $N_{\sigma}$ of particles in the corresponding spin-component (from top to bottom: $({\color{Turquoise}5^F},{\color{Blue}1^F})$, $({\color{Turquoise}4^F},{\color{Blue}2^F})$, $({\color{Turquoise}3^F},{\color{Blue}2^F},{\color{SkyBlue}1^F})$). 
The corresponding symmetries $Y_{\Gamma}$ that were determined in chapter \ref{chap:sym} are associated with each panel. In order to observe the effects of fermionization, the  
density profiles of the same mixtures in the non-interacting cases are shown in the insets. From \cite{Decamp2016}.}
\end{figure}
\begin{figure}\centering
\includegraphics[width=1\linewidth]{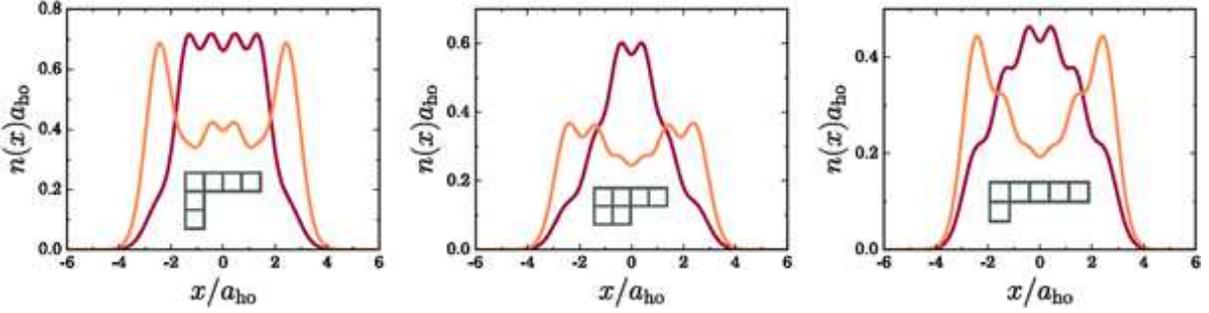}
\caption{\label{dens_bf}Ground state density profiles in units of $a_{ho}^{-1}=\sqrt{m\omega/\hbar}$ for three strongly interacting Bose-Fermi mixtures of $N=6$ particles, normalized 
to the number $N_{\sigma}$ of particles in the corresponding spin-component (from left to right: $({\color{Maroon}3^B},{\color{Orange}3^F})$, $({\color{Maroon}2^B},{\color{Orange}2^F,2^F})$, $({\color{Maroon}2^B,2^B},{\color{Orange}2^F})$). 
For a given mixture, spin-components with the same number of particles and same statistics have the same density profile.
The corresponding symmetries $Y_{\Gamma}$ that were determined in chapter \ref{chap:sym} are associated with each panel. From \cite{Decamp2017}.}
\end{figure}

Let us now analyze exact density profiles $n(x)=G^{(1)}(x,x)$ obtained from Eq.~\eqref{g1sigf} for quantum mixtures with $N=6$ atoms. Results for purely fermionic 
mixtures were published in \cite{Decamp2016}, and for Bose-Fermi mixtures in \cite{Decamp2017}.  Density profiles for the ground state and first excited with a 
different symmetry than the ground state of balanced (respectively imbalanced) fermionic mixtures are given in Fig.~\ref{density_ferm_balanced} and \ref{density_ferm_unbalanced}. 
Ground state density profiles of Bose-Fermi mixtures are given in Fig.~\ref{dens_bf}.

\paragraph{Discussion}

A first observation is that the total density profiles, defined by
\begin{equation}
n(x)=\sum_{\sigma=1}^{\kappa}n_{\sigma}(x),
\end{equation}
are always equal, for the ground state, to the density profile of $N$ spinless fermions (see also Eq.~\eqref{spinlessdensprof}):
\begin{equation}
\label{spinlessdp}
\begin{split}
n_F(x)&=N\int dx_2\ldots dx_N \Psi_F(x,x_2,\ldots,x_N)^2\\
&=\frac{1}{\sqrt{\pi a_0}}\sum_{k=0}^{N-1}\frac{1}{2^kk!}H_k^2(x/a_0)e^{-(x/a_0)^2},
\end{split}
\end{equation}
charaterized by a global parabolic shape and $N$ small density oscillations \cite{Vignolo2000,Kolomeisky2000}. This observation, that was also reported in \cite{massignanNJP},  is a generalization of the equivalence between 
the density profiles of a bosonic Tonks gas and a spinless Fermi gas. 
Consequently, if the mixture is decomposed in $\kappa$ species with equal number of particles per spin-component, the ground-state density profile for every 
spin-component $\sigma$ will verify
\begin{equation}
n_{\sigma}(x)=\frac{1}{\kappa}n_F(x),
\end{equation}
as we can observe in the left panel Fig.~\ref{density_ferm_balanced}. 

Interestingly, as one can see in the right panel of Fig.~\ref{density_ferm_balanced}, it is no longer true in general when considering the total density 
profile of excited states. Instead, we observe that the excited state of $(2^F,2^F,2^F)$, which belongs to the $\Gamma=[3,2,1]$ symmetry class, has the same 
density profile as the ground state density profile of the two-particle component of $(3^F,2^F,1^F)$ (c.f. Fig~\ref{density_ferm_unbalanced}), which belongs 
to the same symmetry class. 
Conversely, the excited states of $(5^F,1^F)$, $(4^F,2^F)$ and $(3^F,2^F,1^F)$, which belong respectively to 
the $[1,1,1,1,1,1]$, $[2,1,1,1,1]$ and $[2,2,2]$ symmetry classes, have similar density profiles than the ground state density profiles of $(6^F)$, $(5^F,1^F)$ and $(3^F,3^F)$ (respectively).
Remark however that the excited state of $(3^F,3^F)$, with symmetry class $\Gamma=[2,2,1,1]$, has not the same density profile as the ground 
state of $(4^F,2^F)$. 
%

The ground state density profiles of imbalanced fermionic mixtures (right panel of Fig~\ref{density_ferm_unbalanced}) and balanced Bose-Fermi mixtures (Fig.~\ref{dens_bf}) have a 
more complex structure than in the balanced fermionic case. In the cases $(5^F,1^F)$ of a polaron and of Bose-Fermi mixtures, we observe a partial spatial 
separation between the polaron (respectively bosonic component(s)) in the center of the trap and the majority component (respectively fermionic component) 
pushed toward the edges. This spatial separation for Bose-Fermi mixtures was predicted by Luttinger theory for homogeneous systems \cite{CazHo03} and by local density approximation 
\cite{Imambekov2006,Imambekov2006b}, and exact diagonalization \cite{Deuretzbacher2017} for harmonically trapped systems. It was also obtained in \cite{Dehkharghani2017} by 
a similar method and DMRG simulations. In the cases of $(4^F,2^F)$ and $(3^F,2^F,1^F)$, the spatial separation of the ground state density profiles is even more complex. In the first case, the system 
displays an alternance between the two components. This can be seen as a realization of an antiferromagnet \cite{Murmann2015}. In the imbalanced three-component system, 
the density profiles are similar to the ones corresponding to the same number of particles in $(4^F,2^F)$ and $(5^F,1^F)$.

In any case, the shape of the density profiles can be seen as a consequence of two complementary phenomena. First, by looking at the insets of Fig.~\ref{density_ferm_balanced} 
and Fig~\ref{density_ferm_unbalanced}, and remembering that the density profile of non-interacting bosons is just a simple central peak, we clearly see the effect 
of fermionization. Indeed, particles that are not subjected to the Pauli principle (like identical bosons or fermions of different spin-components) tend to 
be pushed away in different peaks because of the strong repulsion between them. Second, from the spatial separation discussed in the last paragraph, we observe 
that particles subjected to the Pauli principle, that is fermions belonging to the same spin-component, tend to avoid each other. This can be intuitively understood by analyzing the form of Eq.~\eqref{kvolo}, that we recall here:
\begin{equation}
\label{kvolobis}
K=\sum_{P,Q\in\mathfrak S_N}(a_P-a_Q)^2\alpha_{P,Q},
\end{equation}
where $\alpha_{P,Q}=\alpha_k$ are the nearest-neighbor exchange constants between particles at positions $k$ and $k+1$. Recalling that in the perturbative ansatz we used (section \ref{volansatz}), the ground-state configuration is associated with the maximum value for $K$, and noting that an anti-symmetric exchange corresponds to a zero contribution in Eq.~\eqref{kvolobis}, we can enunciate the following rule:
\vspace{0.5cm}

\begin{center}
\begin{minipage}{0.9\textwidth}
\textit{The spatial configuration of strongly interacting one-dimensional quantum mixtures is such that the number of anti-symmetric exchanges between nearest-neighbors is minimized.
In particular, for the ground-state, the Lieb-Mattis theorem implies that these anti-symmetric exchanges only occur between fermions belonging to the same spin-component.}
\end{minipage}
\end{center}
\vspace{0.5cm}

This fact, together with the parity symmetry of the trap and the fact that the sum of the density profiles is $n_F(x)$, allows to qualitatively predict the shape of the ground-state density profiles.

To conclude this discussion, we have seen that the spatial symmetries investigated in chapter \ref{chap:sym} allow to understand qualitative effects on the shape of the density profiles. We stress, however, that this correspondence is not unequivocal: different symmetry classes can be associated with the same spatial distribution. Thus, the knowledge of the density profile, \textit{e.g.} by an absorption imaging in a cold atom experiment, does not allow to deduce the symmetry class of the system.

\subsubsection{Momentum distributions}
\label{secmd}

\begin{figure}\centering
\includegraphics[width=0.6\linewidth]{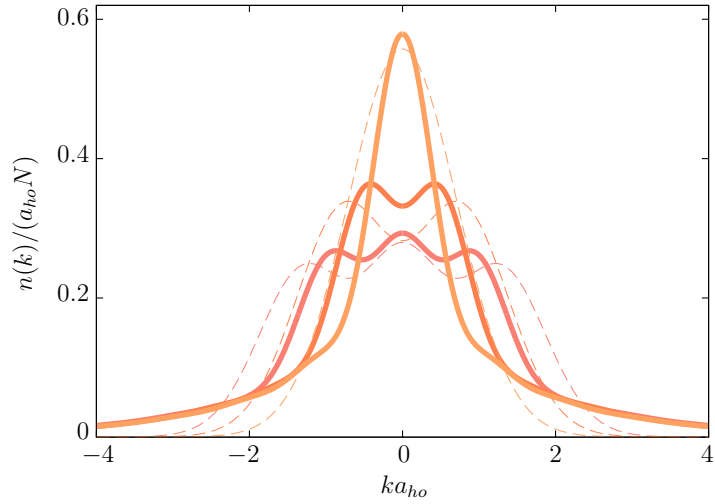}
\caption{\label{md_ferm_balanced}Ground state momentum distributions $n(k)=\kappa n_{\sigma}(k)$ in units of $a_{ho}=\sqrt{\hbar/m\omega}$ for three strongly interacting balanced mixtures of $N=6$ fermions, normalized 
to unity (solid lines, $\kappa={\color{Salmon}2},{\color{Peach}3},{\color{Apricot}6}$.). For a given balanced mixture, the $\kappa$ spin-components have the same momentum distributions. In order to observe the effects of fermionization, the  
momentum distributions of the same mixtures in the non-interacting cases are shown in dashed lines.}
\end{figure}
\begin{figure}\centering
\includegraphics[width=1\linewidth]{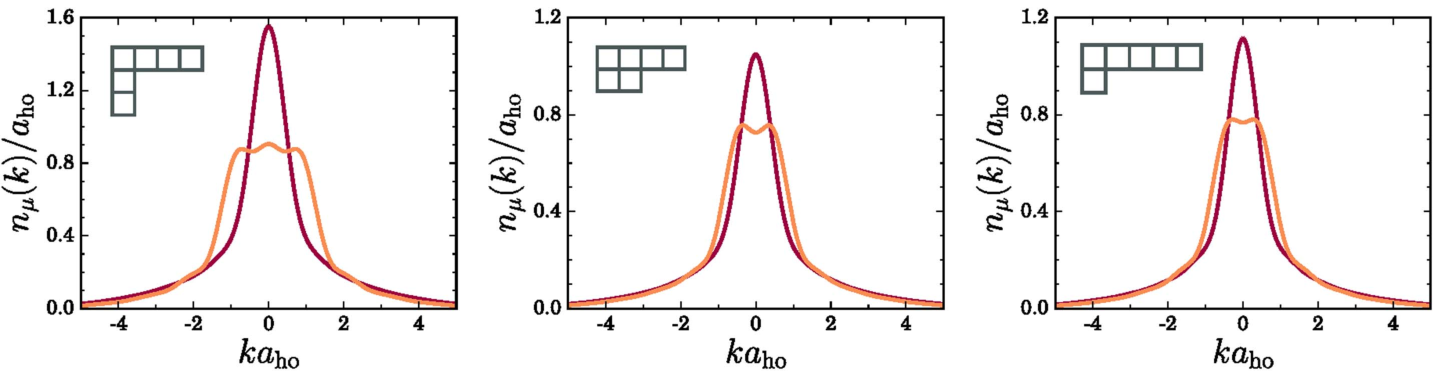}
\caption{\label{mdbosefermi}Ground state momentum distributions $n_{\mu}(k)$ in units of $a_{ho}=\sqrt{\hbar/m\omega}$ for three strongly interacting Bose-Fermi mixtures of $N=6$ particles, normalized 
to the number $N_{\mu}$ of particles in the corresponding spin-component (from left to right: $({\color{Maroon}3^B},{\color{Orange}3^F})$, $({\color{Maroon}2^B},{\color{Orange}2^F,2^F})$, $({\color{Maroon}2^B,2^B},{\color{Orange}2^F})$). 
For a given mixture, spin-components with the same number of particles and same statistics have the same momentum distributions.
The corresponding symmetries $Y_{\Gamma}$ that were determined in chapter \ref{chap:sym} are associated with each panel. From \cite{Decamp2017}.}
\end{figure}

Let us study the momentum distributions $n_{\sigma}(k)=\frac{1}{2\pi}\iint dx dx'~ G_{\sigma}^{(1)}(x,x')e^{-\frac{i}{\hbar}k(x-x')}$, where $G_{\sigma}^{(1)}$ is  as defined in Eq.~\eqref{defg1sigma}. Results for $N=6$ fermionic (resp. Bose-Fermi) 
mixtures were published in \cite{Decamp2016-2} (resp. \cite{Decamp2017}). The corresponding momentum distributions are respectively shown in Fig.~\ref{md_ferm_balanced} 
and Fig.~\ref{mdbosefermi}. Here, we discuss the general qualitative aspects of the momentum distributions --- the asymptotic behaviors will be studied in much details 
in section \ref{sectancont}.

\paragraph{Discussion}

It is a well known fact that the momentum distribution $n_F(k)$ of harmonically trapped spinless non-interacting fermions is equal to its density profile $n_F(x)$ (Eq.~\eqref{spinlessdp})
and is characterized by $N$ peaks \cite{Vignolo2000}. This is a consequence of the duality between $x$ and $k$ in the harmonic oscillator Hamiltonian. 
On the contrary, the momentum distribution $n_B(k)$ of a Tonks gas is characterized by a single central 
peak at k=0 \cite{Papenbrock2003}, showing in passing the limits of the Bose-Fermi mapping discussed in section \ref{tggas}.

In our systems, we observe in Fig.~\ref{mdbosefermi} that the bosonic components in Bose-Fermi mixtures  always display one central peak, while the fermionic 
components have a number of peaks equal to the number of fermions in the considered species, as also observed in \cite{Deuretzbacher2016}. In particular, 
the momentum distribution of the ground state of $(1^F,1^F,1^F,1^F,1^F,1^F)$ is equal to $n_B(k)$ (Fig.~\ref{md_ferm_balanced}). Thus, similarly to what we 
discussed for the density profiles, the exchange symmetry 
and the Pauli principle allow to predict qualitative effects on the momentum distributions of strongly interacting one-dimensional quantum mixtures. Here again, 
however, different exchange symmetries can have similar distributions.

The effect of interactions on momentum distributions is more subtle than on the spatial distributions, 
as one can observe in Fig.~\ref{md_ferm_balanced}. Indeed, the global structure is similar, with 
the fermionized momentum distributions displaying the same number of peaks as the non-interacting 
ones. We observe however a significant reduction of width of these peaks, which 
is dual to the broadening of the density profiles observed in section \ref{dpanalysis}.
Moreover, complementary to this reduction, the large momentum tails are enhanced by interactions. The quantitative 
analysis of these asymptotic behaviors is the subject of section \ref{sectancont}.

\section{Short-range correlations: Tan's contact}
\label{sectancont}

This section is devoted to study of \textit{Tan's contact}, a quantity that governs the large $k$ behavior of the momentum distributions $n(k)$, or equivalently  
the short-range correlations, and therefore goes beyond the Luttinger liquid theory paradigm. After defining this quantity for one-dimensional quantum gases 
and showing how it is universally related to thermodynamic variables of the system in \ref{introtan}, we will study its properties in fermionized one-dimensional 
mixtures in \ref{exacttan}. We will argue, in particular, that Tan's contact is an indicator of the symmetry class of the system. 
Finally, in \ref{scalingstan}, in order to be relevant for actual experiments, we will derive a set of universal scaling laws as functions of interaction, temperature, 
and the transverse confinement.

\subsection{Introduction}
\label{introtan}

\subsubsection{Brief historical review}

In systems interacting via a $\delta$-potential, the momentum distributions $n(k)$  display the following asymptotic algebraic behavior:
\begin{equation}
\label{tancbaeq}
n(k)\underset{k\to\infty}{\sim}\mathcal{C}k^{-4},
\end{equation}
where $\mathcal{C}$ is called the \textit{contact}, or \textit{Tan's contact}. Historically, this behavior was initially highlighted for one-dimensional 
bosonic Tonks gases in \cite{Minguzzi02}, and for any value of the interaction strength in \cite{Olshanii03}. In three-dimensional systems of spin-$\frac{1}{2}$ fermions, 
Tan has also shown that $\mathcal{C}$ is related to thermodynamic quantities of the system --- the so-called \textit{Tan relations} --- such as the dependence of the energy on the scattering length, the quantum 
pressure or the static structure factor \cite{Tan2008a,Tan2008b,Tan2008c} (see also \cite{Braaten2008,Braaten2008b,Zhang2009,Combescot2009}). These results were then extended to two-dimensional \cite{Werner2012,Valiente2011,Valiente2012} 
and one-dimensional \cite{Zwerger2011} spin-$\frac{1}{2}$ gases, and to arbitrary quantum mixtures in two and three dimensions \cite{Werner2012b}, and in one dimension \cite{Patu2017}.

On the experimental side, Tan's contact measurements were performed in three-dimensional strongly interacting fermionic and bosonic ultracold gases, by a direct TOF 
\cite{Stewart2010,Chang2016}, or indirectly by \textit{rf spectroscopy} \cite{Stewart2010,Wild2012,Sagi2012} (exploiting the relation between the contact and the fraction of atoms 
in a given unoccupied state when applying large frequency pulses \cite{Pieri2009,Braaten2008b}) or making use of \textit{Bragg spectroscopy} \cite{Kuhnle2011,Hoinka2013} (exploiting the relation 
between the contact and the structure factor). A precise measurement of Tan's contact in one-dimension remains an experimental challenge because of the enhanced 
fluctuations at large momenta. However, metastable Helium, which allows extremely precise correlation detections (see \textit{e.g.} \cite{Keller2014}), appears to 
be a promising candidate.

\subsubsection{Asymptotic behavior of the momentum distributions in one-dimensional quantum gases}

We now proceed to re-derive Eq.~\eqref{tancbaeq} for one-dimensional quantum mixtures, following \cite{Olshanii03,Patu2017}. 
Their approach is based on the cusp condition (section \ref{secusp}). Other more 
quantum-field-oriented methods rely on the definition of a \textit{generalized function} \cite{Tan2008a,Tan2008b,Tan2008c}, or in a so-called \textit{operator 
product expansion} \cite{Braaten2008,Zwerger2011}.

We consider a system of $N$ one-dimensional particles of coordinates $x_1,\ldots,x_N$ of same mass $m$, subjected to an arbitrary (continuous) external potential $V_{ext}$, 
and then verifying the following Schr\"{o}dinger equation:
\begin{equation}
\left(-\frac{\hbar^2}{2m}\sum_{i=1}^N\frac{\partial^2}{\partial x_i^2}+g_{1D}\sum_{i<j}\delta(x_i-x_j)+V_{ext}(x_1,\ldots,x_N)-E\right)\psi(x_1\ldots,x_N)=0.
\end{equation}
For the moment, we suppose that these particles are identical bosons.

Defining the reduced coordinates $X_{ij}=\frac{x_i+x_j}{2}$ and $x_{ij}=x_i-x_j$ and the reduced masses $M=2m$ and $\mu=m/2$, the $N-$body 
analogue of Eq.~\eqref{cusp2b} implies
\begin{equation}
\label{cuspnbody}
\begin{split}
\psi(x_1,\ldots,x_i,\ldots,x_j,\ldots,x_N)\underset{x_i\to x_j}{=}&\psi(x_1,\ldots,X_{ij},\ldots,X_{ij},\ldots,x_N)\\
&\times\left[1-\frac{|x_{ij}|}{a_{1D}}+\mathcal{O}(x_{ij}^2)\right],
\end{split}
\end{equation}
where $a_{1D}=-\hbar^2/\mu g_{1D}$.

Moreover, we know from Fourier analysis that, for a Lebesgue integrable function $f(x)=|x-x_0|F(x)$ (with $F$ a smooth function), we have \cite{Bleistein2010}:
\begin{equation}
 \int dx ~e^{-ikx}f(x)\underset{k\to\infty}{=}-\frac{2e^{-ikx_0}}{k^2}+\mathcal{O}\left(\frac{1}{|k|^3}\right).
\end{equation}
Then, Eq.~\eqref{cuspnbody} implies
\begin{equation}
\begin{split}
\int dx_1 ~e^{-ikx_1}\psi(x_1,x_2,\ldots,x_N)&\underset{k\to\infty}{\sim}\int dx_1 ~e^{-ikx_1}\sum_{j=2}^N\psi(X_{1j},\ldots,\underset{\uparrow\atop j}{X_{1j}},\ldots,x_N)\left[1-\frac{|x_{ij}|}{a_{1D}}\right]\\
&\underset{k\to\infty}{\sim}\frac{2}{a_{1D}k^2} \sum_{j=2}^Ne^{-ikx_j}\psi(x_j,\ldots,\underset{\uparrow\atop j}{x_j},\ldots,x_N),
\end{split}
\end{equation}
where we have removed one of the terms using the fact that the Fourier transform of a differentiable function with continuous derivative falls to 0 as $o\left(\frac{1}{k^2}\right)$ when $k\to\infty$ \cite{Bracewell1999}. 
Therefore, using $n(k)$'s definition (Eq.~\eqref{mddef}), we find
\begin{equation}
\label{shnouf}
\begin{split}
n(k)&\underset{k\to\infty}{\sim}\frac{2N}{\pi a_{1D}^2k^4}\int dx_2\ldots x_N\sum_{2\le j\le l\le N}e^{-k(x_j-x_l)}
\psi(x_j,\ldots,\underset{\uparrow\atop j}{x_j},\ldots,x_N)\psi(x_l,\ldots,\underset{\uparrow\atop l}{x_l},\ldots,x_N)\\
&\underset{k\to\infty}{\sim}\frac{2N}{\pi a_{1D}^2k^4}\sum_{j=2}^N\int dx_2\ldots x_N~\psi(x_j,\ldots,\underset{\uparrow\atop j}{x_j},\ldots,x_N)^2,
\end{split}
\end{equation}
where again we have neglected the off-diagonal terms  because of the smoothness of the integrand. 

Furthermore, the first-quantized version of $G^{(2)}(x,x')$ (c.f. Eq.~\eqref{g2def2nd}) can be written
\begin{equation}
G^{(2)}(x,x')=\int dx_1\ldots dx_N~ \psi(x_1,\ldots,x_N)^2\sum_{i\neq j}\delta(x-x_i)\delta(x'-x_j).
\end{equation}
Thus, exploiting the permutational symmetry of the integrand, Eq.~\eqref{shnouf} implies Eq.~\eqref{tancbaeq} with
\begin{equation}
\mathcal{C}=\frac{2}{\pi a_{1D}^2}\int dx~G^{(2)}(x,x).
\end{equation}
With this definition, we see that the contact can be interpreted as a measure of the probability of finding two particles in the same region of space.

The generalization to the $\kappa$-component case is straightforward, by restricting the summations on specific components rather than summing over all the 
$N$ particles. More explicitly, for two spin-components $\sigma$ and $\sigma'$, we define
\begin{equation}
G^{(2)}_{\sigma\sigma'}(x,x')=\int dx_1\ldots dx_N~ \psi(x_1,\ldots,x_N)^2\sum_{i\in I_{\sigma},~j\in I_{\sigma'},~i\neq j}\delta(x-x_i)\delta(x'-x_j),
\end{equation}
where $I_{\sigma}$ is the subset of $\{1,\ldots,N\}$ associated with particles of type $\sigma$. Then, it is easy to prove that
\begin{equation}
\label{algbehaviorsigma}
n_{\sigma}(k)\underset{k\to\infty}{\sim}\mathcal{C}_{\sigma}k^{-4},
\end{equation}
where we have
\begin{equation}
\label{defcontsig}
\mathcal{C}_{\sigma}=\frac{2}{\pi a_{1D}^2}\int dx~\sum_{\sigma'=1}^{\kappa}G_{\sigma\sigma'}^{(2)}(x,x),
\end{equation}
which can be seen as a measure of the probability of finding a particle of type $\sigma$ in the same region of space than another particle. Notice that if $\sigma$ is a fermionic component, the term in $\sigma'=\sigma$ cancels out. Moreover, if the mixture is completely balanced, or of it contains 
only two components, all the contacts are equal. In every case, we can define the \textit{total contact} as:
\begin{equation}
\label{deftotalcontact}
\begin{split}
\mathcal{C}_{tot}&=\sum_{\sigma}^{\kappa}\mathcal{C}_{\sigma}\\
&=\frac{2}{\pi a_{1D}^2}\int dx~G^{(2)}(x,x).
\end{split}
\end{equation}

\subsubsection{Tan sweep theorem}

Also known as \textit{Tan adiabatic theorem} in the spin$-\frac{1}{2}$ case, this theorem allows to relate the contact to the derivative of the the energy as a function of the scattering length. It is a simple consequence of the Hellmann-Feynman theorem \cite{Feynman1939}. More precisely, we have:
\begin{equation}
\begin{split}
\frac{\partial E}{\partial a_{1D}}&=\frac{2\hbar^2}{ma_{1D}^2}\int dx_1\ldots dx_N~ \psi(x_1,\ldots,x_N)^2\sum_{i< j}\delta(x_i-x_j)\\
&=\frac{\hbar^2}{ma_{1D}^2}\int dx~G^{(2)}(x,x),
\end{split}
\end{equation}
and therefore, using the definition of the total contact (Eq.~\eqref{deftotalcontact}), we obtain
\begin{equation}
\label{tansweeptot}
\frac{\partial E}{\partial a_{1D}}=\frac{\pi\hbar^2}{2m}~\mathcal{C}_{tot}.
\end{equation}

In order to write the equivalent of Eq.~\eqref{tansweeptot} for $\mathcal{C}_{\sigma}$, we have to modify the definition of our interaction potential, by writing $a_{1D}\delta(x_i-x_j)\equiv a_{\sigma\sigma'}\delta(x_i-x_j)$ when either $(i,j)$ or $(j,i)$ is in $I_{\sigma}\times I_{\sigma'}$. Then, the Hellmann-Feynman theorem implies
\begin{equation}
\frac{\partial E}{\partial a_{\sigma\sigma'}}=\frac{2\hbar^2}{ma_{\sigma\sigma'}^2(1+\delta_{\sigma\sigma'})}\int dx~G^{(2)}_{\sigma\sigma'}(x,x),
\end{equation}
where $\delta_{\sigma\sigma'}$ is the Kronecker symbol, and therefore
\begin{equation}
\label{sweepsigma}
\mathcal{C}_{\sigma}=\frac{m}{\pi\hbar^2}\sum_{\sigma'=1}^{\kappa}(1+\delta_{\sigma\sigma'})\frac{\partial E}{\partial a_{\sigma\sigma'}}.
\end{equation}
Note the term in $\sigma=\sigma'$ is non-zero only in the case where we consider a bosonic component.

\subsection{Exact results in the fermionized limit}
\label{exacttan}


\subsubsection{Exact contact from the perturbative ansatz}

Given Eq.~\eqref{tansweeptot}, we can relate the total contact  in the $g_{1D}\to\infty$ limit to the $K=-\lim_{g_{1D}\to\infty}\frac{\partial E}{\partial g_{1D}^{-1}}$ parameter that we introduced in the perturbative ansatz (section \ref{volansatz}) by the simple relation:
\begin{equation}
\label{ctotk}
\mathcal{C}_{tot}(\infty)=\frac{m^2}{\pi\hbar^4}K.
\end{equation}
Given a solution, characterized by a vector $\vec{a}$ of coefficients (c.f. the defintion of the ansatz in Eq.~\eqref{Psivolo} and the relation \eqref{kvolo} between $K$ and $\vec{a}$), we can be write $\mathcal{C}_{tot}(\infty)$:
\begin{equation}
\label{ctotinfini}
\mathcal{C}_{tot}(\infty)=\frac{m^2}{\pi\hbar^4}\sum_{k=1}^{N-1}\sum_{P\in\mathfrak S_N}(a_P-a_{P(k,k+1)})^2\alpha_k,
\end{equation}
where $(k,k+1)$ is the transposition between particles at positions $k$ and $k+1$ and $\alpha_k$ are the exchange coefficients defined in Eq.~\eqref{alphavolo}.

Moreover, the contact for a given spin-component can be extracted from Eq.~\eqref{sweepsigma} by restricting the second sum in Eq.~\eqref{ctotinfini} over 
the subset $\tilde{\mathfrak{S}}_N(k||\sigma,\sigma')$ of permutations so that the indexes in positions $k$ and $k+1$ correspond to particles belonging to components 
$\sigma$ and $\sigma'$:
\begin{equation}
\tilde{\mathfrak{S}}_N(k||\sigma,\sigma')=\{P\in\mathfrak{S}_N|(P(k),P(k+1))\in I_{\sigma}\times I_{\sigma'}\cup I_{\sigma'}\times I_{\sigma}\}.
\end{equation}
With these notations, we have:
\begin{equation}
\label{csigmainfini}
\mathcal{C}_{\sigma}(\infty)=\frac{m^2}{2\pi\hbar^4}\sum_{\sigma'=1}^{\kappa}(1+\delta_{\sigma\sigma'})\sum_{k=1}^{N-1}\sum_{P\in\tilde{\mathfrak{S}}_N(k||\sigma,\sigma')}(a_P-a_{P(k,k+1)})^2\alpha_k.
\end{equation}

\begin{figure}\centering
\includegraphics[width=0.6\linewidth]{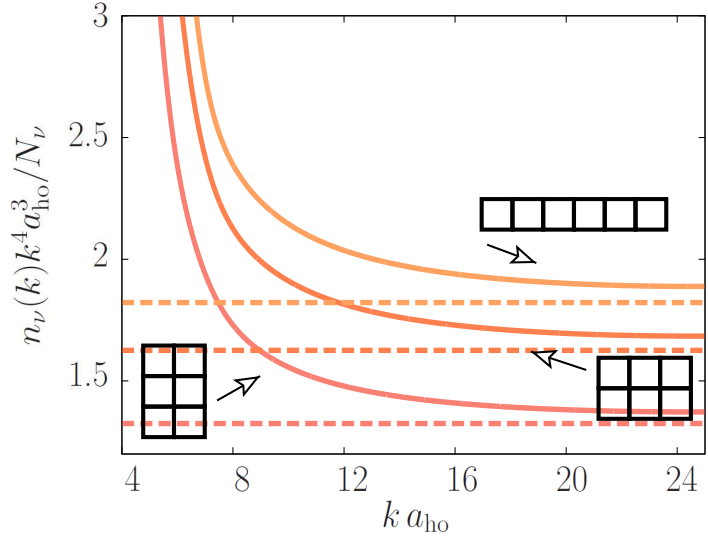}
\caption{\label{tcferm}Ground state $n_{\nu}(k)k^4a^3_{ho}/N_{\nu}$ functions as functions of $ka_{ho}$ where $a_{ho}=\sqrt{\hbar/m\omega}$ for three strongly interacting balanced 
mixtures of $N=6$ fermions (solid lines, $N_{\nu}={\color{Salmon}3},{\color{Peach}2},{\color{Apricot}1}$.). The corresponding symmetries $Y_{\Gamma}$ that were determined in chapter \ref{chap:sym} are associated with each panel. The contacts $\mathcal{C}_{\nu}(\infty)$ computed from Eq.~\eqref{csigmainfini} are shown in dashed lines.}
\end{figure}

\begin{figure}\centering
\includegraphics[width=1\linewidth]{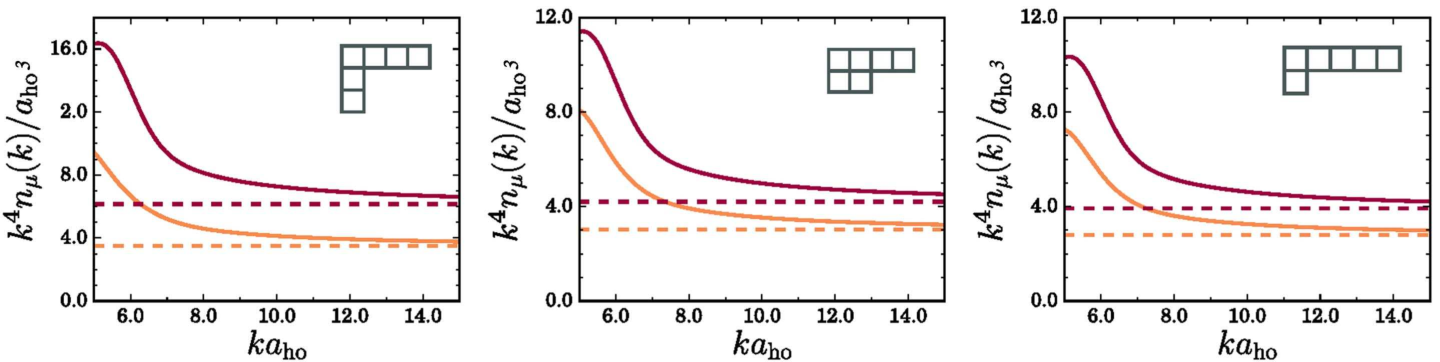}
\caption{\label{tcbf}Ground state $n_{\mu}(k)k^4a^3_{ho}/N_{\nu}$ functions as functions of $ka_{ho}$ where $a_{ho}=\sqrt{\hbar/m\omega}$ for three strongly interacting Bose-Fermi mixtures of $N=6$ particles (from left to right: $({\color{Maroon}3^B},{\color{Orange}3^F})$, $({\color{Maroon}2^B},{\color{Orange}2^F,2^F})$, $({\color{Maroon}2^B,2^B},{\color{Orange}2^F})$). 
For a given mixture, spin-components with the same number of particles and same statistics have the same distributions.
The corresponding symmetries $Y_{\Gamma}$ that were determined in chapter \ref{chap:sym} are associated with each panel.
The contacts $\mathcal{C}_{\mu}(\infty)$ computed from Eq.~\eqref{csigmainfini} are shown in dashed lines. 
From \cite{Decamp2017}.}
\end{figure}

In Figs. \ref{tcferm} and \ref{tcbf}, the $\mathcal{C}_{\sigma}(\infty)$ for balanced fermionic mixtures and Bose-Fermi mixtures (respectively) computed from 
Eq.~\eqref{csigmainfini} are plotted and compared with the $n_{\sigma}(k)k^4$ functions computed in section \ref{secmd}. We observe that the asymptotic behaviors 
verify Eq.~\eqref{algbehaviorsigma}, which comfort us on the consistence of our calculations.

\subsubsection{Discussion: Tan's contact as a symmetry probe}
\label{sectcassp}

In Fig. \ref{tcferm}, we see that the contact of balanced fermionic mixtures increases as a function of the number $\kappa$ of components. This behavior was 
qualitatively observed in the LENS experiment \cite{Pagano2014}. This property is directly related to the global symmetry of the mixture: when the number of 
spin-components increases, the mixture is more an more spatially symmetric. As discussed in chapter \ref{chap:sym}, the (generalized) Lieb-Mattis theorem implies that 
the energy of the system decreases, and more precisely that the energy slope $\partial E/\partial a_{1D}$ increases. Therefore, because of Tan sweep theorem 
(Eq.~\eqref{sweepsigma}), we can conclude that the contact also increases with the number of components. In the fermionized limit, this statement is readily obtained 
by observing Eq.~\eqref{csigmainfini}. Besides, it also allows to understand why the bosonic contact of a balanced Bose-Fermi mixture is 
bigger than the fermionic one, as one can see in Fig. \ref{tcbf}.

Thus, we see that the contact constitutes an experimentally accessible quantity in order to compare spatial symmetries. In fact, 
because of Eq.~\eqref{ctotk} and the fact that the $K$ coefficients label the states of the system, a precise measurement of $\mathcal{C}_{tot}$ via a standard 
time-of-flight technique would allow to determine uniquely the state of the system, and thus its spatial symmetry (c.f. Fig. \ref{fig:symferm}). Besides, as discussed 
in section \ref{introsym}, because of the duality between the spatial symmetry and the spin symmetry, the contact can also be seen as a magnetic structure probe \cite{Decamp2016-2}.

Finally, we stress that our discussion is based on strong assumptions: zero temperature, infinite interactions, purely one-dimensional system. Section 
\ref{scalingstan} is devoted to studying finite corrections to the contact. However, Tan's contact is associated with high momenta. As argued in \cite{Olshanii03}, 
it is then a more robust experimental parameter as compared to low-momentum related quantities, as it is less sensitive to temperature and to residual three-dimensional 
effects in optical lattices (c.f. Fig. \ref{fig:optlat}).

\subsection{Scaling laws for Tan's contact}
\label{scalingstan}

The results presented in this section where published in \cite{Decamp2016-2} (paragraphs \ref{tcfins} and \ref{tcfintemp}) and \cite{Decamp2018} (paragraph \ref{tcincsdbose}).

\subsubsection{Contact at finite interaction strength}
\label{tcfins}

\paragraph{Preliminary discussion: scaling parameter in the harmonic trap}
In this paragraph, we discuss what is the dimensionless parameter we have to consider when going to the thermodynamic limit. 

To do so, let us first analyze the Lieb-Liniger case 
(c.f appendix \ref{secbethe}), whose Hamiltonian is given by
\begin{equation}
\hat{H}_{LL}=\sum_{i=1}^N-\frac{\hbar^2}{2m}\frac{\partial^2}{\partial x_i^2}+g_{1D}\sum_{i<j}\delta(x_i-x_j),
\end{equation}
where the particle coordinates verify $x_i\in\left[0,L\right]$. The thermodynamic limit consists in considering $N,L\to\infty$ with the lineic density $n\equiv N/L$ 
being kept constant. By writing the coordinates in terms of the typical inter-particle distance $y_i\equiv nx_i\in\left[0,N\right]$, it is natural to write:
\begin{equation}
\hat{H}_{LL}=\frac{\hbar^2n^2}{2m}\left[\sum_{i=1}^N-\frac{\partial^2}{\partial y_i^2}+2\gamma\sum_{i<j}\delta(y_i-y_j)\right],
\end{equation}
where 
\begin{equation}
\label{scalingparamll}
\gamma=\frac{mg_{1D}}{\hbar^2n}
\end{equation}
is the dimensionless interaction strength. Then, we see that the total energy per particle will verify, in the thermodynamic limit, the following scaling behavior:
\begin{equation}
\label{scalingbehaviorll}
\frac{E}{N}=\frac{\hbar^2n^2}{2m}e(\gamma),
\end{equation}
with $e(\gamma)$ a dimensionless function, that can be obtained, at least pertubatively, by Bethe ansatz (c.f. section \ref{GS}). For more complicated statistics 
than a one-component bosonic system, the function $e$ will also depend on the single-species polarizations $\{p_{\sigma}\}\equiv \{N_{\sigma}/N\}_{\sigma\in\{1,\ldots,\kappa\}}$. The take-home message 
is that we have factorized $\hat{H}_{LL}$ by the typical energy $\hbar^2n^2/2m$, which corresponds to the Fermi energy in the Tonks thermodynamic limit, in order to obtain the adimensional scaling parameter $\gamma$ 
(Eq.~\eqref{scalingparamll}) and Eq.~\eqref{scalingbehaviorll}.

In the case where the system is trapped by a harmonic potential of frequency $\omega$, we recall that the Hamiltonian is given by
\begin{equation}
\hat{H}=\sum_{i=1}^N\left(-\frac{\hbar^2}{2m}\frac{\partial^2}{\partial x_i^2}+\frac{1}{2}m\omega^2x_i^2\right)+g_{1D}\sum_{i<j}\delta(x_i-x_j).
\end{equation}
Naively, one could think of re-scaling $\hat{H}$ it terms of $\hbar\omega$ for the energy and $x_i/a_0$ for length (where $a_0=\sqrt{\hbar/m\omega}$ is the harmonic 
oscillator length). However, keeping in mind the reasoning that we had in the absence of harmonic potential, we see that we have to factorize by the Fermi 
energy in the Tonks thermodynamic limit, which is given by $\hbar\omega N$. Thus, by re-scaling the spatial coordinates by $y_i\equiv x_i/a_0\sqrt{N}$, we obtain:
\begin{equation}
\hat{H}=\hbar\omega N\left[\sum_{i=1}^N\left(-\frac{\partial^2}{\partial y_i^2}+\frac{1}{2N^2}y_i^2\right)+\alpha_0\sum_{i<j}\delta(y_i-y_j)\right],
\end{equation}
where the adimensional interaction strength relevant for our problem is given by
\begin{equation}
\alpha_0=\frac{a_0mg_{1D}}{\sqrt{N}\hbar^2}=-\frac{2a_0}{\sqrt{N}a_{1D}}.
\end{equation}
Therefore, analogously to Eq.~\eqref{scalingbehaviorll} the total energy per particle in the thermodynamic limit of the harmonically trapped system is of the form (in the multi-component case)
\begin{equation}
\label{scalingen}
\frac{E}{N}=\hbar\omega N f(\alpha_0,\{p_{\sigma}\}),
\end{equation}
with $f$ an adimensional function.

Note that from these simple arguments, together with Tan adiabatic theorem (Eq.~\eqref{tansweeptot}), we already see that Eq.~\eqref{scalingen} implies that 
the total contact $\mathcal{C}_{tot}$ verifies
\begin{equation}
\label{scalingctotalpha}
\mathcal{C}_{tot}(\alpha_0)=\frac{N^{5/2}}{\pi a_0^3}\alpha_0^2\frac{\partial f(\alpha_0,\{p_{\sigma}\})}{\partial \alpha_0}.
\end{equation}
We now proceed to find an explicit expression for this scaling.

\paragraph{A local density approximation approach}

The Local Density Approximation (LDA) consists in stating that when the harmonic potential is sufficiently shallow as compared to the typical length of the system, as it is the case in a typical cold atom 
experiment, we can consider the system as "locally homogeneous". Then, if we have an expression in the homogeneous case for $e(\gamma)$ (Eq.~\eqref{scalingbehaviorll}), we 
can obtain results in the harmonically trapped case by considering the following density functional:
\begin{equation}
\label{energyfunctional}
E[n]=\int n(x) dx~\left(\frac{\hbar^2n(x)^2}{2m}e(\gamma)+\frac{1}{2}m\omega^2x^2-\mu\right),
\end{equation}
where the density profile $n(x)$ is now a function of $x$ that has to be determined, and the chemical potential $\mu$ is a Lagrange multiplier allowing to fix the total number of particles:
\begin{equation}
\label{normmu}
\int n(x) dx=N.
\end{equation}
LDA is a peculiar class of approximation in  density functional theory \cite{Parr1980}. 

Then, we obtain the ground-state density profile by minimizing $E[n]$, i.e. by solving $\delta E[n]/\delta n=0$. This yields
\begin{equation}
\label{baldaeq}
\frac{3}{2}\frac{\hbar^2}{m}e(\gamma)n(x)^2-\frac{g_{1D}}{2}e'(\gamma)n(x)=\mu\left(1-\frac{x^2}{R_{TF}^2}\right),
\end{equation}
where $R_{TF}\equiv \sqrt{2\mu/(m\omega^2)}$ is the Thomas-Fermi radius.

Once the ground-state density profile $n(x)$ is known, it is easy to obtain Tan's contact by combining Eq.~\eqref{energyfunctional} to Tan adiabatic theorem 
(Eq.~\eqref{tansweeptot}):
\begin{equation}
\label{tansweeplda}
\mathcal{C}_{tot}=\frac{g_{1D}^2m^2}{2\pi\hbar^4}\int dx ~n(x)^2e'\left(\frac{m g_{1D}}{\hbar^2n(x)}\right).
\end{equation}

Note that in this paragraph we have considered only one density profile $n(x)$. This corresponds to the case of a single-component bosonic system or of a 
balanced fermionic or bosonic mixture. If different density profiles $n_1,\ldots,n_{\kappa}$ have to be considered, one will have to define a density functional $E[n_1,\ldots,n_{\kappa}]$ 
and the chemical potentials accordingly. This situation is a little bit more intricate, and in this thesis we have only considered the case of a balanced fermionic 
mixture \cite{Decamp2016-2} (c.f. next paragraph), which is relevant for the LENS experiment \cite{Pagano2014}. Our method was adapted to the 
single-component bosonic case in \cite{Lang2017}.

\paragraph{Expression for strongly interacting balanced fermionic mixtures}

In the homogeneous regime \cite{Guan2012}, we dispose of an explicit Laurent expansion for $e(\gamma)$ when $\gamma\to\infty$ in the case of a balanced fermionic mixture. It is given by:
\begin{equation}
\label{expasione}
\begin{split}
e(\gamma)\underset{\gamma\to\infty}{=}\frac{\pi^2}{3}&\left[1-\frac{4Z_1(\kappa)}{\gamma}+\frac{12Z_1(\kappa)^2}{\gamma^2}\right.\\
&\left.-\frac{32}{\gamma^3}\left(Z_1(\kappa)^3-\frac{Z_3(\kappa)\pi^2}{15}\right)+\mathcal{O}\left(\frac{1}{\gamma^4}\right)\right],
\end{split}
\end{equation}
with 
\begin{equation}
Z_1(\kappa)=-\frac{1}{\kappa}\left(\psi\left(\frac{1}{\kappa}\right)+C_{Euler}\right),
\end{equation}
and
\begin{equation}
Z_3(\kappa)=\frac{1}{\kappa^3}\left(\zeta\left(3,\frac{1}{\kappa}\right)-\zeta(3)\right),
\end{equation}
where $C_{Euler}\approx 0.577$ is the Euler constant and $\psi$ and $\zeta$ are respectively the Digamma and Riemann Zeta functions \cite{0486612724}. In section \ref{strcoup}, we provide a sketch of the proof of relation \eqref{expasione}.

The next step is to write the chemical potential $\mu$ and the density profile $n(x)$ as similar $1/\gamma-$expansions whose coefficients are unknown, and plugging 
these expansions into Eqs~\eqref{baldaeq} and \eqref{normmu}. Then, solving it order by order using Eq.~\eqref{expasione} and applying Eq.~\eqref{tansweeplda}, 
we find the strong-coupling expansion for the Tan's contact of a balanced fermionic mixture:
\begin{equation}
\label{ctotalpha}
\begin{split}
\mathcal{C}_{tot}&(\alpha_0)\underset{\alpha_0\to\infty}{=}\frac{N^{5/2}}{\pi a_0^3}\left[\frac{128\sqrt{2}Z_1(\kappa)}{45\pi^2}+\frac{2(315\pi^2-4096)Z_1(\kappa)^2}{81\pi^4\alpha_0}\right.\\
&\left.-\frac{64\sqrt{2}[25(1437\pi^2-14336)Z_1(\kappa)^3+1728\pi^4Z_3(\kappa)]}{14175\pi^6\alpha_0^2}+\mathcal{O}\left(\frac{1}{\alpha_0^3}\right)\right].
\end{split}
\end{equation}
In particular, in the fermionized limit, Eq.~\eqref{ctotalpha} implies
\begin{equation}
\mathcal{C}_{tot}(\infty)=\frac{N^{5/2}}{ a_0^3}\frac{128\sqrt{2}Z_1(\kappa)}{45\pi^3}.
\end{equation}
Notice that these expressions have the same form as Eq.~\eqref{scalingctotalpha}, which was predicted only by physical arguments. 
Besides, by taking the $\kappa\to\infty$ limit of an infinite number of components, since $\lim_{\kappa\to\infty}Z_1(\kappa)=1$, we see that we obtain 
the strong-coupling expansion for the contact in the bosonic case \cite{Olshanii03,Lang2017}. This suppression of the effects of internal degrees of 
freedom when $\kappa\to\infty$, also known as \textit{high-spin bosonization} was also highlighted in \cite{YangYou2011,Guan2012,LIU201484}. Experimentaly, this 
phenomenon was observed by analyzing the collective mode frequencies when increasing the number $\kappa$ of fermionic components up to $\kappa=6$ \cite{Pagano2014}.

The $N^{5/2}-$dependence of the contact was also observed in Monte Carlo numerical simulations \cite{Matveeva2016}, these results thus constituing an 
analytical proof. Moreover, we observe that $\mathcal{C}_{tot}$ is an increasing function of the number of components $\kappa$, as observed in the LENS 
experiment \cite{Pagano2014} and explained qualitatively by symmetry arguments in section \ref{sectcassp}.

\begin{figure}\centering
\includegraphics[width=0.7\linewidth]{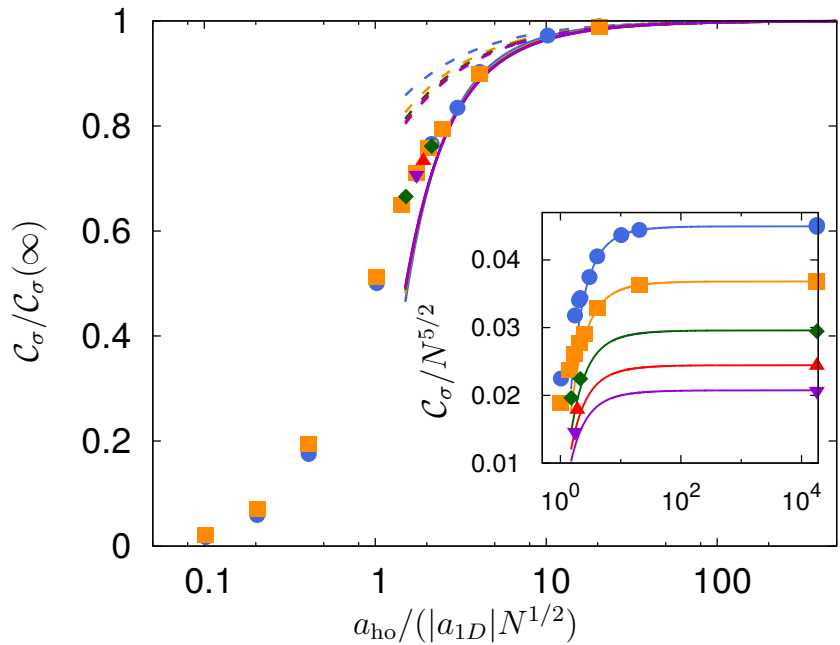}
\caption{\label{ScalingDMRG} Comparison between the strong-coupling expansion (Eq.~\eqref{ctotalpha}) to first and second order in $1/\alpha_0$ (dashed and continuous lines, respectively), and the DMRG simulations performed by Matteo Rizzi and Johannes J\"{u}nemann for the contacts $\mathcal{C}_{\nu}$ of different balanced fermionic mixtures ($N_{\nu}=2$; blue circles: $\kappa=2$, orange squares: $\kappa=3$, green diamonds: $\kappa=4$, red up-triangles: $\kappa=5$, violet down-triangles: $\kappa=6$), as functions of $\alpha_0/2=a_0/(|a_{1D}|N^{1/2})$. By removing the $N^{5/2}$-dependence (inset) and then the asymptotic value of $\mathcal{C}_{\nu}$ (main panel), we see that the data collapse, thus displaying a very weak dependence on the number $\kappa$ of components.  From \cite{Decamp2016-2}.}
\end{figure}

In Fig.~\ref{ScalingDMRG}, we compare the contact strong-coupling expansion of Eq.~\eqref{ctotalpha} with numerical DMRG data obtained by Matteo Rizzi 
and Johannes J\"{u}nemann \cite{Decamp2016-2}. The agreement between the two is extremely good, although we have used strong assumptions in order to perform the LDA. Indeed, our result should, in principle, work only for a very large number of atoms, but we see that it is already very accurate for few-body systems ($4\le N \le 12$). This property, that 
was recently discussed in the bosonic case \cite{2018arXiv180502463R}, suggests that few-body systems, that are hence more accessible for numerical and analytical calculations, already capture the essential behaviors of many-body systems (see also \cite{Lewenstein-Massignan}). Moreover, by re-scaling the contacts by their values at infinite interactions, 
we note that all the plots collapse almost perfectly and therefore display a very weak dependence on $\kappa$.

\subsubsection{Contact at finite temperature}
\label{tcfintemp}

In section \ref{introtan}, when introducing Tan's contact and Tan sweep theorem, we have only considered pure states. However, when temperature is finite, statistical mixtures 
have to be considered. Hopefully, this generalization is pretty straightforward. In particular, 
the finite-temperature version of Eq.~\eqref{sweepsigma}  in the grand canonical ensemble becomes \cite{Patu2017}:
\begin{equation}
\label{tansweeptemp}
\mathcal{C}_{\sigma}(T)=\frac{m}{\pi\hbar^2}\sum_{\sigma'=1}^{\kappa}(1+\delta_{\sigma\sigma'})\left(\frac{\partial \Omega}{\partial a_{\sigma\sigma'}}\right)_{\mu_{\sigma},T},
\end{equation}
where $\Omega$ is the grand potential defined by 
\begin{equation}
\Omega=-k_BT\ln \mathcal{Z},
\end{equation}
with the grand partition function given by
\begin{equation}
\mathcal{Z}=\Tr \left[e^{\left(\sum_{\sigma=1}^{\kappa}\mu_{\sigma} \hat{N_{\sigma}}-\hat{H}\right)/k_BT}\right].
\end{equation}

\paragraph{Virial expansion at high temperatures}

In general, $\Omega$ is hard to compute exactly in our system. However, in the limit of high temperatures, \textit{i.e.}
\begin{equation}
\frac{\hbar\omega}{k_BT}\ll 1,
\end{equation}
it can be expanded in powers of the \textit{fugacity}, which verifies in this limit \cite{Landau1980}
\begin{equation}
z_{\sigma}\equiv e^{\frac{\mu_{\sigma}}{k_BT}}\ll 1.
\end{equation}
Such an expansion is known as a \textit{virial expansion}, and its coefficients can be computed analytically in the strongly repulsive limit \cite{Ho2004,Liu2009,Liu2010}.
It has allowed to compute the high-temperature contact in the fermionized regime in three-dimensional $SU(2)$ systems \cite{Hu2011}, in one-dimensional 
bosonic Tonks gas \cite{Vignolo2013} and $SU(\kappa)$ systems \cite{Decamp2016-2}. In the following, we will first explain how this expansion works in the simplest case of the one-dimensional 
Tonks gas, and then show how it can be extended to other statistics.

In a one-component Bose gas, in the limit of high temperatures, we can forget the $\sigma$ index and write
\begin{equation}
\label{virialz}
\mathcal{Z}\underset{z\to 0}{=}1+zQ_1+z^2Q_2+\mathcal{O}(z^3),
\end{equation}
with $Q_n$ the partial partition function of clusters of size $n$, \textit{i.e.}
\begin{equation}
Q_n=\Tr_n \left[e^{-\hat{H}_n/k_BT}\right].
\end{equation}
By taking the logarithm, we can write $\Omega$ as
\begin{equation}
\Omega\underset{z\to 0}{=}-k_BTQ_1\left(z+b_2z^2\right)+\mathcal{O}(z^3),
\end{equation}
with the \textit{second virial coefficient} given by
\begin{equation}
b_2=(Q_2-Q_1^2/2)/Q_1.
\end{equation}
Then, the one-component version of Eq.~\eqref{tansweeptemp} implies
\begin{equation}
\label{ctbose1}
\begin{split}
\mathcal{C}(T)&=\frac{2m}{\pi\hbar^2}\frac{\partial\Omega}{\partial a_{1D}}\\
&\underset{z\to 0}{=}\frac{2m}{\pi\hbar^2\lambda_{dB}}k_BT~Q_1~c_2z^2+\mathcal{O}(z^3),
\end{split}
\end{equation}
where $\lambda_{dB}=\sqrt{2\pi\hbar^2/mk_BT}$ and the adimensional coefficient $c_2$ is defined by
\begin{equation}
c_2=-\frac{\partial b_2}{\partial(a_{1D}/\lambda_{dB})}.
\end{equation}

In the strongly interacting and high temperature limit, the following relations hold:
\begin{equation}
\label{q1vir}
\begin{split}
Q_1&=\sum_{k=0}^{+\infty}e^{-\frac{\hbar\omega}{k_BT}(n+1/2)}\\
&\sim \frac{k_BT}{\hbar\omega},
\end{split}
\end{equation}
and
\begin{equation}
\label{zvir}
z\sim \frac{N\hbar\omega}{k_BT},
\end{equation}
which can be deduced from
\begin{equation}
N=\int d\epsilon ~\rho_S(\epsilon) \frac{1}{e^{(\epsilon-\mu)/(k_BT)}+1}
\end{equation}
with the density of states given by $\rho_S(\epsilon)=1/(\hbar\omega)$ and by taking the high $T$ limit. Moreover, we have
\begin{equation}
\label{c2virial}
c_2\sim \frac{1}{\sqrt{2}}.
\end{equation}

This last relation is obtained using the fact that 
\begin{equation}
Q_2=Q_1\sum_{\nu}e^{-\epsilon_{\nu}^{rel}/k_BT},
\end{equation}
where $\epsilon_{\nu}^{rel}$ are the energies of the two-body problem. In the harmonic trap, one can prove that \cite{Busch1998}
\begin{equation}
\epsilon_{\nu}^{rel}=\hbar\omega\left(\nu+\frac{1}{2}\right),
\end{equation}
with $\nu$ the solution of the following transcendental equation:
\begin{equation}
\label{transcendentaleq}
f(\nu)\equiv\frac{\Gamma(-\nu/2)}{\Gamma(-\nu/2+1/2)}=\frac{\sqrt{2}a_{1D}}{a_0}.
\end{equation}
In the fermionized regime $a_{1D}$, this implies $\nu\in 2\mathbb{N}+1$. Then, Eq.~\eqref{c2virial} becomes
\begin{equation}
c_2=-\lambda_{dB}\sum_{\nu}\frac{\hbar\omega}{k_BT}\frac{\partial \epsilon_{\nu}^{rel}}{\partial a_{1D}}e^{-\epsilon_{\nu}^{rel}/k_BT},
\end{equation}
and
\begin{equation}
\begin{split}
\frac{\partial \epsilon_{\nu}^{rel}}{\partial a_{1D}}&=\frac{\epsilon_{\nu}^{rel}}{\partial\nu}\frac{\partial\nu}{\partial f}\frac{\partial f}{\partial a_{1D}}\\
&=\frac{\sqrt{2}\hbar\omega}{a_0}\frac{\partial\nu}{\partial f}.
\end{split}
\end{equation}
Using Eq.~\eqref{transcendentaleq} and some algebraic relations on the Euler function $\Gamma$ yields, taking $\nu=2n+1$ with $n\in \mathbb{N}$:
\begin{equation}
\frac{\partial\nu}{\partial f}=\frac{2}{\pi}\frac{\Gamma\left(\frac{3}{2}+n\right)}{n!}.
\end{equation}
Then, we have to evaluate the following sum:
\begin{equation}
\sum_{n\in\mathbb{N}}\frac{\Gamma\left(\frac{3}{2}+n\right)}{n!}e^{-\frac{\hbar\omega}{k_BT}\left(2n+\frac{3}{2}\right)},
\end{equation}
which can be performed exactly and is equal to \cite{Hu2011}
\begin{equation}
\frac{\sqrt{\pi}}{4\left[2\frac{\hbar\omega}{k_BT}\sinh \left(\frac{\hbar\omega}{k_BT}\right)\right]^{3/2}}.
\end{equation}
Thus, taking the $\hbar\omega\ll k_BT$ limit, we readily obtain Eq.~\eqref{c2virial}.

Finally, putting Eqs.~\eqref{q1vir}, \eqref{zvir} and \eqref{c2virial} into Eq.~\eqref{ctbose1}, we find the high-temperature dependence of the Tan's contact of the Tonks gas:
\begin{equation}
\label{ctbg}
\mathcal{C}(T)=\frac{N^2}{\pi^{3/2}a_0^3}\sqrt{\frac{k_BT}{\hbar\omega}}.
\end{equation}
Refinements for this formula can be found in \cite{Yao2018}.

In \cite{Decamp2016-2}, we have generalized Eq.~\eqref{ctbg} to fermionic mixtures. Here, we will describe how to generalize it to Bose-Fermi mixtures, which is straightforward.
Instead of writing the grand partition function as in Eq.~\eqref{virialz}, we write
\begin{equation}
\mathcal{Z}\underset{z\to 0}{=}1+\sum_{\sigma=1}^{\kappa}z_{\sigma}Q_1+\sum_{\sigma,\sigma'}z_{\sigma}z_{\sigma'}Q_{2\sigma\sigma'}+\mathcal{O}(z^3),
\end{equation}
with 
\begin{equation}
z_{\sigma}\sim N_{\sigma}\frac{\hbar\omega}{k_BT}
\end{equation}
in the high temperature and strongly repulsive limit. Note that $Q_{2\sigma\sigma}=0$ if is $\sigma$ is a fermionic component, and otherwise $Q_{2\sigma\sigma'}=Q_2$
as found in the Tonks gas. Then, applying the same procedure than before to the generalized Tan sweep relation (Eq.~\eqref{tansweeptemp}) yields
\begin{equation}
\label{ctmixt}
\mathcal{C}_{\sigma}(T)=\frac{N_{\sigma}}{\pi^{3/2}a_0^3}\sqrt{\frac{k_BT}{\hbar\omega}}\sum_{\sigma'=1}^{\kappa}\chi_{\sigma\sigma'}N_{\sigma'},
\end{equation}
where 
\begin{equation}
\chi_{\sigma\sigma'}=\left\{
\begin{array}{ll}
 0 & \mbox{if } \sigma=\sigma' \mbox{ and  } \sigma \mbox{ is a fermionic component}\\ 1 & \mbox{otherwise}
\end{array}
\right..
\end{equation}
Notice that in the case of a balanced fermionic mixture, Eq.~\eqref{ctmixt} implies
\begin{equation}
\mathcal{C}_{tot}(T)=\frac{N^2}{\pi^{3/2}a_0^3}\sqrt{\frac{k_BT}{\hbar\omega}}\frac{\kappa-1}{\kappa}
\end{equation}
and once again we recover the Tonks result (Eq.~\eqref{ctbg}) in the $\kappa\to\infty$ limit.

\begin{figure}\centering
\includegraphics[width=0.6\linewidth]{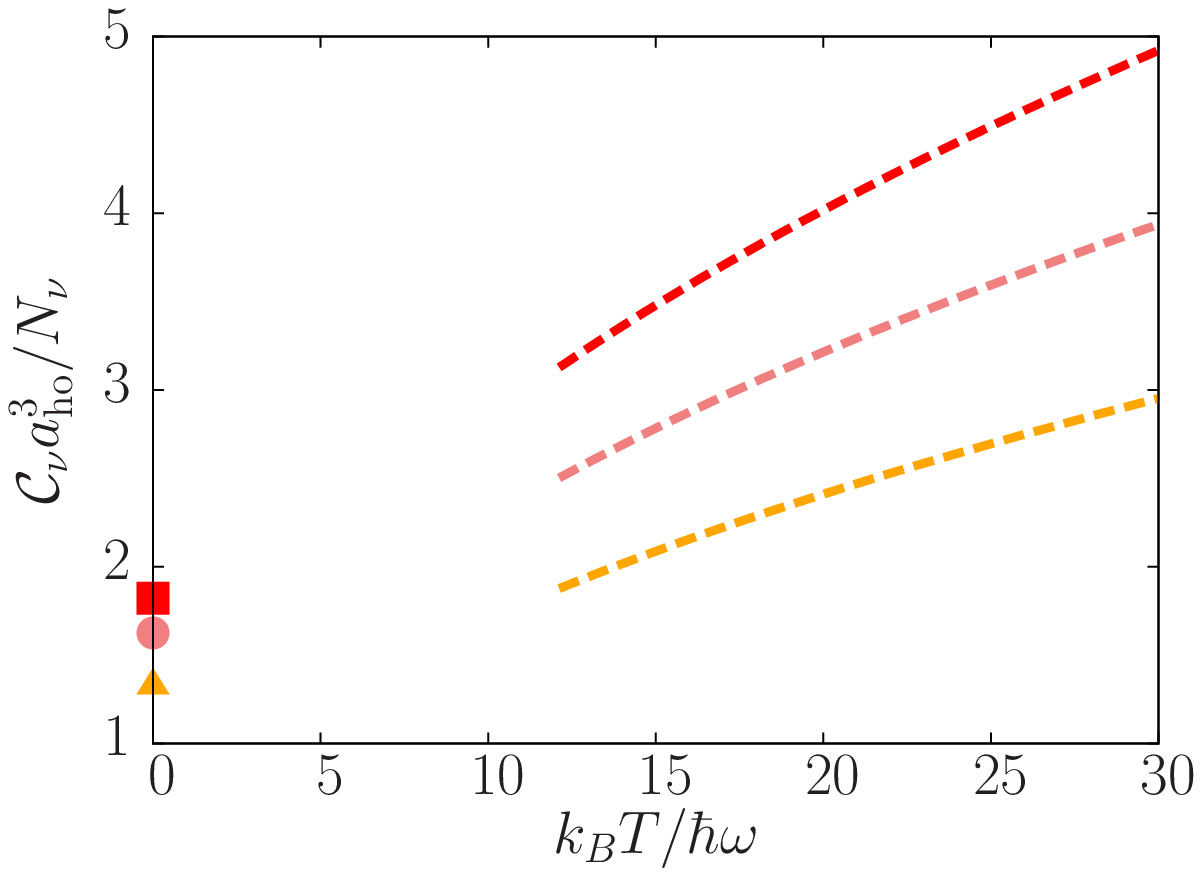}
\caption{\label{momt}Ground state normalized Tan's contacts as functions of $k_BT/\hbar\omega$ for three strongly interacting balanced 
mixtures of $N=6$ fermions (solid lines, $N_{\nu}={\color{Red}3},{\color{Peach}2},{\color{Orange}1}$.). The dashed lines correspond to Eq.~\eqref{ctmixt} 
while the $T=0$ points are given by the exact solution (Eq.~\eqref{csigmainfini}). From \cite{Decamp2016-2}.}
\end{figure}

In Fig.~\ref{momt} are plotted the high-temperature dependencies of balanced fermionic mixtures. The high-temperature curves are given by Eq.~\eqref{ctmixt} and 
the $T=0$ results correspond to the exact solution (Eq.~\eqref{csigmainfini}).

\paragraph{Discussion}

Several interesting facts are worth noticing. First of all, in the fermionized limit $g_{1D}\to\infty$, the contact is an increasing function of temperature. This is a typical one-dimensional effect: 
indeed, one would expect that the correlations (and thus the contact) would be destroyed by the high temperatures. This is indeed the fact for three dimensional 
systems \cite{Hu2011}, but the dimensional constraint implies an opposite behavior for one-dimensional systems (c.f. section \ref{pec1d}). Note that we did not discuss the case 
where interactions are finite, where we expect the presence of maximum, as recently obtained in the bosonic case \cite{Yao2018}. Moreover, we observe that the normalized contact 
is still an increasing function of the number of components, in agreement with the experimental observations \cite{Pagano2014}. Finally, the fact that the 
second virial coefficient is a constant (Eq.~\eqref{c2virial}) implies that the contact at infinite interactions is a \textit{universal} quantity, and thus, as already pointed out in section 
\ref{sectcassp}, a robust experimental observable.

An alternative way of deriving Tan's contact temperature dependence consists in performing, analogously to what we did in section \ref{tcfins}, an LDA on the 
thermal version of the Bethe ansatz equations (See appendix \ref{secbethe}, sections \ref{YYth} and \ref{tbamulti}). This was done recently in \cite{Yao2018} for 
the harmonically trapped Lieb-Liniger gas. This method has the advantage of allowing to express the contact simultaneously as a function of the temperature and the interaction 
strength, and for all temperatures. Nevertheless, extension to multi-component systems appears to be more intricate, given the complexity of the thermodynamic 
Bethe ansatz equations in this case, which involve an infinite set of coupled equations (c.f. Eq.~\eqref{tbaeqmulti}). A promising alternative to these 
equations has however been proposed in \cite{Patu2016} for the Gaudin-Yang model.

\subsubsection{Influence of the transverse confinement}
\label{tcincsdbose}

In section \ref{trapuco}, we have seen that experimentalists are able to construct quasi-1D traps for ultracold atoms by superimposing counter-propagating lasers with 
frequencies $(\omega_x,\omega_y,\omega_z)$ where $\omega_x=\omega_y=\omega_{\perp}$ such that the aspect ratio $\lambda=\omega_z/\omega_{\perp}$  is very small. Therefore,
the energy gap between the transverse ground state and the first excited state is higher than all the typical energies of the system, and the systems behaves  essentially 
like a one-dimensional system oriented in the $z$ direction. Nevertheless, one should not forget that, even if the system in this regime can be described by one-dimensional Hamiltonian, the interactions between 
particles are intrinsically three-dimensional. This can be seen in the derivation and expression of the effective one-dimensional scattering length $a_{1D}$ \cite{Olshanii1998}, 
which is a function of the three-dimensional scattering length $a_{3D}$ (c.f. Eq.~\eqref{a1dexpress}).

Tan's contact, that we defined in Eq.~\eqref{defcontsig}, is a two-body quantity. Then, it depends crucially on the three-dimensional nature of the contact 
interactions. Thus, it is natural to ask ourselves how the contact behaves as a function of $\lambda$, \textit{i.e.} in the \textit{dimensional crossover} 
consisting in allowing progressively the transverse states to be populated. In the following, we describe how this behavior can be studied in the case of a dilute 
Bose gas trapped in a potential of the form $V(x,y,z)=m/2(\omega_{\perp}^2x^2+\omega_{\perp}^2y^2+\omega_z^2z^2)$ with $\lambda\gg 1$, as published \cite{Decamp2018}. This is a bit different from what we studied until now, that is strongly interacting mixtures. However, it constitutes 
a first step toward the characterization of the dimensional crossover behavior of the contact in these systems.

\paragraph{Effective one-dimensional Gross-Pitaevskii equation}

The Gross-Pitaevskii theory is a mean-field theory describing dilute degenerate Bose gases \cite{Gross1961,Pitaevskii1961}. In the $\lambda=\omega_z/\omega_{\perp}\ll 1$ regime, an effective stationary Gross-Pitaevskii equation 
can be written \cite{Leboeuf2001,Gerbier2004,Delgado2006}:
\begin{equation}
  \label{eq_GP}
\left[-\frac{\hbar^2}{2m}\frac{\partial^2}{\partial z^2}+U(z)+\epsilon(n_1)-E\right]\psi(z)=0,   
\end{equation}
where the 1D order parameter $\psi(z)$ is associated to the effective 1D density profile through $n_1(z)=|\psi(z)|^2$, $U(z)$ is an external potential,
and $\epsilon(n_1)$ is given by \cite{Fuchs2003,Gerbier2004}
\begin{equation}
\label{mu}
\epsilon(n_1)=\hbar\omega_{\perp}\sqrt{1+4a_{3D}n_1}.
\end{equation}
The role of this parameter is to describe the effective one-dimensional interactions in the dimensional crossover. Indeed, in the regime where $a_{\perp}n_1\ll 1$, 
$\epsilon(n_1)\simeq \hbar\omega_\perp+2\hbar\omega_\perp a_{3D} n_1$, which corresponds to the standard non-linearity $\propto \psi|\psi|^2$ and thus to the 1D 
equivalent of the 3D Gross-Pitaevskii equation. In this regime, which  is referred as the \textit{mean field regime} (MF) in the following, the transverse wave function is in the ground state of the transverse harmonic oscillator. On the contrary, when $a_{3D}n_1\gg 1$, we have $\epsilon(n_1)\simeq 2\hbar\omega_\perp \sqrt{a_{3D} n_1}$, 
the non-linearity $\propto\psi|\psi|$ is no longer standard and many transverse excited states are populated, thus displaying a Thomas-Fermi profile: this is referred as the \textit{transverse Thomas-Fermi regime} (TTF).

\paragraph{Homogeneous contact from the quantum fluctuations}

We now turn to the calculation of Tan's contact in the homogeneous case, corresponding to $U(z)=0$ in Eq.~\eqref{eq_GP}. In this case the contact is a 
constant function of $z$, and we rather consider the \textit{homogeneous contact}, defined as 
\begin{equation}
\tilde{\mathcal{C}}=\frac{2}{\pi a_{1D}^2}G^{(2)}(0,0).
\end{equation}
In the homogeneous case where the gas is in a box of length $L$, we simply have $\mathcal{C}=L\tilde{\mathcal{C}}$. $\tilde{\mathcal{C}}$ is also called \textit{contact density}.

Note that the correlations $G^{(2)}(0,0)$ can not be deduced from Eq.~\eqref{eq_GP}, which is a mean-field model. In order to do so, we have to take into account 
the quantum fluctuations by a Bogoliubov method \cite{Gerbier2004}. Usually, in a 1D quantum model, one should not only take into account the density fluctuations, 
but also the phase fluctuations. It is indeed these phase fluctuations, related to the collective nature of excitations, which explain for example the impossibility 
of finite-temperature phase transitions in 1D (see section \ref{pec1d} and the MWH theorem). However they are related to the long-range order, or equivalently the 
low-momenta: since the contact is associated to the high-momentum behavior, we can safely neglect them in what follows.

The Bogoliubov expansion in our case is then very similar to its 3D counterpart. The idea is to consider that the system is highly degenerate, with $N_0\equiv N$ particles 
in the ground-state solution $\psi_0$ of Eq.~\eqref{eq_GP}, and add a quantum fluctuation $\delta\hat{\Psi}$ to the field operator of the system \cite{Pitaevskii_book}:
\begin{equation}
\hat{\Psi}(z)=\sqrt{N_0}\psi_0(z)+\sum_{k\neq 0}\psi_k(z)\hat{a}_k,
\end{equation}
where $\hat{a}_k$ annihilates a particle with momentum $\hbar k$ (with an analogous equation for the creation field operator $\hat{\Psi}^{\dagger}$). After expanding 
Eq.~\eqref{eq_GP} on this basis up to the second order in $\delta\hat{\Psi}$, we can perform the celebrated \textit{Bogoliubov transformation} \cite{Bogoliubov1947} which 
consists in diagonalizing the Hamiltonian in terms of the bosonic operator $\hat{b}_k=u_k\hat{a}_k+v_{-k}\hat{a}_{-k}^{\dagger}$, where $u_k$ and $v_k$ obey 
the \textit{Bogoliubov-de Gennes equations}:
\begin{equation}
  \label{eq:bog}
  \hbar\omega 
  \begin{pmatrix}
    u_k \\ v_k
  \end{pmatrix}
  =
  \begin{pmatrix}
    \frac{\hbar^2k^2}{2m}+mc^2 & mc^2 \\ -mc^2 &-\frac{\hbar^2k^2}{2m}-mc^2
  \end{pmatrix}
 \begin{pmatrix}
    u_k \\ v_k
  \end{pmatrix},
\end{equation}
as well as $u_k^2-v_k^2=1$, which is a consequence of the bosonic commutation relations. In Eq.~\eqref{eq:bog}, $c$ is the \textit{effective velocity of sound}, 
and is given in our system by \cite{Stringari1996,Zaremba1998}
\begin{equation}
\label{sndvel}
  mc^2=n_1\epsilon'(n_1).
\end{equation}
Then, solving Eq.~\eqref{eq:bog} yields the famous \textit{Bogoliubov spectrum}, given by
\begin{equation}
\label{bospec}
\hbar\omega(k)=\sqrt{\hbar^2k^2c^2+\left(\frac{\hbar^2k^2}{2m}\right)^2}.
\end{equation}

The momentum distribution for $k\ne 0$ at zero temperature can then been obtained in the Bogoliubov paradigm as:
\begin{equation}
\label{mom}
\begin{split}
  n(k)&=\langle \hat{a}_k^\dagger \hat{a}_k\rangle=v_k^2\\
  &=\frac{\hbar^2k^2/2m+mc^2}{2\hbar\omega(k)}-\frac{1}{2}.
  \end{split}
\end{equation} 
The homogeneous contact is then extracted from Eqs.~\eqref{sndvel}, \eqref{bospec} and \eqref{mom}, giving
\begin{equation}
\label{c1d}
\tilde{\mathcal{C}}=\frac{4}{a_{\perp}^4}\frac{a_{3D}^2n_1^2}{1+4a_{3D}n_1},
\end{equation}
where $a_{\perp}=\sqrt{\hbar/m\omega_{\perp}}$ is the radial harmonic oscillator length.

Notice that the contact goes from a $\mathcal{O}(a_{3D}n_1)$ to a $\mathcal{O}((a_{3D}n_1)^2)$ behavior when going from the TTF to the MF regime.

\paragraph{Comparison with Lieb-Liniger theory}

Here we compare Eq.~\eqref{c1d} to the results obtained from a completely different paradigm, that is the Lieb-Liniger theory (c.f. appendix \ref{secbethe}, section \ref{LL}).

Tan sweep theorem (Eq.~\eqref{sweepsigma}) allows to express the homogeneous contact as a function of the adimensional coupling strength $\gamma=-2/a_{1D}n_1$ and 
the adimensional ground-state energy $e$ (which is obtained solving the Bethe ansatz equations of the Lieb-Liniger model, Eqs~\eqref{norm}, \eqref{fredholm}, \eqref{gsade}):
\begin{equation}
\tilde{\mathcal{C}}=\frac{4n_1^2}{a_{1D}^2}e'(\gamma),
\end{equation}
where $a_{1D}$'s expression if given in Eq.~\eqref{a1dexpress}. Moreover, if $a_{\perp}\gg a_{3D}$, one can consider simultaneously the $\gamma\ll 1$ 
(\textit{i.e.} $a_{1D}n_1\gg 1$) regime and the MF regime $a_{3D}n_1\ll 1$. In this case, Eq.~\eqref{a1dexpress} implies $a_{1D}\approx - a_{\perp}^2/a_{3D}$ and 
the low-coupling expression of $e$ (Eq.~\eqref{lowgamma}) yields
\begin{equation}
\tilde{\mathcal{C}}\sim \frac{4a_{3D}^2n_1^2}{a_{\perp}^4},
\end{equation}
which corresponds exactly to the $a_{3D}n_1\ll 1$ equivalent of $\tilde{\mathcal{C}}$ in Eq.~\eqref{c1d}.

\begin{figure}
\begin{center}
  \includegraphics[width=0.6\linewidth]{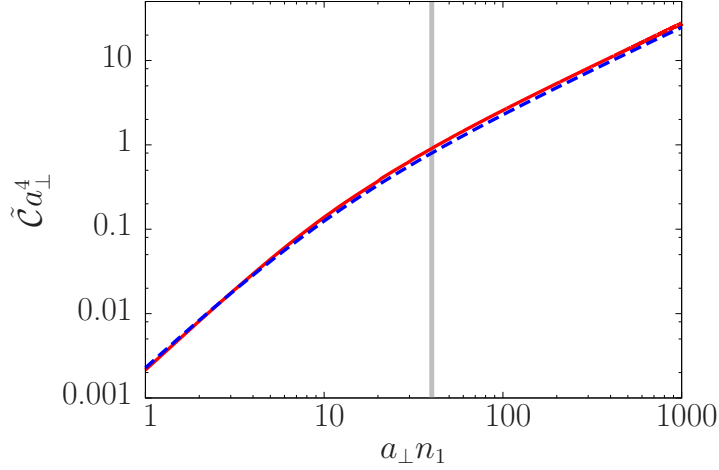}  
  \caption{Homogeneous contact  $\tilde{\mathcal{C}}$ in units of $a_{\perp}^{-4}$ as a function of $a_{\perp}n_1$, with $a_{3D}/a_{\perp}=0.025$. The  dashed blue line is plotted using  
   Eq.~\eqref{c1d} and we obtained the solid red line numerically by the GLL method. The vertical gray line represents 
  $a_{\perp}n_1=1$ and corresponds to the transition from the MF to the TTF regime.
  \label{fig:gllsound}}
  \end{center}
\end{figure}

In the other regimes, it is not possible to compare analytically the one-dimensional Lieb-Liniger theory with our approach. However, a variational gaussian 
ansatz, known as the \textit{generalized Lieb-Liniger theory} (GLL), was proposed in \cite{Salasnich2004,Salasnich2005} in order to take into account the 
effects of the transverse confinement. It consists in supposing that the 3D wave function $\Phi(\vec{r}_1,\ldots,\vec{r}_N)$ (where $\vec{r}_i\equiv (x_i,y_i,z_i)$ 
is the 3D coordinate of particle $i$) is given by:
\begin{equation}
\Phi(\vec{r}_1,\ldots,\vec{r}_N)=\psi(z_1,\ldots,z_N)\prod_{i=1}^N\frac{\exp\left(-\frac{x_i^2+y_i^2}{2\sigma^2}\right)}{\sqrt{\pi}\sigma}.
\end{equation}
Here, the variational parameter $\sigma$ corresponds to the transverse width of the cloud. The lineic total energy $\mathcal{E}$ can then be extracted within 
this hypothesis and is equal to 
\begin{equation}
\label{egll}
\mathcal{E}=\frac{\hbar^2}{2m}n_1^3e\left(\frac{2a_{3D}}{a_{\perp}^2n_1\sigma^2}\right)+\frac{\hbar\omega_{\perp}}{2}n_1\left(\frac{1}{\sigma^2}+\sigma^2\right),
\end{equation}
where $e$ is obtained by solving numerically the Bethe ansatz equations.
Then, $\delta \mathcal{E}/\delta \sigma =0$ yields
\begin{equation}
\label{sigma}
\sigma^4=1+2a_{3D}n_1e'\left(\frac{2a_{3D}}{a_{\perp}^2n_1\sigma^2}\right).
\end{equation}
Once $\mathcal{E}$ is obtained by solving Eqs.~\eqref{egll} and \eqref{sigma} consistently, we can extract the homogeneous contact from Eq.~\eqref{mom} as well as the relation \cite{Stringari1996,Zaremba1998}:
\begin{equation}
c=\sqrt{\frac{n_1}{m}~\frac{\partial^2\mathcal{E}}{\partial n_1^2}}.
\end{equation}
In Fig.~\eqref{fig:gllsound} we have plotted the homogeneous contact obtained by this method and compared it to the analytic result of Eq.~\eqref{c1d}. As one can see, the two methods 
give mutually consistent results.

\paragraph{Trapped one-dimensional contact}

Here we extract the contact $\mathcal{C}$ from the homogeneous contact $\tilde{\mathcal{C}}$ in the realistic case where $U(z)=m\omega_z z^2/2$. To do so, in the same 
spirit as what we did in section \ref{tcfins}, the idea is to perform a LDA. However, in the present case, we dispose of an analytic expression for the 
density profile \cite{Delgado2006}:
\begin{equation}
\begin{split}
\label{densityprofile}
n_1(z)= & \frac{1}{4a_{3D}}\left(\frac{\lambda Z}{a_{\perp}}\right)^2\left[ 1-\left(\frac{z}{Z}\right)^2\right] \\
& +\frac{1}{16a_{3D}}\left(\frac{\lambda Z} {a_{\perp}}\right)^4\left[ 1-\left(\frac{z}{Z}\right)^2\right]^2,
\end{split}
\end{equation}
where $Z$, the axial Thomas-Fermi radius, verifies
\begin{equation}
\frac{1}{15}\left(\frac{\lambda Z}{a_{\perp}}\right)^5+\frac{1}{3}\left(\frac{\lambda Z}{a_{\perp}}\right)^3 =\frac{\lambda N a_{3D}}{a_{\perp}}\equiv\chi_1,
\end{equation}
and can be approximated by \cite{Delgado2008}
\begin{equation}
\frac{\lambda Z}{a_{\perp}}\simeq\left(\frac{1}{(15\chi_1)^{4/5}+\frac{1}{3}}+\frac{1}{57\chi_1+345}+\frac{1}{(3\chi_1)^{4/3}}\right)^{-1/4}.
\end{equation}
The $\chi_1$ parameter characterizes  the transtion between the TTF regime ($\chi_1\gg 1$) and the MF regime ($\chi_1\ll 1$) in the harmonic trap \cite{Stringari2002}.

Then, the LDA approximation implies $\mathcal{C}=\int dz~\tilde{\mathcal{C}}(n_1(z))$, which yields
  \begin{equation}
    \label{ctrap}
    \begin{split}
      \mathcal{C} = & \left\{
        \lambda Z\sqrt{\lambda^2 Z^2 +2a_{\perp}^2}\left(2\lambda^6Z^6+14a_{\perp}^2Z^4\lambda^4+5a_{\perp}^4\lambda^2Z^2-15a_{\perp}^6\right)\right.\\
      & \left. +30a_{\perp}^8
        \operatorname{artanh}\left(\frac{\lambda Z}{\sqrt{\lambda^2Z^2+2a_{\perp}^2}}\right)\right\}\frac{1}{30 a_{\perp}^8
        \lambda\left(\lambda^2 Z^2+2a_{\perp}^2\right)^{3/2}}.
    \end{split}
  \end{equation}
  
  In the TTF and MF regimes, Eq.~\eqref{ctrap} admits respectively the more compact forms:
  \begin{equation}
  \label{iMF}
  \mathcal{C}\underset{\chi_1\gg 1}{\sim} \frac{a_{3D}N}{a_{\perp}^4}\equiv \mathcal{C}_{TTF},
  \end{equation}
and
\begin{equation}
\label{itf}
\mathcal{C}\underset{\chi_1\ll 1}{\sim} \frac{4(3\lambda)^{2/3}}{5a_{\perp}^{14/3}}a_{3D}^{5/3}N^{5/3}.
\end{equation}  
This last expression is, as expected, compatible with the contact obtained by LDA on the weak coupling Lieb-Liniger expansion of the homogeneous contact \cite{Olshanii03,Lang2017}.

\begin{figure}
  \begin{center}
    \includegraphics[width=0.6\linewidth]{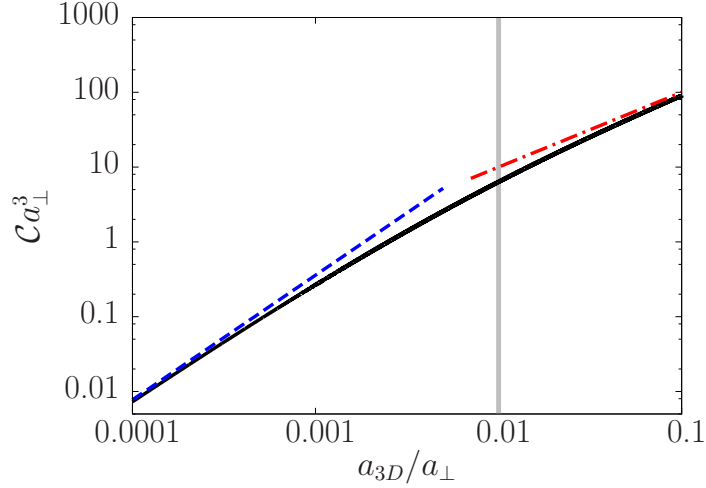}  
    \caption{Tan's contact $\mathcal{C}$ in units of $a_{\perp}^{-3}$ as a function of $a_{3D}/a_{\perp}$. The black curve 
    is plotted using Eq.~(\ref{ctrap}), while the  dashed blue and dot-dashed red curves correspond the MF and TTF expansions from Eqs (\ref{iMF}) and (\ref{itf}), respectively. The  vertical gray line verifies 
    $\chi_1=1$ and is associated with the transition from the MF to the TTF regime. The set of parameters used in these plots is 
    $\left(N=1000,\lambda=0.1\right)$.\label{fig:crossover}}
  \end{center}
\end{figure}

We observe that the contact has clearly distinct scaling behaviors in the TTF regime (where it is linear in $a_{3D}N$) and the MF regime (where $\mathcal{C}\propto (a_{3D}N)^{5/3}$). It can thus be used as an experimental characterization of the dimensional regime in highly anisotropic traps. Notice that, in the TTF regime, which is associated with a three-dimensional system, the contact does not depend on the aspect ratio $\lambda$. In Fig.~\ref{fig:crossover}, we have plotted 
$\mathcal{C}$ as a function of $a_{3D}$.

\begin{figure}
\begin{center}
  \includegraphics[width=0.6\linewidth]{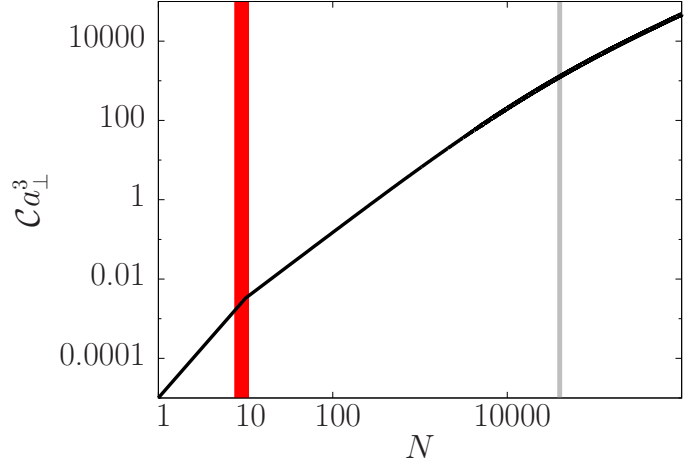}  
  \caption{Tan's contact $\mathcal{C}$ (black line)  in units of $a_{\perp}^{-3}$ as a function of the number of bosons $N$. 
  The thick vertical red line represents
  $\xi_1=N \lambda a_{\perp}^2/a_{3D}^2 = 1$ and corresponds to the transition from the strongly interacting regime to the MF regime. The left part is 
  plotted using Eq.~(\ref{itonks}), the right part
  using Eq.~(\ref{ctrap}). The small shift at the transition is due to corrections to Eq.~(\ref{itonks}) in the intermediate regime (which are not known analytically). The vertical gray
  line corresponds to the transition from the MF regime to the TTF regime at $\chi_1=N\lambda a_{3D}/a_{\perp}=1$.
   The set of parameters used in these plots is
  $\left(\lambda=0.0005,a_{3D}/a_{\perp}=0.05\right)$.\label{fig:contactn}}
  \end{center}
\end{figure}

If we plot the contact as a function of the number of bosons, we can plot on a same graph the TTF, MF and one-dimensional strongly interacting regimes. The relevant parameter in order to characterize the transition from the strongly-interacting regime to the MF regime in the 1D trap is $\xi_1\equiv Na_{1D}^2/a_z^2=N\lambda a_{\perp}^2/a_{3D}^2$ \cite{Stringari2002,PetrShlyap00,Olshanii2001}\footnote{Alternatively, we could have chosen the $\alpha_0$ parameter defined in section \ref{tcfins}.}. The strong-coupling expansion of the trapped bosonic contact, obtained by a similar method as in section \ref{tcfins}, is given by \cite{Lang2017}:
\begin{equation}
\label{itonks}
\mathcal{C}=\frac{N^{\frac{5}{2}}\lambda^{\frac{3}{2}}}{a_{\perp}^3}\left[\frac{256\sqrt{2}}{45\pi^2}+\sqrt{\xi_1}\left(\frac{70}{9\pi^2}-\frac{8192}{81\pi^4}\right)\right].
\end{equation}
The transition between the three regimes is plotted in Fig.~\ref{fig:contactn}.

We have summarized the main results for the bosonic contact in highly elongated traps in table \ref{summary}.
  
\begin{sidewaystable}
\renewcommand{\arraystretch}{2.75}
\begin{center}
\begin{tabular}{|c||c|c|c||c|}
\hline 
Regime & TTF & TTF $\to$ MF & MF &  SI \\\hline\hline
Parameters & $\chi_1\gg 1$ & $\chi_1\simeq 1$ &  $\chi_1\ll 1$ , $\xi_1\gg 1$ &  $\xi_1\ll 1$\\\hline
$n_1(z)$ & $\propto (1-z^2/Z^2)^2$ & $\frac{1}{4a}\left(\frac{\lambda Z}{a_{\perp}}\right)^2\left[ 1-\left(\frac{z}{Z}\right)^2\right]
 +\frac{1}{16a}\left(\frac{\lambda Z} {a_{\perp}}\right)^4\left[ 1-\left(\frac{z}{Z}\right)^2\right]^2$ & $\propto (1-z^2/Z^2)$ & $\propto (1-z^2/Z^2)^{1/2}$\\\hline
$\tilde{\mathcal{C}}$ & $\frac{a_{3D}n_1}{a_{\perp}^4}$ & $\frac{4}{a_{\perp}^4}\frac{a_{3D}^2n_1^2}{1+4a_{3D}n_1}$ & $\frac{4a_{3D}^2n_1^2}{a_{\perp}^4}$ &  $4\pi^2n_1^4\left(\frac{1}{3}+3a_{1D}n_1\right)$ \\\hline
$\mathcal{C}$ & $\frac{a_{3D}N}{a_{\perp}^4}$ & Eq.~(\ref{ctrap}) & 
$\frac{4(3\lambda)^{\frac{2}{3}}}{5}a_{\perp}^{-\frac{14}{3}}a_{3D}^{\frac{5}{3}}N^{\frac{5}{3}}$
& $\frac{N^{\frac{5}{2}}\lambda^{\frac{3}{2}}}{a_{\perp}^3}\left[\frac{256\sqrt{2}}{45\pi^2}+\frac{a_{1D}
\sqrt{N\lambda}}{a_{\perp}}\left(\frac{70}{9\pi^2}-\frac{8192}{81\pi^4}\right)\right]$\\\hline
\end{tabular}\caption{Contacts $\tilde{\mathcal{C}}$ and $\mathcal{C}$ and density profiles in the TTF, MF and 1D strongly interacting (SI) regimes. The transition parameters are given by $\chi_1=N\lambda a/a_{\perp}$ 
and $\xi_1=N\lambda a_{\perp}^2/a^2$. In the intermediate regime $\xi_1\simeq 1$ between MF and SI, the expressions are not known analitycally.}\label{summary}
\end{center}
\end{sidewaystable}  
  
\paragraph{Three-dimensional contact}

As we discussed in section \ref{introtan}, Tan's contact paradigm is not specific to 1D. We can indeed define a 3D contact $\mathcal{C}_{3D}$ analogously to what we did for the 1D contact, which will also verify the (3D) Tan relations. In highly anisotropic traps, it is then natural to wonder how the 1D contact is related to the 3D one. In the quasi-1D regime, which correspond to the MF regime in our case, it can be shown that they are related by a simple geometric factor \cite{Valiente2012,Zhou2017}:
\begin{equation}
\label{c3d1d}
\mathcal{C}_{3D}=\pi a_{\perp}^2 \mathcal{C}.
\end{equation}
Quite remarkably, despite the high non-uniformity of the system in the radial direction, Eq.~\eqref{c3d1d} shows that, in the quasi-1D regime, the system behaves as if it where a cylinder of radius $a_{\perp}$ with a constant lineic contact.

In the TTF regime, the numerous excited states should imply non-negligible non-uniformity effects. Indeed, if we perform a 3D treatment similar to what we just did, \textit{i.e.} by doing a LDA on the homogeneous contact obtained by Bogoliubov theory, one obtain in the 3D Thomas-Fermi regime \cite{Chang2016}:
\begin{equation}
\mathcal{C}_{3D}=\frac{64\pi^2}{7}a_{3D}^2Nn_0,
\end{equation}
where $n_0$ is the atomic density in the center of the trap. Using its expression in highly elongated traps \cite{Baym1996}, we get
\begin{equation}
\mathcal{C}_{3D}=\frac{8\pi}{7}\frac{a_{3D}N}{a_{\perp}^2}\left(\frac{15\lambda Na_{3D}}{a_{\perp}}\right)^{5/2}.
\end{equation}
Moreover, in the TTF regime, the radial Thomas-Fermi radius is given by \cite{Pitaevskii_book}
\begin{equation}
R_{\perp}=2a_{\perp}(an_1(0))^{1/4}.
\end{equation}
Therefore, we have found
\begin{equation}
\begin{split}
\mathcal{C}_{3D}&=\frac{8}{7}\pi R_{\perp}^2\frac{a_{3D}N}{a_{\perp}^2}\\
&=S\mathcal{C}_{TTF},
\end{split}
\end{equation}
where $S\equiv\frac{8}{7}\pi R_{\perp}^2$ is the cross section of the trap up to a numerical factor that can be accounted for the non-uniformity of the system. This last relation is the generalization of Eq.~\eqref{c3d1d} to the TTF regime.
  
\clearemptydoublepage

\clearemptydoublepage
\pagestyle{fancy}
\thispagestyle{empty}
\markboth{Conclusion}{Conclusion}
\chapter*{Conclusion}
\addcontentsline{toc}{chapter}{Conclusion}

This thesis has been devoted to the theoretical study of one-dimensional quantum mixtures, in the experimentally relevant case of particles with short-range and strong repulsions 
trapped in a harmonic potential. These particles are moreover supposed to have equal masses and to be subjected to the same external and interaction potentials regardless of 
their spin-component, which confers highly symmetric properties to the system. This can 
be realized using ultracold atoms with a purely nuclear spin, as it was recently achieved in the groundbreaking experiment of the group of L. Fallani in Florence \cite{Pagano2014}.

In the absence of an external harmonic potential, the system is integrable and one disposes of an extremely powerful theoretical tool, namely the Bethe ansatz, which allows in principle to obtain exact results 
in the system for any range of the interaction strength and temperature. Although integrability is destroyed by the presence of the harmonic potential, 
the system is also exactly solvable in the fermionized limit of infinite repulsions. In chapter \ref{Exactsol}, we have explained how, exploiting a perturbative ansatz, we have obtained  numerically the exact solutions 
for various few-body systems. Our program works in principle for any number of particles and any kind of quantum mixture. Moreover, we have provided a graph 
theory interpretation of the perturbative ansatz. Combined with graph spectral theory, an analysis of our system using this interpretation would be an interesting 
and potentially fruitful perspective.

Once these exact solutions are known exactly, a natural question is to characterize their exchange symmetry. Indeed, although the spatial exchange symmetry 
of particles belonging to the same component is fixed by their bosonic or fermionic nature, the question is more tricky to address when considering particles 
belonging to different components. In chapter \ref{chap:sym}, we have described how we have adapted the so-called class-sum method, which allows to characterize to 
which irreducible representation a wave function belongs, to the perturbative ansatz. We have studied in particular the symmetries of systems with six particles, and analyzed 
the relation between the ordering of the energy levels and the symmetries. We have shown that our system verifies the celebrated Lieb-Mattis theorem, which 
states intuitively that it wants to be as symmetric as possible, whatever the kind of mixture (bosonic, fermionic or mixed). Besides, we were able to compare 
 energy levels that are not comparable with this theorem. Moreover, we have highlighted a connection between this energy ordering and the central characters, 
\textit{i.e.} the eigenvalues of the class-sum operators. If proven, this connection would allow to compare systematically the ordering of energy levels belonging to different 
symmetry class beyond the scope of the Lieb-Mattis theorem. A possible use of this \textit{a priori} knowledge would then be to adapt the perturbative ansatz consequently,
which would considerably reduce the cost of our calculations.

In the last and longest chapter of this thesis (chapter \ref{chap:1bcor}), we have studied the one-body correlations in our system. This quantity is of extreme 
theoretical importance, and can be measured quite easily in a typical ultracold experiment by standard methods such as \textit{in-situ} absorption  imaging and time-of-flight techniques. We have obtained a very general 
formula in order to extract it from a solution obtained by the aforementioned perturbative ansatz. Then, we have analyzed the exact density profiles and momentum 
distributions of various few-body systems, highlighting the effects of the strong interactions and of the symmetry. We have, in particular, obtained simple 
rules in order to guess qualitatively the shape of these profiles from symmetry arguments. Thus, the density profiles are so that particles which are submitted 
to an anti-symmetric exchange (like fermions belonging to the same spin-component) tend to avoid each other, and the momentum distributions contain as many 
peaks as the number of anti-symmetric exchanges. However, although the role of symmetry is  obvious, we have shown that a measurement of the density profile or 
of the central part of the momentum distribution does not allow to extract uniquely the symmetry class of the system.

Then, we have shown that this experimental symmetry characterization can be realized  by a measurement of the so-called Tan's contact, a quantity that 
governs the algebraic asymptotic  behavior of the momentum distributions in short-range interacting quantum gases. Indeed, we have proved that Tan's contact is an increasing function of 
the symmetry, and thus of the number of components, as observed in \cite{Pagano2014}, which makes it a tool in order to compare symmetries. Moreover, provided 
that the experiment is sufficiently accurate, we have shown that a measure of Tan's contact allows to infer uniquely the spatial and spin symmetries, and thus that it can be used 
as a magnetic structure probe. 

The previous results were derived in the perfect case of infinite repulsions and zero temperature. In order to be as relevant as possible for experiment, 
we have then derived scaling laws for Tan's contact. 

First, we have obtained a strong coupling expansion for the contact of a balanced 
fermionic mixture at zero temperature. To do so, we have performed a local density approximation on results obtained in the homogeneous case by Bethe ansatz. Interestingly, 
although it is derived for a large number of particles, this law is in perfect agreement with the exact results we obtained at infinite interactions and with finite-interaction 
DMRG results obtained by our collaborators Matteo Rizzi and Johannes J\"{u}nemann. This agreement is certainly due to a scale invariance in our system, that 
would require a deeper treatment, for instance using the renormalization group theory or conformal field theory. Moreover, we have shown that the dependence we have obtained 
as a function of the number of particles and the number of components is in agreement with previous Monte Carlo simulations \cite{Matveeva2016} and experimental observations \cite{Pagano2014}.

Second, using a virial expansion, we have obtained an expression for the contact at high temperature and infinite interaction that is valid for any kind 
of quantum mixture.  As in the zero temperature case, we have obtained that the contact an increasing function of the number of components. Besides, we have shown that it is also an increasing function of 
temperature, which is a striking consequence of the dimensional constraint. Furthermore,  the contact displays a universal behavior as a function of temperature, since it does not depend explicitly on interactions.
A natural perspective would then be to analyze deeper these universal properties by deriving a scaling law for the contact at finite temperature and finite interaction 
strength simultaneously. This was recently achieved  in the simpler case of a one-component Bose gas, in which the presence of a maximum as a function of the temperature for finite interactions was highlighted \cite{Yao2018}.

Finally, we have taken into account the intrinsically three-dimensional nature of interactions in a realistic quasi-one-dimensional trap by studying the influence 
of the transverse confinement. Using an approach based on Gross-Pitaveskii and Bogoliubov theories combined with a local density approximation, we have obtained 
a scaling law for the contact of a dilute Bose gas in a highly elongated trap as a function of the aspect ratio, three-dimensional scattering length and number of particles. Moreover, 
we have checked that it is compatible with results obtained using Lieb-Liniger theory. We have highlighted that the contact has completely different behaviors in the quasi-one-dimensional 
and three-dimensional cases, and that it can thus be used as a experimental characterization of the dimensional regime. The next step would then be to derive such a scaling 
for more general quantum mixtures as the ones considered earlier in this thesis.

\vspace{0.75cm}

In addition to the perspectives we mentioned above, many compelling questions remain for further prospects. For example, what happens if we consider systems 
with unequal masses? In this case, integrability is broken even in the absence of an external confinement, 
and very little is known on the properties of  such systems, although recent semi-analytical developments using hyperspherical coordinates offer promising 
results for few-body systems \cite{Dehkharghani2016,Harshman2017}. More realistic for the case of Bose-Fermi mixtures, an extension of the results discussed 
in this thesis to these models appears to be a challenging but interesting task.

Furthermore, another interesting issue would be to study the robustness of these features to disorder and quantum chaos by adding a kicked rotor-type potential \cite{Izrailev1990}.
This is motivated by the theoretical observation of $k^{-4}$ tails in the evolution operator of a system of two interacting bosons in an atomic kicked rotor potential \cite{Flach2017}, 
suggesting an interesting analogy with Tan's contact physics. 
In these systems, integrability, correlations, disorder and symmetries would interplay in a non-trivial way, and would possibly allow the occurrence of fascinating many-body features, such as, for example, 
\textit{many-body localization} \cite{Abanin2017}.

\clearemptydoublepage

\appendix
\titlecontents{chapter}
[1pt] 
{\addvspace{3pc}}
{\large\textsc{Appendix \thecontentslabel} --- \bf\sffamily}
{\large \bf\sffamily} 
{\titlerule*[0.75em]{.}\contentspage}
[\addvspace{1pc}]

\part*{Appendices}
\renewcommand{\chaptermark}[1]{\markboth{Annexe \thechapter \ -\ #1}{#1}}

\clearemptydoublepage
\clearemptydoublepage
\pagestyle{fancy}
\thispagestyle{empty}
\chapter{Coordinate Bethe ansatz}
\label{secbethe}

In this appendix, we consider a system of $N$ particles in 1D interacting via the $\delta$-type potential \eqref{upseudo}, without any external potential. This system 
is often referred as a \textit{quantum integrable system}. The definition of integrability in quantum physics is not as clear as in classical physics, but one 
can say, following \cite{Sutherland2004}, that a system is integrable when the scattering events occur without any diffraction. In other words, any scattering 
will simply result in an exchange of the momenta of the particles. In practice, this implies that the system can be solved using a closed set of equations that is 
valid for a wide range of parameters --- without the need of a perturbative method. The usual way to obtain this set of equations is through the so-called 
\textit{Bethe ansatz}, an educated guess for the form of the many-body wave-function that was first used by 
Hans Bethe to solve the 1D Heisenberg XXX model \cite{Bethe1931}. Since then, the number of models that have been solved with this very powerful and elegant method has been 
flourishing, ranging from 1D quantum gases in the continuum \cite{LiebLin,Gaudin1967,Yang67,Sutherland68} and the 1D Hubbard model \cite{Lieb1968}, to 2D classical spin 
chains \cite{Sutherland1967,Baxter1971}, or in more recent theories such as string theory in the context of AdS/CFT correspondence \cite{Amjorn2006}. 
Here we obviously focus on the first category of systems, and we try 
to build an intuitive understanding of this method rather than trying to be exhaustive and perfectly rigorous. After a first general description of the model 
in \ref{model}, we then describe Lieb and Liniger's solution for the interacting Bose gas as a pedagogical example in \ref{LL}. In the third and biggest part 
of this appendix, we analyze the more intricate case of a multi-component system in \ref{multi}\footnote{NB: Although only specific results of this appendix will be used in the main text of this thesis (in particular the Bethe ansatz equations of the Lieb-Liniger model and the strong-coupling expansion of the energy in the multi-component case), we have decided to write it as a self-sufficient introduction to coordinate Bethe ansatz. We indeed have the feeling that a complete set of derivations of this method without the use of involved algebraic tools is missing in the literature, despite its importance in the study of one-dimensional quantum gases.}

\section{Model and first considerations}
\label{model}
We consider a one-dimensional homogeneous (i.e. with no external potential) system of finite size $L$ containing $N$ particles of same masses $m$, interacting via a $\delta$-type 
potential  with an interaction strength $g_{1D}=2c$ with $c>0$. In this section, in order to simplify the expressions, we will write the equations in units of 
$\hbar=2m=1$. The Schr\"{o}dinger equation for the many-body wave-function $\psi(x_1,\ldots,x_N)$, where $x_j\in[0,L]$ is the coordinate of particle $j$, is given by
\begin{equation}
\label{schrodinger}
-\sum_{j=1}^{N}\frac{\partial^2\psi}{\partial x_j^2}+2c\sum_{i<j}\delta(x_i-x_j)\psi=E\psi.
\end{equation}
As explained in section \ref{secusp}, we do not need to specify the symmetry of the many-body wave-function here.
Let us suppose that the $N$ particles are in a given configuration, i.e. that the vector $\{x\}=(x_1,\ldots,x_N)$ is in a given \textit{sector} 
$D_Q=x_{Q1}<x_{Q2}<\cdots<x_{QN}$ of $[0,L]^N$, where $Q=(Q1,\ldots,QN)$ is a permutation of $\{1,\dots,N\}$. 
Eq.~\eqref{schrodinger} can be reduced to a free-wave Helmholtz equation
\begin{equation}
\label{free}
\left(\sum_{j=1}^{N}\frac{\partial^2}{\partial x_j^2}+E\right)\psi=0,
\end{equation}
together with the $N-1$ boundaries conditions, for $j\in\{1,\ldots,N-1\}$:
\begin{equation}
\label{continuity}
\left.\psi\right|_{x_{Q(j+1),Qj}=0^+}=\left.\psi\right|_{x_{Q(j+1),Qj}=0^-}
\end{equation}
and
\begin{equation}
\label{cusp}
\left.\left(\frac{\partial\psi}{\partial x_{Q(j+1)}}-\frac{\partial\psi}{\partial x_{Qj}}\right)\right|_{x_{Q(j+1),Qj}=0^+}-
\left.\left(\frac{\partial\psi}{\partial x_{Q(j+1)}}-\frac{\partial\psi}{\partial x_{Qj}}\right)\right|_{x_{Q(j+1),Qj}=0^-}=2c\left.\psi\right|_{x_{Q(j+1),Qj}},
\end{equation}
with $x_{Q(j+1),Qj}=x_{Q(j+1)}-x_{Qj}$.
Eq.~\eqref{continuity} is simply a continuity equation and Eq.~\eqref{cusp} is the cusp condition \eqref{cusp2b}. Therefore, the \textit{Bethe's hypothesis} consists, in a pretty natural way, in supposing that the solution of
Eq.~\eqref{schrodinger} in the sector $D_Q$ will be given by a combination of plane waves:
\begin{equation}
\label{bethe}
\left.\psi\right|_{\{x\}\in D_Q}=\sum_{P\in\mathfrak S_N} \left\langle Q||P \right\rangle e^{ i\left(x_{Q1}k_{P1}+\cdots+x_{QN}k_{PN}\right)},
\end{equation}
where $\mathfrak S_N$ is the set of all permutations of $\{1,\dots,N\}$ (see section \ref{subsecpermut}) and $\{k\}\in\mathbb{R}^N$ is a set of distinct parameters (pseudo-momenta) associated with the energy 
\begin{equation}
\label{energy}
E=\sum_{j=1}^Nk_j^2.
\end{equation}
Now, all the game will consist in finding the coefficients $\left\langle Q||P \right\rangle$ which satisfy the boundary conditions \eqref{continuity} and
\eqref{cusp} (in order to do that, we will have to have one more set of boundary conditions in $QN-Q1=0$, which is possible if we ask the particles to be on
a circle of size $L$ and impose $\psi$ to be periodic). Although the apparent simplicity of Bethe's hypothesis, the considerations allowing to achieve this 
can be pretty involved. However, if the wave-function $\psi$ is symmetric by exchange of coordinates, as it is the case in the one-component Bose gas, we will have
$\left\langle Q||P \right\rangle=\left\langle Q'||P \right\rangle$ for any $Q,Q'\in\mathfrak S_N$, which will considerably simplify the discussion. This case will be 
studied in section \ref{LL}, and its extension to multi-component systems in section \ref{multi}. In both sections, we will derive the sets of 
\textit{Bethe ansatz equations}, allowing us to extract the ground state and finite-temperature properties in the limit $N,L\to\infty$ with $n=N/L$ keeped constant.

\section{One-component Bose gas: the Lieb-Liniger model}
\label{LL}
\subsection{Deriving the Bethe ansatz equations}
The method employed in this section is pretty much the same as the one used in the original paper of E.H. Lieb and W. Liniger \cite{LiebLin}. As told in section \ref{model}, in the one-component Bose gas we have, for any $Q\in\mathfrak S_N$, $\left.\psi\right|_{\{x\}\in D_Q}=\left.\psi\right|_{\{x\}\in I}$ where
$D_I=x_1<\cdots<x_N$ is the \textit{fundamental sector} of $[0,L]^N$: it is therefore sufficient to obtain the solution \eqref{bethe} in $D_I$ and we can write it:
\begin{equation}
\label{BetheLL}
\left.\psi\right|_{\{x\}\in D_I}=\sum_{P\in\mathfrak S_N}a(P) e^ {i\left(x_1k_{P1}+\cdots+x_Nk_{PN}\right)},
\end{equation}
where $a(P)=\left\langle I||P \right\rangle$ has to satisfy Eqs \eqref{continuity} and \eqref{cusp}. Moreover, the cusp condition \eqref{cusp} simply becomes
\begin{equation}
\label{cuspLL}
\left.\left(\frac{\partial}{\partial x_{j+1}}-\frac{\partial}{\partial x_j}\right)\psi\right|_{x_{j+1}=x_{j}}=c\left.\psi\right|_{x_{j+1}=x_{j}}.
\end{equation}
Considering $P\in\mathfrak S_N$ and $P'=P(j,j+1)$ (i.e. $P$ and $P'$ are the same except $P(j+1)=P'j$ and $Pj=P'(j+1)$), Eqs \eqref{BetheLL} and \eqref{cuspLL}
give
\begin{equation}
\label{phase}
\begin{split}
\frac{a(P')}{a(P)}&=\frac{k_{Pj}-k_{P(j+1)}-ic}{k_{Pj}-k_{P(j+1)}+ic}\\
		  &=-e^{-i\theta(k_{Pj}-k_{P(j+1)})},
		  \end{split}
\end{equation}
with 
\begin{equation}
\label{theta}
\theta\left(k_{Pj}-k_{P(j+1)}\right)=-2\arctan\left(\frac{k_{Pj}-k_{P(j+1)}}{c}\right).
\end{equation}
Thus, the two-body scattering between two identical bosons just
reduces to a phase factor.

\begin{figure}\centering
\begin{tikzpicture}
\draw (0,0) circle(2);
\draw[fill] (0:2) circle(0.1);
\draw[fill] (45:2) circle(0.1);
\draw[fill] (80:2) circle(0.1);
\draw[fill] (110:2) circle(0.1);
\draw[fill] (125:2) circle(0.1);
\draw[fill] (210:2) circle(0.1);
\draw[fill] (290:2) circle(0.1);
\draw[fill] (310:2) circle(0.1);
\draw (1.5,1.5) node[right]{$k_j$};
\draw[>=latex,->] (45:2.7) arc (45:-10:2.7) ;
\end{tikzpicture}

\caption{\label{ringll}Semi-classical interpretation of the Bethe ansatz equations of the Lieb-Liniger model. A boson with pseudo-momentum $k_j$ 
scatters with the $N-1$ other particles and returns to its initial position after a whole turn around the ring. After this operation, the wave-function
will acquire a phase factor $e^{ik_jL}$ due to turn, and a phase factor $(-1)^{N-1}\prod_{l=1,l\neq j}^Ne^{-i\theta(k_j-k_l)}$ due to the two-body scatterings. The periodic
boundary conditions imply that this global phase factor must be equal to 1.}
\end{figure}
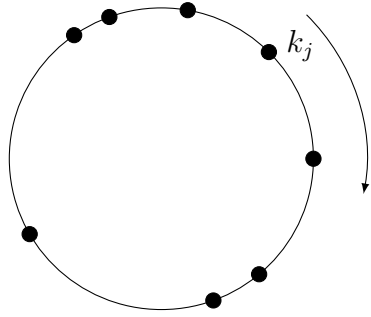

In order to obtain the so-called Bethe ansatz equations, we will have to impose that our system is in fact on a ring of circumference $L$, with periodic boundaries 
conditions for $\psi$. This may appear to be strange at first sight, as it seems to restrict a lot the geometry of the system. However, because we are going to consider eventually 
the thermodynamic limit $N,L\to\infty$, 
the boundary conditions can be chosen adequately without affecting the relevance of the problem (as often in physics, c.f. quantum field theories). More precisely, 
the periodic boundary condition on $\psi$ can be written in $D_I$:
\begin{equation}
\label{pbc}
\psi(x_N-L,x_1,\ldots,x_{N-1})=\psi(x_1,x_2,\dots,x_N).
\end{equation}
If we put this condition into Eq.~\eqref{BetheLL}, given Eq.~\eqref{phase}, it implies that for every $j\in\{1,\dots,N\}$ we have:
\begin{equation}
\label{expbaeLL}
e^{ik_jL}(-1)^{N-1}\prod_{l=1,l\neq j}^Ne^{-i\theta(k_j-k_l)}=1.
\end{equation}
This can be interpreted as the scattering of a particle with pseudo-momentum $k_j$ with the $N-1$ other particles after a whole turn around the 
ring (see Fig. \ref{ringll}). Then, putting Eq.~\eqref{expbaeLL} in a logarithmic form, we obtain a set of $N$ coupled equations, namely the Bethe ansatz 
equations for the Lieb-Liniger model:
\begin{equation}
\label{baeLL}
k_jL=2\pi I_j+\sum_{l=1,l\ne j}^N \theta(k_j-k_l),
\end{equation}
where $I_j$ is an integer when $N$ is odd and a half-odd integer otherwise.

What do the numbers $I_j$ represent? They have to be taken into account in order to remove the ambiguity of the phase when going from Eq.~\eqref{expbaeLL} to
Eq.~\eqref{baeLL}, but they happen to have a deeper physical meaning: they are the \textit{quantum numbers} of the system. In order to understand that, we consider the
Tonks limit $c\to\infty$ of hardcore bosons. In this case, we observe that $\theta(k_j^{\infty}-k_l^{\infty})=0$, so that the Bethe ansatz equations \eqref{baeLL} simply
implies $k_j^{\infty}=\frac{2\pi}{L}I_j$ for any $j\in\{1,\ldots,N\}$. Since we want the $k_j$ to be different (as it is the case in a one-component Fermi gas), we
see that the $I_j$ have to be different. Thus, the $I_j$ label the $k_j$, and a given set of these numbers will completely characterize the state $\{k\}$ of
the system through the Bethe ansatz equations.

\subsection{The ground state}
\label{GS}

Since $\theta(\Delta k)$ monotonically increases with $\Delta k$, it is clear given Eqs.~\eqref{energy} and
\eqref{baeLL} that the energy $E$ of the system is minimized when the set of quantum numbers $I_j$ is as compactly centered around 0 as possible. Therefore,
the ground state of the Lieb-Liniger model is characterized by $I_j\in\{-\frac{N-1}{2},\dots,\frac{N-1}{2}\}$. If we choose the $I_j$ so that 
$I_1<I_2<\cdots<I_N$ we also have
\begin{equation}
\label{distk}
\frac{2\pi I_1}{L}<k_1<\cdots<k_N<\frac{2\pi I_N}{L}.
\end{equation}

We can now consider the thermodynamic limit, where $N,L\to\infty$ with a fixed density $n=\frac{N}{L}$.
In this case, we define the quasi-momentum distribution $\rho(k)$ so that the number of $k$'s
between in a small interval $[k,k+\, d k]$ is $L\rho(k)\, d k$. 
Similarly to Eq.~\eqref{distk}, the $k$'s will distributed  symmetrically
between $q$ and $-q$, with $q$ verifying
\begin{equation}
\label{norm}
\int_{-q}^q \rho(k)\, d k=n.
\end{equation}
Then, the discrete summation over $k$ will become an integral:
\begin{equation}
\sum_{k=1}^N\to L\int_{-q}^q \rho(k)\, d k.
\end{equation}
The thermodynamic equivalent of the quantum number is found noticing that $I_j$
counts the number of $k$'s between 0 and $k_j$ (with the same sign as $k_j$). Therefore,
we can write Eq.~\eqref{baeLL} in the thermodynamic limit:
\begin{equation}
k=2\pi\int_0^k\rho(k')\, d k'+\int_{-q}^q\theta(k-k')\rho(k')\, d k',
\end{equation}
which becomes after differentiation
\begin{equation}
1=2\pi\rho(k)+\int_{-q}^q\theta'(k-k')\rho(k')\, d k'.
\end{equation}
Using the explicit expression for $\theta(\Delta k)$ we get
\begin{equation}
\label{fredholm}
1+2c\int_{-q}^q\frac{\rho(k')\, d k'}{c^2+(k-k')^2}=2\pi\rho(k).
\end{equation}
The ground state energy density $e=\frac{E}{L}$ can then be obtained with 
\begin{equation}
\label{gsade}
e=\int_{-q}^qk^2\rho(k)\, d k.
\end{equation}

Together with Eq.~\eqref{norm}, Eq.~\eqref{fredholm} completely determines the ground
state of the Lieb-Liniger gas in the thermodynamic limit. It is a Fredholm integral equation of the second kind
with a kernel of the form 
\begin{equation}
K(x,y)=\frac{1}{c^2+(x-y)^2}.
\end{equation}
This kernel, and hence Eq.~\eqref{fredholm}, is non singular for any $c>0$, guarantying a 
unique analytic solution. If $c=0$ however it becomes singular. This implies in practice that 
the numerical resolution  of Eq.~\eqref{fredholm} becomes more and more difficult as one 
approaches the $c\to 0$ limit, and that it is very difficult to obtain small $c$ asymptotic expansions 
of physical quantities like the energy. On the contrary, when one approaches the Tonks limit $c\to\infty$,
we have $K(x,y)\approx c^{-2}$, which leads to
\begin{equation}
\label{gsell}
e(\gamma)\underset{\infty}{\sim}\frac{\pi^2}{3}\left(\frac{\gamma}{\gamma+2}\right)^2,
\end{equation}
where $\gamma=\frac{c}{n}$ is the natural interaction parameter of our system. The $\gamma\ll 1$ is a little bit more tricky, but one can check that 
\begin{equation}
\label{lowgamma}
e(\gamma)\underset{0}{\sim}\gamma.
\end{equation}

\subsection{Finite temperature thermodynamics}
\label{YYth}
The so-called \textit{thermodynamic Bethe ansatz} was derived by C.N. Yang and C.P. Yang in \cite{YangYang1969}. They observed that in an excited state, 
the quantum numbers $I_j$ will still be a set of integers/half-odd integers, but not as compact as possible. There will be omitted
lattice sites $J_j$, and equivalently omitted $k$ values. These omitted $k$ values are called \textit{holes}. Then, in the same way than in section \ref{GS},
we can define a density $\rho(k)$ for the number of particles with pseudo-momentum $k$, but also a density $\rho_h(k)$ for the holes. In a similar fashion than
for Eq.~\eqref{fredholm}, we obtain an integral equation for $\rho$ and $\rho_h$:
\begin{equation}
\label{fredholmYY}
1+2c\int_{-\infty}^{\infty}\frac{\rho(k')\, d k'}{c^2+(k-k')^2}=2\pi\left(\rho(k)+\rho_h(k)\right),
\end{equation}
with $\rho$  verifying 
\begin{equation}
\label{normYY}
\int_{-\infty}^{\infty}\rho(k)\, d k=n.
\end{equation}
Note that this time we do not restrict the boundaries of integration. 

Contrary to the case of the ground state, there is here a degeneracy of the quantum states, which implies a non-zero entropy $S$. This degeneracy is given by
\begin{equation}
\frac{[L(\rho+\rho_h)\, d k]!}{[L\rho\, d k]![L\rho_h\, d k]!}\approx 
\exp\left[L\, d k\left((\rho+\rho_h)\ln(\rho+\rho_h)-\rho\ln\rho-\rho_h\ln\rho_h\right)\right].
\end{equation}
Putting the Boltzmann constant equal to 1, the entropy density is then
\begin{equation}
\frac{S}{L}=\int_{-\infty}^{\infty}\left((\rho+\rho_h)\ln(\rho+\rho_h)-\rho\ln\rho-\rho_h\ln\rho_h\right)\, d k.
\end{equation}
We can therefore write the quantum pressure $p=\frac{1}{L}[TS-E+\mu N]$:
\begin{equation}
\label{pressure}
p=\int_{-\infty}^{\infty}\left[T\left((\rho+\rho_h)\ln(\rho+\rho_h)-\rho\ln\rho-\rho_h\ln\rho_h\right)+(\mu-k^2)\rho\right]\, d k.
\end{equation}
Then, minimizing $p$ according to $\rho$ gives with the help of Eq.~\eqref{fredholmYY}:
\begin{equation}
\label{minp}
\ln\left(\frac{\rho_h}{\rho}\right)+\frac{c}{\pi}\int_{-\infty}^{\infty}\frac{\, d k'}{c^2+(k-k')^2}\ln\left(1+\frac{\rho}{\rho_h}\right)
+\frac{1}{T}[\mu-k^2]=0.
\end{equation}
Defining the \textit{pseudo-energy} $\epsilon(k)=T\ln\left(\frac{\rho_h}{\rho}\right)$, Eq.~\eqref{minp} can be re-written as an
integral equation for $\epsilon$:
\begin{equation}
\label{pseudo}
\epsilon(k)=-\mu+k^2-\frac{Tc}{\pi}\int_{-\infty}^{\infty}\frac{\, d k'}{c^2+(k-k')^2}\ln\left(1+e^{-\epsilon(k)/T}\right).
\end{equation}
Using Eqs.~\eqref{fredholmYY}, \eqref{pressure} and \eqref{pseudo} yields a very simple expression for the quantum pressure:
\begin{equation}
\label{pseudopress}
p=\frac{T}{2\pi}\int_{-\infty}^{\infty}\ln\left(1+e^{-\epsilon(k)/T}\right)\, d k.
\end{equation}

Thus, given $\mu$ and $T$, one can solve the integral equation \eqref{pseudo} in order to find $\epsilon$, and then obtain $p$ with Eq.~\eqref{pseudopress}.
The thermodynamic quantities of interest can then be obtained using
\begin{equation}
L\, d p=S\, d T+N\, d \mu.
\end{equation}

Analytics expressions for the Yang-Yang thermodynamics can be found in \cite{Guan2011}.

\section{Extension to the multi-component case}
\label{multi}

\subsection{Scattering operators and consistency}
\label{consistencysec}
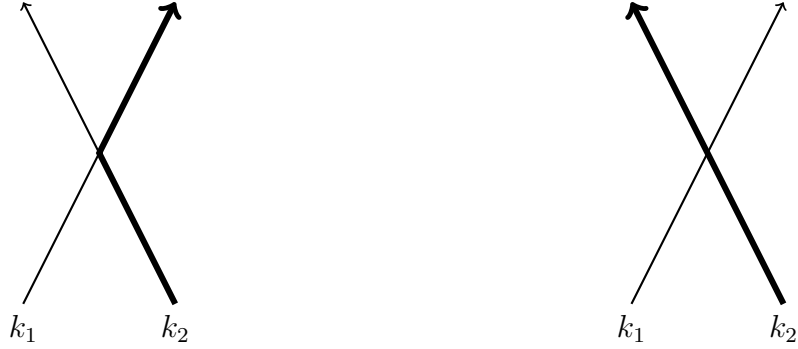
\begin{figure}\centering
\begin{tikzpicture}
\draw[thick,->] (1,0) -- (2,2) -- (1,4);
\draw[line width = 0.8mm,->] (3,0) -- (2,2)--(3,4);
\draw (1,0) node[below]{$k_1$};
\draw (3,0) node[below]{$k_2$};
\draw (9,0) node[below]{$k_1$};
\draw (11,0) node[below]{$k_2$};
\draw[thick,->] (9,0) -- (10,2) -- (11,4) ;
\draw[line width = 0.8mm,->] (11,0) -- (10,2)-- (9,4);
\end{tikzpicture}

\caption{\label{scatter}Two possible scattering diagrams for distinguishable particles. The black particle (thick line) can either be reflected against 
the white one (thin line) or transmitted, corresponding respectively to the left and the right diagram.}
\end{figure}

Let us now consider that the particles are no longer identical. In this case, the scattering between two particles with different internal  states 
(let us say a black one and a white one) will have two possible outcomes: either they will transmit or reflect (see Fig. \ref{scatter}). Equivalently, we can say that they can exchange
their colors, or not. Then, the scattering processes will no longer be described by a phase factor, as it was the case in the Lieb-Liniger model, but they 
will have to be described by a \textit{scattering matrix} $S$.

In order to construct the $S$-matrices of our $N$-body problem, we consider the boundary conditions Eqs.~\eqref{continuity} and \eqref{cusp}, as well as 
$P,P',Q,Q'\in\mathfrak S_N$ with $P'=P(j,j+1)$ and $Q'=Q(j,j+1)$. We then obtain the following relations between the coefficients:
\begin{equation}
\left\langle Q||P \right\rangle+\left\langle Q||P' \right\rangle=\left\langle Q'||P \right\rangle+\left\langle Q'||P' \right\rangle
\end{equation}
for the continuity equation and
\begin{equation}
i(k_{j+1}-k_j)\left(\left\langle Q||P \right\rangle-\left\langle Q||P' \right\rangle+\left\langle Q'||P \right\rangle-\left\langle Q'||P' 
\right\rangle\right)=2c\left(\left\langle Q||P \right\rangle+\left\langle Q||P' \right\rangle\right)
\end{equation}
for the cusp condition. Putting these two relations together brings
\begin{equation}
\label{qp}
\left\langle Q||P' \right\rangle=\frac{ic}{k_{j+1}-k_j-ic}\left\langle Q||P \right\rangle+\frac{k_j-k_{j+1}}{k_j-k_{j+1}+ic}\left\langle Q'||P \right\rangle.
\end{equation}

\begin{figure}\centering
\begin{tikzpicture}
\draw[thick,->] (1,0) -- (5,4);
\draw[thick,->] (5,0) -- (1,4);
\draw[thick,->] (2,0) -- (2,4);
\draw (6,2) node{$=$};three
\draw (1,0) node[below]{$k_1$};
\draw (2,0) node[below]{$k_2$};
\draw (5,0) node[below]{$k_3$};
\draw (7,0) node[below]{$k_1$};
\draw (10,0) node[below]{$k_2$};
\draw (11,0) node[below]{$k_3$};
\draw[thick,->] (7,0) -- (11,4) ;
\draw[thick,->] (11,0) -- (7,4);
\draw[thick,->] (10,0) -- (10,4);
\draw[fill] (10,1) circle(0.05);
\draw[fill] (2,1) circle(0.05);
\draw[fill] (10,3) circle(0.05);
\draw[fill] (2,3) circle(0.05);
\draw[fill] (3,2) circle(0.05);
\draw[fill] (9,2) circle(0.05);
\draw (2,1) node[right]{$S_{12}$};
\draw (2,3) node[right]{$S_{23}$};
\draw (3,2) node[right]{$S_{13}$};
\draw (10,1) node[right]{$S_{23}$};
\draw (10,3) node[right]{$S_{12}$};
\draw (9,2) node[right]{$S_{13}$};
\end{tikzpicture}

\caption{\label{figYB}Diagrammatic interpretation of the Yang-Baxter equations, insuring the consistency of our problem.}
\end{figure}
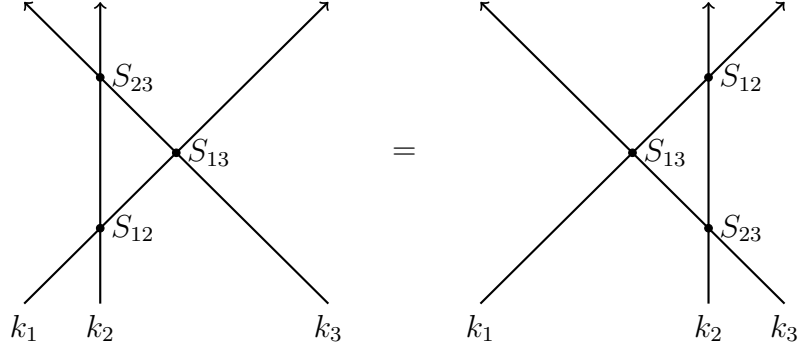

Before interpreting this equation, let us first observe that there are $(N-1)(N!)^2$ equations of this type 
(since $j\in\{1,\dots,N-1\}$ and $P,Q\in\mathfrak S_N$, $\mathfrak S_N$ having $N!$ elements), but there are only $(N!)^2$ coefficients $\left\langle Q||P \right\rangle$. Therefore, 
it is a priori not obvious that these equations are consistent with each other and characterize uniquely the coefficients. We postpone this question of consistency
to a little bit later.

We see in Eq.~\eqref{qp} that $\frac{ic}{k_{j+1}-k_j-ic}\equiv R_{j+1,j}$ is a reflection coefficient ("$Q$ remains $Q$") and $\frac{k_j-k_{j+1}}{k_j-k_{j+1}+ic}\equiv T_{j+1,j}$
is a transmission coefficient ("$Q$ becomes $Q'=Q(j,j+1)$"). Let us organize the $\left\langle Q||P \right\rangle$ coefficients in a $N!\times N!$ matrix, and denote
its columns by $\zeta_P$. Considering $P,P'\in\mathfrak S_N$ with $P'=P(j,l)$, we can write using Eq.~\eqref{qp}
\begin{equation}
\zeta_P=\left(R_{l,j}~\Id+T_{l,j}\hat{P}_{lj}\right)\zeta_{P'},
\end{equation}
where $\Id$ is the identity matrix and $\hat{P}_{lj}$ is the permutation operator of $l$ and $j$, i.e. with 0 coefficients except when the permutations corresponding 
to the coordinates are equal up to a transposition $(j,l)$. Thus, in all generality, considering two particles $a$ and $b$ with momenta $k_j$ and $k_l$, we
can define the scattering operator ${S_{jl}}^{ab}$ by:
\begin{equation}
\label{smatrix}
{S_{jl}}^{ab}=R_{jl}~\Id+T_{jl}\hat{P}_{ab}.
\end{equation}
This is called the \textit{reflection representation} of the scattering matrix, because we are considering the columns $\zeta_P$ and the $Q$ remains unchanged, 
so that the reflection coefficients are on the diagonal. Alternatively, we could have defined the \textit{transmission representation} of the scattering 
matrix ${\tilde{S}_{jl}}^{ab}\equiv \hat{P}_{ab}{S_{jl}}^{ab} =T_{jl}~\Id+R_{jl}\hat{P}_{ab}$ relating $\zeta_{P'}$ to $\hat{P}_{ab}\zeta_{P}$.

As told before, in order to be consistent, the set of $(N-1)(N!)^2$ equations of the form \eqref{qp}, describing two-body scatterings, should lead to a 
unique set of  $\left\langle Q||P \right\rangle$ coefficients. In other words, if we want to obtain  $\left\langle Q'||P' \right\rangle$ from  
$\left\langle Q||P \right\rangle$ where $P,Q,P',Q'\in\mathfrak S_N$, it should not depend on the sequence of two-body scatterings we choose. It is sufficient to check this 
for the three-body case. If we consider an initial state $(ijk)$ and a final state $(kji)$, we want the following diagram to be commutative:
\begin{equation}
\begin{tikzpicture}
  \matrix (m) [matrix of math nodes,row sep=3em,column sep=4em,minimum width=2em]
  {
     (ijk) & (jik) & (jki) \\
     (ikj) & (kij) & (kji)\\};
  \path[-stealth]
    (m-1-1) edge node [left] {$S_{jk}$} (m-2-1)
            edge node [below] {$S_{ij}$} (m-1-2)
    (m-1-2) edge node [below] {$S_{ik}$} (m-1-3)
    (m-2-1) edge node [below] {$S_{ik}$}(m-2-2)
    (m-2-2) edge node [below] {$S_{ij}$} (m-2-3)
    (m-1-3) edge node [left] {$S_{jk}$} (m-2-3)
    
           ;
\end{tikzpicture}
\end{equation}

In terms of the scattering matrices of Eq.~\eqref{smatrix}, the consistency equations then read:
\begin{equation}
\label{Yang-Baxter}
{S_{ij}}^{bc}{S_{ik}}^{ab}{S_{jk}}^{bc}={S_{jk}}^{ab}{S_{ik}}^{bc}{S_{ij}}^{ab}.
\end{equation}
This set of ternary relations constitute the celebrated Yang-Baxter equations \cite{Yang67}. It is easy to check that they are 
verified in our system, confirming 
that our problem is consistent and can be reduced to two-body scatterings. Note that this consistency was trivially verified in the
Lieb-Liniger gas of section \ref{LL}, as the scattering only consisted in a phase change. A graphical interpretation of Eq.~\eqref{Yang-Baxter} is given in
Fig. \ref{figYB}.

Thus, if we obtain $\zeta_I$ for a given set of pseudo-momentum, all the $\zeta_P$'s are determined. We can now suppose that we have the periodic boundary conditions
and do the same procedure as in section \ref{LL}, i.e. we let a particle with pseudo-momentum $k_j$ make a whole turn around the ring (see Fig. \ref{ring}), and we get for all $j\in\{1,\dots,N\}$:
\begin{equation}
\label{bageneral}
e^{ik_jL}\zeta_I={\tilde{S}_{(j+1)j}}^{(j+1)j}\times{\tilde{S}_{(j+2)j}}^{(j+2)j}\times\cdots\times{\tilde{S}_{Nj}}^{Nj}\times{\tilde{S}_{1j}}^{1j}\times\cdots\times{\tilde{S}_{(j-1)j}}^{(j-1)j}\zeta_I.
\end{equation}
In this form, the Bethe ansatz equations are a set of $N$ coupled eigenvalue equations, with the same eigenvector $\zeta_I$. In the thermodynamic limit, this set
of equations is impossible to solve in practice. Hopefully, there are some clever methods to simplify it, when taking into account the fundamental symmetries 
of our multi-component system. The most complete and efficient method to do it is called the \textit{Quantum Inverse Scattering Method} \cite{korepin_bogoliubov_izergin_1993}, but is not very intuitive. 
Rather than expose it here, which would be unnecessary long and involved, we will try to justify the shape of the equations in order to have a intuitive understanding 
of their meaning. We start with the Gaudin-Yang model for the two-component fermionic model in section \ref{GY}, and extend it to the general case in section
\ref{Sutherland}.

\subsection{Two-component fermions: the Gaudin-Yang model}
\label{GY}
\subsubsection{The Bethe-Yang hypothesis}
\label{secbyh}
We consider the case of $N$ spin-$\frac{1}{2}$ particles, divided in $M$  spins down and $N-M$ spins up (these numbers are fixed). This model was solved by Gaudin in \cite{Gaudin1967} and C.N. Yang in \cite{Yang67}.
They noticed that the problem can be separated between the configuration/spin part on the one side and the pseudo-momentum/spatial part where we forget the color/spin
of the particles (i.e. analogous to the Lieb-Liniger case) on the other side. Both sides of the problem are of course coupled through the scattering processes. 
Due to the anti-symmetric nature of same-component fermions, for a given $P\in\mathfrak S_N$, there will be only $C_M^N$ independent coefficients
of the form $\left\langle Q||P \right\rangle$. Then, it is sufficient to know $\left\langle Q||P \right\rangle$ for a given set of positions 
$\{y_1<\cdots<y_M\}\subset\{1,\dots,N\}$ of the $M$ spin downs. Therefore, the quantity of interest for this model, called the \textit{spin wave function} by Yang, can be written:
\begin{equation}
\phi(y_1,\dots,y_M||P)=\epsilon_P\left\langle Q||P \right\rangle,
\end{equation}
where $y_1<\cdots<y_M$ are the positions of the $M$ spins down and $\epsilon_P$ is the signature of the permutation $P$ which takes into account the 
fermionic nature  of the problem. Thus, the spin part of the problem can be seen as a discrete problem of down spins on a chain of length $N$. Similarly to 
the continuous spatial part where each particle has a pseudo-momentum $k_j$, each spin down $y_{\alpha}$ will carry a momentum-like quantity, namely a \textit{rapidity}
$\Lambda_{\alpha}$, which will characterize how the spins will evolve through the scattering processes (this affirmation will be further justified below). Using this observation, Yang made the following generalized Bethe hypothesis
(often called the \textit{Bethe-Yang hypothesis}) for the form of $\phi$:
\begin{equation}
\label{byh}
\phi(y_1,\dots,y_M||P)=\sum_{R\in\mathfrak S_M}b(R)F_P(y_1,\Lambda_1)F_P(y_2,\Lambda_2)\cdots F_P(y_M,\Lambda_M),
\end{equation}
where the  $F_P$ function and the $b(R)$ coefficients remain to be determined, and $\Lambda_1,\dots,\Lambda_M$ are a set of unequal numbers. We summarize these considerations in 
Fig. \ref{fig:Yang2}.

\begin{figure}\centering
\begin{tikzpicture}
\draw (0,0) circle(2);
\draw[fill=white] (0:2) circle(0.1);
\draw[fill=gray] (45:2) circle(0.1);
\draw[fill=white] (80:2) circle(0.1);
\draw[fill=gray] (110:2) circle(0.1);
\draw[fill=white] (125:2) circle(0.1);
\draw[fill=white] (210:2) circle(0.1);
\draw[fill=gray] (290:2) circle(0.1);
\draw[fill=white] (310:2) circle(0.1);
\draw (1.5,1.5) node[right]{$k_j$};
\draw[>=latex,->] (45:2.7) arc (45:-10:2.7) ;
\end{tikzpicture}

\caption{\label{ring}Semi-classical interpretation of the Bethe ansatz equations for the two-component model ($N=8$ and $M=3$). Contrary to the Lieb-Liniger case, the scatterings 
would have to take into account the possibility of a reflection and a transmission, and would therefore be described by $S$-matrices.}
\end{figure}
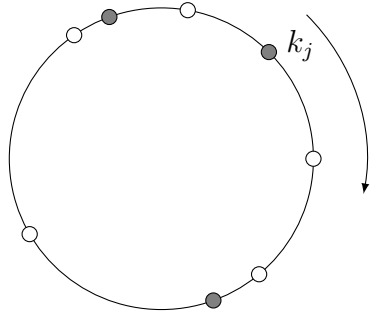

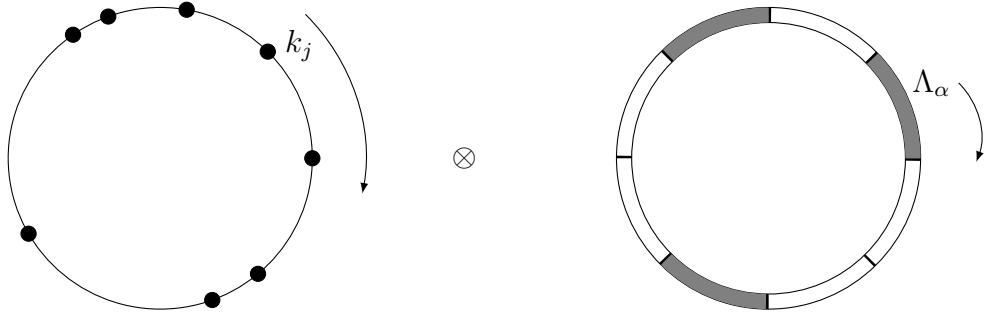
\begin{figure}\centering
\begin{tikzpicture}
\draw (0,0) circle(2);
\draw[fill] (0:2) circle(0.1);
\draw[fill] (45:2) circle(0.1);
\draw[fill] (80:2) circle(0.1);
\draw[fill] (110:2) circle(0.1);
\draw[fill] (125:2) circle(0.1);
\draw[fill] (210:2) circle(0.1);
\draw[fill] (290:2) circle(0.1);
\draw[fill] (310:2) circle(0.1);
\draw (1.5,1.5) node[right]{$k_j$};
\draw[>=latex,->] (45:2.7) arc (45:-10:2.7) ;

\draw (4,0) node{$\otimes$};three

\draw (8,0) circle(2);
\draw (8,0) circle(1.8);
\fill[gray] (8,0) ++(45:2)
arc(45:0:2)
-- ++(0:-0.2)
arc(0:45:1.8)
-- cycle;
\fill[gray] (8,0) ++(135:2)
arc(135:90:2)
-- ++(90:-0.2)
arc(90:135:1.8)
-- cycle;
\fill[gray] (8,0) ++(-90:2)
arc(-90:-135:2)
-- ++(-135:-0.2)
arc(-135:-90:1.8)
-- cycle;
\fill[black] (8,0) ++(-90:2)
arc(-90:-91:2)
-- ++(-91:-0.2)
arc(-91:-90:1.8)
-- cycle;
\fill[black] (8,0) ++(-135:2)
arc(-135:-136:2)
-- ++(-136:-0.2)
arc(-136:-135:1.8)
-- cycle;
\fill[black] (8,0) ++(-180:2)
arc(-180:-181:2)
-- ++(-181:-0.2)
arc(-181:-180:1.8)
-- cycle;
\fill[black] (8,0) ++(-225:2)
arc(-225:-226:2)
-- ++(-226:-0.2)
arc(-226:-225:1.8)
-- cycle;
\fill[black] (8,0) ++(-270:2)
arc(-270:-271:2)
-- ++(-271:-0.2)
arc(-271:-270:1.8)
-- cycle;
\fill[black] (8,0) ++(-315:2)
arc(-315:-316:2)
-- ++(-316:-0.2)
arc(-316:-315:1.8)
-- cycle;
\fill[black] (8,0) ++(0:2)
arc(0:-1:2)
-- ++(-1:-0.2)
arc(-1:0:1.8)
-- cycle;
\fill[black] (8,0) ++(-45:2)
arc(-45:-46:2)
-- ++(-46:-0.2)
arc(-46:-45:1.8)
-- cycle;
\draw (9.75,1) node[right]{$\Lambda_{\alpha}$};
\draw[>=latex,->] (10.5,1) arc (45:-20:1) ;

\end{tikzpicture}
\caption{\label{fig:Yang2}Another way to interpret the two-component model, equivalent to 
Fig. \ref{ring}. Here, the problem is separated between two coupled parts, a spatial part (left), and a spin part (right). The spin part can be seen as $M$ spin downs on a discrete 
chain of length $N$. A rapidity $\Lambda_{\alpha}$ is associated with each spin down.}
\end{figure}

In order to determine $F_P$, let us first consider the two-body problem where $N=2$ and $M=1$. Using previous notations 
(see section \ref{consistencysec}), the scattering process reduces to 
\begin{equation}
\label{two}
\zeta_{(21)}={S_{12}}^{12}\zeta_{(12)}.
\end{equation}
One can check that, for every value of $\Lambda\in\mathbb{R}$, the following set of coefficients verifies Eq.~\eqref{two} \cite{colometatche}:
\begin{equation}
\begin{split}
&\phi(1||(12))=k_1-\Lambda-\frac{ic}{2},\\
&\phi(2||(12))=-(k_2-\Lambda+\frac{ic}{2}).
\end{split}
\end{equation}
We see that the parameter $\Lambda$ is homogeneous to a momentum. If we consider the $N=3$, $M=1$ case, the boundary conditions for $\phi$ can be written
for $j\in{1,2,3}$ and $P,P'\in\mathfrak S_3$ where $P'=P(j,j+1)$:
\begin{equation}
\begin{split}
&\phi(y||P)=\phi(y||P')\quad\text{if}\quad y\neq j,j+1,\\
&\phi(j||P)=-R_{j+1,j}\phi(j||P')+T_{j+1,j}\phi(j+1||P'),\\
&\phi(j+1||P)=-R_{j+1,j}\phi(j+1||P')+T_{j+1,j}\phi(j||P').
\end{split}
\end{equation}
For any $\Lambda,b\in\mathbb{R}$, a solution is given by:
\begin{equation}
\begin{split}
&\phi(1||P)=b(k_{P2}-\Lambda-\frac{ic}{2})(k_{P3}-\Lambda-\frac{ic}{2}),\\
&\phi(2||P)=b(k_{P1}-\Lambda+\frac{ic}{2})(k_{P3}-\Lambda-\frac{ic}{2}),\\
&\phi(3||P)=b(k_{P1}-\Lambda+\frac{ic}{2})(k_{P2}-\Lambda+\frac{ic}{2}).
\end{split}
\end{equation}
Generalizing to $N$ particles with $M=1$ down spin, we have obtained the form of the functions $F_P(y,\Lambda)$:
\begin{equation}
\label{fexpression}
F_P(y,\Lambda)=\prod_{j=1}^{y-1}(k_{Pj}-\Lambda+\frac{ic}{2})\prod_{l=y+1}^N(k_{Pj}-\Lambda-\frac{ic}{2}).
\end{equation}

We are now able to determine $b(R)$. We can write the boundary conditions in the general case in terms of $\bar\phi\equiv\phi(y_1,\ldots,y_M||P)$, for $P,P'\in\mathfrak S_N$ with $P'=P(j,j+1)$ and 
$j\in\{1,\dots,N\}$:
\begin{equation}
\label{boundaryphi}
\begin{split}
&\bar\phi=\phi(y_1,\ldots,y_M||P')\quad\text{if}\quad\forall l,~y_{l}\neq j,j+1\quad\text{or if}\quad\exists l,~y_{l}= j,~y_{l+1}=j+1,\\
&\bar\phi=-R_{j+1,j}\phi(y_1,\dots,y_l,\dots,y_M||P')+T_{j+1,j}\phi(y_1,\dots,y_l+1,\dots,y_M||P')\\&\hspace{10cm}\text{if}\quad y_{l}=j,~y_l+1\neq j+1,\\
&\bar\phi=-R_{j+1,j}\phi(y_1,\dots,y_l,\dots,y_M||P')+T_{j+1,j}\phi(y_1,\dots,y_l-1,\dots,y_M||P')\\&\hspace{10cm}\text{if}\quad y_{l}\neq j,~y_l+1= j+1.
\end{split}
\end{equation}
Then, if we consider $R,R'\in\mathfrak S_M$ with $R'=R(l,l+1)$ and $P,P'\in\mathfrak S_N$ with $P'=P(y_l,y_{l+1})$, we get using Eq.~\eqref{byh}:
\begin{equation}
\begin{split}
b(R)&F_P(y_l,\Lambda_{Rl})F_P(y_l+1,\Lambda_{R(l+1)})+b(R')F_P(y_l,\Lambda_{R(l+1)})F_P(y_l+1,\Lambda_{Rl})\\
&=b(R)F_{P'}(y_l,\Lambda_{Rl})F_{P'}(y_l+1,\Lambda_{R(l+1)})+b(R')F_{P'}(y_l,\Lambda_{R(l+1)})F_{P'}(y_l+1,\Lambda_{Rl}),
\end{split}
\end{equation}
which becomes, using $F$'s expression of Eq.~\eqref{fexpression}:
\begin{equation}
b(R)(\Lambda_{R(l+1)}-\Lambda_{Rl}-ic)=-b(R')(\Lambda_{Rl}-\Lambda_{R(l+1)}-ic).
\end{equation}
In order for this condition to be satisfied in the general case, we can choose:
\begin{equation}
\label{bexpression}
b(R)=\epsilon_R\prod_{j<l}(\Lambda_{Rj}-\Lambda_{Rl}-ic).
\end{equation}

Thus, Eqs.~\eqref{byh}, \eqref{fexpression} and \eqref{bexpression} completely determine the solution for the Gaudin-Yang model as a function of the rapidities 
and pseudo-momenta. We can now apply the periodic boundary conditions in order to determine these parameters.

\subsubsection{Bethe ansatz equations and ground state}
Similarly to the Lieb-Liniger model, we suppose now that the total wave function follows the periodic boundary condition of Eq.~\eqref{pbc}, namely
$\psi(x_N-L,x_1,\ldots,x_{N-1})=\psi(x_1,x_2,\dots,x_N)$. Equivalently, we say that the particle $N$ with momentum $k_j$ will make a whole turn around the ring (c.f.
 Fig. \ref{ring}). Here, we have to distinct two possibilities: the case where particle $N$ is a spin up and the case where it is a spin down. In terms 
 of $\phi$, for $P,P'\in\mathfrak S_N$ where $P'=(PN,P1,P2,\ldots,P(N-1))$, these cases can be written respectively:
 \begin{equation}
 \phi(y_1,\ldots,y_M)||P)e^{ik_{PN}L}=\phi(y_1+1,\ldots,y_M+1)||P'),
 \end{equation}
and
 \begin{equation}
 \phi(y_1,\ldots,y_{M-1},N)||P)e^{ik_{PN}L}=\phi(1,y_1+1,\ldots,y_{M-1}+1)||P').
 \end{equation}
Using $\phi$'s complete expression within the Bethe-Yang hypothesis (Eqs.~\eqref{byh}, \eqref{fexpression} and \eqref{bexpression}) yields
\begin{equation}
\begin{split}
&e^{ik_{PN}L}=\prod_{\beta=1}^M\frac{k_{PN}-\Lambda_{\beta}+\frac{ic}{2}}{k_{PN}-\Lambda_{\beta}-\frac{ic}{2}},\\
&\prod_{j=1}^{N}\frac{k_j-\Lambda_{RM}+\frac{ic}{2}}{k_j-\Lambda_{RM}-\frac{ic}{2}}=-\prod_{\beta=1}^M\frac{\Lambda_{RM}-
\Lambda_{\beta}-ic}{\Lambda_{RM}-\Lambda_{\beta}+ic}.
\end{split}
\end{equation}
Taking the logarithmic form, we obtain the $N+M$ Bethe ansatz equations for the Gaudin-Yang model:
\begin{equation}
\begin{split}
\label{baGY}
&k_jL=2\pi I_j+\sum_{\beta=1}^M\theta(2k_j-2\Lambda_{\beta}),\quad j\in\{1,\ldots,N\},\\
&-\sum_{j=1}^N\theta(2\Lambda_{\alpha}-2k_j)=2\pi J_{\alpha}-\sum_{\beta=1}^M\theta(\Lambda_{\alpha}-\Lambda_{\beta}),\quad\alpha\in\{1,\ldots,M\},
\end{split}
\end{equation}
where the phase $\theta$ is defined as in Eq.~\eqref{theta}, and $I_j,J_{\alpha}$ are the quantum numbers of this model. We can observe the power of Bethe-Yang 
hypothesis by comparing Eqs~\eqref{baGY} and \eqref{bageneral}: here, we have only added $M$ scalar equations as compared to the one-component case!

Just like for the Lieb-Liniger model,
the quantum numbers will define the ground state when $I_j\in\{-\frac{N-1}{2},\dots,\frac{N-1}{2}\}$ and $J_{\alpha}\in\{-\frac{M-1}{2},\dots,\frac{M-1}{2}\}$.
We can then take the thermodynamic limit $N,M,L\to\infty$ keeping $n=\frac{N}{L}$ and $n_{\downarrow}=\frac{M}{L}$ constant, and define the quasi-momentum 
and spin distributions verifying respectively
\begin{equation}
\int_{-q}^q \rho(k)\, d k=n,
\end{equation}
and
\begin{equation}
\int_{-s}^s \sigma(\Lambda)\, d \Lambda=n_{\downarrow}.
\end{equation}
We finally get the integral equations for the ground state:
\begin{equation}
\begin{split}
&1+4c\int_{-s}^s\frac{\sigma(\Lambda)\, d \Lambda}{c^2+4(k-\Lambda)^2}=2\pi\rho(k),\\
&-2c\int_{-s}^s\frac{\sigma(\Lambda')\, d \Lambda'}{c^2+(k-k')^2}+4c\int_{-q}^q\frac{\rho(k)\, d k}{c^2+4(k-\Lambda)^2}=2\pi\sigma(\Lambda).
\end{split}
\end{equation}
The ground state energy density can then be obtained using
\begin{equation}
e=\int_{-q}^qk^2\rho(k)\, d k.
\end{equation}

\subsection{General case: the nested Bethe ansatz}
\label{Sutherland}

\subsubsection{Bethe ansatz equations}

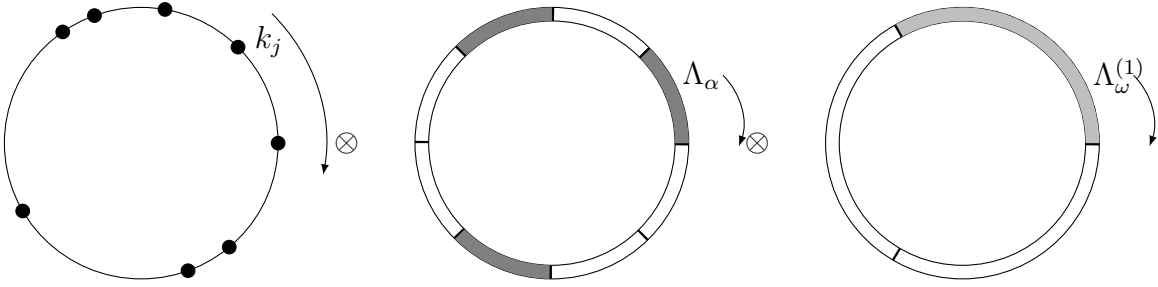
\begin{figure}\centering
\begin{tikzpicture}
\begin{scope}[scale=0.9]
\draw (0,0) circle(2);
\draw[fill] (0:2) circle(0.1);
\draw[fill] (45:2) circle(0.1);
\draw[fill] (80:2) circle(0.1);
\draw[fill] (110:2) circle(0.1);
\draw[fill] (125:2) circle(0.1);
\draw[fill] (210:2) circle(0.1);
\draw[fill] (290:2) circle(0.1);
\draw[fill] (310:2) circle(0.1);
\draw (1.5,1.5) node[right]{$k_j$};
\draw[>=latex,->] (45:2.7) arc (45:-10:2.7) ;

\draw (3,0) node{$\otimes$};three

\draw (6,0) circle(2);
\draw (6,0) circle(1.8);
\fill[gray] (6,0) ++(45:2)
arc(45:0:2)
-- ++(0:-0.2)
arc(0:45:1.8)
-- cycle;
\fill[gray] (6,0) ++(135:2)
arc(135:90:2)
-- ++(90:-0.2)
arc(90:135:1.8)
-- cycle;
\fill[gray] (6,0) ++(-90:2)
arc(-90:-135:2)
-- ++(-135:-0.2)
arc(-135:-90:1.8)
-- cycle;
\fill[black] (6,0) ++(-90:2)
arc(-90:-91:2)
-- ++(-91:-0.2)
arc(-91:-90:1.8)
-- cycle;
\fill[black] (6,0) ++(-135:2)
arc(-135:-136:2)
-- ++(-136:-0.2)
arc(-136:-135:1.8)
-- cycle;
\fill[black] (6,0) ++(-180:2)
arc(-180:-181:2)
-- ++(-181:-0.2)
arc(-181:-180:1.8)
-- cycle;
\fill[black] (6,0) ++(-225:2)
arc(-225:-226:2)
-- ++(-226:-0.2)
arc(-226:-225:1.8)
-- cycle;
\fill[black] (6,0) ++(-270:2)
arc(-270:-271:2)
-- ++(-271:-0.2)
arc(-271:-270:1.8)
-- cycle;
\fill[black] (6,0) ++(-315:2)
arc(-315:-316:2)
-- ++(-316:-0.2)
arc(-316:-315:1.8)
-- cycle;
\fill[black] (6,0) ++(0:2)
arc(0:-1:2)
-- ++(-1:-0.2)
arc(-1:0:1.8)
-- cycle;
\fill[black] (6,0) ++(-45:2)
arc(-45:-46:2)
-- ++(-46:-0.2)
arc(-46:-45:1.8)
-- cycle;
\draw (7.75,1) node[right]{$\Lambda_{\alpha}$};
\draw[>=latex,->] (8.5,1) arc (45:-20:1) ;

\draw (9,0) node{$\otimes$};three

\draw (12,0) circle(2);
\draw (12,0) circle(1.8);
\fill[color=gray!50] (12,0) ++(120:2)
arc(120:0:2)
-- ++(0:-0.2)
arc(0:120:1.8)
-- cycle;
\fill[black] (12,0) ++(0:2)
arc(0:-1:2)
-- ++(-1:-0.2)
arc(-1:0:1.8)
-- cycle;
\fill[black] (12,0) ++(120:2)
arc(120:119:2)
-- ++(119:-0.2)
arc(119:120:1.8)
-- cycle;
\fill[black] (12,0) ++(240:2)
arc(240:239:2)
-- ++(239:-0.2)
arc(239:240:1.8)
-- cycle;

\draw (13.75,1) node[right]{$\Lambda^{(1)}_{\omega}$};
\draw[>=latex,->] (14.5,1) arc (45:-20:1) ;

\end{scope}
\end{tikzpicture}
\caption{\label{fig:sutherland}Semi-classical interpretation of the nested Bethe ansatz equations for the 3-component gas, with five particles of type 1, two particles
of type 2 and one particle of type 3. A pseudo-momentum $k_j$ is associated with each of the eight particles (left), a rapidity $\Lambda_{\alpha}$ with each of the three particles 
particles of type 2 and 3 (center), and a rapidity $\Lambda^{(1)}_{\omega}$ for the particle of type 3 (right).}
\end{figure}

The model of $\kappa$-component fermions was solved by Sutherland in \cite{Sutherland68}. Without entering the details, the idea is to apply successively the Bethe-Yang hypothesis to the 
coefficients of the spin wave-functions, resulting in Bethe ansatz equations of smaller and smaller dimensions (hence the name of \textit{nested Bethe ansatz}).
More precisely, let us consider the case of a three-component model of $N$ particles with $N-M$ particles of type 1, $M-M_1$ particles of type 2 and $M_1$ 
particles of type 3. We can first separate the $N-M$ type-1 particles and the $M$ other, and treat the problem similarly to the Gaudin-Yang model of section
\ref{GY}. Since the $M$ "other particles" are of different type, we will have to write the Bethe-Yang hypothesis of Eq.~\eqref{byh} in a different form, taking 
into account the permutations of the $M$ particles. This is exactly what we did when we went from the Lieb-Liniger model of section \ref{LL} to the Gaudin-Yang 
model of two-component fermions. Thus, if $Q\in\mathfrak S_M$, the spin wave-function $\phi$ for the $M$ particles is written in the (discrete) sector 
$1\le y_{Q1}<\cdots<y_{QM}\le N$:
\begin{equation}
\phi=\sum_{P\in\mathfrak S_M}\left\langle Q||P \right\rangle F(\Lambda_{P1},y_{Q1})\cdots F(\Lambda_{PM},y_{QM}).
\end{equation}
We can then repeat the same procedure as in section \ref{consistencysec}, namely we arrange the coefficients $\left\langle Q||P \right\rangle$ in a
$M!\times M!$ matrix, translate the boundary conditions in terms of scattering matrices and check the consistency of our problem guaranteed by the 
Yang-Baxter equations. Then, we can separate the $M$ particles between the $M-M_1$ type-2 and the $M_1$ type-3 particles and re-apply the Bethe-Yang hypothesis for the coefficients 
$\left\langle Q||P \right\rangle$, introducing rapidities $\Lambda^{(1)}$ for the type-3 particles. Finally, after applying the usual periodic boundary conditions, we find:
\begin{equation}
\label{nested3}
\begin{split}
&e^{ik_jL}=\prod_{\beta=1}^M\frac{k_j-\Lambda_{\beta}+\frac{ic}{2}}{k_j-\Lambda_{\beta}-\frac{ic}{2}},~j\in\{1,\ldots,N\},\\
&\prod_{j=1}^{N}\frac{k_j-\Lambda_{\alpha}+\frac{ic}{2}}{k_j-\Lambda_{\alpha}-\frac{ic}{2}}=-\prod_{\beta=1}^M\frac{\Lambda_{\alpha}-
\Lambda_{\beta}-ic}{\Lambda_{\alpha}-\Lambda_{\beta}+ic}
\prod_{\gamma=1}^{M_1}\frac{\Lambda_{\alpha}-\Lambda^{(1)}_{\gamma}+\frac{ic}{2}}{\Lambda_{\alpha}-\Lambda^{(1)}_{\gamma}-\frac{ic}{2}},~\alpha\in\{1,\ldots,M\},\\
&\prod_{\beta=1}^M\frac{k_j-\Lambda_{\beta}+\frac{ic}{2}}{k_j-\Lambda_{\beta}-\frac{ic}{2}}=-\prod_{\gamma=1}^{M_1}\frac{\Lambda^{(1)}_{\omega}-
\Lambda^{(1)}_{\gamma}-ic}{\Lambda^{(1)}_{\omega}-\Lambda^{(1)}_{\gamma}+ic},~\omega\in\{1,\ldots,M_1\}.
\end{split}
\end{equation}
Long story short, we can again consider the logarithm form of Eq.~\eqref{nested3}, introducing the $N+M+M_1$ quantum numbers $I_j,J_{\alpha}, K_{\omega}$ for the three-component
model, which have to be as compactly centered as possible around the origin when considering the ground state. After that, we can consider the thermodynamic limit
with fixed densities, in a very similar manner as we did before, resulting in three coupled integral equations for the distributions $\rho(k)$, $\sigma(\Lambda)$, and
$\tau(\Lambda^{(1)})$. An illustration of the nested Bethe ansatz is given in Fig. \ref{fig:sutherland}.

Following these ideas, in the case of a $\kappa$-component Fermi gas divided in $[m_1,\dots,m_{\kappa}]$ particles per species, we can define a set of $\kappa$ rapidities,
distributions and integration limits denoted respectively
$k_i$, $\rho_i(k_i)$ and $B_i$. Defining $M_i=\sum_{j=i}^{\kappa}m_j$ for all $i\in\{1,\dots,\kappa\}$, we can write the $\kappa$ coupled integral Bethe ansatz equations for the 
ground state in the thermodynamic limit:
\begin{equation}
\begin{split}
\label{baeSutherland}
& 2\pi\rho_1=1+4c\int_{-B_2}^{B_2}\frac{\rho_2\, d k_2}{c^2+4(k_1-k_2)^2},\\
& \int_{-B_{i+1}}^{B_{i+1}}\frac{4c\rho_{i+1}\, d k_{i+1}}{c^2+4(k_i-k_{i+1})^2}
+\int_{-B_{i-1}}^{B_{i-1}}\frac{4c\rho_{i-1}\, d k_{i-1}}{c^2+4(k_i-k_{i-1})^2}
=2\pi\rho_i+\int_{-B_i}^{B_i}\frac{2c\rho_i\, d k_i'}{c^2+(k_i-k_i')^2}\\&\hspace{10cm}\text{for }~i\in\{2,\ldots,\kappa-1\},\\
&\int_{-B_{\kappa-1}}^{B_{\kappa-1}}\frac{4c\rho_{\kappa-1}\, d k_{\kappa-1}}{c^2+4(k_{\kappa}-k_{\kappa-1})^2}=2\pi\rho_{\kappa}
+\int_{-B_{\kappa}}^{B_{\kappa}}\frac{2c\rho_{\kappa}\, d k_{\kappa}'}{c^2+(k_{\kappa}-k_{\kappa}')^2},
\end{split}
\end{equation}
together with the following normalization conditions:
\begin{equation}
\label{normsuther}
\frac{M_i}{L}=\int_{-B_i}^{B_i}\rho_i\, d k_i,\quad i\in\{1,\ldots,\kappa\},
\end{equation}
and the formula for the ground state energy density:
\begin{equation}
\label{gsensuther}
e=\int_{-B_1}^{B_1}\rho_1k_1^2\, d k_1.
\end{equation}

\subsubsection{Strong-coupling expansion in the balanced fermionic case}
\label{strcoup}

In general, Eqs.~\eqref{baeSutherland} are very hard to solve in practice. In particular, the boundaries $B_i$ are not easy to evaluate. Hopefully, in the balanced case, there are some simplifications that allow to access a strong-coupling expansion for the ground-state energy (Eq.~\eqref{gsensuther}). Here, we provide the ideas for the proof of this expansion, following \cite{Guan2012}.

First, the key property is that the fact that a fermionic mixture is balanced implies that the boundaries verify $B_i\to\infty$ for every $i\ge 2$. Indeed, by integrating Eq.~\eqref{baeSutherland} we obtain for every $i\ge 2$:
\begin{equation}
\int_{-\infty}^{\infty}\rho_i(k)dk=\frac{M_{i-1}}{L}-\frac{M_{i}}{L}+\frac{M_{i+1}}{L}=\frac{M_{i}}{L}
\end{equation}
in the balanced case. It is then clear, by comparing to Eq.~\eqref{gsensuther}, that all boundaries except $B_1$ go to infinity.

Writing the Fourier transform of $\rho_i(k)$ as $\tilde{\rho}_i(\omega)$ and introducing the function $\rho_{1in}(k)\equiv \theta_{[-B_1,B_1]}(k)\rho_1(k)$ where $\theta_{[-B_1,B_1]}$ is the indicator function of $[-B_1,B_1]$, 
we can now Fourier transform Eqs.~\eqref{baeSutherland} and find after some algebra, for every $i\ge 2$:
\begin{equation}
\label{tfrhoi}
\tilde{\rho}_i(\omega)=\frac{\tilde{\rho}_{1in}(\omega)\sinh\left[\frac{1}{2}(\kappa-i+1)|\omega|c\right]}{\sinh\left[\frac{1}{2}\kappa|\omega|c\right]}.
\end{equation}

Then, the condition $c\gg 1$ allows to perform perturbative developments of the kernels of integrals in Eqs.~\eqref{baeSutherland} (c.f. the Lieb-Liniger case, section \ref{GS}). By doing so and using Eqs.~\eqref{tfrhoi}, \eqref{normsuther} and \eqref{gsensuther}, Guan \textit{et al} have obtained:
\begin{equation}
\tilde{\rho}_{1in}(\omega)\approx n-\frac{e\omega^2}{2},\quad\tilde{\rho}_{2}(\omega)\approx \frac{\sinh\left[\frac{1}{2}(\kappa-1)|\omega|c\right]}{\sinh\left[\frac{1}{2}\kappa|\omega|c\right]}\left[n-\frac{e\omega^2}{2}\right].
\end{equation}
Plugging this equation into the first line of Eq.~\eqref{baeSutherland} yields:
\begin{equation}
\label{rho1y}
\rho_1(k)=\frac{1}{2\pi}+\frac{nY_0(k)}{2\pi}-\frac{eY_2(k)}{4\pi}+\mathcal{O}(c^{-4}),
\end{equation}
with
\begin{equation}
Y_{\alpha}(k)=\int d\omega~ \frac{\sinh\left[\frac{1}{2}(\kappa-1)|\omega|c\right]}{\sinh\left[\frac{1}{2}\kappa|\omega|c\right]}\exp(i\omega k-c|\omega|/2)\omega^{\alpha}.
\end{equation}

In the $c\to\infty$ limit, it can be shown that
\begin{equation}
Y_0(k)=\frac{2Z_1(\kappa)}{c}-\frac{2Z_3(\kappa)k^2}{c^3}+\mathcal{O}(c^{-4})
\end{equation}
and
\begin{equation}
Y_2(k)=\frac{4Z_3(\kappa)}{c^3}+\mathcal{O}(c^{-4}),
\end{equation}
with
\begin{equation}
Z_1(\kappa)=-\frac{1}{\kappa}\left(\psi\left(\frac{1}{\kappa}\right)+C_{Euler}\right),
\end{equation}
and
\begin{equation}
Z_3(\kappa)=\frac{1}{\kappa^3}\left(\zeta\left(3,\frac{1}{\kappa}\right)-\zeta(3)\right),
\end{equation}
where $C_{Euler}\approx 0.577$ is the Euler constant and $\psi$ and $\zeta$ are respectively the Digamma and Riemann Zeta functions \cite{0486612724}. Then, plugging Eq.~\eqref{rho1y} into Eqs~\eqref{normsuther} and \eqref{gsensuther} gives, after some calculations, Eq.~\eqref{expasione} of the main text.

\subsection{Thermodynamic Bethe ansatz}
\label{tbamulti}

The thermodynamic Bethe ansatz for multi-component fermions was obtained by Takahashi and Lai for two-component fermions \cite{Takahashi1971,Lai1971} 
and was extended to $\kappa$ components by Schlottmann \cite{Schlottmann1993}. Although the global idea is similar to the Yang-Yang case of section 
\ref{YYth}, that is considering the excited states in terms of particle and hole densities, there is one fundamental difference. Indeed, in section \ref{YYth}
we were only considering real pseudo-momenta: a complex pseudo-momentum means the an exponentially decreasing wave-function as a function of the 
relative distance between particles, which is not the case when $c>0$. When considering multicomponent fermions however, we have introduced spin rapidities $\Lambda$
which have no reason to be real numbers for excited states. The so-called \textit{string hypothesis} assumes that the complex spin-rapidities of excited states form discrete strings 
of arbitrary length $(2m-1)$ where $m\ge 1$:
\begin{equation}
\Lambda_m=\xi_m+i\nu \frac{c}{2}+\delta(L),\quad\nu\in\{-(m-1),\ldots,m-1\},
\end{equation}
where $\xi\in\mathbb{R}$ and $\delta(L)$ vanishes in the thermodynamic limit. The intuitive idea behind this hypothesis is that if a spin rapidity has an imaginary 
part, it will result in exponentially vanishing terms in the Bethe ansatz equations which would have to be compensated by poles in the 
$\frac{\Lambda-k-\frac{ic}{2}}{\Lambda-k+\frac{ic}{2}}$ factors, which is achieved when the complex rapidities are separated by $\frac{ic}{2}$. As compared 
with the Yang-Yang case when  we were going from Eq.~\eqref{fredholm} for the ground state to Eq.~\eqref{fredholmYY} for the excited states, the excited analogous 
of Eq.~\eqref{baeSutherland} will contain sums over strings of arbitrary length for each class of spin rapidities. We adapt the notations of Eq.~\eqref{baeSutherland}
in the following way: $\rho\equiv\rho_1$ and $\sigma_m^{(l)}(\Lambda)$ is the density for strings of length $m$ associated with real rapidities $k_{l+1}$, so 
that 
\begin{equation}
\label{constraints}
\begin{split}
&\frac{N}{L}=\frac{M_1}{L}=\int dk\,\rho(k),\\
&\frac{M_{l+1}}{L}=\sum_{m=1}^{\infty}m\int d\Lambda \,\sigma_m^{(l)}(\Lambda),\quad l\in\{1,\ldots,\kappa-1\}.
\end{split}
\end{equation}
We define the associated hole densities $\rho_h$ and $\sigma_{mh}^{(l)}(\Lambda)$ analogously to section \ref{YYth}, as well as the pseudo-energies 
$\epsilon(k)=T\ln\left(\frac{\rho_h}{\rho}\right)$ and $\varphi_m^{(l)}(\Lambda)=T\ln\left(\frac{\sigma_{mh}^{(l)}}{\sigma_{m}^{(l)}}\right)=T\ln(\eta_m^{(l)})$.
We can then define an entropy density for each class of excitations, and minimize the free energy of the system given the constraints of Eq.~\eqref{constraints}, 
each of them being associated with a Lagrange multiplier $A_l$ ($A_0$ is the chemical potential $\mu$ of section \ref{YYth}). After some (tedious) algebra, 
we find the analogous of the integral equation \eqref{pseudo} in the multicomponent case:
\begin{equation}
\label{tbaeqmulti}
\begin{split}
&\epsilon(k)=k^2-A_0-\frac{T}{\pi}\sum_{m=1}^{\infty}\int d\Lambda \,\frac{\frac{mc}{2}}{(\Lambda-k)^2+\left(\frac{mc}{2}\right)^2}\ln\left(1+\eta_m^{(1)}(\Lambda)^{-1}\right),\\
\ln&\left(1+\eta_m^{(l)}(\Lambda)\right)=-\frac{mA_l}{T}\sum_{n=1}^{\infty}\int d\Lambda' \,D_{mn}(\Lambda-\Lambda')\ln\left(1+\eta_n^{(l)}(\Lambda')^{-1}\right)\\
&-\sum_{n=1}^{\infty}\int d\Lambda' \,C_{mn}(\Lambda-\Lambda')\ln\left(1+\eta_n^{(l+1)}(\Lambda')^{-1}\right)\ln\left(1+\eta_n^{(l-1)}(\Lambda')^{-1}\right)\\&\hspace{8cm}\text{for }~l\in\{1,\ldots,\kappa-1\}\text{ and }~m\in\mathbb{N}^*,\\
\end{split}
\end{equation}
where we have $D_{mn}(\Lambda)=\mathcal{F}\left[ \coth(|\omega c|/2)\{\exp(-|n-m||\omega c|/2)-\exp(-(n+m)|\omega c|/2)\}\right]$ and 
$C_{mn}(\Lambda)=\mathcal{F}\left[\hat D_{mn}(\omega)/(2\cosh(\omega c/2))\right]$ ($\mathcal{F}$ being the Fourier transform operator),  $\eta_1^{(0)}=\exp(\epsilon/T)$ and $\eta_m^{(0)}=\infty$ 
for $m\ge2$,  and $\eta_m^{(\kappa)}=\infty$. We see that in this case, the pseudo-energies are given by an \textit{infinite} set of coupled integral equations! Then, one can 
extract the quantum pressure using Eq.~\eqref{pseudopress}.

\clearemptydoublepage

\clearemptydoublepage
\clearemptydoublepage
\pagestyle{fancy}
\thispagestyle{empty}
\chapter{List of publications}
\label{listpub}

\begin{itemize}
\item \textbf{J. Decamp}, P. Armagnat, B. Fang, M. Albert, A. Minguzzi and P. Vignolo, \textit{Exact density profiles and symmetry classification for strongly interacting multi-component Fermi gases in tight waveguides}, New J. Phys. \textbf{18}, 055011 (2016)
\item \textbf{J. Decamp}, J. J\"{u}nemann, M. Albert, M. Rizzi, A. Minguzzi, P. Vignolo, \textit{High-momentum tails as magnetic-structure probes for strongly correlated $SU(\kappa)$ fermionic mixtures in one-dimensional traps}, Phys. Rev. A \textbf{94}, 053614 (2016)
\item \textbf{J. Decamp}, J. J\"{u}nemann, M. Albert, M. Rizzi, A. Minguzzi, P. Vignolo, \textit{Strongly correlated one-dimensional Bose-Fermi quantum mixtures: symmetry and correlations}, New J. Phys. \textbf{19}, 125001 (2017)
\item \textbf{J. Decamp}, M. Albert, P. Vignolo, \textit{Tan's contact in a cigar-shaped dilute Bose gas}, Phys. Rev. A \textbf{97}, 033611 (2018)
\end{itemize}
\clearemptydoublepage

\clearemptydoublepage
\phantomsection 
\bibliographystyle{ThesisStyle}
\bibliography{Biblio}

\end{document}